\documentstyle{elsart}

\begin{document}
\begin{frontmatter}

\title{Use of Harmonic Inversion Techniques\\
      in Semiclassical Quantization and\\
      Analysis of Quantum Spectra}

\author{J\"org Main}
\address{Institut f\"ur Theoretische Physik I, Ruhr-Universit\"at Bochum,
D-44780 Bochum, Germany}

\vskip 1.5cm
\tableofcontents


\begin{abstract}
Harmonic inversion is introduced as a powerful tool for both the analysis of 
quantum spectra and semiclassical periodic orbit quantization.
The method allows to circumvent the uncertainty principle of the conventional
Fourier transform and to extract dynamical information from quantum spectra
which has been unattainable before, such as bifurcations of orbits,
the uncovering of hidden ghost orbits in complex phase space, and the direct 
observation of symmetry breaking effects.
The method also solves the fundamental convergence problems in semiclassical
periodic orbit theories -- for both the Berry-Tabor formula and Gutzwiller's
trace formula -- and can therefore be applied as a novel technique for 
periodic orbit quantization, i.e., to calculate semiclassical eigenenergies 
from a finite set of classical periodic orbits.
The advantage of periodic orbit quantization by harmonic inversion is the
universality and wide applicability of the method, which will be demonstrated
in this work for various open and bound systems with underlying regular,
chaotic, and even mixed classical dynamics.
The efficiency of the method is increased, i.e., the number of orbits required
for periodic orbit quantization is reduced, when the harmonic inversion 
technique is generalized to the analysis of cross-correlated periodic orbit 
sums.
The method provides not only the eigenenergies and resonances of systems
but also allows the semiclassical calculation of diagonal matrix elements
and, e.g., for atoms in external fields, individual non-diagonal transition
strengths.
Furthermore, it is possible to include higher order terms of the $\hbar$ 
expanded periodic orbit sum to obtain semiclassical spectra beyond the 
Gutzwiller and Berry-Tabor approximation.


\end{abstract}
\end{frontmatter}

\newpage

\section{Introduction}
\subsection{Motivation of semiclassical concepts}
Since the development of quantum mechanics in the early decades of this
century quantum mechanical methods and computational techniques have
become a powerful tool for accurate numerical calculations in atomic
and molecular systems.
The excellent agreement between experimental measurements and quantum
calculations has silenced any serious critics on the fundamental concepts 
of quantum mechanics, and there are nowadays no doubts that quantum mechanics
is the correct theory for microscopic systems.
Nevertheless, there has been an increasing interest in semiclassical theories 
during recent years.
The reasons for the resurgence of semiclassics are the following.
Firstly, the quantum mechanical methods for solving multidimensional, 
non-integrable systems generically imply intense numerical calculations, 
e.g., the diagonalization of a large, but truncated Hamiltonian in a suitably
chosen basis.
Such calculations provide little insight into the underlying dynamics of
the system.
By contrast, semiclassical methods open the way to a deeper understanding 
of the system, and therefore can serve for the interpretation of experimental 
or numerically calculated quantum mechanical data in physical terms.
Secondly, the relation between the quantum mechanics of microscopic systems
and the classical mechanics of the macroscopic world is of fundamental
interest and importance for a deeper understanding of nature.

This relation was evident in the early days of quantum mechanics, when
semiclassical techniques provided the only quantization rules, i.e.,
the WKB quantization of one-dimensional systems or the generalization to
the Einstein-Brillouin-Keller (EBK) quantization \cite{Ein17,Bri26,Kel58}
for systems with $n$ degrees of freedom.
However, the EBK torus quantization is limited to integrable or at least
near-integrable systems.
In non-integrable systems the KAM tori are destroyed \cite{Schu88,Lich83},
a complete set of classical constants of motion does not exist any more, 
and therefore the eigenstates of the quantized system cannot be characterized 
by a complete set of quantum numbers.
The ``breakdown'' of the semiclassical quantization rules for non-regular,
i.e., {\em chaotic} systems was already recognized by Einstein in 1917
\cite{Ein17}.
The failure of the ``old'' quantum mechanics to describe more complicated
systems such as the helium atom \cite{Win92}, and, at the same time,
the development and success of the ``modern'' wave mechanics are the
reasons for little interest in semiclassical theories for several decades.
The connection between wave mechanics and classical dynamics especially
for chaotic systems remained an open question during that period.

\subsubsection{Basic semiclassical theories}
\paragraph*{Chaotic systems: Gutzwiller's trace formula}
The problem was reconsidered by Gutzwiller around 1970 \cite{Gut67,Gut90}.
Although a semiclassical quantization of {\em individual} eigenstates is,
in principle, impossible for chaotic systems, Gutzwiller derived a 
semiclassical formula for the density of states as a whole.
Starting from the exact quantum expression given as the trace of the Green's 
operator he replaced the exact Green's function with its semiclassical 
approximation.
Applying stationary phase approximations he could finally write
the semiclassical density of states as the sum of a smooth part, i.e., 
the Weyl term, and an oscillating sum over all periodic orbits of the 
corresponding classical system.
For this reason, Gutzwiller's theory is also commonly known as 
{\em periodic orbit theory}.
Gutzwiller's semiclassical trace formula is valid for isolated and unstable 
periodic orbits, i.e., for fully chaotic systems with a complete hyperbolic 
dynamics.
Examples of hyperbolic systems are $n$-disk repellers, as models for 
hyperbolic scattering \cite{Cvi89,Eck93,Eck95,Wir92,Wir93,Cvi97,Wir97},
the stadium billiard \cite{Hel84,Tom93,Tan97},
the anisotropic Kepler problem \cite{Gut90},
and the hydrogen atom in magnetic fields at very high energies far above 
the ionization threshold \cite{Han95,Tan96,Schn97}.
Gutzwiller's trace formula is exact only in exceptional cases, e.g., for
the geodesic flow on a surface with constant negative curvature \cite{Gut90}.
In general, the semiclassical periodic orbit sum is just the leading order
term of an infinite series in powers of the Planck constant, $\hbar$.
Methods to derive the higher order contributions of the $\hbar$ expansion
are presented in Refs.\ \cite{Gas93,Alo93,Vat96}.

\paragraph*{Integrable systems: The Berry-Tabor formula}
For integrable systems a semiclassical trace formula was derived by 
Berry and Tabor \cite{Ber76,Ber77}.
The Berry-Tabor formula describes the density of states in terms of the
periodic orbits of the system and is therefore the analogue of Gutzwiller's
trace formula for integrable systems.
The equation is known to be formally equivalent to the EBK torus quantization.
A generalization of the Berry-Tabor formula to near-integrable systems is
given in Refs.\ \cite{Tom95,Ull96}.

\paragraph*{Mixed systems: Uniform semiclassical approximations}
Physical systems are usually neither integrable nor exhibit
complete hyperbolic dynamics.
Generic systems are of mixed type, characterized by the coexistence of
regular torus structures and stochastic regions in the classical Poincar\'e
surface of section.
A typical example is the hydrogen atom in a magnetic field 
\cite{Fri89,Has89,Wat93}, which undergoes a transition from near-integrable 
at low energies to complete hyperbolic dynamics at high excitation.
In mixed systems the classical trajectories, i.e., the periodic orbits,
undergo bifurcations.
At bifurcations periodic orbits change their structure and stability 
properties, new orbits are born, or orbits vanish.
A systematic classification of the various types of bifurcations is possible
by application of {\em normal form theory} \cite{Mey70,Mao92,Sad95,Sad96},
where the phase space structure around the bifurcation point is analyzed 
with the help of classical perturbation theory in local coordinates defined 
parallel and perpendicular to the periodic orbit.
At bifurcation points periodic orbits are neither isolated nor belong to
a regular torus in phase space.
As a consequence the periodic orbit amplitudes diverge in both semiclassical 
expressions, Gutzwiller's trace formula and the Berry-Tabor formula, i.e.,
both formulae are not valid near bifurcations.
The correct semiclassical solutions must simultaneously account for
all periodic orbits which participate at the bifurcation, including
``ghost'' orbits \cite{Kus93,Mai97a} in the complex generalization of phase
space, which can be important near bifurcations.
Such solutions are called {\em uniform semiclassical approximations} and
can be constructed by application of {\em catastrophe theory}
\cite{Pos78,Ber80} in terms of catastrophe diffraction integrals.
Uniform semiclassical approximations have been derived in Refs.\ 
\cite{Alm87,Alm88} for the simplest and generic types of bifurcations and 
in Refs.\ \cite{Mai97a,Scho97a,Scho97b,Mai98a} for nongeneric bifurcations 
with higher codimension and corank.

With Gutzwiller's trace formula for isolated periodic orbits, the Berry-Tabor
formula for regular tori, and the uniform semiclassical approximations for
orbits near bifurcations we have, in principle, the basic equations
for the semiclassical investigation of all systems with regular, chaotic,
and mixed regular-chaotic dynamics.
There are, however, fundamental problems in practical applications of these
equations for the calculation of semiclassical spectra, and the development
of techniques to overcome these problems has been the objective of intense 
research during recent years.

\subsubsection{Convergence problems of the semiclassical trace formulae}
The most serious problem of Gutzwiller's trace formula is that for chaotic
systems the periodic orbit sum does not converge in the physically interesting
energy region, i.e., on and below the real axis, where the eigenstates of 
bound systems and the resonances of open systems are located, respectively.
For chaotic systems the sum of the absolute values of the periodic orbit 
terms diverges exponentially because of the exponential proliferation of
the number of orbits with increasing periods.
The convergence problems are similar for the quantization of regular systems
with the Berry-Tabor formula, however, the divergence is algebraic instead
of exponential in this case.

Because of the technical problems encountered in trying to extract 
eigenvalues directly from the periodic orbit sum, Gutzwiller's trace formula 
has been used in many applications mainly for the interpretation of 
experimental or theoretical spectra of chaotic systems.
For example the periodic orbit theory served as basis for the development
of {\em scaled-energy spectroscopy}, a method where the classical dynamics
of a system is fixed within long-range quantum spectra \cite{Hol88,Mai91}.
The Fourier transforms of the scaled spectra exhibit peaks at positions
given by the classical action of the periodic orbits.
The action spectra can therefore be interpreted directly in terms of the 
periodic orbits of the system, and the technique of scaled-energy spectroscopy
is now well established, e.g., for investigations of atoms in external fields
\cite{Mai94a,Vel93,Del94,Rai94,Cou94,Cou95a,Cou95b,Neu97,Spe97,Hag98}.
The resolution of the Fourier transform action spectra is restricted by
the uncertainty principle of the Fourier transform, i.e., the method allows
the identification of usually short orbits in the low-dense part of the 
action spectra.
A fully resolved action spectrum would require an infinite length of the 
original scaled quantum spectrum.

Although periodic orbit theory has been very successful in the
interpretation of quantum spectra, the extraction of individual eigenstates 
and resonances directly from periodic orbit quantization remains of 
fundamental interest and importance.
As mentioned above the semiclassical trace formulae diverge at the physical 
interesting regions.
They are convergent, however, at complex energies with positive imaginary
part above a certain value, viz.\ the entropy barrier.
Thus the problem of revealing the semiclassical eigenenergies and resonances
is closely related to finding an analytic continuation of the trace
formulae to the region at and below the real axis.
Several refinements have been introduced in recent years in order to 
transform the periodic orbit sum in the physical domain of interest to a
conditionally convergent series by, e.g., using symbolic dynamics and the 
cycle expansion \cite{Cvi89,Cvi88,Art90}, the Riemann-Siegel look-alike 
formula and pseudo-orbit expansions \cite{Ber90,Kea92,Ber92}, surface of 
section techniques \cite{Bog92a,Bog92b}, and heat-kernel regularization
\cite{Sie91,Aur92a,Aur92b}.
These techniques are mostly designed for systems with special properties,
e.g., the cycle expansion requires the existence and knowledge of a symbolic
code and is most successful for open systems, while the Riemann-Siegel
look-alike formula and the heat-kernel regularization are restricted to
bound systems.
Until now there is no {\em universal} method, which allows periodic orbit
quantization of a large variety of bound and open systems with an underlying
regular, mixed, or chaotic classical dynamics.

\subsection{Objective of this work}
The main objective of this work is the development of novel methods for
$(a)$ the analysis of quantum spectra and $(b)$ periodic orbit quantization.
The conventional action or recurrence spectra obtained by Fourier 
transformation of finite range quantum spectra suffer from the fundamental 
resolution problem because of the uncertainty principle of the Fourier 
transformation.
The broadening of recurrence peaks usually prevents the detailed analysis
of structures near bifurcations of classical orbits.
We will present a method which allows, e.g., the detailed analysis of
bifurcations and symmetry breaking and the study of higher order $\hbar$
corrections to the periodic orbit sum.
For periodic orbit quantization the aim is the development of a universal 
method for periodic orbit quantization which does not depend on special 
properties of the system, such as the existence of a symbolic code
or the applicability of a functional equation.
As will be shown both problems can be solved by applying methods
for {\em high resolution spectral analysis} \cite{Mar87}.
The computational techniques for high resolution spectral analysis have
recently been significantly improved \cite{Wal95,Man97a,Man97b},
and we will demonstrate that the state of the art methods are a powerful 
tool for the semiclassical quantization and analysis of dynamical systems.

\subsubsection{High precision analysis of quantum spectra}
Within the semiclassical approximation of Gutzwiller's trace formula
or the Berry-Tabor formula the density of states of a scaled quantum
spectrum is a superposition of sinusoidal modulations as a function of 
the energy or an appropriate scaling parameter.
The frequencies of the modulations are determined by the periods, i.e.,
the classical action of orbits.
The amplitudes and phases of the oscillations depend on the stability 
properties of the periodic orbits and the Maslov indices.
To extract information about the classical dynamics from the quantum spectrum
it is therefore quite natural to Fourier transform the spectrum from
``energy'' domain to ``time'' (or, for scaled spectra, ``action'') domain
\cite{Hol88,Mai91,Mai94a,Mai86,Win86,Win87a,Win87b,Dan95,Hue95,Wei98}.
The periodic orbits appear as peaks at positions given by the periods
of orbits, and the peak heights exhibit the amplitudes of periodic orbit
contributions in the semiclassical trace formulae.
The comparison with classical calculations allows the interpretation of 
quantum spectra in terms of the periodic orbits of the corresponding 
classical system.
However, the analyzed experimental or theoretical quantum spectra are usually
of finite length, and thus the sinusoidal modulations are truncated when
Fourier transformed.
The truncation implies a broadening of peaks in the time domain spectra
because of the uncertainty principle of the Fourier transform.
The widths of the recurrence peaks are determined by the total length of the
original spectrum.
The uncertainty principle prevents until now precision tests of the 
semiclassical theories because neither the peak positions (periods of orbits)
nor the amplitudes can be obtained from the time domain spectra to high
accuracy.
Furthermore, the uncertainty principle implies an overlapping of broadened 
recurrence peaks when the separation of neighboring periods is less than
the peak widths.
In this case individual periodic orbit contributions cannot be revealed
any more.
This is especially a disadvantage, e.g., when following, in quantum spectra 
of systems with mixed regular-chaotic dynamics, the bifurcation tree of
periodic orbits.
All orbits involved in a bifurcation, including complex ``ghost'' orbits
\cite{Kus93,Mai97a,Mai98a}, have nearly the same period close to the 
bifurcation point.
The details of the bifurcations, especially of catastrophes with higher 
codimension or corank \cite{Pos78,Ber80}, cannot be resolved with the help 
of the conventional Fourier transform.
The same is true for the effects of symmetry breaking, e.g., breaking of
the cylindrical symmetry of the hydrogen atom in crossed magnetic and
electric fields \cite{Neu97} or the ``temporal symmetry breaking'' 
\cite{Spe97,Hag98} of atoms in oscillating fields.
The symmetry breaking should result in a splitting of peaks in the
Fourier transform recurrence spectra.
Until now this phenomenon could only be observed indirectly 
via constructive and destructive interference effects of orbits 
with broken symmetry \cite{Neu97,Spe97}.

To overcome the resolution problem in the conventional recurrence spectra
and to achieve a high resolution analysis of quantum spectra it is first
of all necessary to mention that the uncertainty principle of the Fourier
transformation is not a fundamental restriction to the resolution of the
time domain recurrence spectra, and is not comparable to the fundamental 
Heisenberg uncertainty principle in quantum mechanics.
This can be illustrated for the simple example of a sinusoidal function with
only one frequency.
The knowledge of the signal at only three points is sufficient to recover,
although not uniquely, the three unknown parameters, viz.\ the frequency,
amplitude and phase of the modulation.
Similarly, a superposition of $N$ sinusoidal functions can be recovered,
in principle, from a set of at least $3N$ data points $(t_i,f_i)$.
If the data points are exact the spectral parameters $\{\omega_k,A_k,\phi_k\}$
can also be exactly determined from a set of $3N$ {\em nonlinear} equations
\begin{equation}
f_i=\sum_{k=1}^N A_k\sin(\omega_kt_i-\phi_k) \; ,
\end{equation}
i.e., there is no uncertainty principle.
The main difference between this procedure and the Fourier transformation 
is that we use the linear superposition of sinusoidal functions as an ansatz 
for the functional form of our data, and the frequencies, amplitudes, and
phases are adjusted to obtain the best approximation to the data points.
By contrast, the Fourier transform is not based on any constraints.
In case of a discrete Fourier transform (DFT) the frequencies are chosen
on a (usually equidistant) grid, and the amplitudes are determined from
a {\em linear} set of equations, which can be solved numerically very
efficiently, e.g., by fast Fourier transform (FFT).

As mentioned above the high precision spectral analysis requires the
numerical solution of a {\em nonlinear} set of equations, and the availability
of efficient algorithms has been the bottleneck for practical applications
in the past.
Several techniques have been developed \cite{Mar87}, however, most of them
are restricted -- for reasons of storage requirements, computational time, 
and stability of the algorithm -- to signals with quite low number 
of frequencies.
By contrast, the number of frequencies (periodic orbits) in quantum spectra
of chaotic systems is infinite or at least very large, and applied methods
for high precision spectral analysis must be able to handle signals with
large numbers of frequencies.
This requirement is fulfilled by the method of {\em harmonic inversion by
filter-diagonalization}, which was recently developed by Wall and Neuhauser
\cite{Wal95} and significantly improved by Mandelshtam and Taylor
\cite{Man97a,Man97b}.
The decisive step is to recast the nonlinear set of equations as a generalized
eigenvalue problem.
Using an appropriate basis set the generalized eigenvalue equation is solved
with the filter-diagonalization method, which means that the frequencies of
the signal can be determined within small frequency windows and only small
matrices must be diagonalized numerically even when the signal is composed
of a very large number of sinusoidal oscillations.

We will introduce harmonic inversion as a powerful tool for the high precision
analysis of quantum spectra, which allows to circumvent the uncertainty
principle of the conventional Fourier transform analysis and to extract 
previously unattainable information from the spectra.
In particular the following items are investigated:
\begin{itemize}
\item
High precision check of semiclassical theories:
The analysis of spectra allows a direct quantitative comparison of
the quantum and classical periodic orbit quantities to many significant
digits.
\item
Uncovering of periodic orbit bifurcations in quantum spectra, and the
verification of ghost orbits and uniform semiclassical approximations.
\item
Direct observation of symmetry breaking effects.
\item
Quantitative interpretation of the differences between quantum and 
semiclassical spectra in terms of the $\hbar$ expansion of the periodic 
orbit sum.
\end{itemize}
Results will be presented for various physical systems, e.g., the hydrogen 
atom in external fields, the circle billiard, and the three disk scattering 
problem.

\subsubsection{Periodic orbit quantization}
The analysis of quantum spectra by harmonic inversion and, e.g., the
observation of ``ghost'' orbits, symmetry breaking effects, or higher
order $\hbar$ corrections to the periodic orbit contributions provides
a deeper understanding of the relation between quantum mechanics and the
underlying classical dynamics of the system.
However, the inverse procedure, i.e., the calculation of semiclassical
eigenenergies and resonances directly from classical input data is of 
at least the same importance and even more challenging.
As mentioned above the periodic orbit sums suffer from fundamental convergence
problems in the physically interesting domain and much effort has been
undertaken in recent years to overcome these problems
\cite{Cvi89,Cvi88,Art90,Ber90,Kea92,Ber92,Bog92a,Bog92b,Sie91,Aur92a,Aur92b}.
Although many of the refinements which have been introduced are very
efficient for a specific model system or a class of systems, they all suffer
from the disadvantage of non-universality.
The cycle expansion technique \cite{Cvi89,Cvi88,Art90}
requires a completely hyperbolic dynamics and the existence of a symbolic code.
The method is most efficient only for open systems, e.g., for three-disk
or $n$-disk pinball scattering 
\cite{Cvi89,Eck93,Eck95,Wir92,Wir93,Cvi97,Wir97}.
By contrast, the Riemann-Siegel look-alike formula and pseudo orbit expansion
of Berry and Keating \cite{Ber90,Kea92,Ber92} can only be applied for
bound systems.
The same is true for surface of section techniques \cite{Bog92a,Bog92b} and
heat-kernel regularization \cite{Sie91,Aur92a,Aur92b}.
We will introduce high resolution spectral analysis, and in particular 
harmonic inversion by filter-diagonalization as a novel and {\em universal} 
method for periodic orbit quantization.
Universality means that the method can be applied to open and bound systems
with regular and chaotic classical dynamics as well.

Formally the semiclassical density of states (more precisely the 
semiclassical response function) can be written as the Fourier transform 
of the periodic orbit recurrence signal
\begin{equation}
 C^{\rm sc}(s)=\sum_{\rm po}{\cal A}_{\rm po}\delta(s-s_{\rm po}) \; ,
\end{equation}
with ${\cal A}_{\rm po}$ and $s_{\rm po}$ the periodic orbit amplitudes
and periods (actions), respectively.
If all orbits up to infinite length are considered the Fourier transform
of $C^{\rm sc}(s)$ is equivalent to the non-convergent periodic orbit sum.
For chaotic systems a numerical search for all periodic orbits
is impossible and, furthermore, does not make sense because the periodic
orbit sum does not converge anyway.
On the other hand, the truncation of the Fourier integral at finite maximum 
period $s_{\rm max}$ yields a smoothed spectrum only \cite{Win88}.
However, the low resolution property of the spectrum can be interpreted
as a consequence of the uncertainty principle of the Fourier transform.
We can now argue in an analogous way as in the previous section that the
uncertainty principle can be circumvented with the help of high resolution 
spectral analysis, and propose the following procedure for periodic orbit
quantization by harmonic inversion.

Let us assume that the periodic orbits with periods $0<s<s_{\rm max}$ are 
available and the semiclassical recurrence function $C^{\rm sc}(s)$ has been 
constructed.
This signal can now be harmonically inverted and thus adjusted to the
functional form of the quantum recurrence function $C^{\rm qm}(s)$, which
is a superposition of sinusoidal oscillations with frequencies given by
the quantum mechanical eigenvalues of the system.
The frequencies obtained by harmonic inversion of $C^{\rm sc}(s)$ should
therefore be the semiclassical approximations to the exact eigenenergies.
The universality and wide applicability of this novel quantization scheme
follows from the fact that the periodic orbit recurrence signal $C^{\rm sc}(s)$
can be obtained for a large variety of systems because the calculation does
not depend on special properties of the system, such as boundness,
ergodicity, or the existence of a symbolic dynamics.
For systems with underlying regular or chaotic classical dynamics the 
amplitudes ${\cal A}_{\rm po}$ of periodic orbit contributions are directly
obtained from the Berry-Tabor formula \cite{Ber76,Ber77} and Gutzwiller's 
trace formula \cite{Gut90}, respectively.

As mentioned above the harmonic inversion technique requires the knowledge 
of the signal $C^{\rm sc}(s)$ up to a finite maximum period $s_{\rm max}$.
The efficiency of the quantization method strongly depends on the signal 
length which is required to obtain a certain number of eigenenergies.
In chaotic systems periodic orbits proliferate exponentially with increasing
period and usually the orbits must be searched numerically.
It is therefore highly desirable to use the shortest possible signal for
periodic orbit quantization.
The efficiency of the method can be improved if not just the single signal
$C^{\rm sc}(s)$ is harmonically inverted but additional classical
information obtained from a set of smooth and linearly independent observables
is used to construct a semiclassical cross-correlated periodic
orbit signal.
The cross-correlation function can be analyzed with a generalized harmonic
inversion technique and allows calculating semiclassical eigenenergies from
a significantly reduced set of orbits or alternatively to improve the accuracy
of spectra obtained with the same set of orbits.

However, the semiclassical eigenenergies deviate -- apart from a few 
exceptions, e.g., the geodesic flow on a surface with constant negative 
curvature \cite{Gut90} -- from the exact quantum mechanical eigenvalues.
The reason is that Gutzwiller's trace formula and the Berry-Tabor formula
are only the leading order terms of the $\hbar$ expansion of periodic orbit
contributions \cite{Gas93,Alo93,Vat96}.
It will be shown how the higher order $\hbar$ corrections of the periodic 
orbit sum can be used within the harmonic inversion procedure to improve,
order by order, the semiclassical accuracy of eigenenergies, i.e., to obtain 
eigenvalues beyond the Gutzwiller and Berry-Tabor approximation.

\subsection{Outline}
The manuscript is organized as follows.
In Chapter \ref{analysis} the high precision analysis of quantum spectra is
discussed.
After general remarks on Fourier transform recurrence spectra in Section
\ref{FT_spectra} we introduce in Section \ref{Circumvent} harmonic inversion 
as a tool to circumvent the uncertainty principle of the conventional Fourier 
transformation \cite{Mai97c}.
In Section \ref{prec_check} a precision check of the periodic orbit theory 
is demonstrated by way of example of the hydrogen atom in a magnetic field.
For the hydrogen atom in external fields we furthermore illustrate that
dynamical information which has been unattainable before can be extracted
from the quantum spectra.
In Section \ref{ghost_sec} we investigate in detail the quantum manifestations
of bifurcations of orbits related to a hyperbolic umbilic catastrophe
\cite{Mai98a,Wil97} and a butterfly catastrophe \cite{Mai97a,Mai97b}, and in 
Section \ref{sym_breaking} we directly uncover effects of symmetry breaking 
in the quantum spectra.
In Section \ref{hbar_sec} we analyze by way of example of the circle billiard
and the three disk system the difference between the exact quantum and the
semiclassical spectra.
It is shown that the deviations between the spectra can be quantitatively 
interpreted in terms of the next order $\hbar$ contributions of the
$\hbar$ expanded periodic orbit sum.

In Chapter \ref{po_quant} we propose harmonic inversion as a novel technique
for periodic orbit quantization \cite{Mai97d,Mai98b}.
The method is introduced in Section \ref{zeta_sec} for a mathematical model,
viz.\ the calculation of zeros of Riemann's zeta function.
This model shows a formal analogy with the semiclassical trace formulae, and 
is chosen for the reasons that no extensive numerical periodic orbit search 
is necessary and the results can be directly compared to the exact zeros
of the Riemann zeta function.
In Section \ref{po_quant_sec} the method is derived for the periodic 
orbit quantization of physical systems and is demonstrated in Section 
\ref{three_disk_sec} for the example of the three disk scattering system
with fully chaotic (hyperbolic) classical dynamics \cite{Mai97d,Mai98b}, 
and in Section \ref{mixed_sec} for the hydrogen atom in a magnetic field as 
a prototype example of a system with mixed regular-chaotic dynamics 
\cite{Mai98f}.
In Section \ref{cross_corr_po_sums} the efficiency of the method is improved
by a generalization of the technique to the harmonic inversion of
cross-correlated periodic orbit sums, which allows to significantly reduce
the number of orbits required for the semiclassical quantization \cite{Mai98c},
and in Section \ref{hbar_sec2} we derive the concept for periodic orbit
quantization beyond the Gutzwiller and Berry-Tabor approximation by harmonic 
inversion of the $\hbar$ expansion of the periodic orbit sum \cite{Mai98d}.
The methods of Sections \ref{cross_corr_po_sums} and \ref{hbar_sec2} are
illustrated in Section \ref{circle_billiard} for the example of the 
circle billiard \cite{Mai98c,Mai98d,Wei99a,Wei99b}.
Finally, in Section \ref{photo_sec} we demonstrate the semiclassical
calculation of individual transition matrix elements for atoms in 
external fields \cite{Mai98g}.

Chapter \ref{conclusion} concludes with a summary and outlines possible
future applications, e.g., the analysis of experimental spectra and the
periodic orbit quantization of systems without scaling properties.

Computational details of harmonic inversion by filter-diagonalization are 
presented in Appendix \ref{harm_inv_app}, and the calculation and asymptotic 
behavior of some catastrophe diffraction integrals is discussed in Appendix 
\ref{diffr_int_app}.

\section{High precision analysis of quantum spectra}
\label{analysis}
\setcounter{equation}{0}
Semiclassical {\em periodic orbit} theory \cite{Gut67,Gut90} and 
{\em closed orbit} theory \cite{Du87,Du88,Bog88a,Bog89} have become the key 
for the interpretation of quantum spectra of classically chaotic systems.
The semiclassical spectra at least in low resolution are given as the sum
of a smooth background and a superposition of modulations whose amplitudes,
frequencies, and phases are solely determined by the closed or periodic 
orbits of the classical system.
For the interpretation of quantum spectra in terms of classical orbits it 
is therefore most natural to obtain the recurrence spectrum by
Fourier transforming the energy spectrum to the time domain.
Each closed or periodic orbit should show up as a sharp $\delta$-peak
at the recurrence time (period), provided, first, the classical recurrence 
times do not change within the whole range of the spectrum and, second, 
the Fourier integral is calculated along an infinite energy range.
Both conditions are usually not fulfilled.
However, the problem regarding the energy dependence of recurrence times
can be solved in systems possessing a classical scaling property by 
application of scaling techniques.
The second condition is never fulfilled in practice, i.e., the length of 
quantum spectra is always restricted either by experimental limitations or, 
in theoretical calculations, by the growing dimension of the Hamiltonian 
matrix which has to be diagonalized numerically.
The given length of a quantum spectrum determines the resolution of the
quantum recurrence spectrum due to the uncertainty principle, 
$\Delta E \cdot \Delta T \sim \hbar$, when the conventional Fourier transform 
is used.
Only those closed or periodic orbits can be clearly identified quantum
mechanically which appear as isolated non-overlapping peaks in the quantum 
recurrence spectra.
This is especially not the case for orbits which undergo bifurcations at
energies close to the bifurcation point.
As will be shown the resolution of quantum recurrence spectra can be 
significantly improved beyond the limitations of the uncertainty principle
of the Fourier transform by methods of high resolution spectral analysis.

In the following we first review the conventional analysis of quantum
spectra by Fourier transformation and discuss on the example of the
hydrogen atom in a magnetic field the achievements and limitations of
the Fourier transform recurrence spectra.
In Section \ref{Circumvent} we review methods of 
{\em high resolution spectral analysis} which can serve to circumvent the 
uncertainty principle of the Fourier transform.
In Sections \ref{prec_check} to \ref{hbar_sec} the harmonic inversion 
technique is applied to calculate {\em high resolution recurrence spectra} 
beyond the limitations of the uncertainty principle of the Fourier 
transformation from experimental or theoretical quantum spectra of 
{\em finite length}.
The method allows to reveal information about the dynamics of the system
which is completely hidden in the Fourier transform recurrence spectra.
In particular, it allows to identify real orbits with nearly degenerate 
periods, to detect complex ``ghost'' orbits which are of importance 
in the vicinity of bifurcations \cite{Kus93,Mai97a,Mai98a}, and to 
investigate higher order $\hbar$ corrections of the periodic orbit 
contributions \cite{Gas93,Alo93,Vat96}.

\subsection{Fourier transform recurrence spectra}
\label{FT_spectra}
According to {\em periodic orbit} theory \cite{Gut67,Gut90} the semiclassical
density of states can be written as the sum of a smooth background 
$\varrho_0(E)$ and oscillatory modulations induced by the periodic orbits, 
\begin{equation}
\label{rho_sc_E}
 \varrho(E) = \varrho_0(E) + {1\over\pi} {\rm Re} \,
 \sum_{\rm po} \sum_{r=1}^\infty A_{\rm po,r}(E) e^{irS_{\rm po}(E)/\hbar} \; ,
\end{equation}
with $A_{\rm po,r}(E)$ the amplitudes of the modulations and the classical 
actions $S_{\rm po}(E)$ of a {\em primitive} periodic orbit (po) determining 
the frequencies of the oscillations.
Linearizing the action around $E=E_0$ yields
\begin{equation}
 S(E) \approx S(E_0) + \left.{dS\over dE}\right|_{E_0}(E-E_0)
  = S(E_0) + T_0 (E-E_0) \; ,
\label{S_expan}
\end{equation}
with $T_0$ the time period of the orbit at energy $E_0$.
When Eq.\ (\ref{S_expan}) is inserted in (\ref{rho_sc_E}) the semiclassical 
density of states is locally given as a superposition of sinusoidal 
oscillations, and it might appear that the problem of identifying the 
amplitudes $A_{\rm po}$ and time periods $T_{\rm po}$ that contribute to 
the quantum spectrum $\varrho(E)$ can be solved by Fourier transforming 
$\varrho(E)$ to the time domain, i.e., each periodic orbit should show
up as a $\delta$-peak at the recurrence time $t=T_0$.
However, fully resolved recurrence spectra can be obtained only if the two
following conditions are fulfilled.
First, the Fourier integral is calculated along an infinite energy range,
and, second, the classical recurrence times do not change within the whole 
range of the spectrum.
The first condition is usually not fulfilled when quantum spectra are 
obtained from an experimental measurement or a numerical quantum calculation.
In that case the Fourier transformation is restricted to a finite energy
range and the resolution of the time domain spectrum is limited by
the uncertainty principle of the Fourier transform.
It is the main objective of this Chapter to introduce high resolution methods
for the spectral analysis which allow to circumvent the uncertainty principle
of the Fourier transform and to obtain fully resolved recurrence spectra
from the analysis of quantum spectra with finite length.
However, the second condition is usually not fulfilled either.
Both the amplitudes and recurrence times of periodic orbits are in general
nontrivial functions of the energy $E$, and, even worse, the whole phase 
space structure changes with energy when periodic orbits undergo bifurcations.
For the interpretation of quantum spectra in terms of periodic orbit
quantities it is therefore in general not appropriate to analyze the 
frequencies of long ranged spectra $\varrho(E)$.
The problems due to the energy dependence of periodic orbit quantities
have been solved by the development and application of scaling techniques.

\paragraph*{Scaling techniques}
Many systems possess a classical scaling property in that the classical 
dynamics does not depend on an external scaling parameter $w$ and the action
of trajectories varies linearly with $w$.
Examples are systems with homogeneous potentials, billiard systems, or
the hydrogen atom in external fields.
In billiard systems the shapes of periodic orbits are solely determined
by the geometry of the borders, and the classical action depends on the
length $L_{\rm po}$ of the periodic orbit,
$S_{\rm po}=\hbar kL_{\rm po}$, with $k$ the wave number.
For a particle with mass $m$ moving in a billiard system it is therefore 
most appropriate to take the wave number as the scaling parameter, i.e.,
$w=k=\sqrt{2mE}/\hbar$.
For a system with a homogeneous potential $V({\bf q})$ with
$V(c{\bf q})=c^\alpha V({\bf q})$ only the size of periodic orbits but not
their shape changes with varying energy $E$.
Introducing a scaling parameter as a power of the energy, 
$w=E^{1/\alpha+1/2}$, 
the classical action of a periodic orbit is obtained as 
$S_{\rm po}(E)=(E/E_0)^{1/\alpha+1/2}S_{\rm po}(E_0)$, 
with $E_0$ being a reference energy, i.e., the action depends linearly
on a scaling parameter defined as $w=(E/E_0)^{1/\alpha+1/2}$.
For example the Coulomb potential is a homogeneous potential with $\alpha=-1$
and the bound Coulomb spectrum at negative energies ($E_0=-1/2$) is 
transformed by the scaling procedure to a simple spectrum with equidistant 
lines at $w=n$ with $n=1,2,\dots$ the principal quantum number.
For atoms in external magnetic and electric fields the shape of periodic
orbits changes if the field strengths are fixed and only the energy is varied.
However, these systems possess a scaling property if both the energy and the
field strengths are simultaneously varied.
Details for the hydrogen atom in a magnetic field will be given below.
The scaling parameter plays the role of an inverse effective Planck constant,
$w=\hbar_{\rm eff}^{-1}$.
In theoretical investigations it is even possible to apply scaling techniques
to general systems with non-homogeneous potentials if the Planck constant
is formally used as a variable parameter.
Non-homogeneous potentials are important, e.g., in molecular dynamics, and
the genaralized scaling technique has been applied in Ref.\ \cite{Mai97e}
to analyze quantum spectra of the HO$_2$ molecule.

When the scaling technique is applied to a quantum system the scaling
parameter $w$ is quantized, i.e., bound systems exhibit sharp lines
at real eigenvalues $w=w_k$ and open systems have resonances related to 
complex poles $w=w_k$ of the scaled Green's function $G^+(w)$.
By varying the scaled energy a direct comparison of the quantum recurrence
spectra with the bifurcation diagram of the underlying classical system
is possible \cite{Hol88,Mai94a}.

The semiclassical approximation to the scaled spectrum is given by
\begin{equation}
\label{rho_sc}
 \varrho(w) = \varrho_0(w) + {1\over\pi} {\rm Re} \,
 \sum_{\rm po} \sum_{r=1}^\infty A_{\rm po,r} e^{irs_{\rm po} w} \; ,
\end{equation}
with
\begin{equation}
   s_{\rm po}
 = \oint_{\rm po} {\tilde{\bf p}}d{\tilde{\bf q}}
 = {1\over w} \oint_{\rm po} {\bf p}d{\bf q}
\end{equation}
the scaled action of a {primitive} periodic orbit.
The vectors $\tilde{\bf q}$ and $\tilde{\bf p}$ are the coordinates
and momenta of the scaled Hamiltonian.
In contrast to Eq.\ \ref{rho_sc_E} the periodic orbit sum (\ref{rho_sc})
is a superposition of sinusoidal oscillations as a function of the
scaling parameter $w$.
Therefore the scaled spectra $\varrho(w)$ can be Fourier transformed along 
arbitrarily long ranges of $w$ to generate Fourier transform recurrence 
spectra of in principle arbitrarily high resolution, i.e., yielding sharp
$\delta$-peaks at the positions of the scaled action $s=s_{\rm po}$ of 
periodic orbits.
The high resolution analysis of quantum spectra in the following sections is
possible only in conjunction with the application of the scaling technique.

The periodic orbit amplitudes $A_{\rm po,r}$ in Gutzwiller's trace formula 
(\ref{rho_sc}) for scaled systems are given by
\begin{eqnarray}
 A_{\rm po,r} = {s_{\rm po}\over\sqrt{|{\rm det}(M_{\rm po}^r-I)|}} \,
                e^{-ir{\pi\over 2}\mu_{\rm po}} \; ,
\label{A_po_r}
\end{eqnarray}
with $M_{\rm po}$ and $\mu_{\rm po}$ the monodromy matrix and Maslov index 
of the primitive periodic orbits, respectively, and $r$ the repetition number
of orbits.
The monodromy matrix ${\bf M}$ is the stability matrix restricted to 
deviations perpendicular to a periodic orbit after period time $T$. 
We here discuss systems with two degrees of freedom.
If $\delta q(0)$ is a small deviation perpendicular to the orbit in
coordinate space at time $t=0$ and $\delta p(0)$ an initial deviation in 
momentum space, the corresponding deviations at time $t=T$ are related to
the monodromy matrix \cite{Bog89,Bog88b}:
\begin{equation}
   \left(\begin{array}{c} \delta q(T) \\
                          \delta p(T) \end{array} \right)
 = {\bf M} \left( \begin{array}{c} \delta q(0) \\
                                   \delta p(0)\end{array} \right)
 = \left(\begin{array}{c} m_{11}~m_{12} \\ 
                          m_{21}~m_{22} \end{array} \right)
           \left( \begin{array}{c} \delta q(0) \\
                                   \delta p(0)\end{array} \right) \; .
\end{equation}
To compute ${\bf M}$ one considers an initial deviation solely
in coordinate space to obtain the matrix elements $m_{11}$ and $m_{21}$, and 
an initial deviation solely in momentum space to obtain $m_{12}$ and $m_{22}$.
In practice a linearized system of differential equations obtained by 
differentiating Hamilton's equations of motion with respect to the 
phase space coordinates is numerically integrated.
The Maslov index $\mu_{\rm po}$ increases by one every time the trajectory 
passes a conjugate point or a caustic.
Therefore the amplitudes $A_{\rm po,r}$ are complex numbers containing 
phase information determined by the Maslov indices of orbits.
The classical actions $s_{\rm po}$ are usually real numbers, although
they can be complex in general.
Non-real actions $s_{\rm po}$ indicate ``ghost'' orbits
\cite{Kus93,Mai97a,Mai98a} which exist in the complex continuation of 
the classical phase space.

As mentioned above the problem of identifying $A_{\rm po}$ and 
$s_{\rm po}$ that contribute to the quantum spectrum $\varrho(w)$ 
can in principle be solved by Fourier transforming $\varrho(w)$ to the 
action domain,
\begin{equation}
 C(s) = {1\over 2\pi}\int_{w_1}^{w_2} \left[\varrho(w)-\varrho_0(w)\right]
        e^{-isw}dw \; .
\label{C_FT}
\end{equation}
If the quantum spectrum $\varrho(w)$ has infinite length and the periodic
orbits (i.e., the actions $s_{\rm po}$) are real the Fourier transform of
Eq. \ref{rho_sc} indeed results in a fully resolved recurrence spectrum
\begin{eqnarray}
 C(s) &=& {1\over 2\pi}\int_{-\infty}^{+\infty}
          \left[\varrho(w)-\varrho_0(w)\right] e^{-isw}dw  \nonumber \\
      &=& {1\over 2\pi}\sum_{\rm po} \sum_{r=1}^\infty
           \left( A_{\rm po,r}\delta(s-rs_{\rm po})
                + A_{\rm po,r}^\ast\delta(s+rs_{\rm po}) \right)  \; .
\label{Cs}
\end{eqnarray}
The periodic orbits are identified as $\delta$-peaks in the recurrence 
spectrum at positions $s=\pm s_{\rm po}$ (and their repetitions at 
$s=\pm rs_{\rm po}$) and the recurrence strengths are given as the amplitudes 
$A_{\rm po,r}$ (Eq.\ \ref{A_po_r}) of Gutzwiller's trace formula.
However, for finite range spectra the $\delta$-functions in Eq.\ \ref{Cs}
must be replaced apart from a phase factor with
\[
 \delta(s) \longrightarrow {\sin((w_2-w_1)s/2)\over \pi s} \; .
\]
Unfortunately, the recurrence peaks are broadened by the uncertainty 
principle of the Fourier transform and furthermore the function 
$\sin((w_2-w_1)s/2)/\pi s$ has side-peaks which are not related to
recurrences of periodic orbits but are solely an undesirable effect of the
sharp cut of the Fourier integral at $w_1$ and $w_2$.
In complicated recurrence spectra it can be difficult or impossible to
separate the physical recurrences and the unphysical side-peaks.
The occurrence of side-peaks can be avoided by multiplying the spectra
with a window function $h(w)$, i.e.\
\begin{equation}
 C(s) = {1\over 2\pi}\int_{w_1}^{w_2} h(w) \left[\varrho(w)-\varrho_0(w)\right]
        e^{-isw}dw  \; ,
\end{equation}
where $h(w)$ is equal to one at the center of the spectrum and decreases 
smoothly close to zero at the borders $w_1$ and $w_2$ of the Fourier integral.
An example is a Gaussian window
\[
 h(w) = \exp\left(-{(w-w_0)^2\over 2\sigma^2}\right)
\]
centered at $w_0=(w_1+w_2)/2$ and with sufficiently chosen width 
$\sigma\approx (w_2-w_1)/6$.
The Gaussian window suppresses unphysical side-peaks in the recurrence spectra,
however, the decrease of the side-peaks is paid by an additional broadening of 
the central peak.
Various other types of window functions $h(w)$ have been used as a compromise 
between the least uncertainty of the central recurrence peak and the optimal
suppression of side-peaks.
However, the fundamental problem of the uncertainty principle of the Fourier
transform cannot be solved with any window function $h(w)$.

As a first example for the analysis of quantum spectra we introduce
the hydrogen atom in a magnetic field (for reviews see 
\cite{Fri89,Has89,Wat93}) given by the Hamiltonian [in atomic units, 
magnetic field strength $\gamma=B/(2.35 \times 10^5\, {\rm T})$, 
angular momentum $L_z=0$]
\begin{equation}
\label{Hamfkt}
 H = {1\over2}{\bf p}^2 - {1\over r} + {1\over8}\gamma^2 (x^2+y^2) \; .
\end{equation}
As mentioned above this system possesses a scaling property.
Introducing scaled coordinates, $\tilde{\bf r} = \gamma^{2/3}{\bf r}$,
and momenta, $\tilde{\bf p} = \gamma^{-1/3}{\bf p}$,
the classical dynamics of the scaled Hamiltonian
\begin{equation}
\label{Hamfkt_scal}
 \tilde H = {1\over2}\tilde{\bf p}^2 - {1\over \tilde r}
 + {1\over8}(\tilde x^2+\tilde y^2) = E\gamma^{-2/3} \; .
\end{equation}
does not depend on two parameters, the energy $E$ and the magnetic field 
strength $\gamma$, but solely on the scaled energy 
\begin{equation}
 \tilde E = E\gamma^{-2/3} \; .
\end{equation}
The classical action of the trajectories scales as
\begin{equation}
 S_{\rm po} = s_{\rm po} \gamma^{-1/3} = s_{\rm po}w \; .
\end{equation}

\begin{figure}[b]
\vspace{13.5cm}
\includegraphics{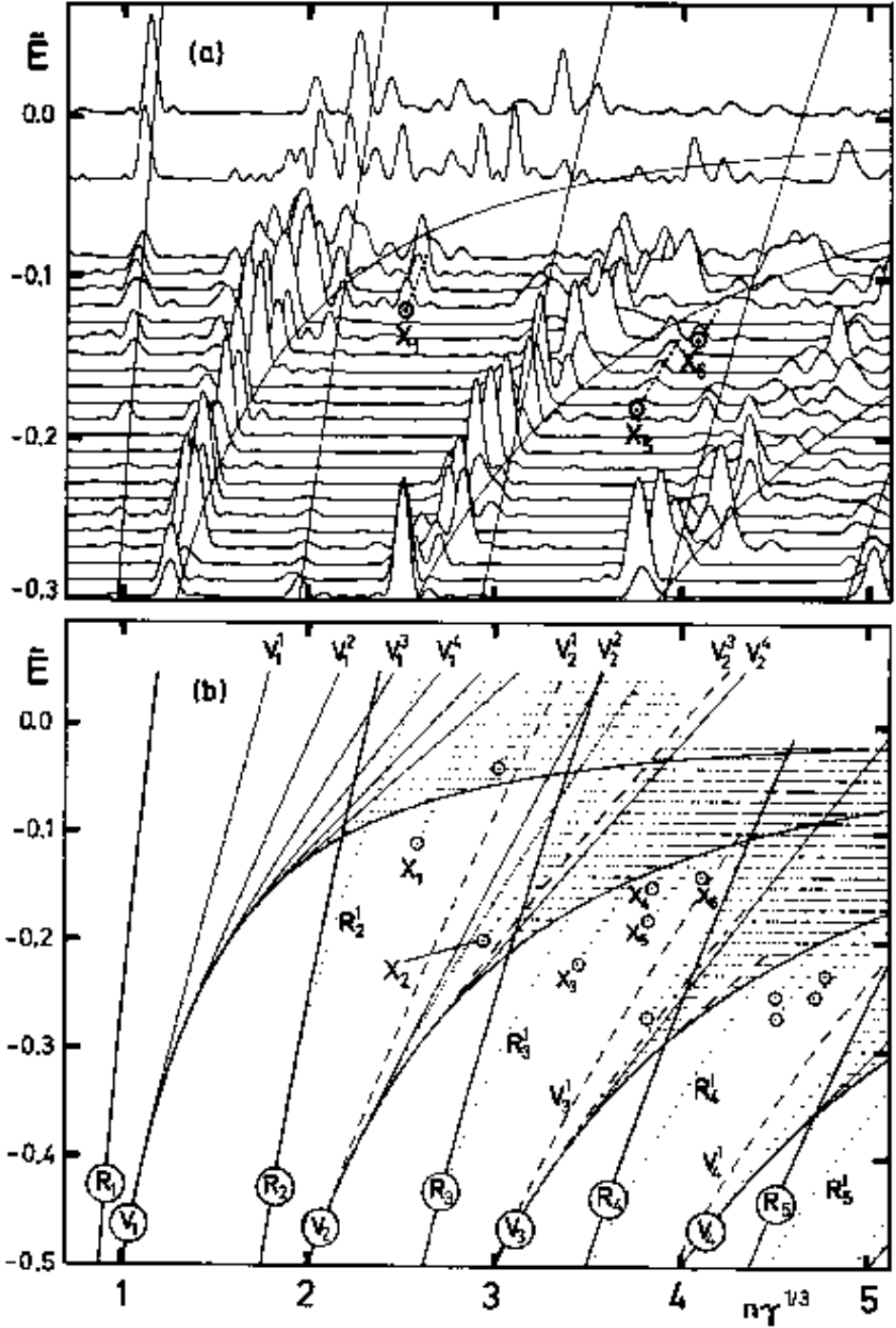}
\caption{\label{fig01} 
(a) Experimental recurrence spectra of the hydrogen atom in a magnetic field
in overlay form. Even-parity, $m=0$ final state.
(b) Scaled action of closed classical orbits through origin.
(From Ref.\ [45].)
}
\end{figure}

\begin{figure}[b]
\vspace{13.5cm}
\includegraphics{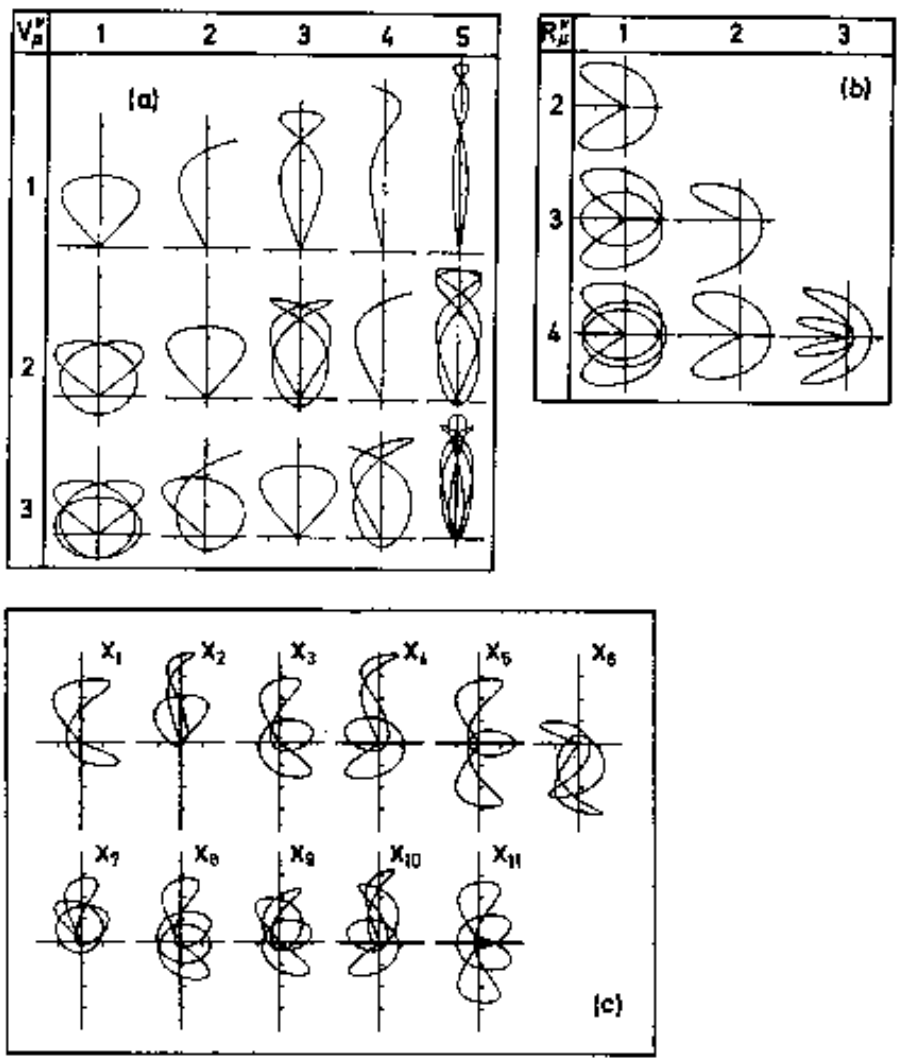}
\caption{\label{fig02} 
Closed classical orbits of the hydrogen atom in a magnetic field in
$(\rho,z)$ projection.
(a) Vibrator, (b) rotator, and (c) ``exotic'' orbits.
(From Ref.\ [46].)
}
\end{figure}

Based on the scaling relations of the classical Hamiltonian the 
experimental technique of {\em scaled energy spectroscopy} was developed 
\cite{Hol88,Mai91,Mai94a}.
Experimental spectra on the hydrogen atom at constant scaled energy 
have been measured by varying the magnetic field strength linearly on a 
scale $w\equiv\gamma^{-1/3}$, adjusting simultaneously the energy $E$ 
(via the wave length of the exciting laser light) so that the scaled energy 
$\tilde E = E\gamma^{-2/3}$ is kept constant at a given value.
The spectra have been Fourier transformed in the experimentally accessible
range $36\le\gamma^{-1/3}\le 50$.
Experimental recurrence spectra at scaled energies $-0.3\le\tilde E\le 0$
are presented in Fig.\ \ref{fig01}a.
The overlay exhibits a well-structured system of clustered branches of
resonances in the scaled energy-action plane.
For comparison Fig.\ \ref{fig01}b presents the energy-action spectrum of
the closed classical orbits, i.e.\ the scaled action of orbits as a function
of the energy.
As can be seen the clustered branches of resonances in the experimental
recurrence spectra (Fig.\ \ref{fig01}a) well resemble the classical 
bifurcation tree of closed orbits in Fig.\ \ref{fig01}b, although a
one to one comparison between recurrence peaks and closed orbits is
limited by the finite resolution of the quantum recurrence spectra.
Closed orbits bifurcating from the $\mu$th repetition of the orbit parallel 
to the magnetic field axis are called ``vibrators'' $V_\mu^\nu$, and orbits 
bifurcating from the $\mu$th repetition of the perpendicular orbit are 
called ``rotators'' $R_\mu^\nu$ in Fig.\ \ref{fig01}.
Orbits $X_\mu$ are created mainly in tangent bifurcations.
The graphs of some closed orbits are given in Fig.\ \ref{fig02}.
A detailed comparison of the peak heights of resonances in the experimental
recurrence spectra with the semiclassical amplitudes obtained from closed 
orbit theory can be found in Ref.\ \cite{Mai94a}.

\begin{figure}[b]
\vspace{13.5cm}
\includegraphics{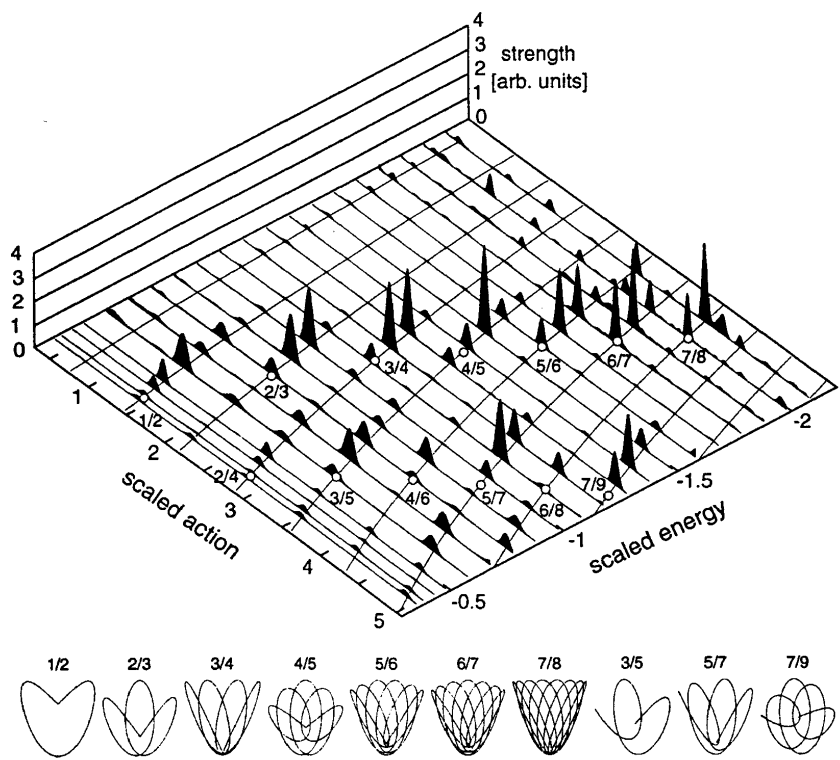}
\caption{\label{fig03} 
Experimental recurrence spectra of lithium in an electric field in overlay 
form.
The curves mark the scaled action of the parallel orbit and its repetitions
as a function of scaled energy.
The open circles mark the bifurcation points of closed orbits, whose
shapes are shown along the bottom.
(From Ref.\ [52].)
}
\end{figure}

Scaled energy spectroscopy has become a well established method to
investigate the dynamics of hydrogenic and nonhydrogenic atoms in external
fields.
Recurrence spectra of helium in a magnetic field are presented in
Ref.\ \cite{Vel93,Del94}.
Recurrence peaks of the helium atom which cannot be identified by hydrogenic
closed orbits have been explained by scattering of the highly excited
electron at the ionic core \cite{Del94,Dan95,Hue95}.
Rubidium has been studied in crossed electric and magnetic fields in
Ref.\ \cite{Rai94} and unidentified recurrence peaks have been interpreted
in terms of classical core scattering \cite{Wei98}.
The Stark effect on lithium has been investigated by the MIT group
\cite{Cou94,Cou95a,Cou95b}.
For atoms in an electric field $F$ the scaling parameter is $w\equiv F^{-1/4}$,
and $EF^{-1/2}$ is the scaled energy.
Experimental recurrence spectra of lithium in an electric field and the
corresponding closed orbits are presented in Fig.\ \ref{fig03}.
Strong recurrence peaks occur close to the bifurcations of the parallel
orbit and its repetitions marked by open circles in Fig.\ \ref{fig03}.
The new orbits created in bifurcations have almost the same action as the
corresponding return of the parallel orbit, and, similar as in Fig.\ 
\ref{fig01} for the hydrogen atom in a magnetic field, the small splittings 
of recurrence peaks are not resolved in the experimental recurrence spectra.

\begin{figure}[b]
\vspace{9.0cm}
\includegraphics{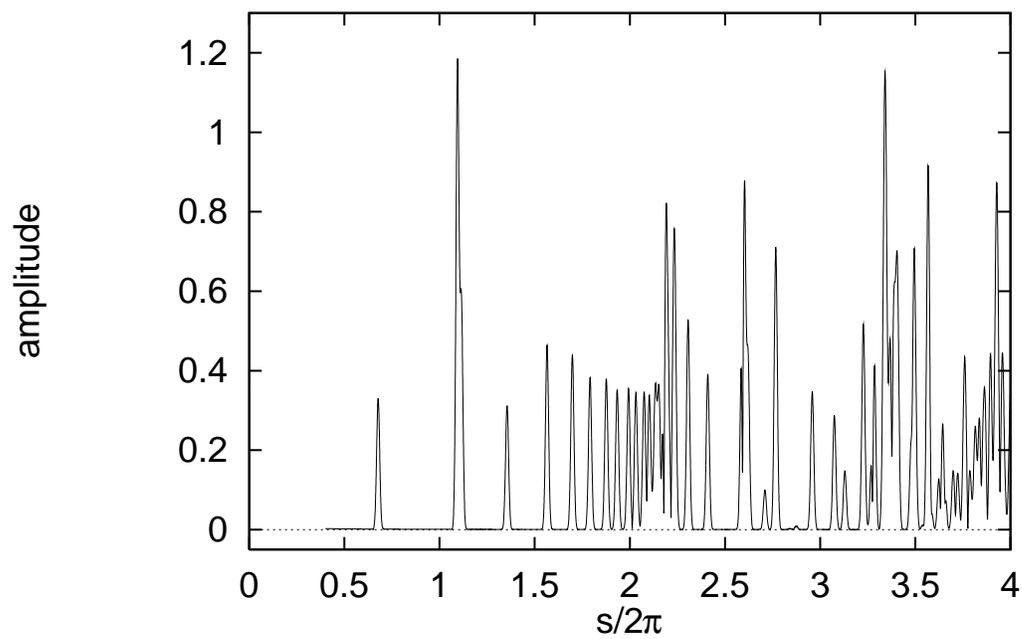}
\caption{\label{fig04} 
Recurrence spectrum (Fourier transform) for the density of states of the 
hydrogen atom in a magnetic field at scaled energy 
$\tilde E=E\gamma^{-2/3}=-0.1$.
}
\end{figure}

It is important to note that the finite resolution of the experimental
recurrence spectra in Figs.\ \ref{fig01} and \ref{fig03} is not caused by
the finite bandwidth of the exciting laser but results as discussed above
from the finite length of the Fourier transformed scaled spectra and the
uncertainty principle of the Fourier transform.
This can be illustrated by analyzing a numerically computed spectrum instead
of an experimentally measured spectrum.
We study the hydrogen atom in a magnetic field at constant scaled energy 
$\tilde E=-0.1$.
At this energy the classical dynamics is completely chaotic and all periodic
orbits are unstable.
We calculated 9715 states in the region $w<140$ by numerical diagonalization
of the Hamiltonian matrix in a complete basis set.
For details of the quantum calculations see, e.g., Ref.\ \cite{Mai94b}.
In Fig.\ \ref{fig04} the quantum density of states is analyzed by the
conventional Fourier transform.
To get rid of unphysical side-peaks the spectrum was multiplied with a 
Gaussian function $h(w)$ with width $\sigma$ chosen in accordance with 
the total length of the quantum spectrum.
Fig.\ \ref{fig04} clearly exhibits recurrence peaks which can be related
to classical periodic orbits.
However, the widths of the peaks are approximately $\Delta s/2\pi=0.03$,
and it is impossible to determine the periods of the classical orbits
to higher accuracy than about $0.01$ from the Fourier analysis of the 
quantum spectrum.
Recurrence peaks of orbits with similar periods overlap, as can be clearly
seen around $s/2\pi=2.1$, and at least guessed, e.g., at $s/2\pi=1.1$, or 
$s/2\pi=2.6$.
A precise determination of the amplitudes is impossible especially for the
overlapping peaks.
Furthermore, the Fourier transform does not allow to distinguish between 
real and ghost orbits.
In the following we will demonstrate that the quality of recurrence
spectra can be significantly improved by application of state of the art
methods for high resolution spectral analysis.

\subsection{Circumventing the uncertainty principle}
\label{Circumvent}
Instead of using the standard Fourier analysis, to extract the amplitudes 
and actions we propose to apply methods for high resolution spectral analysis.
The fundamental difference between the Fourier transform and high resolution
methods is the following.
Assume a complex signal $C(t)$ given on an equidistant grid of points
$t_n=t_0 + n\Delta t$
as a superposition of exponential functions, i.e.
\begin{equation}
 C(t_n) \equiv c_n = \sum_{k} d_k e^{-it_n\omega_k} \; ,
   \quad n=1,2,\dots ,N \; .
\label{C_s_n}
\end{equation}
In the Discrete Fourier Transform (DFT) $N$ real frequencies
\begin{equation}
 \omega_k = {2\pi k\over N\Delta t}  \; , \quad k=1,2,\dots ,N
\end{equation}
are fixed and evenly spaced.
The $N$ complex amplitudes $d_k$ of the Fourier transform are determined 
by solving a {\em linear} set of equations, yielding
\begin{equation}
 d_k = {1\over{N}}\sum_{n=1}^N c_n e^{it_n\omega_k} \; , \quad 
  k=1,2,\dots ,N  \; .
\label{DFT}
\end{equation}
The sums in Eq.\ \ref{DFT} can be calculated numerically very efficiently, 
e.g., by the Fast Fourier Transform algorithm (FFT).
The resolution $\Delta\omega$ of the Fourier transform is controlled by the
total length $N\Delta t$ of the signal,
\begin{equation}
 \Delta\omega = {2\pi\over N\Delta t} \; ,
\end{equation}
which is the spacing between the grid points in the frequency domain.
Only those spectral features that are separated from each other by more
than $\Delta w$ can be resolved.
This is referred to as the ``uncertainty principle'' of the Fourier transform.
By contrast, both the amplitudes $d_k$ {and} the frequencies $\omega_k$ 
in Eq.\ \ref{C_s_n} are free parameters when methods for high resolution 
spectral analysis are applied.
Because the frequencies $\omega_k$ are free adjusting parameters they are 
allowed to appear very close to each other and therefore the resolution is
practically infinite.
Using the data points of the signal $C(t)$ at $t=t_n$ the set of in general
complex parameters $\{d_k,\omega_k\}$ are given as the solution of a 
{\em nonlinear} set of equations.
Unfortunately, the nonlinear set of equations does not have a solution in
closed form similar to Eq.\ \ref{DFT} for the discrete Fourier transform.
In fact, the numerical calculation of the parameters $\{d_k,\omega_k\}$ from
the nonlinear set of equations is the central and nontrivial problem of all 
methods for high resolution spectral analysis.

The numerical harmonic inversion of a given signal like Eq.\ \ref{C_s_n} 
is a fundamental problem in physics, electrical engineering and many other 
diverse fields.
The problem has already been addressed (in a slightly different form)
in the 18th century by Baron de Prony, who converted the nonlinear set
of equations (\ref{C_s_n}) to a linear algebra problem.
There are several approaches related to the Prony method used for a high
resolution spectral analysis of short time signals, such as the modern
versions of the Prony method, MUSIC (MUltiple SIgnal Classification) and 
ESPRIT (Estimation of Signal Parameters via Rotational Invariance Technique)
\cite{Mar87,Roy91,Gra92}.
As opposed to the Fourier transform these methods are not related to a
linear (unitary) transformation of the signal and are highly nonlinear
by their nature.
However, the common feature present in most versions of these methods is
converting the nonlinear fitting problem to a linear algebraic one.
Note also that while ESPRIT uses exclusively linear algebra, the Prony
method or MUSIC require some additional search in the frequency domain
which makes them less efficient.
To our best knowledge none of these nonlinear methods is able to handle a
signal (\ref{C_s_n}) that contains ``too many'' frequencies as they lead 
to unfeasibly large and typically ill-conditioned linear algebra problems 
\cite{Man97b}.
This is especially the case for the analysis of quantum spectra because
such spectra cannot be treated as ``short signals'' and contain a high
number of frequencies given by (see below) the number of periodic orbits 
of the underlying classical system.
Decisive progress in the numerical techniques for harmonic inversion has
recently been achieved by Wall and Neuhauser \cite{Wal95}.
Their method is conceptually based on the original filter-diagonalization
method of Neuhauser \cite{Neu90} designed to obtain the eigenspectrum of
a Hamiltonian operator in any selected small energy range.
The main idea is to associate the signal $C(t)$ (Eq.\ \ref{C_s_n}) with
an autocorrelation function of a suitable dynamical system,
\begin{equation}
 C(t) = \left(\Phi_0, e^{-i\hat\Omega t}\Phi_0\right) \; ,
\label{C_s_auto}
\end{equation}
where the brackets define a complex symmetric inner product (i.e., no
complex conjugation).
Eq.\ \ref{C_s_auto} establishes an equivalence between the problem of
extracting information from the signal $C(t)$ with the one of diagonalizing
the evolution operator $\exp(-i\hat\Omega)$ of the underlying dynamical
system.
The frequencies $\omega_k$ of the signal $C(t)$ are the eigenvalues of the 
operator $\hat\Omega$, i.e.
\begin{equation}
 \hat\Omega|\Phi_k) = \omega_k|\Phi_k) \; ,
\label{omega_eq}
\end{equation}
and the amplitudes $d_k$ are formally given as
\begin{equation}
 d_k = \left(\Phi_0,\Phi_k\right)^2 \; .
\end{equation}
After introducing an appropriate basis set, the operator $\exp(-i\hat\Omega)$
can be diagonalized using the method of {\em filter-diagonalization}
\cite{Wal95,Man97a,Man97b}.
Operationally this is done by solving a small generalized eigenvalue problem
whose eigenvalues yield the frequencies in a chosen window.
Typically, the numerical handling of matrices with approximate dimension
$100 \times 100$ is sufficient even when the number of frequencies in the
signal $C(s)$ is about $10000$ or more.
The knowledge of the operator $\hat\Omega$ itself is not required as for
a properly chosen basis the matrix elements of $\hat\Omega$ can be expressed
only in terms of the signal $C(t)$.
The advantage of the filter-diagonalization technique is its numerical
stability with respect to both the length and complexity (the number
and density of the contributing frequencies) of the signal.
Details of the method of harmonic inversion by filter-diagonalization
are given in Ref.\ \cite{Man97b} and in Appendix \ref{harm_inv}.

We now want to apply harmonic inversion as a tool for the high precision
analysis of quantum spectra.
In quantum calculations the bound state spectrum is given as a sum of 
$\delta$-functions,
\begin{equation}
 \varrho^{\rm qm}(w) = \sum_k \delta(w-w_k) \; .
\end{equation}
Instead of using the Fourier transform (Eq.\ \ref{C_FT}) we want to
adjust $\varrho^{\rm qm}(w)$ to the functional form of Gutzwiller's
semiclassical trace formula (\ref{rho_sc}), which can be written as
\begin{equation}
   \varrho^{\rm sc}(w)
 = \varrho_0(w) + {1\over 2\pi} \sum_{\rm po} \sum_{r=1}^\infty
   \left\{ A_{\rm po,r} e^{irs_{\rm po} w} +
           A_{\rm po,r}^\ast e^{-irs_{\rm po}^\ast w} \right\} \; .
\label{rho_sc2}
\end{equation}
The fluctuating part of the semiclassical density of states (\ref{rho_sc2})
has exactly the functional form of the signal $C(t)$ in Eq.\ \ref{C_s_n}
with $t$ replaced by the scaling parameter $w$.
The amplitudes and frequencies $\{d_k,\omega_k\}$ in Eq.\ \ref{C_s_n} are 
the amplitudes and scaled actions $\{A_{\rm po,r}/2\pi,-rs_{\rm po}\}$
of the periodic orbit contributions and their conjugate pairs
$\{A_{\rm po,r}^\ast/2\pi,rs_{\rm po}^\ast\}$.
In order to obtain $\varrho^{\rm qm}(w)$ on an evenly spaced grid the spectrum
is regularized by convoluting it with a narrow Gaussian function having 
the width $\sigma \ll 1/s_{\rm max}$, where $s_{\rm max}$ is the 
scaled action of the longest orbit of interest.
The regularized density of states reads
\begin{equation}
   \varrho_\sigma^{\rm qm}(w)
 = \sum_k \delta_\sigma(w-w_k)
 = {1\over\sqrt{2\pi}\sigma} \sum_k e^{-{(w-w_k)^2/ 2\sigma^2}} \; ,
\label{rho_sigma}
\end{equation}
and is the starting point for the harmonic inversion procedure.
The step width for the discretization of $\varrho_\sigma^{\rm qm}(w)$
is typically chosen as $\Delta w\approx\sigma/3$.
The convolution of $\varrho^{\rm qm}(w)$ with a Gaussian function does 
not effect the frequencies, i.e., the values obtained for the scaled actions 
$s_{\rm po}$ of the periodic orbits, but just results in a small damping 
of the amplitudes
\begin{equation}
   A_{\rm po,r} \longrightarrow A_{\rm po,r}^{(\sigma)}
 = A_{\rm po,r} e^{-s_{\rm po}^2\sigma^2/2} \; .
\end{equation}
For open systems the density of states is given by
\begin{equation}
 \varrho^{\rm qm}(w) = -{1\over\pi}\, {\rm Im} \, \sum_k {1\over w-w_k} \; ,
\label{rho_qm_open}
\end{equation}
with complex resonances $w_k$.
If the minimum of the resonance widths $\Gamma_k=-2\,{\rm Im}\, w_k$ is 
larger than the step width $\Delta w$, there is no need to convolute 
$\varrho^{\rm qm}(w)$ with a Gaussian function and the density of states 
can directly be analyzed by harmonic inversion in the same way as for bound 
systems.

\subsection{Precision check of the periodic orbit theory}
\label{prec_check}
\begin{figure}[b]
\vspace{13.5cm}
\includegraphics{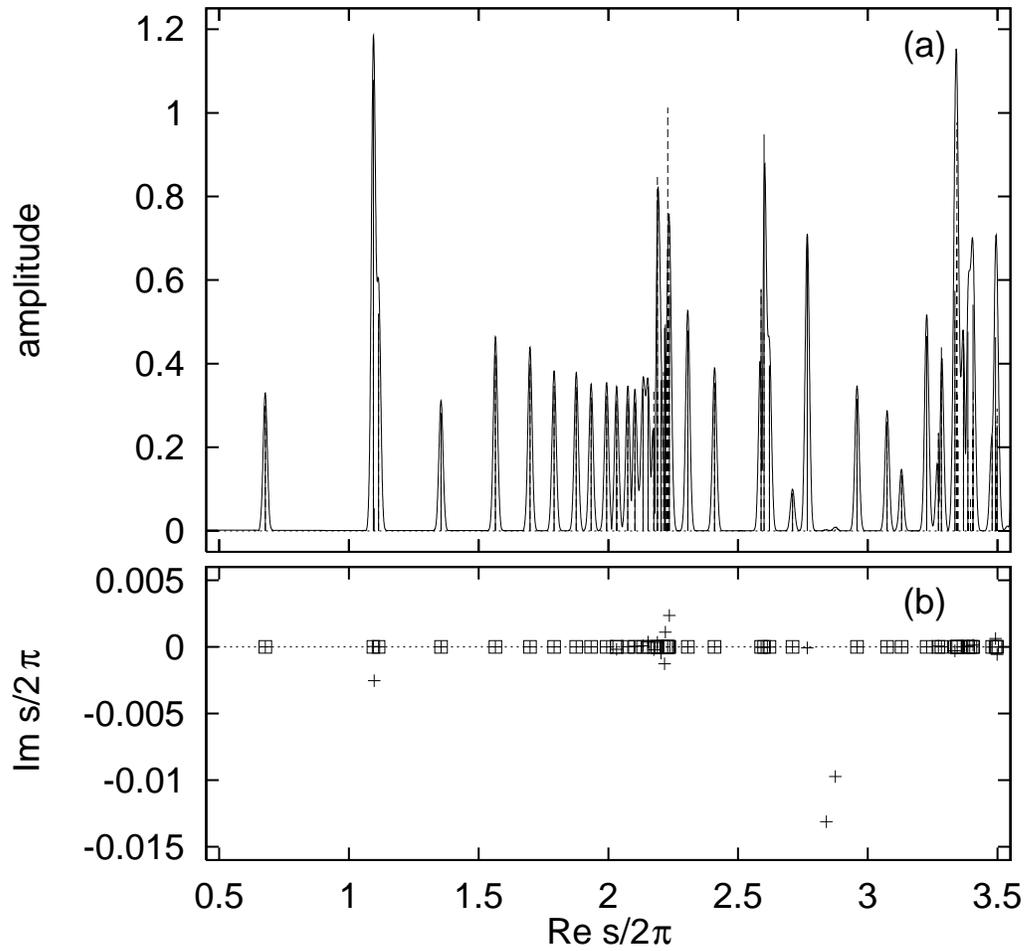}
\caption{\label{fig05} 
(a) Recurrence spectrum for the density of states of the hydrogen atom in a 
magnetic field at scaled energy $\tilde E=E\gamma^{-2/3}=-0.1$.
Smooth line: Conventional Fourier transform.
Solid stick spectrum: High resolution quantum recurrence spectrum.
Dashed sticks (hardly visible under solid sticks): Recurrence 
spectrum from semiclassical {\em periodic orbit} theory.
(b) Complex actions. Crosses and squares are the quantum and classical
results, respectively.
}
\end{figure}

\begin{figure}[b]
\vspace{9.9cm}
\includegraphics{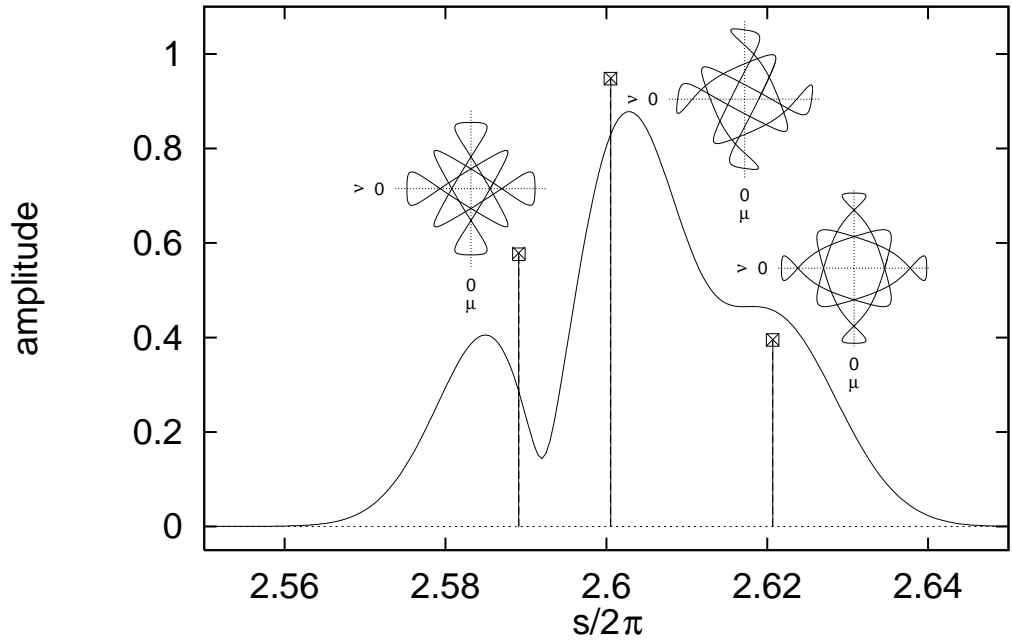}
\caption{\label{fig06} 
Recurrence spectrum for the density of states of the hydrogen atom in a 
magnetic field at scaled energy $\tilde E=E\gamma^{-2/3}=-0.1$.
Smooth line: Conventional Fourier transform.
Solid stick spectrum and crosses: High resolution quantum recurrence spectrum.
Dashed sticks (hardly visible under solid sticks) and squares: Recurrence 
spectrum from semiclassical {\em periodic orbit} theory.
The recurrence peaks are identified by periodic orbits drawn in 
semiparabolical coordinates $\mu=(r+z)^{1/2}$, $\nu=(r-z)^{1/2}$.
(From Ref.\ [79].)
}
\end{figure}

As a first application of harmonic inversion for the high resolution analysis
of quantum spectra we investigate the hydrogen atom in a magnetic field
given by the Hamiltonian (\ref{Hamfkt_scal}) and compare the results of the
harmonic inversion method to the conventional Fourier transform presented
in Fig.\ \ref{fig04}.
We analyze the density of states (\ref{rho_sigma}) with $\sigma=0.0015$
at constant scaled energy $\tilde E=-0.1$.
From the discussion of the harmonic inversion technique and especially
Eq.\ \ref{omega_eq} it follows that the frequencies in 
$\varrho_\sigma^{\rm qm}(w)$, i.e., the actions of periodic orbits, are not
obtained from a continuous frequency spectrum (or a spectrum defined on an 
equidistant grid of frequencies) but are given as {\em discrete} and in
general complex eigenvalues.
Actions with imaginary part significantly below zero indicate ``ghost''
orbit contributions to the quantum spectrum.
The actions $s_{\rm po}^{\rm qm}/2\pi$ and absolute values of the 
amplitudes $|A^{\rm qm}|$ are given in the first three columns of
Table \ref{table1}.
The last two columns present the classical actions $s_{\rm po}^{\rm cl}/2\pi$
of the periodic orbits and the absolute values of the semiclassical amplitudes
\begin{eqnarray}
   |A^{\rm cl}|
 = {s_{\rm po}^{\rm prim}/2\pi\over\sqrt{|{\rm det}(M_{\rm po}-I)|}} \; ,
\label{A_po}
\end{eqnarray}
with $s_{\rm po}^{\rm prim}$ the action of the {\em primitive} periodic
orbit, i.e., the first recurrence of the orbit.
A graphical comparison between the Fourier transform, the high resolution 
quantum recurrence spectrum, and the semiclassical recurrence spectrum is 
presented in Fig.\ \ref{fig05}.
The crosses in Fig.\ \ref{fig05}b are the (complex) actions obtained by
the harmonic inversion of the quantum mechanical density of states, and
the squares are the actions of the (real) classical periodic orbits.
The crosses and squares are in excellent agreement with a few exceptions, 
e.g., around $s/2\pi\approx 2.2$ and $s/2\pi\approx 2.8$ which will be
discussed later.
The amplitudes of the periodic orbit contributions are illustrated in
Fig.\ \ref{fig05}a.
The solid sticks are the amplitudes obtained by harmonic inversion of 
the quantum spectrum and the dashed sticks (hardly visible under solid sticks)
present the corresponding semiclassical results.
For comparison, the conventional Fourier transform recurrence spectrum is
drawn as a solid line.
To visualize more clearly the improvement of the high resolution recurrence
spectrum compared to the conventional Fourier transform a small part of
the recurrence spectrum (Fig.\ \ref{fig05}a) around $s/2\pi=2.6$ is enlarged 
in Fig.\ \ref{fig06}.
The smooth line is the absolute value of the conventional Fourier transform.
Its shape suggests the existence of at least three periodic orbits but
obviously the recurrence spectrum is not completely resolved.
The results of the high resolution spectral analysis are presented as sticks
and crosses at the positions defined by the scaled actions 
$s_{\rm po}^{\rm qm}$ with peak heights $|A^{\rm qm}|$.
Note that the positions of the peaks are considerably shifted with respect to
the maxima of the conventional Fourier transform.
To compare the quantum recurrence spectrum with Gutzwiller's periodic orbit 
theory the semiclassical results are presented as dashed sticks and squares in
Fig.\ \ref{fig06}.
For illustration the shapes of periodic orbits are also shown 
(in semiparabolic coordinates $\mu=\sqrt{r+z}$, $\nu=\sqrt{r-z}$).
For these three orbits the agreement between the semiclassical and 
the high resolution quantum recurrence spectrum is nearly perfect, 
deviations are within the stick widths.
The relative deviations between the quantum mechanical and classical actions 
and amplitudes are given in Table \ref{table2}.

The excellent agreement between the classical periodic orbit data and the
quantum mechanical results obtained by harmonic inversion of the density
of states may appear to be in contradiction to the fact that Gutzwiller's 
trace formula is an approximation, i.e., only the leading term of the 
semiclassical $\hbar$ expansion of the periodic orbit sum.
The small deviations given in Table \ref{table2} are certainly due to 
numerical limitations and do not indicate effects of higher order $\hbar$
corrections.
The reason is that the higher order $\hbar$ contributions of the periodic
orbits do not have the functional form of Eq.\ \ref{rho_sc2} as a linear
superposition of exponential functions of the scaling parameter $w$.
As the harmonic inversion procedure adjusts the quantum mechanical density
of states $\varrho^{\rm qm}(w)$ to the ansatz (\ref{rho_sc2}) with free 
parameters $s_{\rm po}$ and $A_{\rm po}$, the {\em exact} parameters 
$s_{\rm po}$ and $A_{\rm po}$ of the classical periodic orbits should
provide the optimal adjustment to the quantum spectrum within the lowest
order $\hbar$ approximation.
However, the harmonic inversion technique can also be used to reveal
the higher order $\hbar$ contributions of the periodic orbits in the 
quantum spectra, as will be demonstrated in Section \ref{hbar_sec}.

The quantum mechanical high resolution recurrence spectrum is not in good
agreement with the classical calculations around $s/2\pi\approx 2.2$
(see crosses and squares in Fig.\ \ref{fig05}b).
This period is close to the scaled action 
$s_{\rm po}/2\pi =(-2\tilde E)^{-1/2}=2.2361$ of the parallel orbit, which 
undergoes an infinite series of bifurcations with increasing energy 
\cite{Win87c}.
If orbits are too close to bifurcations they can no longer be treated
as {\em isolated} periodic orbits, which in turn is the basic assumption
for Gutzwiller's semiclassical trace formula (\ref{rho_sc}).
The disagreement between the crosses and squares in Fig.\ \ref{fig05} 
around $s/2\pi\approx 2.2$ thus indicates the breakdown of the semiclassical 
ansatz (\ref{rho_sc}) for non-isolated orbits.
The structures around $s/2\pi\approx 2.8$ can also not be explained by
real classical periodic orbits.
They are related to uniform semiclassical approximations and complex 
ghost orbits as will be discussed in the next section.

\subsection{Ghost orbits and uniform semiclassical approximations}
\label{ghost_sec}
Gutzwiller's periodic orbit theory (\ref{rho_sc}) is valid for isolated
periodic orbits where the determinant $\det(M_{\rm po}^r-I)$, i.e.,
the denominator in Eq.\ \ref{A_po_r} is nonzero.
However, the periodic orbit amplitudes $A_{{\rm po},r}$ diverge close to
the bifurcation points of orbits where $M_{\rm po}^r-I$ has a vanishing
determinant.
To remove the unphysical singularity from the semiclassical expressions
all periodic orbits which participate at the bifurcation must be
considered simultaneously in a {\em uniform} semiclassical approximation
\cite{Alm87,Alm88}.
The uniform solutions can be constructed with the help of 
{\em catastrophe theory} \cite{Pos78,Ber80}, and have been studied, e.g.,
for the kicked top \cite{Kus93,Scho97a,Scho97b,Scho97c} and the hydrogen
atom in a magnetic field \cite{Mai97a,Mai98a}.
In the vicinity of bifurcations ``ghost'' orbits, i.e., periodic orbits in
the complex continuation of phase space can be very important.
In general the ghost orbits have real or complex actions, $s_{\rm po}$.
As can be shown from the asymptotic expansions of the uniform semiclassical
approximations \cite{Mai97a,Mai98a} those ghosts with positive imaginary
part of the action, ${\rm Im}\, s_{\rm po}>0$, are of physical relevance.
They contribute as
\[
 A_{\rm po}\, e^{i\, {\rm Re}\, s_{\rm po}w}
           \, e^{-\, {\rm Im}\, s_{\rm po}w}
\]
to Gutzwiller's periodic orbit sum (\ref{rho_sc}), i.e., the modulations
of the ghost orbits are exponentially damped with increasing scaling 
parameter $w$.

Here we will only present a brief derivation of uniform semiclassical 
approximations.
For details we refer the reader to the literature
\cite{Kus93,Mai97a,Scho97a,Scho97b,Mai98a,Scho97c}.
Our main interest is to demonstrate how bifurcations and ghost orbits can 
directly be uncovered in quantum spectra with the help of the harmonic 
inversion technique and the high resolution recurrence spectra.
Note that a detailed investigation of these phenomena is in general 
impossible with the Fourier transform because the orbits participating at 
the bifurcation have nearly the same period and thus cannot be resolved in 
the conventional recurrence spectra.
We will discuss two different types of bifurcations by way of example of
the hydrogen atom in a magnetic field:
The hyperbolic umbilic and the butterfly catastrophe.

\begin{figure}[b]
\vspace{13.5cm}
\includegraphics{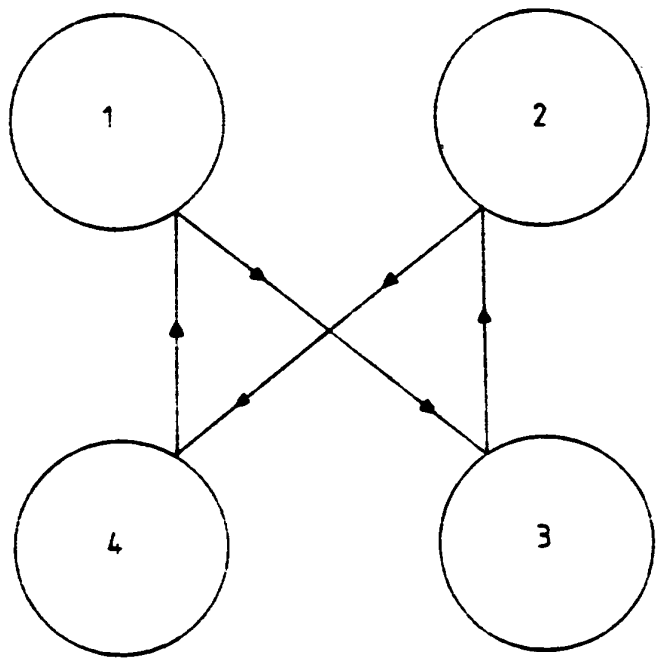}
\caption{\label{fig07} 
Schematic view of the four-disk scattering problem.
(From Ref.\ [102].)
}
\end{figure}

In the following section we will adopt the symbolic code of 
Ref.\ \cite{Eck90} for the nomenclature of periodic orbits.
Introducing scaled semiparabolic coordinates 
$\mu=\gamma^{1/3}\sqrt{r+z}$ and $\nu=\gamma^{1/3}\sqrt{r-z}$ the
scaled Hamiltonian of the hydrogen atom in a magnetic field reads
\begin{equation}
 h = {1\over 2}(p_\mu^2+p_\nu^2) - \tilde E(\mu^2+\nu^2)
   + {1\over 8}(\mu^4\nu^2+\mu^2\nu^4) = 2 \; .
\end{equation}
As pointed out in \cite{Eck90} the effective potential
$V(\mu,\nu)=-\tilde E(\mu^2+\nu^2)+(\mu^4\nu^2+\mu^2\nu^4)/8$
is bounded for $\tilde E>0$ by four hyperbolas.
When the smooth potential is replaced with hard walls periodic orbits can
be assigned with the same ternary symbolic code as orbits of the four-disk
scattering problem (see Fig.\ \ref{fig07}).
In a first step a ternary alphabet \{0AC\} is introduced.
The symbol 0 labels scattering to the opposite disk, and the symbols
C and A label scattering to the neighboring disk in clockwise or
anticlockwise direction, respectively.
The orbit in Fig.\ \ref{fig07} is coded 0C0A.
The ternary code can be made more efficient with respect to the exchange
symmetry of the C and A symbol if it is redefined in the following way.
The Cs and As are replaced with the sympol {\tt +}, if consecutive letters
ignoring the 0s are equal and they are replaced with the symbol {\tt -},
if consecutive letters differ.
With this redefinition the orbit shown in Fig.\ \ref{fig07} is coded
{\tt 0-0-}, or because of its periodicity even simpler {\tt 0-}.
For the hydrogen atom in a magnetic field there is a one to one 
correspondence between the periodic orbits and the ternary symbolic code
at energies $\tilde E>\tilde E_c=+0.3287$ \cite{Han95,Schn97}.
Below the critical energy orbits undergo bifurcations and the code is
not unique.
However, nearly all unstable periodic orbits can be uniquely assigned
with this code even at negative energies.

\subsubsection{The hyperbolic umbilic catastrophe}
\label{umbilic_sec}
\begin{figure}[b]
\vspace{13.5cm}
\includegraphics{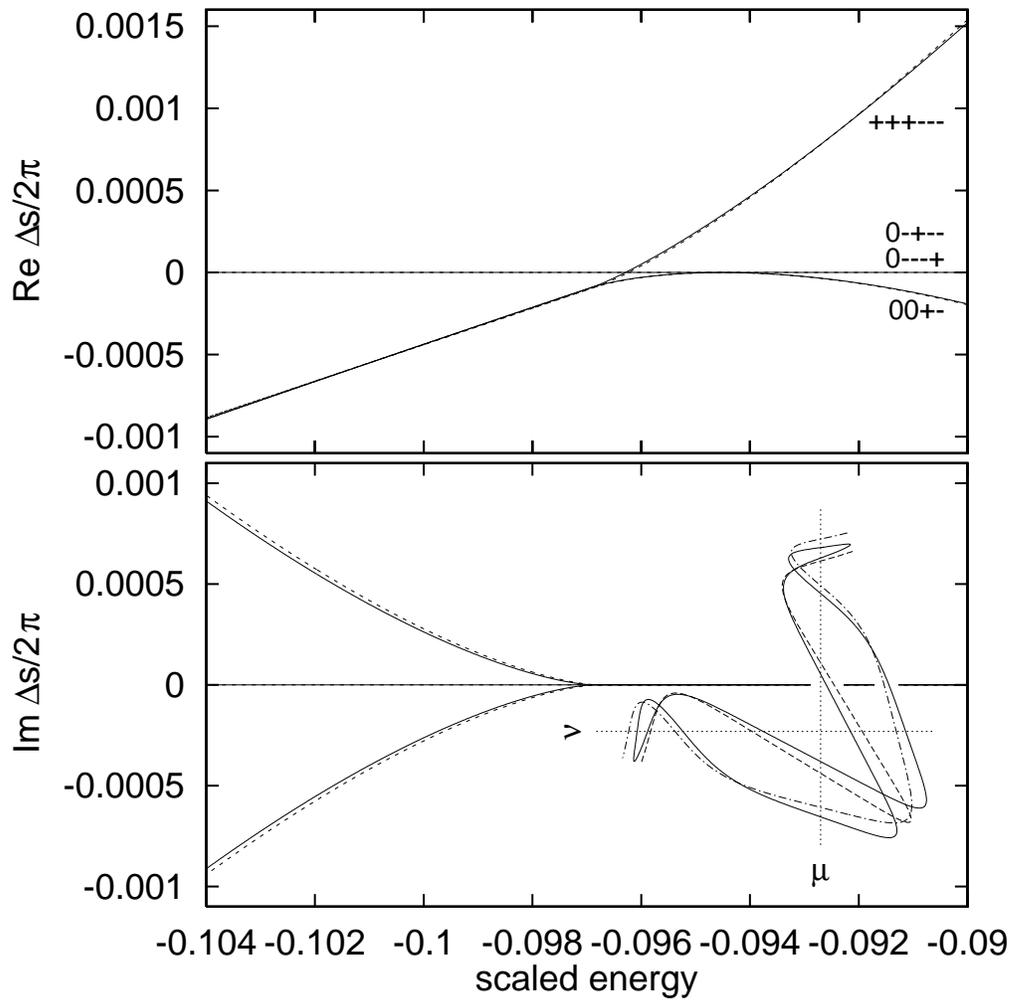}
\caption{\label{fig08} 
Difference $\Delta s$ between the classical action of the 
four periodic orbits involved in the bifurcations.
Dashed lines: Analytical fits related to the hyperbolic umbilic catastrophe.
Inset: Graphs of periodic orbits {\tt 0-+--} and its time reversal {\tt 0---+}
(solid line), {\tt 00+-} (dashed line), and {\tt +++---} (dash-dotted line)
drawn in semiparabolical coordinates $\mu=(r+z)^{1/2}$, $\nu=(r-z)^{1/2}$.
(From Ref.\ [44].)
}
\end{figure}

\begin{figure}[b]
\vspace{13.5cm}
\includegraphics{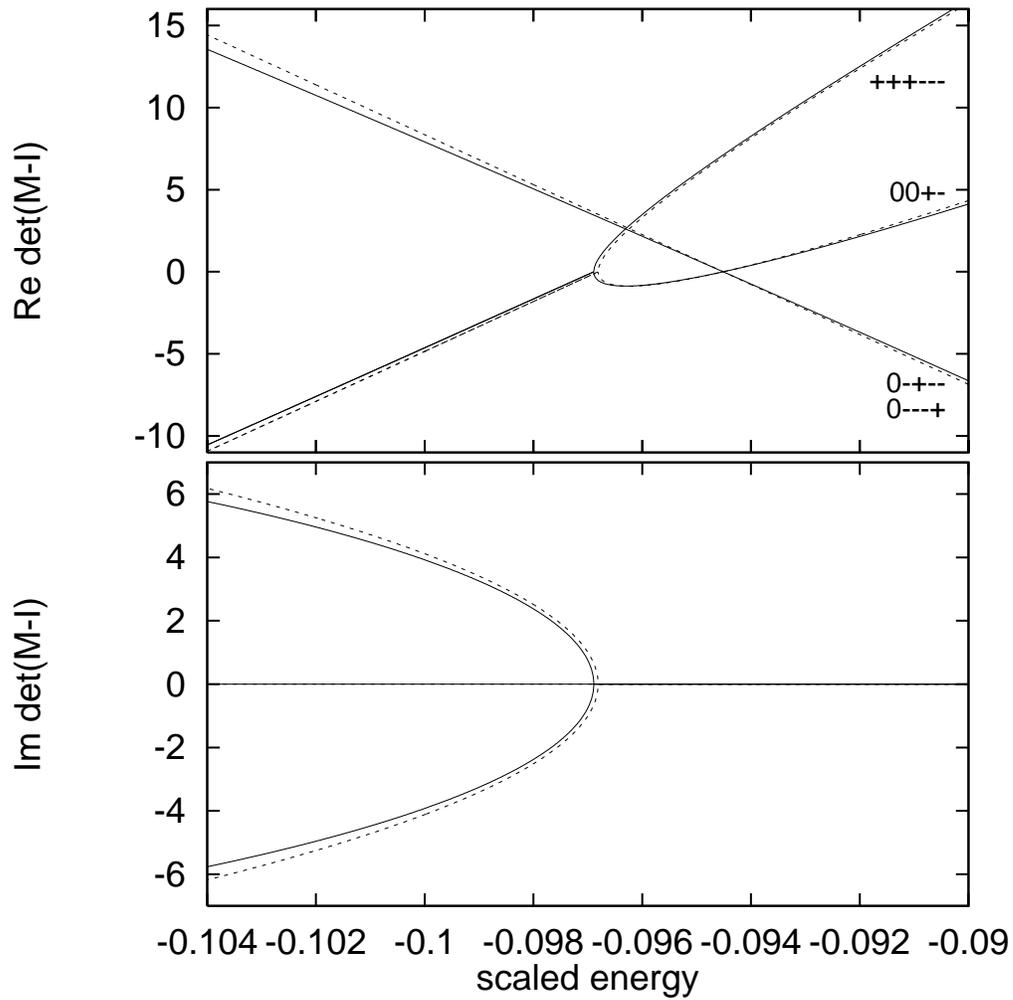}
\caption{\label{fig09} 
Same as Fig.\ \ref{fig08} but for the determinant $\det(M-I)$ of the 
periodic orbits.
(From Ref.\ [44].)
}
\end{figure}

As a first example we investigate the structure at $s_{\rm po}/2\pi=2.767$ 
in the recurrence spectrum of the hydrogen atom in a magnetic field at
constant scaled energy $\tilde E=E\gamma^{-2/3}=-0.1$
(see Fig.\ \ref{fig05} and Table \ref{table1}).
At this energy no periodic orbit with an action close to 
$s_{\rm po}/2\pi=2.767$ does exist.
However, the strong recurrence peak in Fig.\ \ref{fig05}a indicates a
near bifurcation.
This bifurcation and the corresponding semiclassical approximation have
been studied in detail in Refs.\ \cite{Mai98a,Wil97}.
Four periodic orbits are created through two nearby bifurcations around the 
scaled energy $\tilde E \approx -0.096$ where we search for both real and 
complex ``ghost'' orbits.
For the nomenclature of the real orbits we adopt the symbolic code of Ref.\
\cite{Eck90} as explained above.
At scaled energy $\tilde E_b^{(1)}=-0.09689$, the two orbits {\tt 00+-} and 
{\tt +++---} are born in a tangent bifurcation.
At energies $\tilde E<\tilde E_b^{(1)}$, a prebifurcation ghost orbit and
its complex conjugate exist in the complex continuation of the phase space.
Orbit {\tt 00+-} is born unstable, and turns stable at the slightly higher
energy $\tilde E_b^{(2)} = -0.09451$.
This is the bifurcation point of two additional orbits, {\tt 0-+--} and its
time reversal {\tt 0---+}, which also have ghost orbits as predecessors.
The graphs of the real orbits at energy $E=0$ are shown as insets in Fig.\ 
\ref{fig08},
and the classical periodic orbit parameters are presented as solid lines in 
Figs.\ \ref{fig08} and \ref{fig09}.
Fig.\ \ref{fig08} shows the difference in scaled action between the orbits.
The action of orbit {\tt 0-+--} (or its time reversal {\tt 0---+}), which is 
real also for its prebifurcation ghost orbits, has been taken as the 
reference action.
The uniform semiclassical approximation for the four orbits involved in the
bifurcation can be expressed in terms of the diffraction integral of a
hyperbolic umbilic catastrophe
\begin{equation}
 \Psi(x,y) = \int_{-\infty}^{+\infty}dp \int_{-\infty}^{+\infty}dq
   e^{i\Phi(p,q;x,y)}
\label{Psi_def}
\end{equation}
with
\begin{equation}
 \Phi(p,q;x,y) = p^3 + q^3 + y(p+q)^2 + x(p+q) \; .
\end{equation}
For our convenience the function $\Phi(p,q;x,y)$ slightly differs from
the standard polynomial of the hyperbolic umbilic catastrophe given in 
Ref.\ \cite{Ber80} but the diffraction integral (\ref{Psi_def}) can be 
easily transformed to the standard representation.
The four stationary points of the integral (\ref{Psi_def}) are readily 
obtained from the condition $\nabla\Phi=0$ as
\begin{equation}
 p_0 = -q_0 = \pm\sqrt{-x/3}
 \Rightarrow \Phi(p_0,q_0;x,y) = 0
\label{stat_points_a}
\end{equation}
and
\begin{eqnarray}
 p_0 = q_0 &=& -{2\over 3}y \pm\sqrt{{4\over 9}y^2-{x\over 3}} \Rightarrow
   \nonumber \\
 \Phi(p_0,q_0;x,y) &=& {4\over 3}y\left({8\over 9}y^2-x\right)
   \mp 4\left({4\over 9}y^2-{x\over 3}\right)^{3/2} \; .
\label{stat_points_b}
\end{eqnarray}
The function $\Phi(p_0,q_0;x,y)$ must now be adapted to the classical action 
of the four periodic orbits, i.e., 
$\Delta S = w\Delta s \approx \Phi(p_0,q_0;x,y)$,
which is well fulfilled for
\begin{equation}
 x = a w^{2/3} \left(\tilde E - \tilde E_b^{(2)}\right) \; ; \quad
 y = b w^{1/3} \; ,
\label{xy_def}
\end{equation}
and constants $a=-5.415$, $b=0.09665$, as can be seen from the dashed lines 
in Fig.\ \ref{fig08}.
Note that the agreement holds for both the real and complex ghost orbits.

The next step to obtain the uniform solution is to calculate the diffraction 
integral (\ref{Psi_def}) within the stationary phase approximation.
For $\tilde E > \tilde E_b^{(2)}$ there are four real stationary points
$(p_0,q_0)$ (see Eqs.\ \ref{stat_points_a} and \ref{stat_points_b}), and 
after expanding $\Phi(p,q;x,y)$ around the stationary points up to second 
order in $p$ and $q$, the diffraction integral becomes the sum of
Fresnel integrals, viz.\
\begin{equation}
 \Psi(x,y) \stackrel{x\ll 0}{\sim} {2\pi\over\sqrt{-3x}} + \sum_{+,-}
 {\pi e^{i\left[{4\over 3}y\left({8\over 9}y^2-x\right)\mp
 4\left({4\over 9}y^2-{x\over 3}\right)^{3/2}\pm{\pi\over 2}\right]} \over
 \sqrt{(4y^2-3x)\mp 2y\sqrt{4y^2-3x}}} \, .
\label{Psi_asym}
\end{equation}
The terms of Eq.\ \ref{Psi_asym} can now be compared to the standard
periodic orbit contributions (\ref{A_po}) of Gutzwiller's trace formula.
In our example the first term is related to the orbit {\tt 0-+--} (with a
multiplicity factor of 2 for its time reversal {\tt 0---+}), and the other 
two terms are related to the orbits {\tt 00+-} and {\tt +++---} for the upper 
and lower sign, respectively.
The phase shift in the numerators describe the differences of the action 
$\Delta S$ and of the Maslov index $\Delta\mu=\mp 1$ relative to the 
reference orbit {\tt 0-+--}.
The denominators are, up to a factor $cw^{1/3}$, with $c=0.1034$, the 
square root of $|\det(M-I)|$, with $M$ the stability matrix.
Fig.\ \ref{fig09} presents the comparison for the determinants obtained from 
classical periodic orbit calculations (solid lines) and from Eqs.\
\ref{xy_def} and \ref{Psi_asym} (dashed lines).
The agreement is very good for both the real and complex ghost orbits,
similar to the agreement found for $\Delta s$ in Fig.\ \ref{fig08}.
The constant $c$ introduced above determines the normalization of the
uniform semiclassical approximation for the hyperbolic umbilic bifurcation
which is finally obtained as
\cite{Mai98a,Wil97}
\begin{eqnarray}
   {\cal A}_{\rm uniform}(\tilde E,w)
 = (c/\pi) s_0 w^{1/3} \, \Psi\left(aw^{2/3} (\tilde E - \tilde E_b^{(2)}),
   bw^{1/3}\right) e^{i[s_0 w-{\pi\over 2}\mu_0]} \; , \nonumber \\
\label{A_uniform}
\end{eqnarray}
with $s_0$ and $\mu_0$ denoting the orbital action and Maslov index of the 
reference orbit {\tt 0-+--}, and the constants $a$, $b$, and $c$ as given 
above.

\begin{figure}[b]
\vspace{16.0cm}
\includegraphics{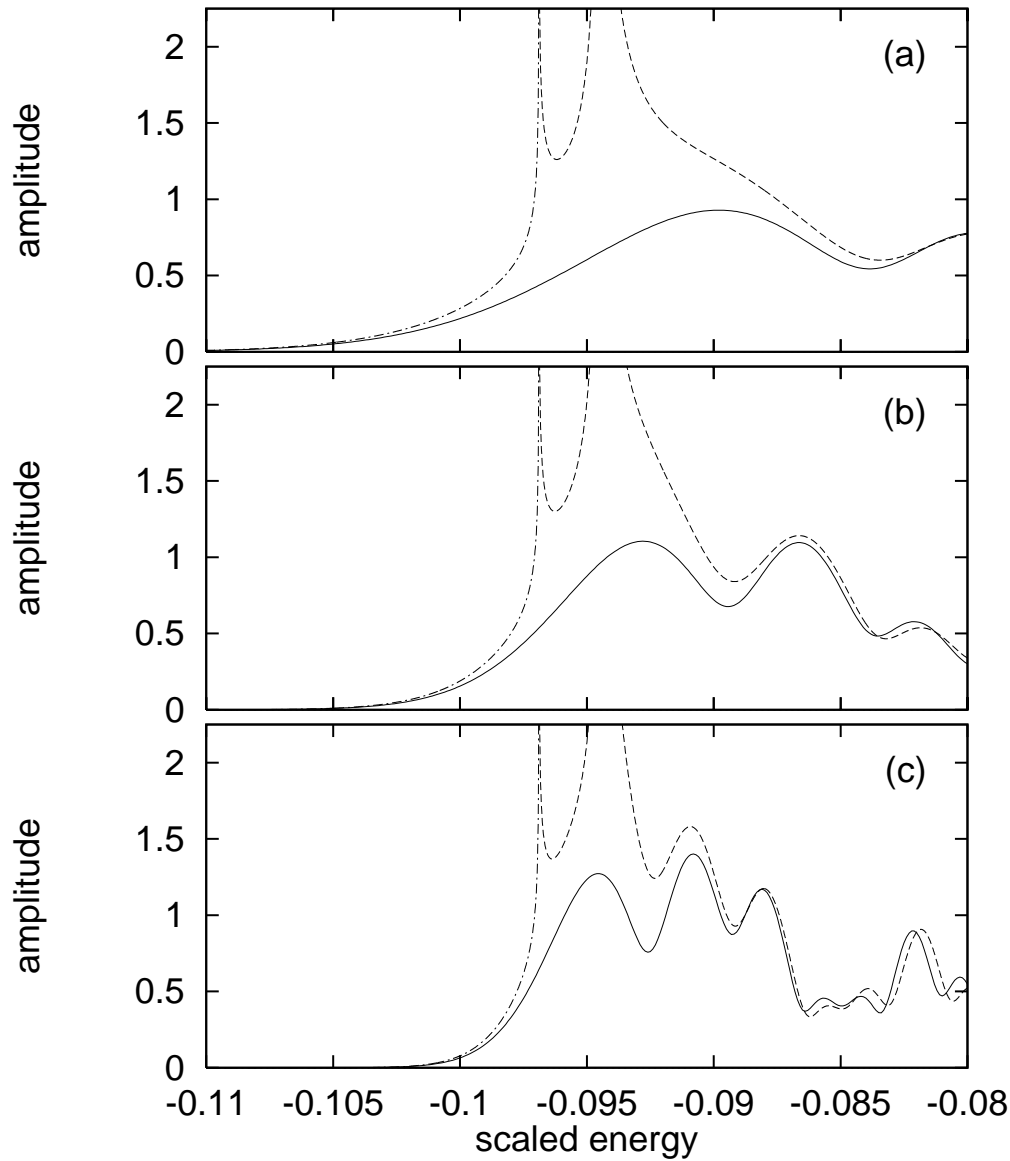}
\caption{\label{fig10} 
Semiclassical amplitudes (absolute values) for magnetic field strength 
(a) $\gamma=10^{-7}$, (b) $\gamma=10^{-8}$, and (c) $\gamma=10^{-9}$
in units of the time period $T_0$.
Dashed line: Amplitudes of the standard semiclassical trace formula.
Dash-dotted line: Ghost orbit contribution.
Solid line: Uniform approximation of the hyperbolic umbilic catastrophe.
(From Ref.\ [44].)
}
\end{figure}

The comparison between the amplitudes (\ref{A_po}) of the conventional 
semiclassical trace formula for isolated periodic orbits and the uniform 
approximation (\ref{A_uniform}) for the hyperbolic umbilic catastrophe is 
presented in Fig.\ \ref{fig10} at the magnetic field strengths 
$\gamma=10^{-7}$, $\gamma=10^{-8}$, and $\gamma=10^{-9}$.
For graphical purposes we suppress the highly oscillatory part resulting 
from the function $\exp[i(S_0/\hbar-{\pi\over2}\mu_0)]$ by plotting the 
absolute value of ${\cal A}(\tilde E,w)$ instead of the real part.
The dashed line in Fig.\ \ref{fig10} is the superposition of the isolated 
periodic orbit contributions from the four orbits involved in the bifurcations.
The modulations of the amplitude are caused by the constructive and 
destructive interference of the real orbits at energies 
$\tilde E > \tilde E_b^{(2)}$ and are most pronounced at low magnetic
field strength (see Fig.\ \ref{fig10}c).
The amplitude diverges at the two bifurcation points.
For the calculation of the uniform approximation (\ref{A_uniform}) we
numerically evaluated the catastrophe diffraction integral (\ref{Psi_def})
using a more simple and direct technique as described in \cite{Uze83}.
Details of our method which is based on Taylor series expansions 
are given in Appendix \ref{hyp_umb_app}.
The solid line in Fig.\ \ref{fig10} is the uniform approximation 
(\ref{A_uniform}).
It does not diverge at the bifurcation points but decreases exponentially 
at energies $\tilde E < \tilde E_b^{(1)}$.
At these energies no real orbits exist and the amplitude in the standard 
formulation would be zero when only real orbits are considered.
However, the exponential tail of the uniform approximation (\ref{A_uniform})
is well reproduced by a ghost orbit \cite{Mai97a,Kus93} with positive 
imaginary part of the complex action.
As can be shown, the asymptotic expansion of the diffraction integral
(\ref{Psi_def}) has, for $x\gg 0$, exactly the form of Eq.\ \ref{A_po} but 
with complex action $s$ and determinant $\det(M-I)$.
The ghost orbit contribution is shown as dash-dotted line in Fig.\ 
\ref{fig10}.

\begin{figure}[b]
\vspace{9.0cm}
\includegraphics{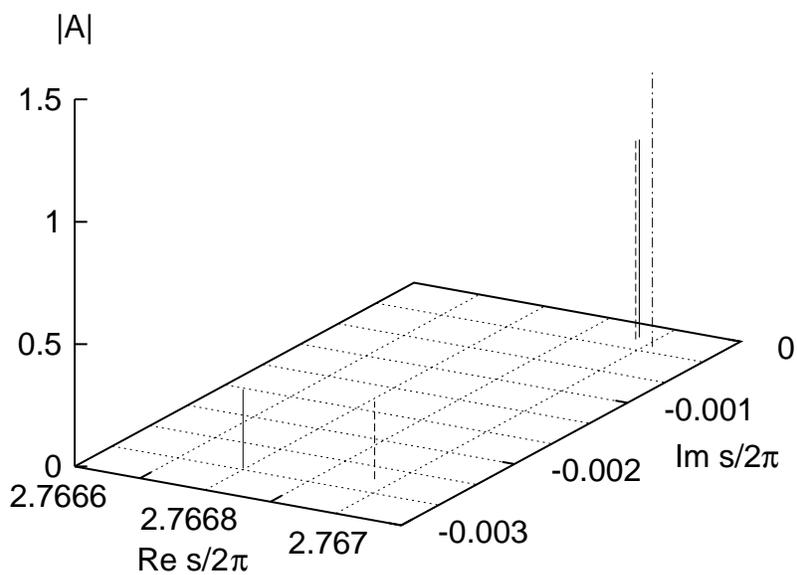}
\caption{\label{fig11} 
High resolution recurrence spectra at scaled energy $\tilde E=-0.1$.
Solid peaks: Part of the quantum recurrence spectra.
Dash-dotted peak: Classical ghost orbit contribution.
Dashed peaks: Uniform approximation of the hyperbolic umbilic catastrophe.
(From Ref.\ [44].)
}
\end{figure}

To verify the hyperbolic umbilic catastrophe in the quantum spectrum we
analyze all three, the exact quantum spectrum, the uniform semiclassical
approximation (\ref{A_uniform}), and Gutzwiller's periodic orbit formula
for isolated periodic orbits by means of the harmonic inversion technique
at scaled energy $\tilde E=-0.1$, which is slightly below the two bifurcation
points.
The part of the complex action plane which is of interest for the hyperbolic
umbilic catastrophe is presented in Fig.\ \ref{fig11}.
The two solid peaks mark the positions $s_k/2\pi$ and the absolute values 
of amplitudes $|A_k|$ obtained from the quantum spectrum.
As mentioned above, at this energy only one classical ghost orbit is of 
physical relevance and marked as dash-dotted peak in Fig.\ \ref{fig11}.
The position of that peak is in good agreement with the quantum result
but the amplitude is enhanced.
This enhancement is expected for isolated periodic orbit contributions near 
bifurcations which become singular exactly at the bifurcation points.
The harmonic inversion analysis of the uniform approximation (\ref{A_uniform}) 
at constant scaled energy $\tilde E=-0.1$ in the same range $0<w<140$ 
is presented as dashed peaks in Fig.\ \ref{fig11}.
The two peaks agree well with the quantum results for both the complex
actions and amplitudes.
The enhancement of the ghost orbit peak and the additional non-classical 
peak observed in the quantum spectrum are therefore clearly identified as 
artifacts of the bifurcation, i.e., the hyperbolic umbilic catastrophe.

\subsubsection{The butterfly catastrophe: Uncovering the ``hidden'' ghost orbits}
\label{bfly_sec}
We now investigate the butterfly catastrophe, which is of importance, e.g.,
in photoabsorption spectra of the hydrogen atom in a magnetic field.
In contrast to the density of states the photoabsorption spectra for
dipole transitions from a low-lying initial state to highly excited final
states can be measured experimentally \cite{Hol88,Mai91,Mai94a}.
The semiclassical photoabsorption cross-section is obtained by
{\em closed orbit} theory \cite{Du87,Du88,Bog88a,Bog89} as the superposition
of a smooth background and sinusoidal modulations induced by closed orbits
starting at and returning back to the nucleus.
Although the derivation of the semiclassical oscillator strength for
dipole transitions ({\em closed orbit} theory) differs from the derivation
of the semiclassical density of states ({\em periodic orbit} theory)
the final results have a very similar structure and therefore spectra can 
be analyzed in the same way by conventional Fourier transform or the high 
resolution harmonic inversion technique.
In closed orbit theory the semiclassical oscillator strength is given by
\begin{equation}
 f = f^0 + f^{\rm osc}
\end{equation}
with $f^0$ the oscillator strength of the field free hydrogen atom 
at energy $E=0$ and
\begin{eqnarray}
 f^{\rm osc} &=& 2(E_f-E_i) w^{-1/2}\, {\rm Im}\, \sum_{{\rm cl.o.}~k}
  {2(2\pi)^{3/2}\over \sqrt{|\tilde m_{12}^{(k)}|}}
  \sqrt{\sin\vartheta_{i,k}\vartheta_{f,k}}
 \nonumber \\  &\times&
  {\cal Y}_m(\vartheta_{i,k}){\cal Y}_m(\vartheta_{f,k})
  \exp\left\{i\left(s_k w + {1\over 2}m\tilde T_k - {\pi\over 2}\mu_k^0
     + {\pi\over 4}\right)\right\}
\label{f_osc}
\end{eqnarray}
the fluctuating part of the oscillator strength.
In (\ref{f_osc}) $E_f$ and $E_i$ are the energies of the final and initial
state, $m$ is the magnetic quantum number, $s_k$, $\tilde T_k=\gamma T_k$, 
and $\mu_k^0$ are the scaled action, scaled time, and the Maslov index of 
closed orbit $k$, $\vartheta_i$ and $\vartheta_f$ are the starting and 
returning angle of the orbit with respect to the magnetic field axis, and 
$\tilde m_{12}$ is an element of the scaled monodromy matrix.
The angular functions ${\cal Y}_m(\vartheta)$ depend on the initial state
and polarization of light (see Appendix \ref{Ym_app}).
For more details see Refs.\ \cite{Mai94a,Du88,Bog89}.
The fluctuating terms (\ref{f_osc}) of the photoabsorption cross section are
sinusoidal functions of the scaling parameter $w=\gamma^{-1/3}$ despite a
factor of $w^{-1/2}$.
To obtain the same functional form as Eq.\ \ref{rho_sc2} for the density
of states, which is required for harmonic inversion, we multiply the 
oscillator strength $f$ by $w^{1/2}$ for both the semiclassical and quantum 
mechanical photoabsorption spectra.

\begin{figure}[b]
\vspace{11.5cm}
\includegraphics{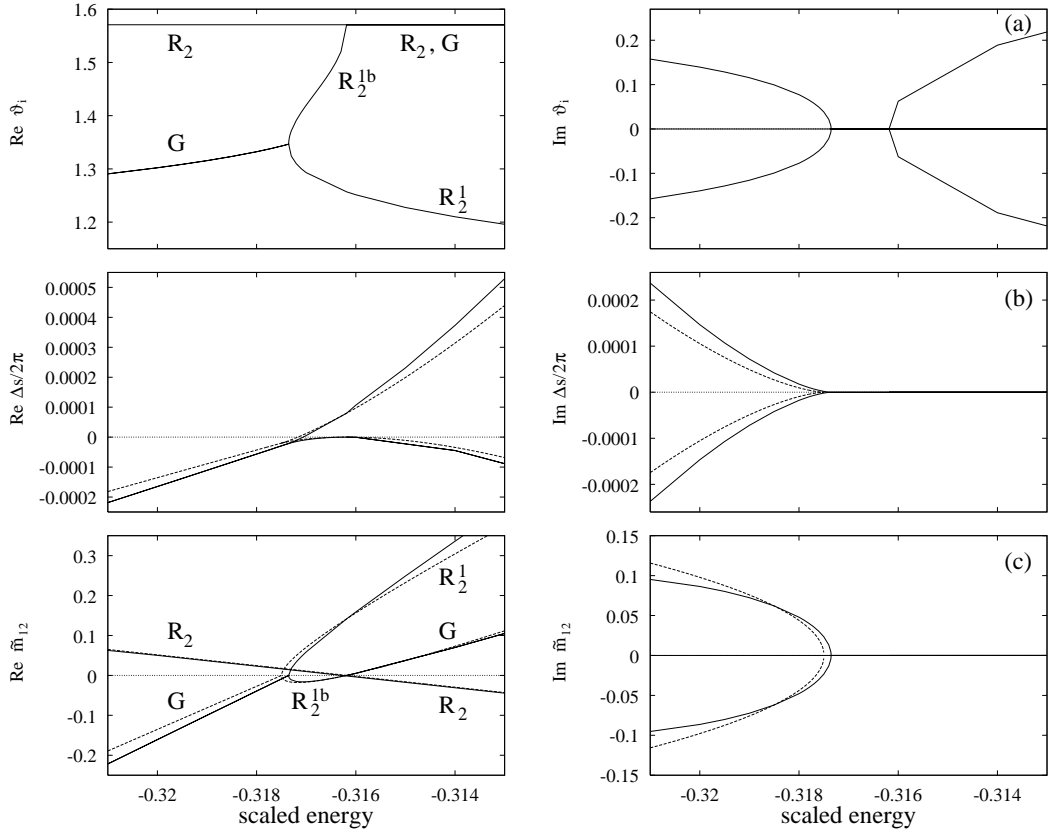}
\caption{\label{fig12} 
$(a)$ Real and imaginary part of starting angle $\vartheta_i$ for closed orbits
related to the bifurcating scenario of the (period doubled) perpendicular
orbit $R_2$.
$(b)$ Difference $\Delta s/2\pi$ between the classical action of the 
(period doubled) perpendicular orbit $R_2$ and real and ghost orbits 
bifurcating from it.
$(c)$ Monodromy matrix element $\tilde m_{12}$ of the perpendicular orbit 
$R_2$ and orbits bifurcating from it.
Dashed lines: Analytical fits (see text).
(From Ref.\ [37].)
}
\end{figure}

In analogy to Gutzwiller's trace formula Eq.\ \ref{f_osc} for photoabsorption 
spectra is valid only for isolated orbits and diverges at bifurcations of 
orbits, where $\tilde m_{12}$ is zero.
Near bifurcations the closed orbit contributions in (\ref{f_osc}) must be
replaced with uniform semiclassical approximations, which have been studied
in detail for the fold, cusp, and butterfly catastrophe in Refs.\ 
\cite{Mai97a,Mai97b}.
Here we restrict the discussion to the butterfly catastrophe, which is
especially interesting because of the existence of a ``hidden'' ghost orbit,
which can be uncovered in the photoabsorption spectrum of the hydrogen atom
in a magnetic field with the help of the harmonic inversion technique.
As an example we investigate real and ghost orbits related to the period
doubling of the perpendicular orbit $R_1$.
(For the closed orbits we adopt the nomenclature of Refs.\ \cite{Hol88,Mai91},
see also Fig.\ \ref{fig02}.)
This closed orbit bifurcation is more complicated because
various orbits with similar periods undergo two different elementary types 
of bifurcations at nearly the same energy.
The structure of bifurcations and the appearance of ghost orbits can clearly 
be seen in the energy dependence of the starting angles $\vartheta_i$ in
Fig.\ \ref{fig12}a.
Two orbits $R_2^1$ and $R_2^{1b}$ are born in a saddle node bifurcation at
$\tilde E_b^{(1)}=-0.31735345$, $\vartheta_i=1.3465$.
Below the bifurcation energy we find 
an associated ghost orbit and its complex conjugate.
Orbit $R_2^{1b}$ is real only in a very short energy interval 
($\Delta\tilde E \approx 0.001$), and is then involved in the next 
bifurcation at $\tilde E_b^{(2)}=-0.31618537$, $\vartheta_i=\pi/2$.
This is the period doubling bifurcation of the perpendicular orbit $R_1$, 
which exists at all energies ($\vartheta_i=\pi/2$ in Fig.\ \ref{fig12}a).
The real orbit $R_2^{1b}$ separates from $R_1$ at energies {\em below} 
the bifurcation point, i.e.\ a real orbit vanishes with increasing energy.
Consequently, associated ghost orbits are expected at energies {\em above} 
the bifurcation, i.e.\ $\tilde E > \tilde E_b^{(2)}$, and indeed such 
``postbifurcation'' ghosts have been found.
Its complex starting angles are also shown in Fig.\ \ref{fig12}a.
The energy dependence of scaled actions, or, more precisely, the difference
$\Delta s$ with respect to the action of the period doubled 
perpendicular orbit $R_2$,
is presented in Fig.\ \ref{fig12}b (solid lines), and the graph for the
monodromy matrix element $\tilde m_{12}$ is given in Fig.\ \ref{fig12}c.
It can be seen that the actions and the monodromy matrix elements of the ghost 
orbits related to the saddle node bifurcation of $R_2^1$ become complex at 
$\tilde E < \tilde E_b^{(1)}$, while these parameters remain real 
for the postbifurcation ghosts at $\tilde E > \tilde E_b^{(2)}$.
The two bifurcations are so closely adjacent that formulae for the saddle 
node bifurcation and the period doubling are no reasonable approximation to 
$\Delta s(\tilde E)$ and $\tilde m_{12}(\tilde E)$ in the neighborhood
of the bifurcations.
However, both functions can be fitted well by the more complicated formulae
\cite{Mai97a}
\begin{equation}
 \Delta s = 
 k \left(\tilde\sigma\left({\tilde E}-{\tilde E}_b^{(2)}\right)
  + {2\over3}\left\{1\pm\left[\tilde\sigma\left({\tilde E}-{\tilde E}_b^{(2)}
   \right)+1\right]^{3/2}\right\}\right)
\label{R2_DS}
\end{equation}
and
\begin{eqnarray}
\label{R2_m12}
\tilde m_{12} &=& -\tilde M \left({\tilde E}-{\tilde E}_b^{(2)}\right) ~~~~
    ({\rm orbit}~R_2) \nonumber \\
\tilde m_{12} &=& 4\tilde M \left({\tilde E}-{\tilde E}_b^{(2)}\right)
  + {4\tilde M\over \tilde\sigma}\left[1\pm\sqrt{\tilde\sigma
      \left({\tilde E}-{\tilde E}_b^{(2)}\right)+1}\right] \\
  &&    (R_2^1, R_2^{1b},~ {\rm and~ ghosts}) \nonumber
\end{eqnarray}
with $k=3.768\cdot10^{-4}$, $\tilde\sigma=763.6$, and $\tilde M=13.52$
(see dashed lines in Figs.\ \ref{fig12}b and \ref{fig12}c).
Note that Eqs.\ (\ref{R2_DS}) and (\ref{R2_m12}) describe the complete 
scenario for the real and the ghost orbits including both the saddle node 
and period doubling bifurcations.
We also mention that orbits with angles $\vartheta_i\ne\vartheta_f$ have
to be counted twice because they correspond to different orbits when 
traversed in either direction, and therefore  a total number of {\em five} 
closed orbits, including ghosts, is considered here in the bifurcation
scenario around the period doubling of the perpendicular orbit.

\begin{figure}[b]
\vspace{16.0cm}
\includegraphics{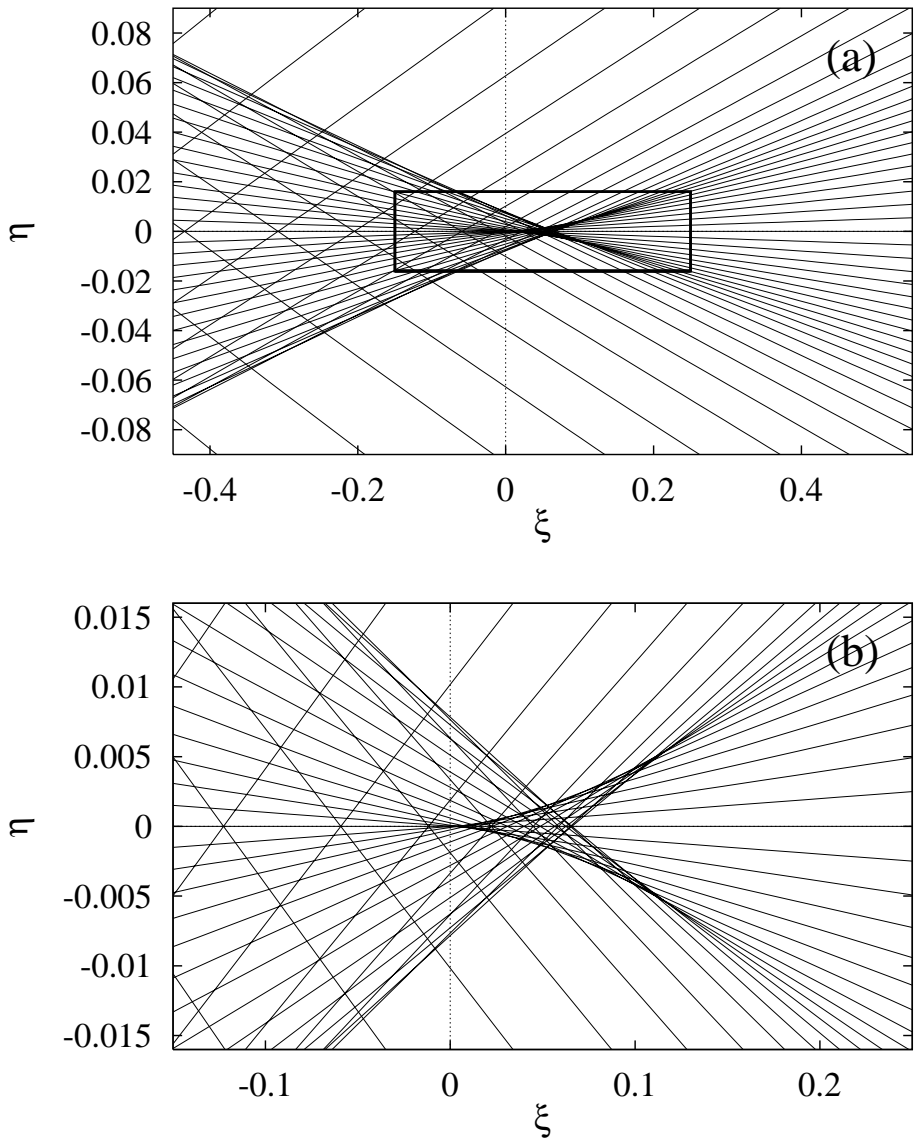}
\caption{\label{fig13} 
$(a)$ Butterfly catastrophe of returning orbits (in rotated semiparabolic 
coordinates) related to the bifurcation of the perpendicular orbit at
$\tilde E_b=-0.31618537$.
$(b)$ Magnification of the marked region close to the nucleus.
There are one, three, or five orbits returning to each point $(\xi,\eta)$
in coordinate space.
(From Ref.\ [37].)
}
\end{figure}

The bunch of trajectories forming the butterfly is given by the 
Hamilton-Jacobi equations (with $p_\xi=\partial S / \partial \xi$ and 
$p_\eta=\partial S / \partial \eta$)
\begin{eqnarray}
 3dp_\eta^5 + 2cp_\xi^2p_\eta^3 
   + p_\xi^4\left[(\xi-\xi_0)p_\eta-\eta p_\xi\right] & = & 0  \nonumber \\
 p_\xi^2 + p_\eta^2 & = & 4 \; ,
\label{Ham_Jacobi_bfly}
\end{eqnarray}
where $\xi$ and $\eta$ are rotated semiparabolic coordinates
\begin{eqnarray}
 \xi  &=& \mu \cos{\vartheta_f\over2} + \nu \sin{\vartheta_f\over2} 
       =  \sqrt{2r} \cos{\vartheta-\vartheta_f\over2} \\
 \eta &=& \nu \cos{\vartheta_f\over2} - \mu \sin{\vartheta_f\over2}
       =  \sqrt{2r} \sin{\vartheta-\vartheta_f\over2} \; ,
\end{eqnarray}
so that the $\xi$ and $\eta$ axes are now parallel and perpendicular to the 
returning orbit.
The parameters $c$, $d$, and $\xi_0$ in Eq.\ \ref{Ham_Jacobi_bfly} will be 
specified later.
With $p_\eta/p_\xi\approx a\Delta\vartheta_i$ we obtain
\begin{equation}
 \eta(\xi) = 3a^5d(\Delta\vartheta_i)^5 + 2a^3c(\Delta\vartheta_i)^3
     + a(\Delta\vartheta_i)(\xi-\xi_0) \; .
\label{eta_bfly}
\end{equation}
The butterfly is illustrated in Fig.\ \ref{fig13}.
Depending on the number of real solutions $\Delta\vartheta_i$ of
Eq.\ (\ref{eta_bfly}) there exist one, three, or five orbits returning
to each point $(\xi,\eta)$.
The different regions are separated by caustics.

The uniform semiclassical approximations for closed orbits near bifurcations
can in general be written as
\begin{eqnarray}
 f_k^{\rm osc} &=& 2(E-E_i) \, \sqrt{\sin\vartheta_i\sin\vartheta_f} 
        {\cal Y}_m(\vartheta_i) {\cal Y}_m(\vartheta_f) \nonumber \\
  && \times {\rm Im} \left\{{\cal A}\exp
         \left(i\left[S_m^k-{\pi\over2}\mu^k+{\pi\over4}\right]\right)\right\}
  \, ,
\label{f_osc_ansatz}
\end{eqnarray}
where the complex amplitude $\cal A$ is defined implicitly by the diffraction
integral \cite{Mai97a,Mai97b}
\begin{eqnarray}
\label{A_def}
 I(r) &\equiv&
  \sum_{\lambda={\rm out,in}} \int_0^\pi 
     {\exp\left(i\left[\Delta S^{\lambda, k}(r,\vartheta)
           - {\pi\over 2}\Delta\mu^{\lambda, k}\right]\right) \over
        \sqrt{\left|\det\left({\partial(\mu,\nu) \over 
        \partial(\tau,\vartheta_i)}\right)^{\lambda,k} \right|}} \, 
        d\vartheta \nonumber \\
 &=&  \; {\cal A} \; \times \; {\cos\left(\sqrt{8r}-{3\over4}\pi\right) \over
               {2\pi \, (2r)^{1/4}}} \; .
\end{eqnarray}
To find a uniform semiclassical approximation for the butterfly catastrophe
we have to solve the Hamilton-Jacobi equations (\ref{Ham_Jacobi_bfly}) 
at least in the vicinity of the central returning orbit and to insert
the action $S(r,\vartheta)$ into Eq.\ (\ref{A_def}).
For the classical action we obtain
\begin{eqnarray}
 \Delta S^\lambda(r,\vartheta) =
    \pm \sqrt{8r} + {1\over4}\xi_0(\vartheta-\vartheta_f)^2
      + {1\over16}c(\vartheta-\vartheta_f)^4
      - {1\over64}d(\vartheta-\vartheta_f)^6  \; , \nonumber \\
\label{S_bfly}
\end{eqnarray}
and for the determinant in the denominator of (\ref{A_def}) we find
$\pm a\sqrt{8r}$ in the limit $r\gg0$ and $\vartheta\approx\vartheta_f$.
Summing up in (\ref{A_def}) the contributions of the incoming and the outgoing 
orbit (with Maslov indices $\Delta\mu^\lambda$ as for the cusp) we obtain 
the integral $I(r)$ and the amplitude $\cal A$ for the butterfly catastrophe 
$(t\equiv\vartheta-\vartheta_f)$:
\begin{eqnarray}
 I(r)
 &=& 2^{1/4} r^{-1/4} |a|^{-1/2} \exp{(-i{\pi\over4})}
       \cos\left(\sqrt{8r}-{3\over4}\pi\right) \nonumber \\
 &\times& \int_{-\infty}^{+\infty} \exp\left(i
       \left[(\xi_0/4)t^2 - (c/16)t^4 - (d/64)t^6\right]\right)\, dt \\
 \Rightarrow {\cal A}
 &=& 2^{5/2} \pi |a|^{-1/2} \exp{(-i{\pi\over4})} d^{-1/6}~
   \Psi\left(-d^{-1/3}\xi_0,-cd^{-2/3}\right)  \quad ,
\label{A_bfly}
\end{eqnarray}
where
\begin{equation}
 \Psi(x,y) \equiv \int_{-\infty}^{+\infty} \exp{\left(-i\left[xt^2
     + yt^4 + t^6\right]\right)}\, dt
\end{equation}
is an analytic function in both variables $x$ and $y$.
Its numerical calculation and asymptotic properties are discussed in Appendix
\ref{bfly_app}.
The uniform result for the oscillatory part of the transition strength
now reads
\begin{eqnarray}
 f^{\rm osc} &=& 2(E-E_i) \, \sqrt{\sin\vartheta_i\sin\vartheta_f} 
        {\cal Y}_m(\vartheta_i) {\cal Y}_m(\vartheta_f)
        2^{5/2} \pi |a|^{-1/2} d^{-1/6}  \nonumber \\ 
 &\times& 
       {\rm Im} \left\{\exp\left(i\left[S_m^0-{\pi\over2}\mu^0\right]\right)
        \Psi\left(-d^{-1/3}\xi_0,-cd^{-2/3}\right)\right\} \quad .
\label{f_osc_bfly_a}
\end{eqnarray}
It is very illustrative to study the asymptotic behavior of the uniform
approximation (\ref{f_osc_bfly_a}) as we obtain, on the one hand, the relation
between the parameters $a$, $c$, $d$, and $\xi_0$ and  the actions and the 
monodromy matrix elements of closed classical orbits, and, on the other hand,
the role of
complex ghost orbits related to this type of bifurcation is revealed.
In the following we discuss both limits $\xi_0 \gg 0$, i.e.\ scaled energy
$\tilde E \gg \tilde E_b^{(2)}$, and $\xi_0 \ll 0$, i.e.\ $\tilde E \ll 
\tilde E_b^{(1)}$.

\paragraph*{Asymptotic behavior at scaled energy $\tilde E\gg\tilde E_b^{(2)}$}
Applying Eq.\ (\ref{psi_asy_minus}) from Appendix \ref{bfly_app} to the 
$\Psi$-function in the uniform approximation (\ref{f_osc_bfly_a}), we obtain 
the asymptotic formula for $\xi_0 \gg 0$ 
\begin{eqnarray}
 f^{\rm osc} &=& 2(E-E_i) \, \sqrt{\sin\vartheta_i\sin\vartheta_f} 
        {\cal Y}_m(\vartheta_i) {\cal Y}_m(\vartheta_f)~
              2(2\pi)^{3/2} |a\xi_0|^{-1/2}   \nonumber \\
 &\times& \left\{\sin\left(S_m^0-{\pi\over2}\mu^0+{\pi\over4}\right)
 + \left[1+{c^2\over3d\xi_0}\left(1+\sqrt{(3d/c^2)\xi_0+1}\right)\right]^{-1/2}
     \right.  \nonumber \\
 &\times& \left.\sin\left(S_m^0
  + {c^3\over9d^2}\left\{{3d\over c^2}\xi_0 + {2\over3}
    \left[1+\left(1+{3d\over c^2}\xi_0\right)^{3/2}\right]\right\}
    - {\pi\over2}(\mu^0+1)+{\pi\over4}\right)
  \right\} \; . \nonumber \\
\label{f_osc_bfly_asy1}
\end{eqnarray}
Comparing with the solutions for isolated closed orbits (\ref{f_osc})
we can identify the contributions of three real closed orbits.
The classical action of orbit 1 is $S_m^0$, its Maslov index is $\mu^0$ and
the monodromy matrix element $m_{12}$ is given by
\begin{equation}
 m_{12}^{(1)} = -a\xi_0 \equiv -\gamma^{-1/3} \tilde M
  (\tilde E - \tilde E_b) \,
\label{m12_1_bfly}
\end{equation}
where the parameter $\tilde M$ can be determined by closed orbit 
calculations (see Eqs.\ \ref{R2_DS} and \ref{R2_m12}).
Orbits 2 and 3 are symmetric with respect to the $z=0$ plane and have the
same orbital parameters, i.e.\ Maslov index $\mu^0+1$ and the classical action
and the monodromy matrix element
\begin{eqnarray}
 S^{(2,3)} &=& S_m^0 + \Delta S = S_m^0 +
    {c^3\over9d^2}\left\{{3d\over c^2}\xi_0 + {2\over3}
    \left[1+\left({3d\over c^2}\xi_0+1\right)^{3/2}\right]\right\}
\label{S_2_bfly}  \\
 m_{12}^{(2,3)} &=&  4a\xi_0
    \left[1+{c^2\over3d\xi_0}\left(1+\sqrt{(3d/c^2)\xi_0+1}\right)\right]
        \; .
\label{m12_2_bfly}
\end{eqnarray}
In the example of the bifurcation of orbit $R_2$ at scaled energy
$\tilde E_b^{(2)}=-0.31618537$, orbit 1 is the orbit $R_2$ perpendicular to 
the magnetic field axis, while orbits 2 and 3 can be identified with $R_2^1$
traversed in both directions ($\vartheta_{i,3}=\pi-\vartheta_{i,2}$).
With the help of Eqs.\ (\ref{m12_1_bfly}) to (\ref{m12_2_bfly}) and classical 
scaling properties of the action and the monodromy matrix the parameters 
$a$, $c$, $d$, and $\xi_0$ in the uniform approximation (\ref{f_osc_bfly_a}) 
can now be expressed completely in terms of closed-orbit parameters
$k$, $\tilde\sigma$, and $\tilde M$ (see Eqs.\ \ref{R2_DS} and \ref{R2_m12}),
\begin{eqnarray}
\label{bfly_param}
 |a|^{-1/2}d^{-1/6} &=& 3^{1/6} \gamma^{1/18} k^{1/3}
      \left(\tilde\sigma / \tilde M\right)^{1/2}  \nonumber \\
 d^{-1/3}\xi_0 &=& 3^{1/3} k^{2/3} \gamma^{-2/9} \tilde\sigma 
      \left(\tilde E - \tilde E_b^{(2)}\right)  \\
 cd^{-2/3} &=& (9k)^{1/3} \gamma^{-1/9}  \; ,  \nonumber
\end{eqnarray}
and the uniform approximation for the butterfly catastrophe finally reads
\begin{eqnarray}
 f^{\rm osc} &=& 2(E-E_i) \, \sqrt{\sin\vartheta_i\sin\vartheta_f}
        {\cal Y}_m(\vartheta_i) {\cal Y}_m(\vartheta_f)
  \nonumber \\  &\times& 
  \pi \gamma^{1/18} 3^{1/6}k^{1/3}(32\tilde\sigma/\tilde M)^{1/2} \;
 {\rm Im} \bigg\{\exp\left(i\left[S_m^0-{\pi\over2}\mu^0\right]\right)
  \nonumber \\  &\times& 
 \Psi\left(-3^{1/3}k^{2/3}\gamma^{-2/9}\tilde\sigma(\tilde E-\tilde E_b^{(2)}),
           -(9k)^{1/3}\gamma^{-1/9}\right)\bigg\}  \; .
\label{f_osc_bfly_b}
\end{eqnarray}
In the classical analysis complex ghost orbits were
discovered both below and above the bifurcation energy.
At $\tilde E>\tilde E_b^{(2)}$ they have the property that the classical action
and the monodromy matrix remain real, although coordinates and momenta in phase
space are complex.
These ghost orbits do not appear in the asymptotic expansion 
(\ref{f_osc_bfly_asy1}) of the uniform approximation (\ref{f_osc_bfly_a}),
and therefore, in analogy with the cusp catastrophe (see \cite{Mai97a}), 
they do not have a physical meaning.
The situation is different at energy $\tilde E<\tilde E_b^{(1)}$ where a
``hidden ghost'' with physical meaning will be revealed in the following.

\paragraph*{Asymptotic behavior and ``hidden ghost'' at scaled energy 
$\tilde E \ll \tilde E_b^{(1)}$}
At scaled energies below the bifurcation point we can apply the asymptotic
formula (\ref{psi_asy_plus}) from Appendix \ref{bfly_app} to the 
$\Psi$-function in the uniform approximation (\ref{f_osc_bfly_a}) and obtain
\begin{eqnarray}
 f^{\rm osc} &=& 2(E-E_i) \, \sqrt{\sin\vartheta_i\sin\vartheta_f} 
        {\cal Y}_m(\vartheta_i) {\cal Y}_m(\vartheta_f)~
              2(2\pi)^{3/2} |a\xi_0|^{-1/2}
   \nonumber \\ &\times&
  \Bigg\{\sin\left(S_m^0-{\pi\over2}(\mu^0+1)+{\pi\over4}\right) \nonumber \\
  &+& {\rm Im} \Bigg\{ \left[1+{c^2\over3d\xi_0}
             \left(1-i\sqrt{-(3d/c^2)\xi_0-1}\right)\right]^{-1/2}
    \exp \Bigg( i\Bigg[ S_m^0 + {c^3\over9d^2}  \nonumber \\
  &\times& \left\{{3d\over c^2}\xi_0 + {2\over3}
    \left[1+i\left(-{3d\over c^2}\xi_0-1\right)^{3/2}\right]\right\}
    - {\pi\over2}\mu^0+{\pi\over4}\Bigg] \Bigg)
  \Bigg\} \Bigg\} \; .
\label{f_osc_bfly_asy2}
\end{eqnarray}
The first term in Eq.\ (\ref{f_osc_bfly_asy2}) can be identified as a real 
orbit with the same classical action and monodromy matrix element as in 
(\ref{f_osc_bfly_asy1}), but with a Maslov index increased by one.
The second term in (\ref{f_osc_bfly_asy2}) is a ghost orbit contribution
resulting from a superposition of two closed orbits with complex action 
and monodromy matrix element, 
\begin{eqnarray}
 S^{(2,3)} &=& S_m^0 +
    {c^3\over9d^2}\left\{{3d\over c^2}\xi_0 + {2\over3}
    \left[1+i\left(-{3d\over c^2}\xi_0-1\right)^{3/2}\right]\right\}
\label{S_3_bfly}  \\
 m_{12}^{(2,3)} &=&  4a\xi_0
   \left[1+{c^2\over3d\xi_0}\left(1-i\sqrt{-(3d/c^2)\xi_0-1}\right)\right] \; ,
\label{m12_3_bfly}
\end{eqnarray}
traversed in both directions.
The positive imaginary part of the classical action results in an exponential
damping of the ghost orbit contribution to the oscillator strength amplitude
with decreasing energy similar to the situation at a fold catastrophe.
In contrast to the fold catastrophe (see Ref.\ \cite{Mai97a}) and to the 
hyperbolic umbilic catastrophe discussed in Section \ref{umbilic_sec} the 
ghost orbit related to a butterfly catastrophe is always accompanied by a 
real orbit with almost the same classical action.
Because the contribution of the real orbit is not exponentially damped its
amplitude at energies where the asymptotic formula (\ref{f_osc_bfly_asy2})
is valid is much stronger than the ghost contribution.
Therefore we call the ghost orbit in (\ref{f_osc_bfly_asy2}) a
``hidden ghost'' \cite{Mai97a,Mai97b}.
Note that classically the complex conjugate of the hidden ghost orbit also 
exists.
The negative imaginary part of its classical action would result in an
unphysical exponential increase of amplitude with decreasing energy, and
consequently the complex conjugate ghost orbit does not appear in the 
asymptotic formula (\ref{f_osc_bfly_asy2}).
In Ref.\ \cite{Mai97a} it was admitted that it might be rather difficult 
to find evidence for hidden ghost orbits in the Fourier transform of
experimental or theoretical scaled energy spectra.
However, with the harmonic inversion technique we are now able to uncover
the hidden ghosts in high resolution quantum recurrence spectra.

\begin{figure}[b]
\vspace{13.5cm}
\includegraphics{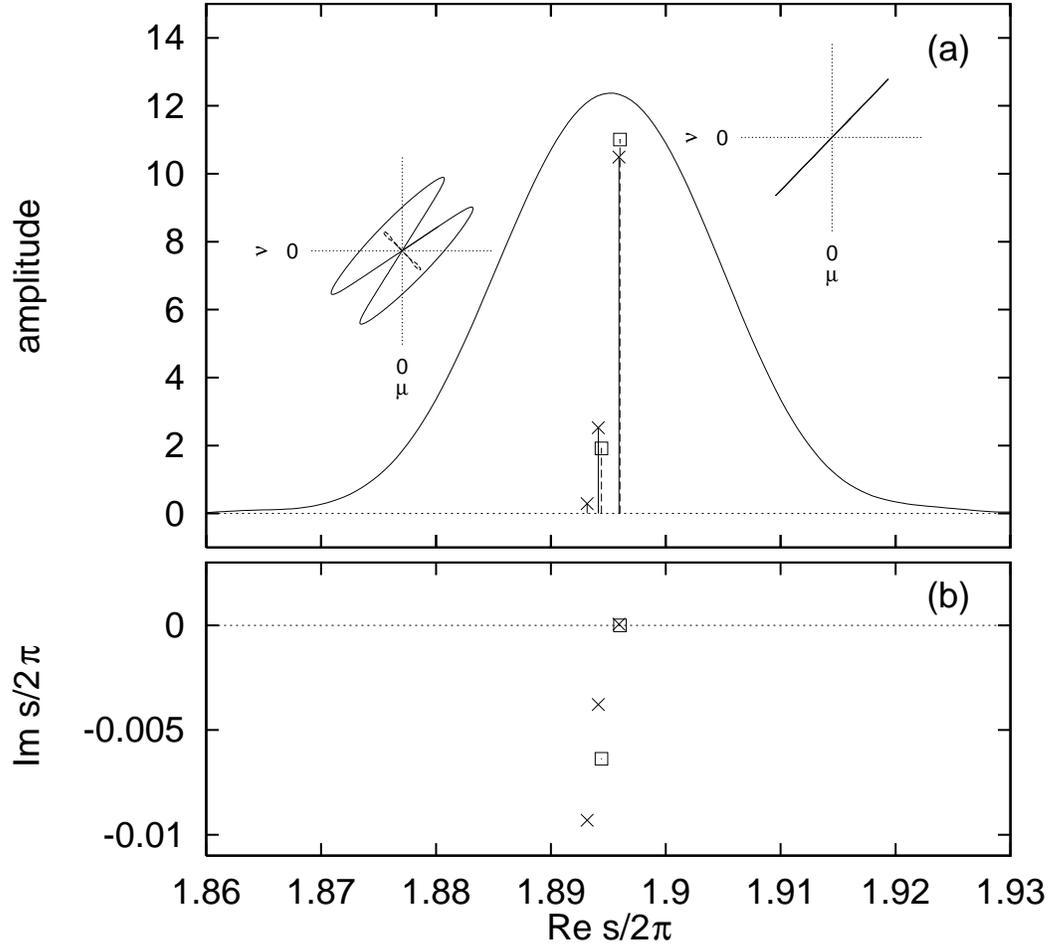}
\caption{\label{fig14} 
(a) Recurrence spectrum for the photoabsorption cross section of the 
hydrogen atom in a magnetic field at scaled energy 
$\tilde E=E\gamma^{-2/3}=-0.35$.
Transition $|2p0\rangle \to |m^{\pi_z}=0^+\rangle$.
Smooth line: Conventional Fourier transform.
Solid stick spectrum and crosses: High resolution quantum recurrence spectrum.
Dashed sticks and squares: Recurrence spectrum from semiclassical 
{\em closed orbit} theory.
The two strongest recurrence peaks are identified by a real and complex 
ghost orbit which are presented as insets.
The solid and dashed lines in the insets are the real and imaginary part 
in semiparabolical coordinates $\mu=(r+z)^{1/2}$, $\nu=(r-z)^{1/2}$.
(b) Complex actions. Crosses and squares are the quantum and classical
results, respectively.
(From Ref.\ [79].)
}
\end{figure}

\paragraph*{Uncovering the hidden ghost orbit}
To uncover the hidden ghost orbit related to the period doubling bifurcation
of the perpendicular orbit in photoabsorption spectra of the hydrogen
atom in a magnetic field we calculated the quantum spectrum at constant
scaled energy $\tilde E=-0.35$, which is sufficiently far below the two
bifurcation energies around $\tilde E=-0.317$ so that the real orbit $R_2$
and the prebifurcation ghost orbit of $R_2^1$ are approximately isolated
orbits.
We calculated 2823 transitions from the initial state $|2p0\rangle$ to final 
states with magnetic quantum number $m=0$ in the region $w=\gamma^{-1/3}<100$.
The scaled photoabsorption spectrum was analyzed by conventional Fourier
transform and by the high resolution harmonic inversion technique.
The interesting part of the recurrence spectrum in the region
$1.86<s/2\pi<1.93$ is presented in Fig.\ \ref{fig14}.
The conventional Fourier transform (smooth line) has a maximum
at $s/2\pi=1.895$ which is roughly twice the period of the perpendicular 
orbit but does not give any hint on the existence of a ghost orbit.
The key points of the harmonic inversion analysis are that the resolution
of the recurrence spectrum is not restricted by the uncertainty principle of 
the Fourier transform and that the method supplies {\em complex} frequencies 
$s^{\rm qm}$ of the analyzed quantum spectrum $f^{\rm qm}(w)$, which 
can be interpreted as complex actions $s_{\rm po}^{\rm cl}$ of ghost orbits.
The high resolution spectral analysis uncovers one real and two complex
actions $s^{\rm qm}$ around $s/2\pi=1.9$, which are marked as crosses 
in Fig.\ \ref{fig14}b.
They can be compared to actions of real and complex classical orbits.
We find two closed orbits, i.e., the period doubling of the perpendicular
orbit with scaled action $s_{\rm po}^{\rm cl}/2\pi=1.896011$ and a ghost
orbit with scaled action $s_{\rm po}^{\rm cl}/2\pi=1.894401-i0.006372$,
which are marked as squares in Fig.\ \ref{fig14}b.
The shapes of the real and complex closed orbits are presented as insets
in Fig.\ \ref{fig14}a (in semiparabolic coordinates $\mu, \nu$).
The real part of the ghost orbit (solid line) is similar to the shape of
orbit $R_2^1$ which is created as a real orbit at much higher energy 
$\tilde E=-0.317$.
The actions and also the amplitudes in the quantum and classical recurrence
spectrum (see crosses and squares in Fig.\ \ref{fig14}) agree very well
for the two closed orbit recurrences.
The deviation between the imaginary parts of the actions of the quantum
and classical ghost orbit can be explained by the fact that the orbits
are only approximately isolated at energy $\tilde E=-0.35$.
The correct semiclassical formula is the uniform semiclassical approximation
(\ref{f_osc_bfly_b}) of the butterfly catastrophe.
This interpretation is supported by the occurrence of a third complex
resonance with large imaginary part of the action and small amplitude
in the quantum recurrence spectrum (Fig.\ \ref{fig14}), which has no 
classical analogue.
The non-classical peak is similar to the occurrence of a non-classical
peak in Fig.\ \ref{fig11} for the hyperbolic umbilic catastrophe and is 
clearly an artifact of the near bifurcation.
The uncovering of the hidden ghost orbit, which obviously is impossible
with the conventional Fourier transform (solid line in Fig.\ \ref{fig14})
demonstrates that harmonic inversion is a very powerful tool for the analysis
of quantum spectra, which can reveal structures and information from the 
spectra that have been unattainable before.

\subsection{Symmetry breaking}
\label{sym_breaking}
Methods for high resolution spectral analysis are also helpful for a direct
observation of symmetry breaking effects in quantum spectra.
In many cases physical systems possess symmetries, e.g., a cylindrical
symmetry around a fixed axis.
In such situations the closed or periodic orbits of the classical system
appear in continuous families.
All members of the family have the same stability parameters and periods, 
i.e., they are observed as one peak in the Fourier transform recurrence 
spectrum of the quantum mechanical density of states or the transition 
spectrum.
The dynamics is profoundly changed when the symmetry is broken by a (weak)
external perturbation.
The behavior of the dynamics and the corresponding semiclassical theory is 
described in Ref.\ \cite{Cre96}.
In general, out of a continuous family of orbits only two closed or periodic 
orbits survive the symmetry breaking.
Recently, quantum manifestations of symmetry breaking have been observed 
experimentally for atoms in external fields:
The cylindrical symmetry of the hydrogen atom in a magnetic field was
broken in crossed magnetic and electric fields \cite{Neu97}, and a
``temporal symmetry breaking'' was studied on lithium atoms in an oscillating
electric field \cite{Spe97,Hag98}.

\begin{figure}[b]
\vspace{13.5cm}
\includegraphics{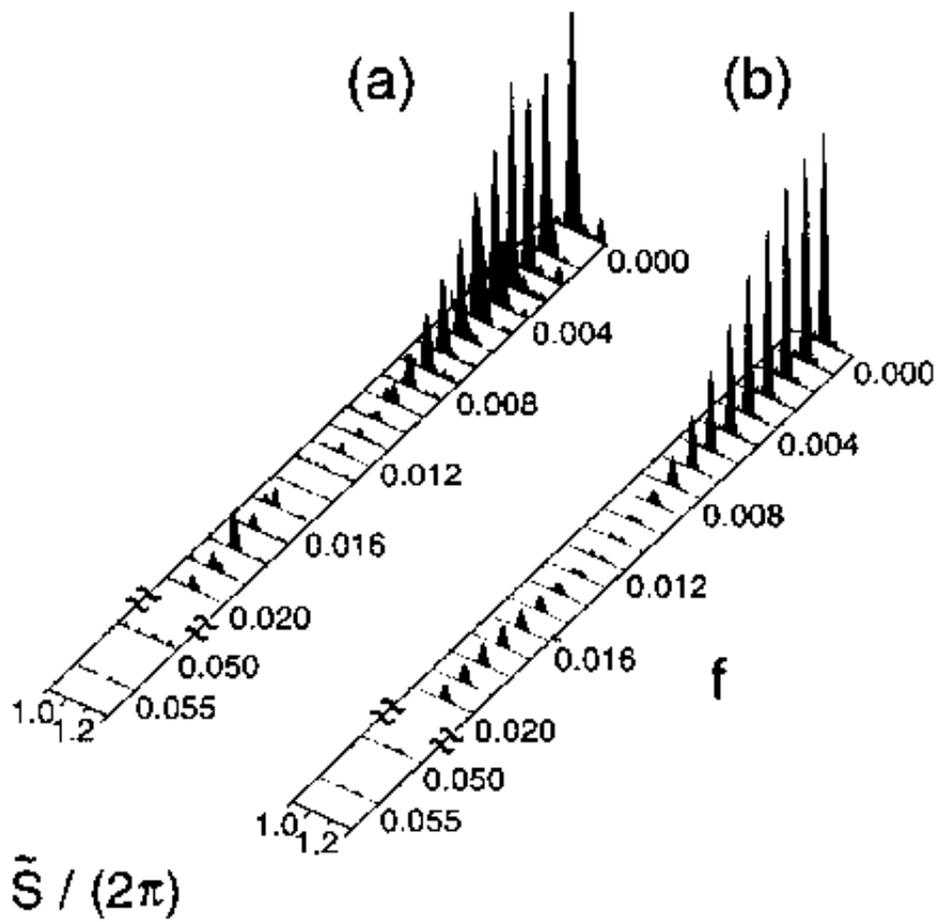}
\caption{\label{fig15} 
Segments of (a) experimental and (b) theoretical scaled action spectra
for the hydrogen atom in crossed magnetic and electric fields at
constant scaled energy $\tilde E=-0.15$.
(From Ref.\ [54].)
}
\end{figure}

\begin{figure}[b]
\vspace{13.5cm}
\includegraphics{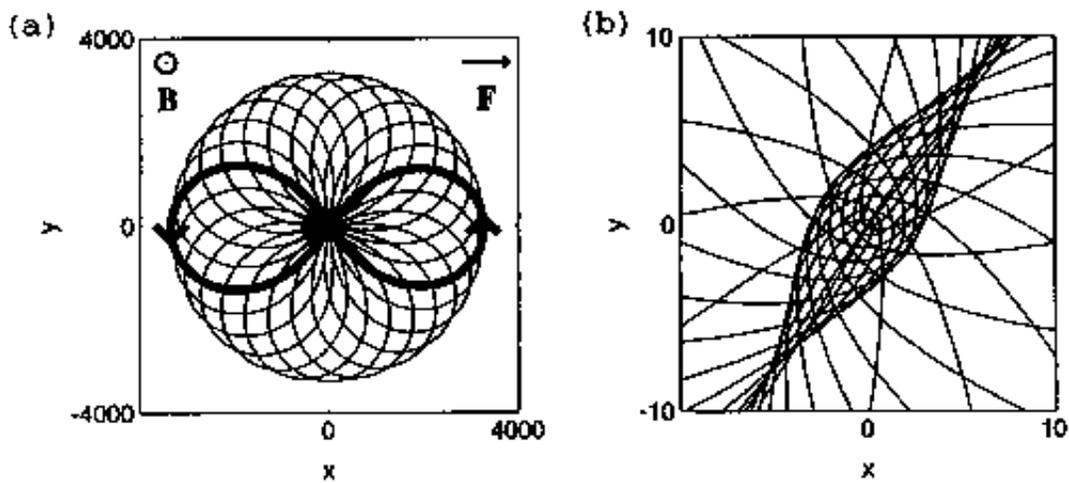}
\caption{\label{fig16} 
(a) Classical trajectories for the hydrogen atom in crossed fields in the
plane perpendicular to the magnetic field axis.
Scaled energy $\tilde E=-0.15$ and field strength $f=0.012$.
(b) Closeup of the returning part of the trajectories as they approach
the nucleus.
(From Ref.\ [54].)
}
\end{figure}

However, the expected splitting of recurrence peaks cannot be observed 
directly because of the finite resolution of the Fourier transform.
This can be seen in Fig.\ \ref{fig15} for the hydrogen atom in crossed fields.
Fig.\ \ref{fig15}a presents segments of the experimental recurrence spectra
at constant scaled energy $\tilde E=-0.15$ and scaled electric field 
strengths $0\le f=F\gamma^{-4/3}\le 0.055$.
$F$ is the electric field strength in atomic units 
$(F_0=5.14\times 10^9\,{\rm V/cm})$.
For comparison Fig.\ \ref{fig15}b shows the theoretical recurrence spectra 
obtained from semiclassical closed orbit theory \cite{Neu97}.
Spectra have been Fourier transformed in the range 
$34.0\le w=\gamma^{-1/3}\le 61.7$.
The recurrence structure corresponds to the classical orbit in the plane 
perpendicular to the magnetic field axis drawn in Fig.\ \ref{fig16}.
At vanishing electric field trajectories starting at the origin are exactly 
closed at each return to the nucleus (thin line in Fig.\ \ref{fig16}a).
For nonzero electric field, only two orbits with slightly different periods
(shown with heavy lines in Fig.\ \ref{fig16}a) return exactly to the origin.
A closeup of the returning part of the trajectories as they approach the
nucleus at scaled energy $\tilde E=-0.15$ and field strength $f=0.012$
is given in Fig.\ \ref{fig16}b.
The two closed orbits return to the nucleus diagonally; all others follow
near-parabolic paths near the nucleus.
In Fig.\ \ref{fig15} no splitting of the recurrence peak can be observed
due to the finite resolution of the Fourier transform spectra.
However, the amplitude of the peak changes with increasing electric
field strength, and indeed in Ref.\ \cite{Neu97} the symmetry breaking 
was identified indirectly by the constructive and destructive interference 
of the two orbits resulting in a Bessel function type modulation of 
amplitudes of recurrence peaks as a function of the strength of the 
symmetry breaking perturbation.

We now want to apply harmonic inversion to directly uncover 
the splitting of recurrence peaks when symmetries are broken.
As in Ref.\ \cite{Neu97} we investigate the hydrogen atom 
in crossed magnetic and electric fields.
Schr\"odinger's equation for hydrogen in crossed fields reads
\begin{equation}
   \left[ {1\over2}{\bf p}^2 - {1\over r} + {1\over 2}\gamma L_z 
 + {1\over8}\gamma^2 (x^2+y^2) + Fx \right] \Psi
 = E \Psi \; ,
\label{Schr_eq_crossed}
\end{equation}
with $\gamma$ and $F$ the magnetic and electric field strength 
(in atomic units), and $E$ the energy.
The eigenvalue problem (\ref{Schr_eq_crossed}) was solved numerically
for fixed external fields $\gamma$ and $F$ in Refs.\ \cite{Mai94b,Mai92}.
To study the effects of symmetry breaking on the quantum spectra we
want to use the scaling properties of the classical system and to analyze
spectra at constant scaled energy $\tilde E =E\gamma^{-2/3}$ and scaled 
electric field $f=F\gamma^{-4/3}$.
Eigenvalues are obtained for the scaling parameter $w=\gamma^{-1/3}$.
Introducing dilated coordinates $\bar{\bf r}=\gamma^{2/3}{\bf r}$
Eq.\ \ref{Schr_eq_crossed} reads
\begin{equation}
 w^2 \left[\bar {\bf p}^2\right]\Psi + w \left[L_z\right]\Psi
 + \left[{1\over 4}(\bar x^2+\bar y^2)+2f\bar x - 2\tilde E
 - {2\over \bar r}\right]\Psi = 0 \; .
\label{Schr_eq_crossed_scal}
\end{equation}
The peculiarity of (\ref{Schr_eq_crossed_scal}) is that it is a
quadratic (instead of linear) eigenvalue equation for the scaling 
parameter $w$, which cannot be solved straightforwardly with standard
diagonalization routines for linear eigenvalue problems.
To solve the quadratic Schr\"odinger equation (\ref{Schr_eq_crossed_scal})
we use the following technique.
We write
\begin{equation}
 w^2 = \lambda = \lambda_0 + \Delta\lambda \; ,
\end{equation}
with $\lambda_0$ fixed and assume that $\Delta\lambda$ is small.
Now the scaling parameter $w$ in the para\-magnetic term of 
(\ref{Schr_eq_crossed_scal}) can be approximated by
\begin{equation}
 w = \sqrt{\lambda}
 \approx \sqrt{\lambda_0} + {1\over 2\sqrt{\lambda_0}} \Delta\lambda \; .
\label{w_approx}
\end{equation}
Replacing $w$ in (\ref{Schr_eq_crossed_scal}) with this approximation yields
a generalized linear eigenvalue problem for $\Delta\lambda$,
\begin{eqnarray}
 & & \left[\lambda_0\bar {\bf p}^2 + \sqrt{\lambda_0}L_z 
   + {1\over 4}(\bar x^2+\bar y^2)+2f\bar x - 2\tilde E
   - {2\over \bar r}\right]\Psi \nonumber \\
 &=& -\Delta\lambda\left[\bar {\bf p}^2 + {1\over 2\sqrt{\lambda_0}}L_z\right]
    \Psi \; ,
\label{Schr_eq_crossed_appr}
\end{eqnarray}
which can be solved numerically with the Spectral Transformation Lanczos
Method (STLM) \cite{Eri80}.
The numerical details for the diagonalization of the linearized
Schr\"odinger equation (\ref{Schr_eq_crossed_appr})
in a complete set of Sturmian type basis functions 
are similar to the diagonalization of Eq.\ \ref{Schr_eq_crossed} at constant 
field strengths as described in \cite{Mai94b}.
It is convenient to choose identical parameters for both $\lambda_0$ and
the center of the spectral transformation, so that the STLM method provides
eigenvalues in the local neighborhood of $\lambda\approx\lambda_0$, i.e.,
$\Delta\lambda$ is small.
However, the obtained eigenvalues are still not very precise because of the
approximation (\ref{w_approx}).
In a second step we therefore apply perturbation theory and use the 
eigenvectors $|\Psi\rangle$ of Eq.\ \ref{Schr_eq_crossed_appr} to
solve the quadratic equation (\ref{Schr_eq_crossed_scal}) separately
for each eigenvalue, $w_n$, in the corresponding one-dimensional subspace,
i.e.
\begin{eqnarray}
 &&      w_n^2\, \langle\Psi_n|\bar {\bf p}^2|\Psi_n\rangle 
       + w_n\, \langle\Psi_n|L_z|\Psi_n\rangle \nonumber \\
 && {} + \langle\Psi_n|{1\over 4}(\bar x^2+\bar y^2)+2f\bar x
       - 2\tilde E - {2\over \bar r}|\Psi_n\rangle = 0 \; .
\label{Schr_eq_crossed_pert}
\end{eqnarray}
By this procedure the accuracy of the eigenvalues is significantly improved.
An alternative method for the exact solution of the quadratic eigenvalue
equation (\ref{Schr_eq_crossed_scal}) is described in Ref.\ \cite{Spr96}.

\begin{figure}[b]
\vspace{9.9cm}
\includegraphics{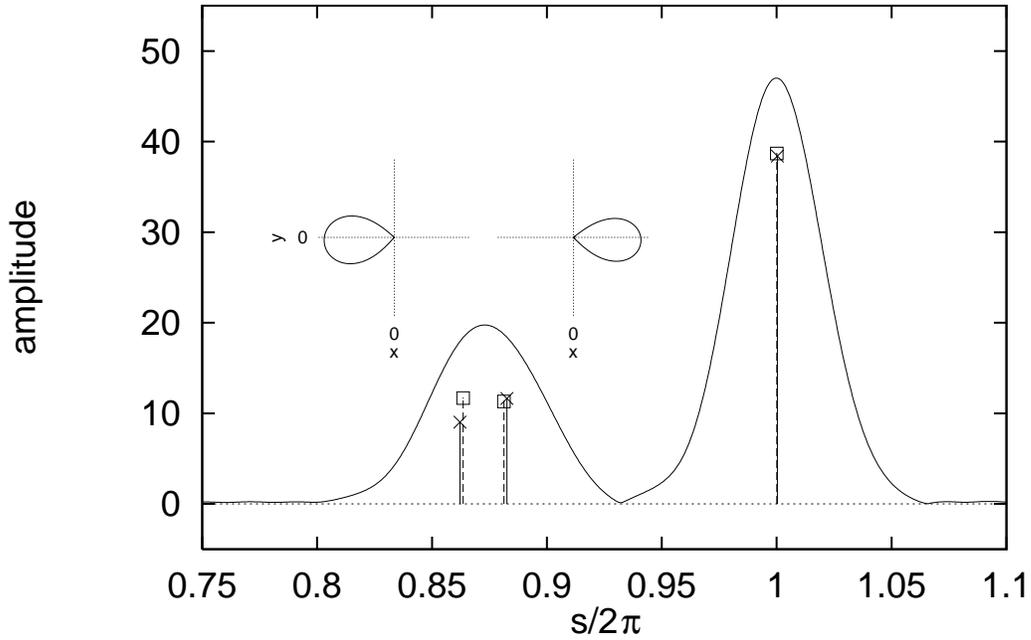}
\caption{\label{fig17} 
Recurrence spectrum for the photoabsorption cross section of the 
hydrogen atom in crossed magnetic and electric fields at scaled energy 
$\tilde E=E\gamma^{-2/3}=-0.5$ and field strength 
$f=F\gamma^{-4/3}=0.02$.
Transition $|2p0\rangle \to |\pi_z=+1\rangle$.
Smooth line: Conventional Fourier transform.
Solid stick spectrum and crosses: High resolution quantum recurrence spectrum.
Dashed sticks and squares: Recurrence spectrum from semiclassical 
{\em closed orbit} theory.
The two recurrence peaks around $s/2\pi=0.87$ are identified by 
closed orbits in the $(x,y)$ plane presented as insets in the figure. 
(From Ref.\ [79].)
}
\end{figure}

We now investigate the symmetry breaking in the hydrogen atom at constant
scaled energy $\tilde E=-0.5$.
Without electric field ($f=0$) the two shortest closed orbit recurrences 
are the perpendicular orbit at $s/2\pi=0.872$ and the parallel orbit at 
$s/2\pi=1$.
When the cylindrical symmetry is broken by a crossed electric field, all
three-dimensional orbits should split into two nearby peaks.
The only exception is the orbit parallel to the magnetic field axis, which
does not appear as a continuous family of closed orbits.
To verify the symmetry breaking in quantum spectra we calculated the  
photoabsorption spectrum (transitions from the initial state 
$|2p0\rangle$ to the final states with even $z$-parity) up to $w=50$
for the crossed field atom at constant scaled energy $\tilde E=-0.5$ and 
scaled field strength $f=0.02$.
The interesting part of the resulting recurrence spectrum obtained by both
the conventional Fourier transform and the high resolution harmonic inversion
technique are presented in Fig.\ \ref{fig17}.
The conventional Fourier transform (smooth line) shows two peaks around
$s/2\pi=0.87$ (the perpendicular orbit) and $s/2\pi=1$ (the parallel orbit).
However, none of the peaks is split.
The high resolution recurrence spectrum obtained by harmonic inversion
is drawn as solid sticks and crosses in Fig.\ \ref{fig17} and clearly
exhibits a splitting of the recurrence peak of the perpendicular orbit
at $s/2\pi\approx 0.87$.
For comparison the semiclassical recurrence spectrum is presented as
dashed sticks and squares.
In the plane perpendicular to the magnetic field axis two closed orbits
have been found with slightly different classical actions 
$s^{\rm cl}/2\pi=0.864$ and $s^{\rm cl}/2\pi=0.881$.
The shapes of these orbits are illustrated as insets in Fig.\ \ref{fig17}.
As can be seen the semiclassical and the high resolution quantum recurrence
spectrum are in excellent agreement, i.e., the symmetry breaking in crossed
fields has been directly uncovered by harmonic inversion of the quantum 
mechanical photoabsorption spectrum.

\subsection{$\hbar$ expansion of the periodic orbit sum}
\label{hbar_sec}
In the previous sections we have introduced harmonic inversion as a
powerful method for the high resolution analysis of quantum spectra,
which allows a direct and quantitative comparison of the quantum spectra
with semiclassical theories.
However, the excellent agreement to many significant digits for both the 
periods and amplitudes of quantum mechanical recurrence peaks with the 
semiclassical periodic orbit contributions (see Section \ref{prec_check})
may be surprising for the following reason.
Periodic orbit theory is exact only for a special class of systems, e.g., the 
geodesic motion on a surface with constant negative curvature \cite{Gut90}.
In general, the semiclassical periodic orbit sum is only the leading order
contribution of an infinite series in powers of the Planck constant
\cite{Gas93,Alo93,Vat96}.
Therefore it might be expected that only the high resolution analysis of
{\em semiclassical} spectra yields perfect agreement with periodic
orbit theory while the harmonic inversion of {\em quantum} spectra should
show small but noticeable deviations from the semiclassical recurrence
spectra.
As will be shown in the following,
the absence of such deviations is related to the functional form of the
$\hbar$ expansion of the periodic orbit sum and special properties of the
harmonic inversion method.

In scaling systems, where the classical action of periodic orbits scales as 
\[
S_{\rm po}/\hbar=s_{\rm po}/\hbar_{\rm eff}=s_{\rm po}w \; ,
\] 
the scaling parameter plays the role of an effective Planck constant, i.e.
\[
 w \equiv \hbar_{\rm eff}^{-1} \; ,
\]
and the $\hbar$ expansion of the periodic orbit sum can therefore be written
as a power series in $w^{-1}$.
The fluctuating part of the semiclassical response function is given by
\begin{equation}
   g(w)
 = \sum_{n=0}^\infty g_n(w)
 = \sum_{n=0}^\infty {1\over w^{n}} \sum_{\rm po} 
   {\cal A}_{\rm po}^{(n)} e^{is_{\rm po}w} \; .
\label{g_hbar_series}
\end{equation}
The ${\cal A}_{\rm po}^{(n)}$ are the complex amplitudes of the $n^{\rm th}$
order periodic orbit contributions including phase information from the 
Maslov indices.
When quantum spectra are analyzed in the semiclassical regime, i.e., at
sufficiently high scaling parameter $w$, the higher order $\hbar$ correction
terms $(n\ge 1)$ of the series (\ref{g_hbar_series}) are certainly small
compared to the zeroth order terms.
However, the reason why the higher order contributions are not uncovered
by the harmonic inversion method is that only the zeroth order $(n=0)$ terms
in (\ref{g_hbar_series}) fulfill the ansatz (\ref{C_s_n}) required for
harmonic inversion, i.e., they are exponential functions with constant 
frequencies and amplitudes.
The higher order terms $(n\ge 1)$ have amplitudes decreasing $\sim w^{-n}$
with increasing scaling parameter, and thus do not fulfill the ansatz 
(\ref{C_s_n}).
Therefore only the zeroth order amplitudes ${\cal A}_{\rm po}^{(0)}$ can be
obtained as converged parameters from the high resolution harmonic inversion
analysis of the quantum spectra.
The terms with $n\ge 1$ have similar properties as weak ``noise'' \cite{Man97b}
and are separated by the harmonic inversion method from the ``true'' signal.
In other words, the zeroth order approximation of the periodic orbit sum
(\ref{g_hbar_series}) with amplitudes ${\cal A}_{\rm po}^{(0)}$ given by
Gutzwiller's trace formula is the best fit to the quantum spectra within
the given ansatz as a linear superposition of exponential functions of the
scaling parameter $w$, and the corresponding parameters are obtained from
the harmonic inversion procedure.

However, the higher order terms of the $\hbar$ expansion (\ref{g_hbar_series})
can be revealed by harmonic inversion as will be demonstrated in the
following.
The periodic orbit terms ${\cal A}_{\rm po}^{(n)}$ can be obtained provided
that the quantum spectrum and the $(n-1)^{\rm st}$ order eigenvalues 
$w_{k,n-1}$ are given.
We can then calculate the difference between the exact quantum mechanical
and the $(n-1)^{\rm st}$ order response function
\begin{equation}
   g^{\rm qm}(w) - \sum_{j=0}^{n-1} g_j(w)
 = \sum_{j=n}^\infty g_j(w)
 = \sum_{j=n}^\infty {1\over w^{j}} \sum_{\rm po} 
   {\cal A}_{\rm po}^{(j)} e^{is_{\rm po}w} \; .
\label{Dg_n}
\end{equation}
The leading order terms in Eq.\ \ref{Dg_n} are $\sim w^{-n}$, i.e.,
multiplication with $w^n$ yields
\begin{equation}
   w^n\left[g^{\rm qm}(w) - \sum_{j=0}^{n-1} g_j(w)\right]
 = \sum_{\rm po} {\cal A}_{\rm po}^{(n)} e^{is_{\rm po}w}
   + {\cal O}\left({1\over w}\right) \; .
\label{G_n}
\end{equation}
In Eq.\ \ref{G_n} we have restored the functional form (\ref{C_s_n}), i.e.,
a linear superposition of exponential functions of $w$.
The harmonic inversion of the function (\ref{G_n}) will now provide
the periods $s_{\rm po}$ and the $n^{\rm th}$ order amplitudes 
${\cal A}_{\rm po}^{(n)}$ of the $\hbar$ expansion (\ref{g_hbar_series}).
We will illustrate the method on two different examples, i.e., the circle
billiard and the three disk scattering problem.

\subsubsection{The circle billiard}
\label{circle_analysis}
The circle billiard is an integrable and even separable bound system, and 
has been chosen here mainly for the sake of simplicity, since all the
relevant physical quantities, i.e., the quantum and semiclassical
eigenenergies, and the periodic orbits can easily be obtained.
In polar coordinates $(\rho,\phi)$ and after separation of the $\phi$-motion
Schr\"odinger's equation reads
\begin{equation}
   -{\hbar^2\over 2M}\left({1\over\rho}{\partial\over\partial\rho}\rho
    {\partial\over\partial\rho}-{m^2\over\rho^2}\right)\Psi_m(\rho)
 = E\Psi_m(\rho) \; ,
\label{Schr_eq_circle}
\end{equation}
with $M$ the mass of the particle and $m$ the angular momentum quantum number.
The wave functions must fulfill the boundary condition $\Psi_m(R)=0$ with
$R$ the radius of the circle billiard.
Defining
\[
 E = {\hbar^2k^2\over 2M}
\]
and after substitution of $z\equiv k\rho$ Schr\"odinger's equation
(\ref{Schr_eq_circle}) is transformed into the differential equation for
the Bessel functions \cite{Abr65}
\begin{equation}
 z^2 J_m''(z) + z J_m'(z) + (z^2-m^2) J_m(z) = 0 \; .
\label{Bessel_eq}
\end{equation}
The quantum mechanical eigenvalues $k_{n,m}^{\rm qm}$ are obtained from the 
boundary condition $J_m(k_{n,m}^{\rm qm}R)=0$ as zeros of the Bessel functions.
We calculated numerically the first 32469 eigenenergies of the circle billiard
in the region $k_{n,m}R<510$.
Note that states with $m\ne 0$ are twofold degenerate.

The semiclassical eigenenergies are obtained from an EBK torus quantization
or, after separation of the $\phi$-motion, even more simply from the
one-dimensional WKB quantization of the radial motion in the centrifugal
potential, i.e.
\begin{equation}
   2\int_{m/k}^R p_\rho d\rho
 = 2\hbar k \int_{m/k}^R \sqrt{1-(m/k\rho)^2}d\rho
 = 2\pi\hbar \left(n+{3\over 4}\right) \; ,
\label{circle_WKB_cond}
\end{equation}
with $n=0,1,2,\dots$ the radial quantum number.
The r.h.s.\ of Eq.\ \ref{circle_WKB_cond} takes into account the
correct boundary conditions of the semiclassical wave functions at the
classical turning points.
The zeroth order semiclassical eigenenergies are finally obtained from
the quantization condition
\begin{equation}
 kR\sqrt{1-(m/kR)^2} - |m|\arccos{|m|\over kR} = \pi\left(n+{3\over 4}\right)
  \; .
\label{circle_EBK}
\end{equation}
The semiclassical spectrum has been calculated with the help of 
Eq.\ \ref{circle_EBK} in the same region $k_{n,m}^{\rm sc}R<510$ as the 
exact quantum spectrum.

For the comparison of the spectra with periodic orbit theory we need to
calculate the periodic orbits and their physical quantities.
For the circle billiard all quantities are obtained analytically.
In the following we choose $R=1$.
The periodic orbits of the circle billiard are those orbits for which the 
angle between two bounces is a rational multiple of $2\pi$, i.e., the periods 
$\ell_{\rm po}$ are obtained from the condition
\begin{equation}
 \ell_{\rm po} = 2m_r \sin \gamma \; ,
\end{equation}
with 
\[
\gamma\equiv\pi {m_\phi\over m_r} \; ,
\]
$m_\phi=1,2,\dots$ the number of turns of the orbit around the origin, 
and $m_r=2m_\phi,2m_\phi+1,\dots$ the number of reflections at the boundary 
of the circle.
Periodic orbits with $m_r\ne 2m_\phi$ can be traversed in two directions
and thus have multiplicity 2.
Because the classical dynamics of the circle billiard is regular the 
Berry-Tabor formula \cite{Ber76} must be applied instead of Gutzwiller's
trace formula for the calculation of the semiclassical density of states.
We obtain
\begin{equation}
   {1\over\sqrt{k}} g(k)
 = {1\over\sqrt{k}} \sum_{n=0}^\infty g_n(k)
 = \sum_{n=0}^\infty {1\over k^{n}} \sum_{\rm po} 
   {\cal A}_{\rm po}^{(n)} e^{i\ell_{\rm po}k} \; ,
\label{g_hbar_series_circ}
\end{equation}
which basically differs from Eq.\ \ref{g_hbar_series} by a factor of
$k^{-1/2}$ on the l.h.s.\ of (\ref{g_hbar_series_circ}).
[Note that for billiard systems the scaling parameter is the absolute value of
the wave vector, $w\equiv k=|{\bf p}|/\hbar$, and the action is proportional 
to the length of the orbit, $S_{\rm po}=\hbar k\ell_{\rm po}$.]
The zeroth order periodic orbit amplitudes obtained from the Berry-Tabor
formula read
\begin{equation}
   {\cal A}_{\rm po}^{(0)}
 = \sqrt{\pi\over 2}{\ell_{\rm po}^{3/2}\over m_r^2}
   e^{-i({\pi\over 2}\mu_{\rm po}+{\pi\over 4})}  \; ,
\label{A0_circ}
\end{equation}
with $\mu_{\rm po}=3m_r$ the Maslov index.
For the calculation of the first order periodic orbit amplitudes
${\cal A}_{\rm po}^{(1)}$ in Eq.\ \ref{g_hbar_series_circ} we adopt the 
method of Alonso and Gaspard \cite{Alo93}.
After a lengthy calculation we finally obtain
\begin{equation}
   {\cal A}_{\rm po}^{(1)}
 = {1\over 2}\sqrt{\pi m_r} \, 
   \underbrace{5-2\sin^2\gamma\over 3\sin^{3/2}\gamma}_{f(\gamma)}
   e^{-i({\pi\over 2}\mu_{\rm po}-{\pi\over 4})}  \; .
\label{A1_circ}
\end{equation}
A detailed derivation of Eqs.\ \ref{A0_circ} and \ref{A1_circ} will be given 
elsewhere \cite{Wei99a,Wei99b}.

\begin{figure}[t]
\vspace{7.5cm}
\includegraphics{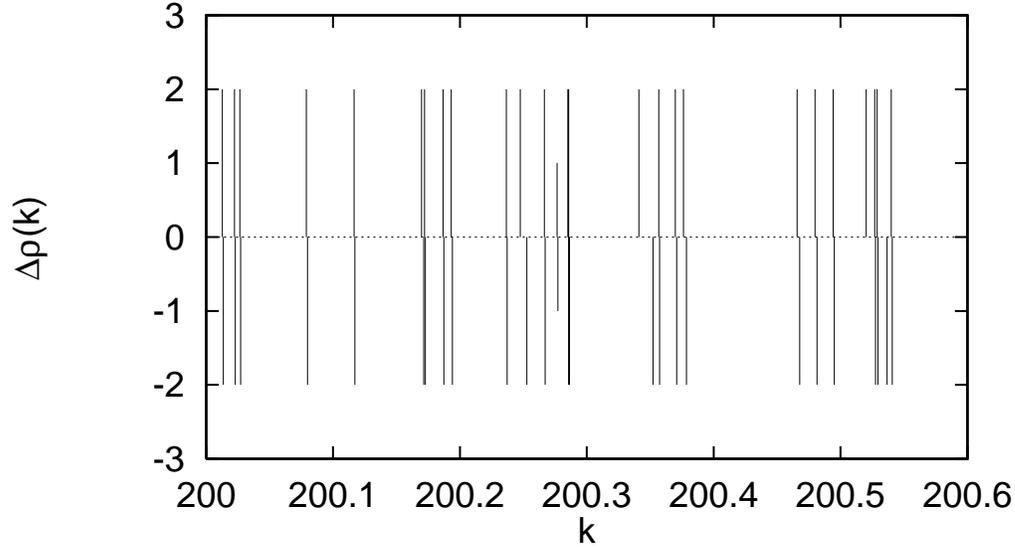}
\caption{\label{fig18} 
Part of the difference spectrum
$\Delta\varrho(k)=\varrho^{\bf qm}(k)-\varrho^{\rm sc}(k)$
between the quantum and the semiclassical density of states for the
circle billiard with radius $R=1$.
}
\end{figure}

\begin{figure}[b]
\vspace{7.8cm}
\includegraphics{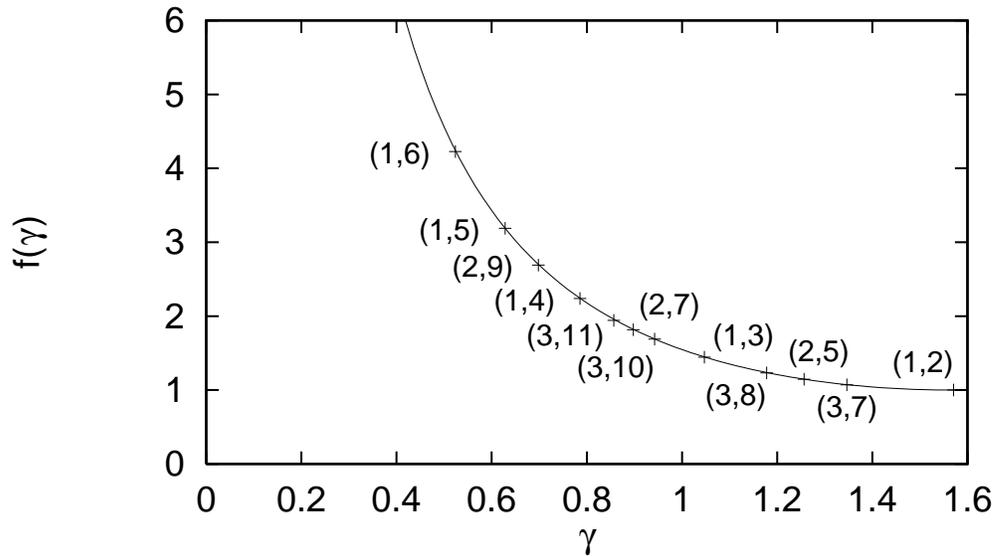}
\caption{\label{fig19} 
First order $\hbar$ correction terms of the semiclassical periodic orbit sum
for the circle billiard.
Solid line: Semiclassical Theory.
Crosses: Harmonic inversion analysis of the difference spectrum
$k^{1/2}[\varrho^{\rm qm}(k)-\varrho^{\rm sc}(k)]$.
The periodic orbits are marked by the numbers $(m_\phi,m_r)$.
}
\end{figure}

When the quantum mechanical density of states, or, more precisely, the
spectrum $k^{-1/2}\varrho^{\rm qm}(k)$ is analyzed by the harmonic inversion
method the periodic orbit quantities $\ell_{\rm po}$ and 
${\cal A}_{\rm po}^{(0)}$ are obtained to very high precision \cite{Wei99a}.
However, we are now interested in the $\hbar$ expansion of the periodic
orbit sum and want to verify the first order corrections
${\cal A}_{\rm po}^{(1)}$ directly in the quantum spectrum.
We therefore analyze the difference spectrum 
$\Delta\varrho(k)=\varrho^{\rm qm}(k)-\varrho^{\rm sc}(k)$
between the quantum and the semiclassical density of states.
A small part of this spectrum at $k\approx 200$ is presented 
in Fig.\ \ref{fig18}.
The absolute values of the peak heights mark the multiplicities of the states.
To restore the functional form which is required for the harmonic inversion
procedure the difference spectrum was multiplied by $\sqrt{k}$, and the
resulting signal $\sqrt{k}\Delta\varrho(k)$ was analyzed in the region
$100<k<500$.
The results are presented in Fig.\ \ref{fig19}.
For a direct comparison with Eq.\ \ref{A1_circ} we transform for each
periodic orbit the obtained period $\ell_{\rm po}$ and the first order 
amplitude ${\cal A}_{\rm po}^{(1)}$ into the quantities
\[
 \gamma=\arcsin{\ell_{\rm po}\over 2m_r} \quad {\rm and} \quad 
 f(\gamma)\equiv {2\over\sqrt{\pi m_r}}\, |{\cal A}_{\rm po}^{(1)}| \; .
\]
These quantities are plotted as crosses in Fig.\ \ref{fig19}.
The periodic orbits are marked by the numbers $(m_\phi,m_r)$.
The solid line
\[
 f(\gamma) = {5-2\sin^2\gamma\over 3\sin^{3/2}\gamma}
\]
is the result of Eq.\ \ref{A1_circ}.
As can be seen the theoretical curve and the crosses obtained by harmonic
inversion of the difference between the quantum and semiclassical density 
of states are in excellent agreement.

The semiclassical accuracy of the zeroth order eigenenergies of the circle
billiard has been discussed in Refs.\ \cite{Pro93,Boa94} in terms of, e.g., 
the average error in units of the mean level spacing.
The analysis presented here provides a more physical interpretation of
the deviations between the quantum and semiclassical eigenenergies in
terms of the higher order corrections of the $\hbar$ expanded
periodic orbit sum (\ref{g_hbar_series}).
The direct application of the series (\ref{g_hbar_series}) for the
semiclassical quantization beyond the Gutzwiller and Berry-Tabor
approximation will be discussed in Section \ref{hbar_sec2}.

\subsubsection{The three disk scattering system}
\label{three_disk_analysis}
\begin{figure}[b]
\vspace{17.5cm}
\includegraphics{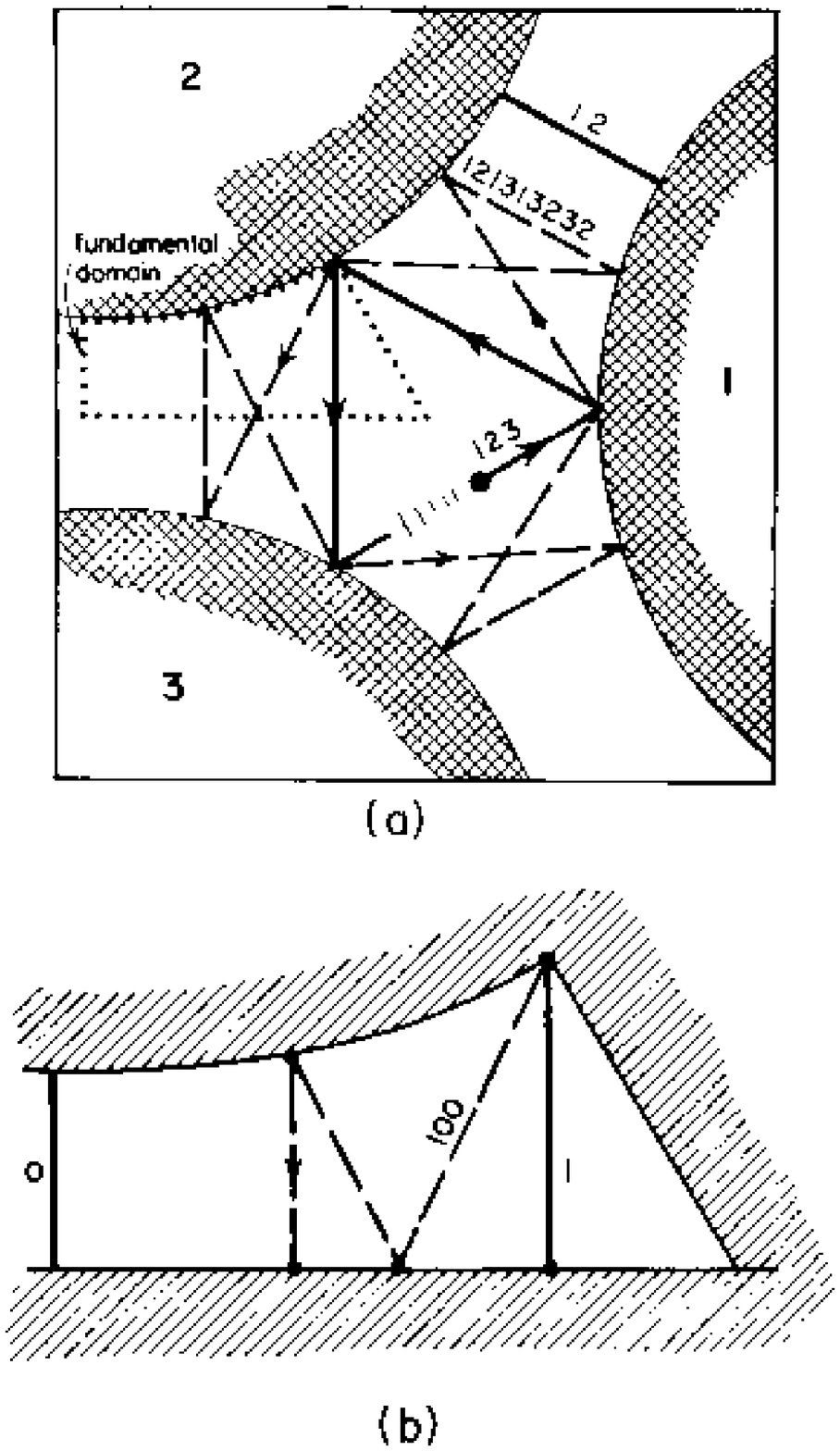}
\caption{\label{fig20} 
The scattering geometry for the three-disk system.
(a) The three disks with 12, 123, and 121313232 cycles indicated.
(b) The fundamental domain, i.e., a wedge consisting of a section of a disk,
two segments of symmetry axis acting as straight mirror walls, and an
escape gap. The above cycles restricted to the fundamental domain are now
the 0, 1, and 100 cycle.
(From Ref.\ [9].)
}
\end{figure}

As a second example for the investigation of the $\hbar$ expansion of the 
periodic orbit sum we now consider a billiard system consisting of three
identical hard disks with unit radius $R=1$, displaced from each other by
the same distance, $d$.
The classical, semiclassical, and quantum dynamics of this scattering system 
has been studied by Gaspard and Rice \cite{Gas89}.
In recent years the system has served as a prototype model for periodic orbit 
quantization by cycle expansion techniques 
\cite{Cvi89,Eck93,Eck95,Cvi97,Wir97}.
If the disks are separated by a distance $d>2.0481419$ there is a one-to-one 
identity between the periodic orbits and a symbolic code, whereas for 
$d<2.04821419$ pruning of orbits sets in \cite{Han93}.
The geometry of the three disk scattering system is shown in Fig.\ \ref{fig20}.
The symbolic code of a periodic orbit is obtained by numbering of the disks,
'1', '2', and '3', and by bookkeeping the reflections of the orbit at
the three disks.
E.g., $\overline{121313232}$ is the symbolic code of the primitive orbit 
with cycle length $n=9$ drawn with a dashed line in Fig.\ \ref{fig20}.
The bar, which is often omitted, indicates the periodicity of the orbit.
The three disk scattering system is invariant under the symmetry operations
of the group $C_{3v}$, i.e., three reflections at symmetry lines and two 
rotations by $2\pi/3$ and $4\pi/3$.
Periodic orbits have $\sigma_v$ symmetry if they are invariant under 
reflections and $C_3, C_3^2$ symmetry if they are invariant under rotations.
After symmetry decomposition the periodic orbits in the fundamental domain
(see Fig.\ \ref{fig20}b) can be classified by a binary symbolic code of 
symbols '0' and '1', where each '0' represents a change between clockwise
and anticlockwise scattering in the original three disk system \cite{Cvi89}.
The symbolic code of the orbit 121313232 in Fig.\ \ref{fig20}a restricted
to the fundamental domain is 100100100 or because of its periodicity even 
simpler 100.
The quantum resonances are also classified by symmetries.
Resonances with $A_1$ ($A_2$) symmetry are symmetric (antisymmetric) under
reflections at a symmetry line, and resonances with $E$ symmetry are
invariant (up to a complex phase factor $c$ with $c^3=1$) under rotations 
by $2\pi/3$ and $4\pi/3$.
The resonances in the $E$ subspace are twofold degenerate.
In the following we analyze resonances with $A_1$ symmetry.

\begin{figure}[b]
\vspace{17.5cm}
\includegraphics{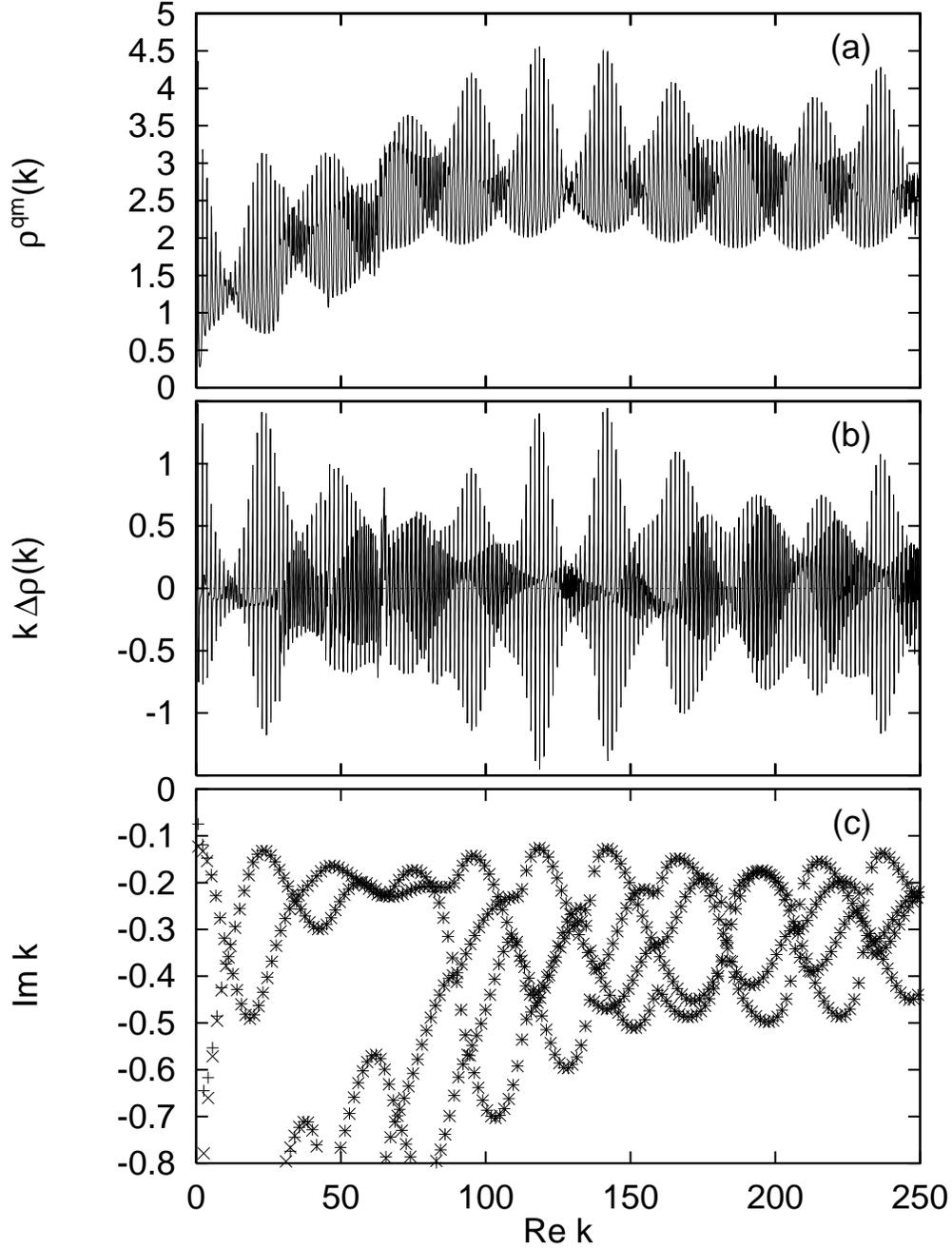}
\caption{\label{fig21} 
Spectra of the three disk scattering system ($A_1$ subspace) with radius 
$R=1$ and distance $d=6$ between the disks.
(a) Fluctuating part of the quantum mechanical density of states.
(b) Difference spectrum $k[\varrho^{\rm qm}(k)-\varrho^{\rm sc}(k)]$ 
between the quantum and semiclassical density of states.
(c) Quantum $(+)$ and semiclassical $(\times)$ $A_1$ resonances of the 
Green's function.
}
\end{figure}

The three disk scattering system does not have any bound states.
However, the fluctuating part of the density of states can be written as
\begin{equation}
 \varrho(k) = -{1\over\pi}\, {\rm Im}\, \sum_n {1\over k-k_n}
\end{equation}
with $k_n$ the complex resonances (poles) of the Green's function.
For distance $d=6$ the quantum mechanical $A_1$ resonances have been 
calculated by Wirzba \cite{Cvi97,Wir97,Wir98} in the region 
$0<{\rm Re}~k<250$, and are plotted with the $+$ symbols 
in Fig.\ \ref{fig21}c.
The corresponding fluctuating part of the quantum density of states is shown
in Fig.\ \ref{fig21}a.
This spectrum can be analyzed by harmonic inversion to extract the lengths
(periods) of the orbits and the zeroth order amplitudes.
However, as in the previous section for the circle billiard we here want to
investigate the difference between the quantum and semiclassical spectrum.
The semiclassical $A_1$ resonances have been calculated by Wirzba using the
$12^{\rm th}$ order in the curvature expansion of the Gutzwiller-Voros
zeta-function \cite{Cvi97,Wir97,Wir98} and are plotted by the crosses
$(\times)$ in Fig.\ \ref{fig21}c.
The differences between the quantum and semiclassical resonances are usually
much smaller than the size of the symbols in Fig.\ \ref{fig21}c and are
visible only at small values of ${\rm Re}~k$.
However, the deviations become clearly pronounced in Fig.\ \ref{fig21}b,
which presents the difference between the quantum and zeroth order 
semiclassical density of states 
\begin{equation}
   k[\varrho^{\rm qm}(k)-\varrho^{\rm sc}(k)]
 = -{1\over\pi}\, {\rm Im}\, 
   \sum_{\rm po} {\cal A}_{\rm po}^{(1)} e^{i\ell_{\rm po}k}
   + {\cal O}\left({1\over k}\right) \; .
\label{Drho_3disk}
\end{equation}
Eq.\ \ref{Drho_3disk} implies that by harmonic inversion of the spectrum 
in Fig.\ \ref{fig21}b we can extract the first order amplitudes 
${\cal A}_{\rm po}^{(1)}$ of the $\hbar$ expansion of the periodic orbit sum.
In Table \ref{table3} we present the ratio of the zeroth and first order
amplitudes, $|{\cal A}_{\rm po}^{(1)}/{\cal A}_{\rm po}^{(0)}|^{\rm hi}$
obtained by harmonic inversion of the spectra in Fig.\ \ref{fig21}a
and \ref{fig21}b for all periodic orbits up to cycle length 5 in the
symmetry reduced symbolic code \cite{Cvi89}.
These quantities have been calculated by Alonso and Gaspard \cite{Alo93} 
and we present their results 
$|{\cal A}_{\rm po}^{(1)}/{\cal A}_{\rm po}^{(0)}|^{\rm cl}$ in Table
\ref{table3} for comparison.
In Ref.\ \cite{Alo93} the periodic orbits have not been symmetry reduced 
and the set of orbits chosen by Alonso and Gaspard is complete only up to 
cycle length 3 in the symmetry reduced symbolic code.
However, Table \ref{table3} clearly illustrates a very good agreement
between the amplitudes obtained by the harmonic inversion analysis of 
spectra and the results of Ref.\ \cite{Alo93}.

We have here obtained information about the first order terms of the
$\hbar$ expansion of the periodic orbit sum from the analysis of the
quantum and the lowest (zeroth) order semiclassical resonances.
In general, the quantum spectrum and its $n^{\rm th}$ order semiclassical
approximation are required to extract information about the $(n+1)^{\rm st}$
order terms of the $\hbar$ expansion series (\ref{g_hbar_series}).

\section{Periodic orbit quantization by harmonic inversion}
\label{po_quant}
\setcounter{equation}{0}
Since the development of {\em periodic orbit theory} by Gutzwiller 
\cite{Gut67,Gut90} it has become a fundamental question as to how individual
semiclassical eigenenergies and resonances can be obtained from periodic
orbit quantization for classically chaotic systems.
A major problem is the exponential proliferation of the number of periodic 
orbits with increasing period, resulting in a divergence of 
Gutzwiller's trace formula at real energies and below the real axis, where 
the poles of the Green's function are located.
The periodic orbit sum is a Dirichlet series
\begin{equation}
 g(w) = \sum_n A_n e^{i s_n w} \; ,
\label{g}
\end{equation}
where the parameters $A_n$ and $s_n$ are the amplitudes and periods (actions)
of the periodic orbit contributions.
In most applications Eq.\ \ref{g} is absolutely convergent only in the region 
${\rm Im}~w > c > 0$ with $c$ the entropy barrier of the system, while the 
poles of $g(w)$, i.e., the bound states and resonances, are located on and 
below the real axis, ${\rm Im}~w \le 0$.
Thus, to extract individual eigenstates, the semiclassical trace formula 
(\ref{g}) has to be analytically continued to the region of the quantum poles.
Up to now no general procedure is known for the analytic continuation of a
non-convergent Dirichlet series of the type of Eq.\ \ref{g}.
All existing techniques are restricted to special situations.
For bound and ergodic systems the semiclassical eigenenergies can be 
extracted with the help of a functional equation and the mean staircase 
function (Weyl term), resulting in a Riemann-Siegel look-alike formula 
\cite{Ber90,Kea92,Ber92}.
Alternative semiclassical quantization conditions based on a semiclassical
representation of the spectral staircase \cite{Aur92a,Aur92b} and
derived from a quantum version of a classical Poincar\'e map \cite{Bog92b}
are also restricted to bound and ergodic systems.
For systems with a symbolic dynamics the periodic orbit sum (\ref{g}) can
be reformulated as an infinite Euler product, which can be expanded
in terms of the cycle length of the symbolic code.
If the contributions of longer orbits are shadowed by the contributions
of short orbits the cycle expansion technique can remarkably improve the
convergence properties of the series and allows to extract the bound states
and resonances of bound and open systems, respectively
\cite{Cvi89,Eck93,Eck95,Art90}.
A combination of the cycle expansion technique with a functional equation 
for bound systems has been studied by Tanner et al.\ \cite{Tan91}.
However, the existence of a simple symbolic code is restricted to very few
systems, and cycle expansion techniques cannot be applied, e.g., to the general
class of systems with mixed regular-chaotic classical dynamics.

In this chapter we present a general technique for the analytic continuation 
and the extraction of poles of a non-convergent series of the type of 
Eq.\ \ref{g}.
The method is based on {harmonic inversion} by filter-diagonalization.
The advantage of the method is that it does not depend on special 
properties of the system such as ergodicity or the existence
of a symbolic dynamics for periodic orbits.
It does not even require the knowledge of the mean staircase function, i.e.,
the Weyl term in dynamical systems.
The only assumption we have to make is that the analytic continuation of the
Dirichlet series $g(w)$ (Eq.\ \ref{g})
is a linear combination of poles $(w-w_k)^{-1}$, which is exactly the
functional form of, e.g., a quantum mechanical response function with
real and complex parameters $w_k$ representing the bound states and 
resonances of the system, respectively.
To demonstrate the general applicability and accuracy of our method we will 
apply it to three systems with completely different properties, first the 
zeros of the Riemann zeta function \cite{Edw74,Tit86}, as a mathematical 
model for a bound system, second the three disk scattering system as a 
physical example of an open system with classically chaotic dynamics, and
third the circle billiard as an integrable system.

As pointed out by Berry \cite{Ber86} the density of zeros of Riemann's zeta 
function can be written, in formal analogy with Gutzwiller's semiclassical
trace formula, as a non-convergent series, where the ``periodic orbits'' are
the prime numbers.
A special property of this system is the existence of a functional equation
which allows the calculation of Riemann zeros via the Riemann-Siegel formula 
\cite{Edw74,Tit86,Ber86,Odl90}.
An analogous functional equation for quantum systems with an underlying 
chaotic (ergodic) classical dynamics has served as the basis for the 
development of a semiclassical quantization rule for bound ergodic systems
\cite{Ber90,Kea92,Ber92}.
The Riemann zeta function has also served as a mathematical model to study
the statistical properties of level distributions \cite{Odl90,Boh84,Bog95}.
We will demonstrate in Section \ref{zeta_sec} that harmonic inversion can 
reveal the Riemann zeros with extremely high accuracy and with just prime 
numbers as input data.
The most important advantage of our method is, however, its wide 
applicability, i.e., it can be generalized in a straightforward way to 
non-ergodic bound or open systems, and the procedure for periodic orbit
quantization by harmonic inversion will be discussed in Section 
\ref{po_quant_sec}.

As an example of periodic orbit quantization of a physical system 
we investigate in Section \ref{three_disk_sec} the three disk scattering 
problem, which is an open and non-ergodic system.
Its classical dynamics is purely hyperbolic, and the periodic 
orbits can be classified by a complete binary symbolic code.
This system has served as the prototype for the development of 
cycle expansion techniques \cite{Cvi89,Eck93,Eck95}.
When applying the harmonic inversion technique to the three disk scattering
system we will highlight the general applicability of our method by not 
having to make use of its symbolic dynamics in any way.

The power of the method will be illustrated in Section \ref{mixed_sec}
on a challenging physical system which has not been solved previously
with any semiclassical quantization technique.
The hydrogen atom in a magnetic field shows a transition from near-integrable
to chaotic classical dynamics with increasing excitation energy.
We apply periodic orbit quantization by harmonic inversion to the hydrogen
atom in the mixed regular-chaotic regime.

An important question is the efficiency of methods for periodic orbit
quantization, i.e., the number of periodic orbits required for the 
calculation of a certain number of poles of the response function $g(w)$.
It is evident that methods invoking special properties of a given system
may be remarkably efficient.
E.g., the Riemann-Siegel type formulae \cite{Ber90,Kea92,Ber92} require
the periodic orbits up to a maximum period which is by about a factor of
four shorter compared to the required signal length for the harmonic 
inversion technique discussed in Section \ref{po_quant_sec}.
We therefore propose in Section \ref{cross_corr_po_sums} an extension of
the harmonic inversion method to the harmonic inversion of 
{\em cross-correlated} periodic orbit sums.
The method uses additional semiclassical information obtained from a set
of linearly independent smooth observables and allows to significantly
reduce the number of periodic orbits and thus to improve the efficiency 
of periodic orbit quantization by harmonic inversion.

Periodic orbit theory yields exact eigenenergies only in exceptional
cases, e.g., for the geodesic motion on the constant negative curvature 
surface \cite{Gut90}.
As already discussed in Section \ref{hbar_sec},
Gutzwiller's periodic orbit sum is, in general, just the leading order 
term of an infinite series in powers of the Planck constant.
Methods for the calculation of the higher order periodic orbit 
contributions were developed in \cite{Gas93,Alo93,Vat96}.
In Section \ref{hbar_sec2} we demonstrate how the higher order $\hbar$
corrections of the periodic orbit sum can be used to improve the accuracy
of the semiclassical eigenenergies, i.e., to obtain eigenenergies beyond
the standard semiclassical approximation of periodic orbit theory.

Both methods, the harmonic inversion of cross-correlated periodic orbit sums,
which allows to significantly reduce the required number of periodic orbits
for semiclassical quantization, and the calculation of eigenenergies
beyond the lowest order $\hbar$ approximation will be illustrated in Section
\ref{circle_billiard} by way of example of the circle billiard.
Furthermore, for this system we will calculate semiclassically the 
diagonal matrix elements of various operators.
The circle billiard is an integrable system and therefore the Berry-Tabor
formula for integrable systems \cite{Ber76,Ber77} is valid in this case 
rather than Gutzwiller's trace formula.
However, periodic orbit quantization by harmonic inversion can be applied
for both regular and chaotic systems as well which demonstrates the 
universality and wide applicability of the method.

Finally, in Section \ref{photo_sec} we will calculate semiclassically
the photoabsorption spectra of atoms in external fields.
Applying a combination of closed orbit theory \cite{Du87,Du88,Bog88a,Bog89},
the cross-correlation approach of Section \ref{cross_corr_po_sums}, and
the harmonic inversion method we obtain {\em individual} non-diagonal 
transition matrix elements between low-lying initial states and strongly
perturbed Rydberg states of the magnetized hydrogen atom.

\subsection{A mathematical model: Riemann's zeta function}
\label{zeta_sec}
Our goal is to introduce our method for periodic orbit quantization
by harmonic inversion using, as an example, the well defined problem of 
calculating zeros of the Riemann zeta function.
There are essentially two advantages of studying the zeta function instead
of a ``real'' physical bound system.
First, the Riemann analogue of Gutzwiller's trace formula is exact,
as is the case for systems with constant negative curvature 
\cite{Gut90,Aur92b},
whereas the semiclassical trace formula for systems with plane geometry 
is correct only to first order in $\hbar$.
This allows a direct check on the precision of the method.
Second, no extensive periodic orbit search is necessary for the calculation
of Riemann zeros, as the only input data are just prime numbers.
It is not our intention to introduce yet another method for computing
Riemann zeros, which, as an objective in its own right, can be accomplished
more efficiently by specific procedures.
Rather, in our context the Riemann zeta function serves primarily as a 
mathematical model to illustrate the power of our technique when applied
to bound systems.

\subsubsection{General remarks}
Before discussing the harmonic inversion method we start with recapitulating 
a few brief remarks on Riemann's zeta function necessary for our purposes.
The hypothesis of Riemann is that all the non-trivial zeros of the analytic
continuation of the function
\begin{equation}
   \zeta(z)
 = \sum_{n=1}^\infty n^{-z}
 = \prod_p \left(1-p^{-z}\right)^{-1} \; , 
   \quad ({\rm Re}~ z>1,~ p: {\rm primes})
\label{zeta_def}
\end{equation}
have real part $1\over 2$, so that the values $w=w_k$, defined by
\begin{equation}
 \zeta\left({1\over 2}-iw_k\right) = 0,
\end{equation}
are all real or purely imaginary \cite{Edw74,Tit86}.
The Riemann staircase function for the zeros along the line
$z={1\over 2}-iw$, defined as
\begin{equation}
 N(w) = \sum_{k=1}^\infty \Theta(w-w_k),
\label{Riemann_staircase}
\end{equation}
i.e.\ the number $N(w)$ of 
zeros with $w_k<w$, can be split \cite{Edw74,Tit86,Ber86} into a smooth part,
\begin{eqnarray}
   \bar N(w)
 &=& {1\over\pi}\arg\Gamma\left({1\over4}+{1\over2}iw\right)
   -{w\over 2\pi}\ln\pi + 1  \nonumber \\
 &=& {w\over 2\pi} \left(\ln \left\{{w\over 2\pi}\right\}-1\right)
   + {7\over 8} + {1\over 48\pi w} - {7\over 5760\pi w^3}+{\cal O}(w^{-5}) \; ,
\label{N_bar}
\end{eqnarray}
and a fluctuating part,
\begin{equation}
 N_{\rm osc}(w) = - {1\over \pi} \lim_{\eta\to 0} \, {\rm Im} \, 
  \ln \zeta\left({1\over 2}-i(w+i\eta)\right) \; .
\label{N_osc_def}
\end{equation}
Substituting the product formula (\ref{zeta_def}) (assuming that it can 
be used when ${\rm Re}~z={1\over 2}$) into (\ref{N_osc_def})
and expanding the logarithms yields
\begin{equation}
 N_{\rm osc}(w) = - {1\over \pi} \, {\rm Im} \, \sum_p \sum_{m=1}^\infty
 {1\over mp^{m/2}} \, e^{iwm\ln(p)} \; .
\end{equation}
Therefore the density of zeros along the line $z={1\over 2}-iw$ can
formally be written as
\begin{equation}
   \varrho_{\rm osc}(w)
 = \frac{dN_{\rm osc}}{dw}
 = -{1\over \pi} \, {\rm Im} \, g(w)
\label{rho_zeta}
\end{equation}
with the response function $g(w)$ given by the series 
\begin{equation}
g(w) = i\sum_p \sum_{m=1}^\infty \frac{\ln(p)}{p^{m/2}} \, e^{iwm\ln(p)} \ ,
\label{g_nc}
\end{equation}
which converges only for ${\rm Im}\, w > {1\over 2}$.
Obviously Eq.\ \ref{g_nc} is of the same type as the response function
(\ref{g}), with the entropy barrier $c={1\over2}$, i.e., Eq.\ \ref{g_nc} 
does not converge on the real axis, where the Riemann zeros are located.
The mathematical analogy between the above equation and Gutzwiller's
periodic orbit sum
\begin{equation}
 \varrho_{\rm osc}(E) \approx -{1\over \pi} \, {\rm Im} \, 
 \sum_{\rm po} {\cal A}_{\rm po} \, e^{iS_{\rm po}},
\label{rho_po}
\end{equation}
with ${\cal A}_{\rm po}$ the amplitudes and $S_{\rm po}$ the classical 
actions (including phase information) of the periodic orbit contributions,
was already pointed out by Berry \cite{Ber86,Ber90}.
For the Riemann zeta function the primitive periodic orbits have to be
identified with the primes $p$, and the integer $m$ formally counts the
``repetitions'' of orbits.
The ``amplitudes'' and ``actions'' are then given by 
\begin{eqnarray}
\label{Apm}
 {\cal A}_{pm}&=& i{\ln(p)\over p^{m/2}} \; , \\
\label{Spm}
 S_{pm}&=&mw\ln(p) \; .
\end{eqnarray}
Both equation (\ref{g_nc}) for the Riemann zeros and 
-- for most classically chaotic physical systems -- 
the periodic orbit sum (\ref{rho_po}) do not converge. 
In particular, zeros of the zeta function, or semiclassical 
eigenstates, cannot be obtained directly using these expressions.
The problem is to find the analytic continuation of these equations to the
region where the Riemann zeros or, for physical systems, the eigenenergies
and resonances, are located.
Eq.\ \ref{g_nc} is the starting point for our introduction and discussion
of the harmonic inversion technique for the example of the Riemann zeta
function.
The generalization of the method to periodic orbit quantization 
(Eq.\ \ref{rho_po}) in Section \ref{po_quant_sec} will be straightforward.

Although Eq.\ \ref{g_nc} is the starting point for the harmonic inversion
method, for completeness we quote the Riemann-Siegel formula,
which is the most efficient approach to computing Riemann zeros.
For the Riemann zeta function it follows from a functional equation
\cite{Edw74} that the function
\begin{equation}
 Z(w) = \exp\left\{-i\left[\arg \Gamma\left({1\over 4}+{1\over 2}iw\right)
-{1\over 2}w\ln\pi\right]\right\}
        \zeta\left({1\over 2}-iw\right)
\end{equation}
is real, and even for real $w$.
The asymptotic representation of $Z(w)$ for large $w$,
\begin{eqnarray}
 Z(w) &=& -2\sum_{n=1}^{{\rm Int}[\sqrt{w/2\pi}]}
 \frac{\cos\{\pi\bar N(w)-w\ln n\}}{n^{1/2}} \nonumber \\
 &-& (-1)^{{\rm Int}[\sqrt{w/2\pi}]}\left({2\pi\over w}\right)^{1\over4} \;
   {\cos\left(2\pi\left(t^2-t-{1/16}\right)\right)
   \over\cos\left(2\pi t\right)} + \dots \; ,
\label{Riemann_Siegel}
\end{eqnarray}
with $t=\sqrt{w/2\pi}-{\rm Int}[\sqrt{w/2\pi}]$
is known as the Riemann-Siegel formula and has been employed (with several 
more correction terms) in effective methods for computing 
Riemann zeros \cite{Odl90}.
Note that the principal sum in (\ref{Riemann_Siegel}) has discontinuities
at integer positions of $\sqrt{w/2\pi}$, and therefore the Riemann zeros
obtained from the principal sum are correct only to about 1 to 15 percent
of the mean spacing between the zeros.
The higher order corrections to the principal Riemann-Siegel sum remove,
one by one, the discontinuities in successive derivatives of $Z(w)$ at the
truncation points and are thus essential to obtaining accurate numerical 
results.
An alternative method for improving the asymptotic representation of $Z(w)$
by smoothing the cutoffs with an error function and adding higher order 
correction terms is presented in \cite{Ber92}.
An analogue of the functional equation for bound and ergodic dynamical
systems has been used as the starting point to develop a ``rule for quantizing 
chaos'' via a ``Riemann-Siegel look-alike formula'' \cite{Ber90,Kea92,Ber92}.
This method is very efficient as it requires the least number of periodic
orbits, but unfortunately it is restricted to ergodic systems on principle
reasons, and cannot be generalized either to systems with regular or mixed 
classical dynamics or to open systems.
By contrast, the method of harmonic inversion does not have these restrictions.
We will demonstrate that Riemann zeros can be obtained directly from the 
"ingredients" of the non-convergent response function (\ref{g_nc}), i.e., 
the set of values $A_{pm}$ and $S_{pm}$, thus avoiding the use of the 
functional equation, the Riemann-Siegel formula, the mean staircase function 
(\ref{N_bar}), or any other special property of the zeta function.
The comparison of results in Section \ref{zeta_results} will show that the 
accuracy of our method goes far beyond the Riemann-Siegel formula 
(\ref{Riemann_Siegel}) without higher order correction terms.
The main goal of this section is to demonstrate that because of the formal 
equivalence between Eqs.\ (\ref{g_nc}) and (\ref{rho_po}) our method 
can then be applied to periodic orbit quantization of dynamical
systems in Section \ref{po_quant_sec} without any modification.

\subsubsection{The ansatz for the Riemann zeros}
\label{Riemann_ansatz}
To find the analytic continuation of Eq.\ (\ref{g_nc}) in the region
${\rm Im}~ w<{1\over 2}$ we essentially wish to fit $g(w)$ to
its exact functional form, 
\begin{equation}
 g_{\rm ex}(w) = \sum_k {d_k \over w-w_k+i0} \; ,
\label{g_ex}
\end{equation}
arising from the definition of the Riemann staircase 
(\ref{Riemann_staircase}).
The ``multiplicities'' $d_k$ in Eq.\ \ref{g_ex} are formally fitting 
parameters, which here should all be equal to 1.
It is hard to directly adjust the non-convergent (on the real axis)
series $g(w)$ to the form of $g_{\rm ex}(w)$.
The first step towards the solution of the problem is to carry out the 
adjustment for the Fourier components of the response function,
\begin{equation}
  C(s) = {1\over 2\pi}\int_{-\infty}^{+\infty} g(w) e^{-isw}dw
= i\sum_p \sum_{m=1}^\infty \frac{\ln(p)}{p^{m/2}} \, \delta(s-m\ln(p)) \, ,
\label{C_nc}
\end{equation}
which after certain regularizations (see below) is a well-behaved function of 
$s$.
Due to the formal analogy with the results of periodic orbit theory 
(see Eqs.\ \ref{Apm} and \ref{Spm}), $C(s)$ can be interpreted as the 
recurrence function for the Riemann zeta function, with the recurrence 
positions $s_{pm}=S_{pm}/w=m\ln(p)$ and recurrence strengths 
of periodic orbit returns $A_{pm}=i\ln(p)p^{-m/2}$.
The exact functional form which now should be used to adjust $C(s)$ is 
given by 
\begin{equation}
   C_{\rm ex}(s)
 = {1\over 2\pi}\int_{-\infty}^{+\infty} g_{\rm ex}(w) e^{-isw}dw
 = -i \sum_{k=1}^\infty d_k e^{-iw_ks} \; .
\label{C_ex}
\end{equation}
$C_{\rm ex}(s)$ is a superposition of sinusoidal functions with
frequencies $w_k$ given by the Riemann zeros and amplitudes $d_k=1$.
(It is convenient to use the word ``frequencies'' for $w_k$ 
referring to the sinusoidal form of $C(s)$. We will also use the word 
``poles'' in the context of the response function $g(w)$.)

Fitting a signal $C(s)$ to the functional form of Eq.\ \ref{C_ex} with,
in general, both complex frequencies $w_k$ and amplitudes $d_k$ is known as
{harmonic inversion}, and has already been introduced in Section 
\ref{Circumvent} for the high resolution analysis of quantum spectra.
The harmonic inversion analysis is especially non-trivial if the number 
of frequencies in the signal $C(s)$ is large, e.g., more than a thousand. 
It is additionally complicated by the fact that the
conventional way to perform the spectral analysis by studying
the Fourier spectrum of $C(s)$ will bring us back to analyzing the 
non-convergent response function $g(w)$ defined in Eq.\ \ref{g_nc}.
Until recently the known techniques of spectral analysis \cite{Mar87} would 
not be applicable in the present case, and it is the filter-diagonalization 
method \cite{Wal95,Man97a,Man97b} which has turned the harmonic inversion 
concept into a general and powerful computational tool.

The signal $C(s)$ as defined by Eq.\ \ref{C_nc} is not yet suitable for
the spectral analysis. 
The next step is to regularize $C(s)$ by convoluting it with a Gaussian 
function to obtain the smoothed signal,
\begin{eqnarray}
     C_\sigma(s)
 &=& {1\over \sqrt{2\pi}\sigma} \int_{-\infty}^{+\infty}
       C(s')e^{-(s-s')^2/2\sigma^2}ds' \nonumber \\
 &=& {i\over \sqrt{2\pi}\sigma} \sum_p \sum_{m=1}^\infty
       {\ln (p) \over p^{m/2}} \, e^{-(s-m\ln(p))^2 / 2\sigma^2}
\label{C_sigma}
\end{eqnarray}
that has to be adjusted to the functional form of the corresponding 
convolution of $C_{\rm ex}(s)$.
The latter is readily obtained by substituting $d_k$ in Eq.\ \ref{C_ex}
by the damped amplitudes,
\begin{equation}
 d_k \to d_k^{(\sigma)} = d_k \, e^{-w_k^2\sigma^2/2} \; .
\label{d_sigma}
\end{equation}
The regularization (\ref{C_sigma}) can also be interpreted as a cut of an 
infinite number of high frequencies in the signal which is of fundamental 
importance for numerically stable harmonic inversion.
Note that the convolution with the Gaussian function is no approximation,
and the obtained frequencies $w_k$ and amplitudes $d_k$ corrected by Eq.\
\ref{d_sigma} are still exact, i.e., do not depend on $\sigma$.
The convolution is therefore not related to the Gaussian smoothing devised
for Riemann zeros in \cite{Del66} and for quantum mechanics in \cite{Aur88},
which provides low resolution spectra only.

The next step is to analyze the signal (\ref{C_sigma}) by harmonic inversion.
The concept of harmonic inversion by filter-diagonalization has already been
explained in Section \ref{Circumvent} and the technical details are given in
Appendix \ref{harm_inv}.
Note that even though the derivation of Eq.\ \ref{C_sigma} assumed that the 
zeros $w_k$ are on the real axis, the analytic properties of $C_\sigma(s)$ 
imply that its representation by Eq.\ \ref{C_sigma} includes not only the 
non-trivial real zeros, but also all the trivial ones, 
$w_k=-i(2k+{1\over 2})$, $k=1,2,\dots$, which are purely imaginary.
The general harmonic inversion procedure does not require the frequencies 
to be real. 
Both the real and imaginary zeros $w_k$ will be obtained as the eigenvalues 
of the non-Hermitian generalized eigenvalue problem, Eq.\ \ref{eq:generalized}
in Appendix \ref{harm_inv}.

\subsubsection{Numerical results}
\label{zeta_results}
\begin{figure}[b]
\vspace{15.0cm}
\includegraphics{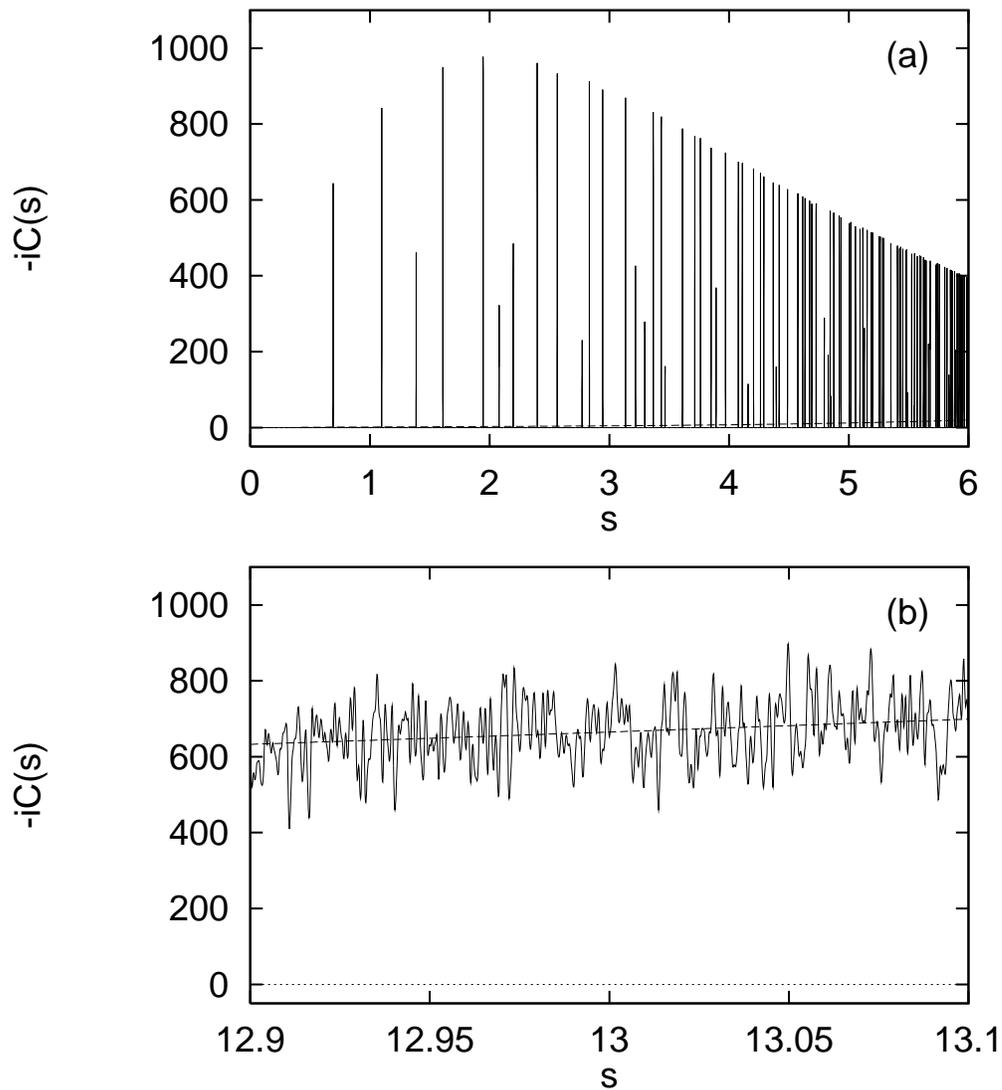}
\caption{\label{fig22} 
``Recurrence'' function $-iC_\sigma(s)$ for the Riemann zeros which has 
been analyzed by harmonic inversion. 
(a) Range $0\le s \le 6$, (b) short range around $s=13$.
The $\delta$-functions have been convoluted by a Gaussian function with width 
$\sigma=0.0003$. Dashed line: Smooth background $\bar C(s)=ie^{s/2}$ 
resulting from the pole of the zeta function at $w=i/2$.
(From Ref.\ [83].)
}
\end{figure}

For a numerical demonstration we construct the signal $C_\sigma(s)$ using 
Eq.\ \ref{C_sigma} in the region $s<\ln(10^6)=13.82$ from the first 78498 
prime numbers and with a Gaussian smoothing width $\sigma=0.0003$.
Parts of the signal are presented in Fig.\ \ref{fig22}.
Up to $s\approx 8$ the Gaussian approximations to the $\delta$-functions
do essentially not overlap (see Fig.\ \ref{fig22}a) whereas for $s\gg 8$ 
the mean spacing $\Delta s$ between successive $\delta$-functions becomes 
much less than the Gaussian width $\sigma=0.0003$ and the signal fluctuates 
around the mean $\bar C(s)=ie^{s/2}$ (see Fig.\ \ref{fig22}b).
From this signal we were able to calculate about 2600 Riemann zeros
to at least 12 digit precision.
For the small generalized eigenvalue problem (see Eq.\ \ref{eq:generalized}
in Appendix \ref{harm_inv}) we used matrices with dimension $J<100$.
Some Riemann zeros $w_k$, the corresponding amplitudes $d_k$, and the 
estimated errors $\varepsilon$ (see Eq.\ \ref{eps_def} in Appendix 
\ref{harm_inv}) are given in Tables \ref{table4} and \ref{table5}.
The pole of the zeta function yields the smooth background 
$\bar C(s)=ie^{s/2}$ of the signal $C(s)$ (see the dashed line in 
Fig.\ \ref{fig22}).
Within the numerical error the Riemann zeros are real and the amplitudes
are consistent with $d_k=1$ for non-degenerate zeros.
To fully appreciate the accuracy of our harmonic inversion technique we note
that zeros obtained from the principal sum of the Riemann-Siegel formula 
(\ref{Riemann_Siegel}) deviate by about 1 to 15 percent of the mean spacing
from the exact zeros.
Including the first correction term in (\ref{Riemann_Siegel}) the 
approximations to the first five zeros read 
$w_1=14.137$, $w_2=21.024$, $w_3=25.018$, $w_4=30.428$, and $w_5=32.933$,
which still significantly deviates from the exact values 
(see Table \ref{table4}).
Considering even higher order correction terms the results will certainly
converge to the exact zeros.
However, the generalization of such higher order corrections to ergodic
dynamical systems is a nontrivial task and requires, e.g., the knowledge
of the terms in the Weyl series, i.e., the mean staircase function after the 
constant \cite{Ber92,Kea94}.
The perfect agreement of our results for the $w_k$ with the exact Riemann 
zeros to full numerical precision is remarkable and clearly demonstrates 
that harmonic inversion by filter-diagonalization is a very powerful and 
accurate technique for the analytic continuation and the extraction of 
poles of a non-convergent series such as Eq.\ \ref{g}.

A few $w_k$ have been obtained (see Table \ref{table6}) which are definitely 
not located on the real axis.
Except for the first at $w=i/2$ they can be identified with the trivial real 
zeros of the zeta function at $z=-2n$; $n=1,2,\dots$
In contrast to the nontrivial zeros with real $w_k$, the numerical accuracy 
for the trivial zeros decreases rapidly with increasing $n$.
The trivial zeros $w_n=-i(2n+{1\over2})$ are the analogue of resonances
in open physical systems with widths increasing with $n$.
The fact that the trivial Riemann zeros are obtained emphasizes the
universality of our method and demonstrates that periodic orbit quantization 
by harmonic inversion can be applied not only to closed but to open systems 
as well.
The decrease of the numerical accuracy for very broad resonances is a 
natural numerical consequence of the harmonic inversion procedure 
\cite{Man97a,Man97b}.
The value $w=i/2$ in Table \ref{table6} is special because in this case the amplitude
is negative, i.e., $d_k=-1$.
Writing the zeta function in the form \cite{Tit86}
\begin{equation}
 \zeta({1\over 2}-iw) = C \prod_k (w-w_k)^{d_k} A(w,w_k)
\end{equation}
where $C$ is a constant and $A$ a regularizing function which ensures
convergence of the product, integer values $d_k$ are the multiplicities of 
{\em zeros}.
Therefore it is reasonable to relate negative integer values with the 
multiplicities of {\em poles}.
In fact, $\zeta(z)$ has a simple pole at $z={1\over 2}-iw=1$ consistent with
$w=i/2$ in Table \ref{table6}.

\subsubsection{Required signal length}
\label{Req_signal_len}
\begin{figure}[b]
\vspace{9.0cm}
\includegraphics{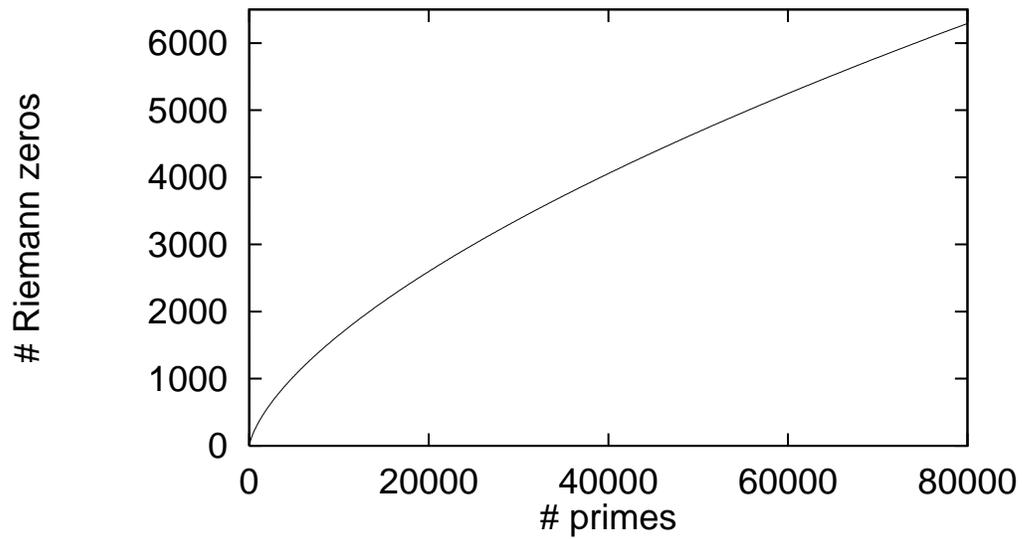}
\caption{\label{fig23} 
Estimated number of converged zeros of the Riemann zeta function, which 
can be obtained by harmonic inversion for given number of primes $p$.
(From Ref.\ [83].)
}
\end{figure}

We have calculated Riemann zeros by harmonic inversion of the signal
$C_\sigma(s)$ (Eq.\ \ref{C_sigma}) which uses prime numbers as input.
The question arises what are the requirements on the signal $C_\sigma(s)$,
in particular what is the required signal length.
In other words, how many Riemann zeros (or semiclassical eigenenergies)
can be converged for a given set of prime numbers (or periodic orbits,
respectively)?
The answer can be directly obtained from the requirements on the harmonic
inversion technique.
In general, the required signal length $s_{\rm max}$ for harmonic inversion 
is related to the average density of frequencies $\bar\varrho(w)$ 
by \cite{Man97b}
\begin{equation}
 s_{\rm max} \approx 4\pi\bar\varrho(w) \; ,
\label{s_max}
\end{equation}
i.e., $s_{\rm max}$ is about two times the Heisenberg length 
\begin{equation}
s_{\rm Heisenberg} \equiv 2\pi\bar\varrho(w) \; .
\end{equation}
From Eq.\ \ref{s_max} the required number of primes (or periodic orbits)
can be directly estimated as 
$\{\#~{\rm primes}~p~|~\ln p < s_{\rm max}\}$ or 
$\{\#~{\rm periodic~orbits}~|~s_{\rm po} < s_{\rm max}\}$.
For the special example of the Riemann zeta function the required number
of primes to have a given number of Riemann zeros converged can be estimated 
analytically.
With the average density of Riemann zeros derived from (\ref{N_bar}),
\begin{equation}
   \bar\varrho(w) = \frac{d\bar N}{dw}
 = {1\over 2\pi}\ln\left({w\over 2\pi}\right) \; ,
\end{equation}
we obtain
\begin{equation}
 s_{\rm max} = \ln(p_{\rm max}) = 2\ln\left({w\over 2\pi}\right)
 \Rightarrow p_{\rm max} = \left({w\over 2\pi}\right)^2 \; .
\end{equation}
The number of primes with $p<p_{\rm max}$ can be estimated from the
prime number theorem
\begin{equation}
   \pi(p_{\rm max}) \sim \frac{p_{\rm max}}{\ln(p_{\rm max})}
 = \frac{(w/2\pi)^2}{2\ln(w/2\pi)} \; .
\end{equation}
On the other hand the number of Riemann zeros as a function of $w$ is
given by Eq.\ (\ref{N_bar}).
The estimated number of Riemann zeros which can be obtained by harmonic 
inversion from a given set of primes is presented in Fig.\ \ref{fig23}.
For example, about 80 zeros $(w<200)$ can be extracted 
from the short signal $C_\sigma(s)$ with $s_{\rm max}=\ln(1000)=6.91$
(168 prime numbers) in agreement with the estimates given above.
Obviously, in the special case of the Riemann zeta function the efficiency 
of our method cannot compete with that of the Riemann-Siegel formula method
(\ref{Riemann_Siegel}) where the number of terms is given by
$n_{\rm max} = {\rm Int}\,[\sqrt{w/2\pi}]$ and, e.g., 5 terms in Eq.\
\ref{Riemann_Siegel} would be sufficient to calculate good approximations
to the Riemann zeros in the region $w<200$.
Our primary intention is to introduce harmonic inversion by way of example
of the zeros of the Riemann zeta function as a universal tool for
periodic orbit quantization, and not to use it as an alternative method
for solving the problem of finding most efficiently zeros of the Riemann
zeta function.
For the semiclassical quantization of bound and ergodic systems a functional
equation can be invoked to derive a Riemann-Siegel look-alike quantization
condition \cite{Ber90,Kea92,Ber92}.
In this case the required number of periodic orbits can be estimated from 
the condition 
\begin{equation}
s_{\rm max} \approx \pi\bar\varrho(w) = {1\over 2} s_{\rm Heisenberg} \; ,
\end{equation}
which differs by a factor of 4 from the required signal length (\ref{s_max}) 
for harmonic inversion.
Obviously, the limitation (\ref{s_max}) is unfavorable for periodic orbit
quantization because of the exponential proliferation of the number of orbits 
in chaotic systems, and methods for a significant reduction of the signal 
length $s_{\rm max}$ are highly desirable.
In fact, this can be achieved by an extension of the harmonic inversion 
technique to cross-correlation signals.
We will return to this problem in  Section \ref{cross_corr_po_sums}.

\subsection{Periodic orbit quantization}
\label{po_quant_sec}
As mentioned in Section \ref{zeta_sec} the basic equation (\ref{g_nc}) used
for the calculation of Riemann zeros has the same mathematical
form as Gutzwiller's semiclassical trace formula.
Both series, Eq.\ \ref{g_nc} and the periodic orbit sum (\ref{rho_po}), 
suffer from similar convergence problems in that they are absolutely
convergent only in the complex half-plane outside the region where the
Riemann zeros, or quantum eigenvalues, respectively, are located.
As a consequence, in a direct summation of periodic orbit contributions 
smoothing techniques must be applied resulting in low resolution spectra
for the density of states \cite{Win88,Aur88}.
To extract individual eigenstates the semiclassical trace formula has to
be analytically continued to the region of the quantum poles, and this
was the subject of intense research during recent years.

\begin{figure}[b]
\vspace{17.5cm}
\includegraphics{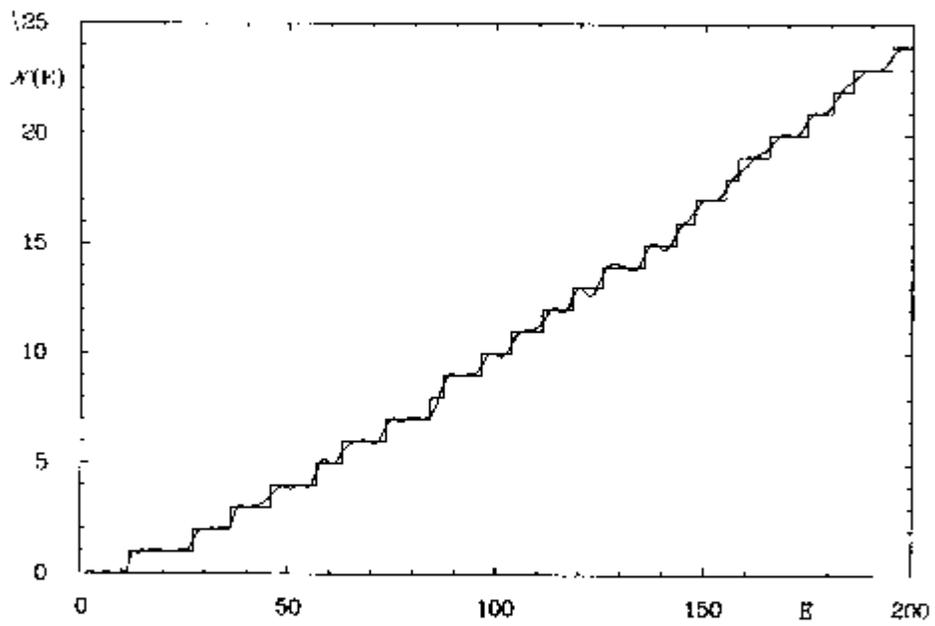}
\caption{\label{fig24} 
The spectral staircase ${\cal N}^+(E)$ and its semiclassical approximation
${\cal N}^+_{\rm sc}(E)$ for the hyperbolic billiard.
${\cal N}^+_{\rm sc}(E)$ was calculated using 101265 periodic orbits.
(From Ref.\ [65].)
}
\end{figure}

\begin{figure}[b]
\vspace{17.5cm}
\includegraphics{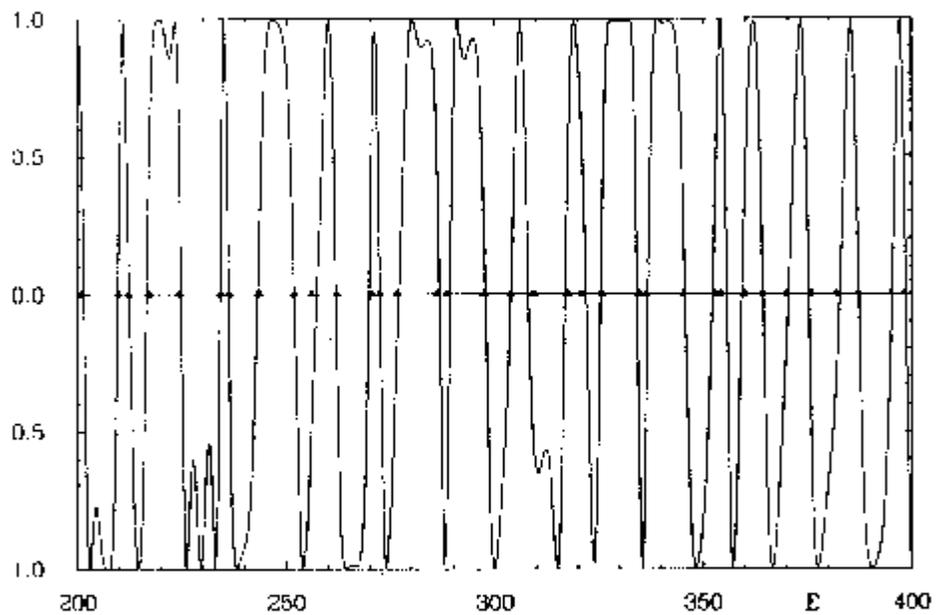}
\caption{\label{fig25} 
The function $\cos\{\pi{\cal N}^+_{\rm sc}(E)\}$ for the hyperbolic billiard.
${\cal N}^+_{\rm sc}(E)$ was evaluated as in Fig.\ \ref{fig24}.
The triangles mark the positions of the true quantum mechanical energies.
(From Ref.\ [65].)
}
\end{figure}

For strongly chaotic bound systems a semiclassical quantization condition
based on a semiclassical representation of the spectral staircase 
\begin{equation}
 {\cal N}(E) = \sum_n\Theta(E-E_n)
\end{equation}
has been developed by Aurich et al.\ in Refs.\ \cite{Aur92a,Aur92b}.
They suggest to replace the spectral staircase ${\cal N}(E)$ with a smooth
semiclassical approximation ${\cal N}_{\rm sc}(E)$, and to evaluate the
semiclassical eigenenergies from the quantization condition
\begin{equation}
 \cos\left\{\pi{\cal N}_{\rm sc}(E)\right\} = 0  \; .
\label{Aur_quant:eq}
\end{equation}
For the even parity states of the hyperbola billiard the exact spectral
staircase ${\cal N}^+(E)$ and the semiclassical approximation 
${\cal N}^+_{\rm sc}(E)$ evaluated from 101265 periodic orbits are shown 
in Fig.\ \ref{fig24}.
The function $\cos\left\{\pi{\cal N}^+_{\rm sc}(E)\right\}$ obtained from
the smooth spectral staircase ${\cal N}^+_{\rm sc}(E)$ in Fig.\ \ref{fig24}
is presented in Fig.\ \ref{fig25}.
According to the quantization condition (\ref{Aur_quant:eq}) the zeros
of this function are the semiclassical energies, and are in good agreement 
with the true quantum mechanical energies marked by triangles in Fig.\ 
\ref{fig25}.
The quantization condition (\ref{Aur_quant:eq}) was also successfully 
applied to the motion of a particle on various Riemann surfaces with
constant negative curvature, e.g., Artin's billiard or the Hadamard-Gutzwiller
model \cite{Aur92a,Aur92b}.
Another technique for {\em bound} and {\em ergodic} systems is to apply an 
approximate functional equation and generalize the Riemann-Siegel formula 
(\ref{Riemann_Siegel}) to dynamical zeta functions \cite{Ber90,Kea92,Ber92}.
The Riemann-Siegel look-alike formula has been applied, e.g., for the 
semiclassical quantization of the hyperbola billiard \cite{Kea94}.

These quantization techniques cannot be applied to {\em open} systems.
However, if a symbolic dynamics for the system exists, i.e., if the
periodic orbits can be classified with the help of a complete symbolic code,
the dynamical zeta function, given as an infinite Euler product over entries 
from classical periodic orbits, can be expanded in terms of the cycle length 
of the orbits \cite{Cvi89,Art90}.
The {\em cycle expansion} series is rapidly convergent if the contributions 
of long orbits are approximately shadowed by contributions of short orbits.
The cycle expansion technique has been applied, e.g., to the three disk
scattering system \cite{Cvi89,Eck93,Eck95} (see also Section 
\ref{three_disk_sec}), the three body Coulomb system \cite{Win92,Ezr91}, 
and to the hydrogen atom in a magnetic field \cite{Tan96}.

\begin{figure}[b]
\vspace{17.5cm}
\includegraphics{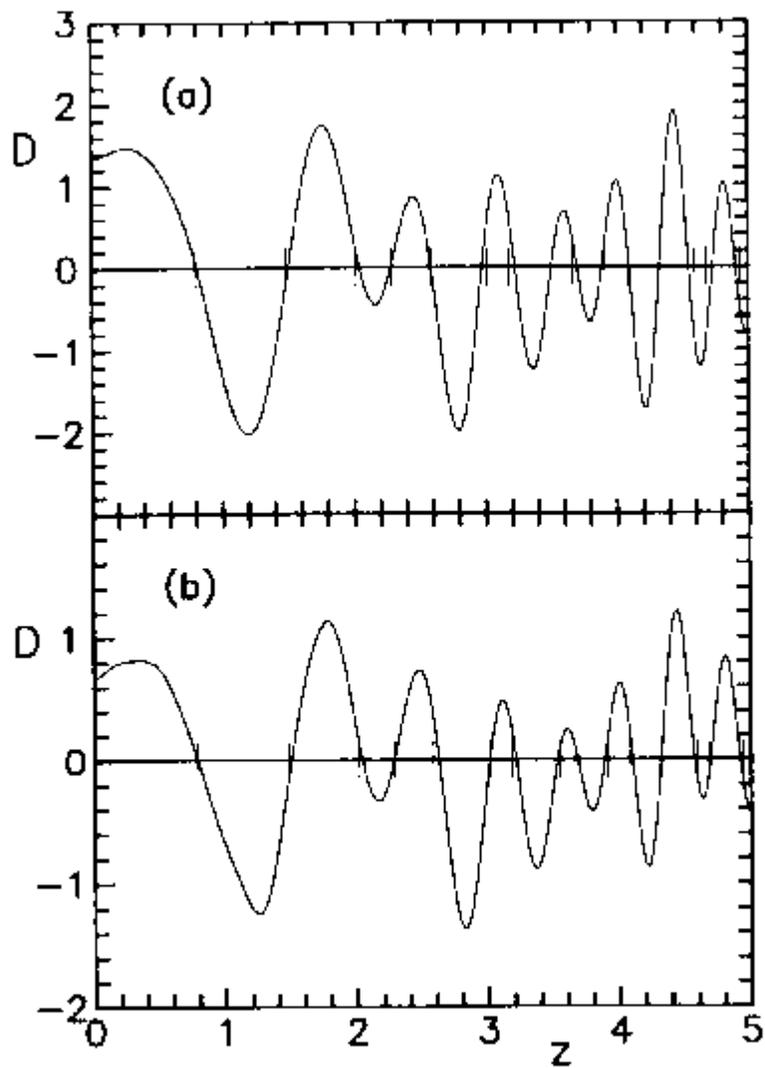}
\caption{\label{fig26} 
The cycle-expanded functional determinant for the $m^\pi=0^+$ subspace
of the anisotropic Kepler problem.
The quantum eigenvalues are marked with vertical bars on the real axis.
(a) All orbits up to length 4 (8 in number) included.
(b) All orbits up to length 8 (71) included.
(From Ref.\ [114].)
}
\end{figure}

\begin{figure}[b]
\vspace{17.5cm}
\includegraphics{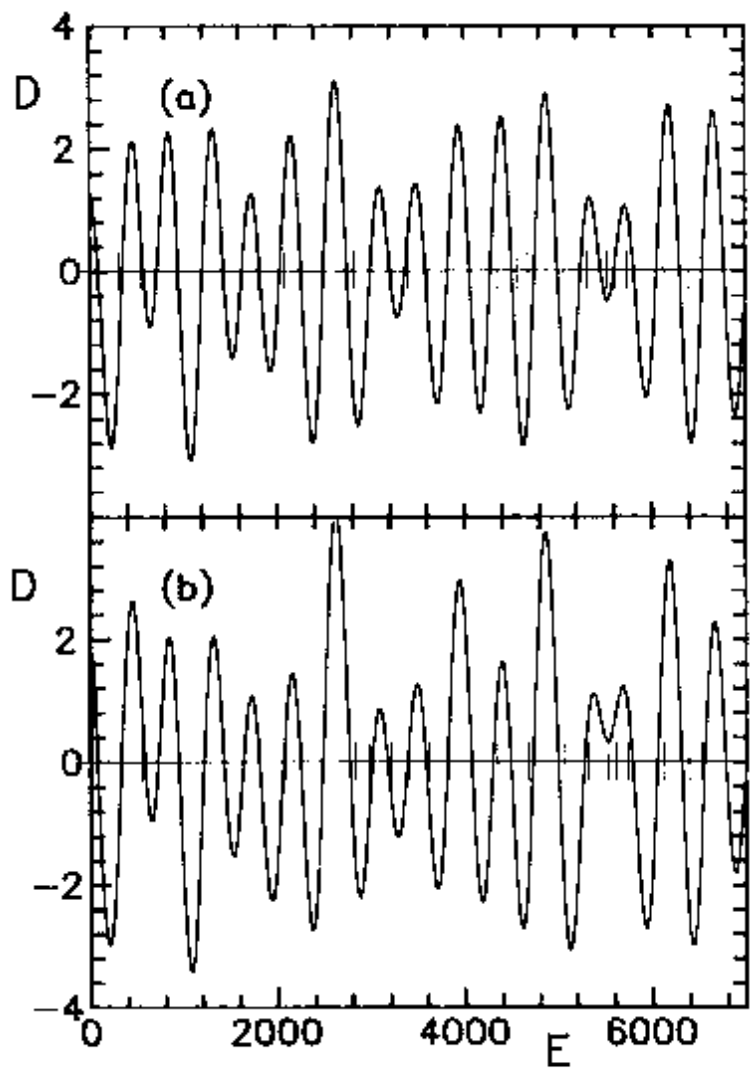}
\caption{\label{fig27} 
The cycle-expanded functional determinant for the closed three disk
billiard (disk radius $R=1$).
The vertical bars mark the exact quantum eigenvalues.
(a) All orbits up to length 2 (3 in number) included.
(b) All orbits up to length 3 (5) included.
(From Ref.\ [114].)
}
\end{figure}

It turns out that the cycle expansion of dynamical zeta functions converges
very slowly for bound systems.
Therefore, a combination of the cycle expansion method with a functional 
equation has been developed by Tanner et al.\ \cite{Tan91,Tan92}.
Applying a functional equation to the dynamical zeta function they conclude
that the zeros of the real expression
\begin{equation}
 D(E) = e^{-i\pi\bar{\cal N}(E)}Z(E) + e^{i\pi\bar{\cal N}(E)}Z^\ast(E) \; ,
\end{equation}
with $\bar{\cal N}(E)$ the mean spectral staircase and $Z(E)$ the cycle
expanded dynamical zeta function should be semiclassical approximations
to the eigenvalues.
The semiclassical functional determinant $D(E)$ is illustrated in Fig.\
\ref{fig26} for the anisotropic Kepler problem \cite{Gut90} and in
Fig.\ \ref{fig27} for the closed three disk billiard, i.e., the system
shown in Fig.\ \ref{fig20} with touching disks.
The vertical bars in Figs.\ \ref{fig26} and \ref{fig27} mark the exact 
eigenvalues.

The existence of a complete symbolic dynamics is more the exception than 
the rule, and therefore cycle expansion techniques cannot be applied, 
in particular, for systems with mixed regular-chaotic classical dynamics.
In this section we apply the same technique that we used for the calculation 
of Riemann zeros to the calculation of semiclassical eigenenergies 
and resonances of physical systems by harmonic inversion of Gutzwiller's 
periodic orbit sum for the propagator.
The method only requires the knowledge of all orbits up to a sufficiently 
long but finite period and does neither rely on an approximate semiclassical
functional equation, nor on the existence of a symbolic code for the orbits.
The universality of the method will therefore allow the investigation of a 
large variety of systems with an underlying chaotic, regular, or even mixed
classical dynamics.
The derivation of an expression for the recurrence function to be harmonically
inverted is analogous to that in Section \ref{Riemann_ansatz}.

\subsubsection{Semiclassical density of states}
Following Gutzwiller \cite{Gut67,Gut90} the semiclassical response function
for chaotic systems is given by
\begin{equation}
 g^{\rm sc}(E) = g^{\rm sc}_0(E)
   + \sum_{\rm po} {\cal A}_{\rm po} e^{iS_{\rm po}} \; ,
\label{g_sc}
\end{equation}
where $g^{\rm sc}_0(E)$ is a smooth function and the $S_{\rm po}$ and 
${\cal A}_{\rm po}$ are the classical actions and weights (including phase 
information given by the Maslov index) of periodic orbit contributions.
Eq.\ (\ref{g_sc}) is also valid for integrable \cite{Ber76,Ber77} and
near-integrable \cite{Tom95,Ull96} systems but with different expressions 
for the amplitudes ${\cal A}_{\rm po}$.
It should also be possible to include complex ``ghost'' orbits 
\cite{Kus93,Mai97a} and uniform semiclassical approximations
\cite{Alm87,Mai98a} close to bifurcations of periodic orbits
in the semiclassical response function (\ref{g_sc}).
The eigenenergies and resonances are the poles of the response function
but, unfortunately, its semiclassical approximation (\ref{g_sc}) does
not converge in the region of the poles, whence the problem is the analytic 
continuation of $g^{\rm sc}(E)$ to this region.

In the following we make the (weak) assumption that the classical system has 
a scaling property (see Section \ref{FT_spectra}), i.e., the shape of periodic
orbits does not depend on the scaling parameter, $w$, and the classical action
scales as 
\begin{equation}
 S_{\rm po} = ws_{\rm po} \; .
\label{S_po}
\end{equation}
Examples of scaling systems are billiards \cite{Cvi89,Hel84}, 
Hamiltonians with homogeneous potentials \cite{Mar89,Tom91}, 
Coulomb systems \cite{Win92}, or the hydrogen atom in external magnetic 
and electric fields \cite{Mai94a,Tan96}.
Eq.\ \ref{S_po} can even be applied for non-scaling, e.g., molecular systems
if a generalized scaling parameter $w\equiv\hbar_{\rm eff}^{-1}$ is introduced
as a new dynamical variable \cite{Mai97e}.
Quantization yields bound states or resonances, $w_k$, for the scaling 
parameter.
In scaling systems the semiclassical response function $g^{\rm sc}(w)$ can be 
Fourier transformed easily to obtain the semiclassical trace of the propagator
\begin{equation}
   C^{\rm sc}(s) 
 = {1 \over 2\pi} \int_{-\infty}^{+\infty} g^{\rm sc}(w) e^{-isw} dw
 = \sum_{\rm po} {\cal A}_{\rm po} \delta\left(s-s_{\rm po}\right) \; .
\label{C_sc}
\end{equation}
The signal $C^{\rm sc}(s)$ has $\delta$-peaks at the positions of the 
classical periods (scaled actions) $s=s_{\rm po}$ of periodic orbits and 
with peak heights (recurrence strengths) ${\cal A}_{\rm po}$, i.e., 
$C^{\rm sc}(s)$ is Gutzwiller's periodic orbit recurrence function.
Consider now the quantum mechanical counterparts of $g^{\rm sc}(w)$ and
$C^{\rm sc}(w)$ taken as the sums over the poles $w_k$ of the Green's 
function,
\begin{equation}
 g^{\rm qm}(w) = \sum_k {d_k \over w-w_k+i0} \; ,
\label{g_qm}
\end{equation}
\begin{equation}
   C^{\rm qm}(s)
 = {1\over 2\pi} \int_{-\infty}^{+\infty} g^{\rm qm}(w) e^{-isw} dw
 = -i\sum_k d_k e^{-i w_k s} \; ,
\label{C_qm}
\end{equation}
with $d_k$ being the multiplicities of resonances, i.e., $d_k=1$ for 
non-degenerate states.
In analogy with the calculation of Riemann zeros from Eq.\ (\ref{C_sigma})
the frequencies, $w_k$, and amplitudes, $d_k$, can be extracted by
harmonic inversion of the signal $C^{\rm sc}(s)$ after convoluting it
with a Gaussian function, i.e.,
\begin{equation}
   C_\sigma^{\rm sc}(s) 
 = {1\over \sqrt{2\pi}\sigma}
   \sum_{\rm po} {\cal A}_{\rm po} e^{(s-s_{\rm po})^2/2\sigma^2} \; .
\label{C_sc_sigma}
\end{equation}
By adjusting $C_\sigma^{\rm sc}(s)$ to the functional form of Eq.\ \ref{C_qm},
the frequencies, $w_k$, can be interpreted as the semiclassical approximation 
to the poles of the Green's function in (\ref{g_qm}).
Note that the harmonic inversion method described in Appendix \ref{harm_inv} 
allows studying signals with complex frequencies $w_k$ as well.
For open systems the complex frequencies can be interpreted as semiclassical 
resonances. 
Note also that the $w_k$ in general differ from the exact quantum eigenvalues
because Gutzwiller's trace formula (\ref{g_sc}) is an approximation, 
correct only to the lowest order in $\hbar$.
Therefore the diagonalization of small matrices 
(Eq.\ \ref{eq:generalized} in Appendix \ref{harm_inv})
does not imply that the results of the periodic orbit quantization are more
``quantum'' in any sense than those obtained, e.g., from a cycle expansion
\cite{Cvi89,Art90}.
However, the harmonic inversion technique also allows the calculation of
higher order $\hbar$ corrections to the periodic orbit sum, and we will
return to this problem in Section \ref{hbar_sec2}.

\subsubsection{Semiclassical matrix elements}
The procedure described above can be generalized in a straightforward manner
to the calculation of semiclassical diagonal matrix elements 
$\langle\psi_k|\hat A|\psi_k\rangle$ of a smooth Hermitian operator
$\hat A$.
In this case we start from the quantum mechanical trace formula 
\begin{equation}
 g_A^{\rm qm}(w) = {\rm tr} \, G^+ \hat A
 = \sum_k {\langle\psi_k|\hat A|\psi_k\rangle \over w-w_k+i0} \; ,
\label{gA_qm}
\end{equation}
which has the same functional form as (\ref{g_qm}), but with 
$d_k=\langle\psi_k|\hat A|\psi_k\rangle$ instead of $d_k=1$.
For the quantum response function $g_A^{\rm qm}(w)$ (Eq.\ \ref{gA_qm})
a semiclassical approximation has been derived in \cite{Wil88,Eck92}, 
which has the same form as Gutzwiller's trace formula (\ref{g_sc}) but
with amplitudes
\begin{equation}
 {\cal A}_{\rm po} = -i {A_p e^{-i{\pi\over 2}\mu_{\rm po}}
 \over \sqrt{|\det(M_{\rm po}-I)|}}
\label{A_po,A}
\end{equation}
where $M_{\rm po}$ is the monodromy matrix and $\mu_{\rm po}$ the Maslov
index of the periodic orbit, and
\begin{equation}
 A_p = \int_0^{s_p} A({\bf q}(s),{\bf p}(s)) ds
\label{A_cl}
\end{equation}
is the classical integral of the observable $A$ over {\em one} period $s_p$ 
of the {\em primitive} periodic orbit.
Note that ${\bf q}(s)$ and ${\bf p}(s)$ are functions of the classical action
instead of time for scaling systems \cite{Boo95}.
Gutzwiller's trace formula for the density of states is obtained with
$\hat A$ being the identity operator, i.e., $A_p=s_p$.
When the semiclassical signal $C^{\rm sc}(s)$ (Eq.\ \ref{C_sc}) with
amplitudes ${\cal A}_{\rm po}$ given by Eqs.\ \ref{A_po,A} and \ref{A_cl}
is analyzed with the method of harmonic inversion the frequencies and 
amplitudes obtained are the semiclassical approximations to the eigenvalues 
$w_k$ and matrix elements $d_k=\langle\psi_k|\hat A|\psi_k\rangle$,
respectively.

\subsection{The three disk scattering system}
\label{three_disk_sec}
Let us consider the three disk scattering system which has served as a model
for periodic orbit quantization by cycle expansion techniques 
\cite{Cvi89,Eck93,Eck95,Wir92,Wir93,Cvi97,Wir97}.
For a brief introduction of the system, its symbolic dynamics and symmetries 
we refer the reader to Section \ref{three_disk_analysis}.
The starting point for the cycle expansion technique is to rewrite
Gutzwiller's trace formula as a dynamical zeta function \cite{Rue87}
\begin{equation}
 1/\xi = \prod_p (1-t_p) \; ,
\label{dyn_zeta_def}
\end{equation}
with $p$ indicating the primitive periodic orbits,
\begin{equation}
 t_p \equiv {1\over\sqrt{\Lambda_p}}\, e^{i[s_pw-{\pi\over 2}\mu_p]} \; ,
\label{tp_def}
\end{equation}
$\mu_p$ the Maslov index, and $\Lambda_p$ the largest eigenvalue of the 
monodromy matrix.
[The dynamical zeta function, Eqs.\ \ref{dyn_zeta_def} and \ref{tp_def},
is based on an approximation. By contrast, the Gutzwiller-Voros zeta function 
\cite{Gut88,Vor88} is exactly equivalent to Gutzwiller's trace formula.
See Refs.\ \cite{Cvi97,Wir97} for more details.]
The semiclassical eigenvalues $w$ are obtained as zeros of the dynamical
zeta function (\ref{dyn_zeta_def}).
However, the convergence problems of the infinite product (\ref{dyn_zeta_def})
are similar to the convergence problems of Gutzwiller's periodic orbit sum.
The basic observation for the three disk system with sufficiently large
distance $d$ was that the periodic orbit quantities of long orbits can be
approximated by the periodic orbit quantities of short orbits, i.e.,
for two orbits with symbolic codes $p_1$ and $p_2$ we have
\begin{eqnarray}
 s_{p_1p_2} &\approx& s_{p_1} + s_{p_2} \nonumber \\
 \Lambda_{p_1p_2} &\approx& \Lambda_{p_1} \Lambda_{p_2} \\
 \mu_{p_1p_2} &=& \mu_{p_1} + \mu_{p_2} \; , \nonumber
\label{po_relations}
\end{eqnarray}
and thus
\begin{equation}
 t_{p_1p_2} \approx t_{p_1} t_{p_2} \; .
\end{equation}
This implies that in the cycle expansion of the dynamical zeta function
(\ref{dyn_zeta_def}) the contributions of long periodic orbits are shadowed
by the contributions of short orbits.
E.g., for the three disk scattering system the cycle expansion up to cycle
length $n=3$ reads
\begin{equation}
 1/\xi = 1 - t_0 -  t_1 - [t_{01}-t_0 t_1]
       - [t_{001} - t_0 t_{01} + t_{011} - t_{01} t_1] - \dots
\label{cycle_exp}
\end{equation}
The terms $t_0$ and $t_1$ are the fundamental contributions, while
the terms in brackets are the curvature corrections, ordered by cycle length,
and can rapidly decrease with increasing cycle length.
With the cycle expansion (\ref{cycle_exp}) the dynamical zeta function
(\ref{dyn_zeta_def}) is analytically continued to the physically important
region below the real axis, where the resonances are located.
For $d=6$ semiclassical resonances were calculated by application
of the cycle expansion technique including all (symmetry reduced)
periodic orbits up to cycle length $n=13$ \cite{Eck95,Cvi97,Wir97}.

\begin{figure}[b]
\vspace{18.0cm}
\includegraphics{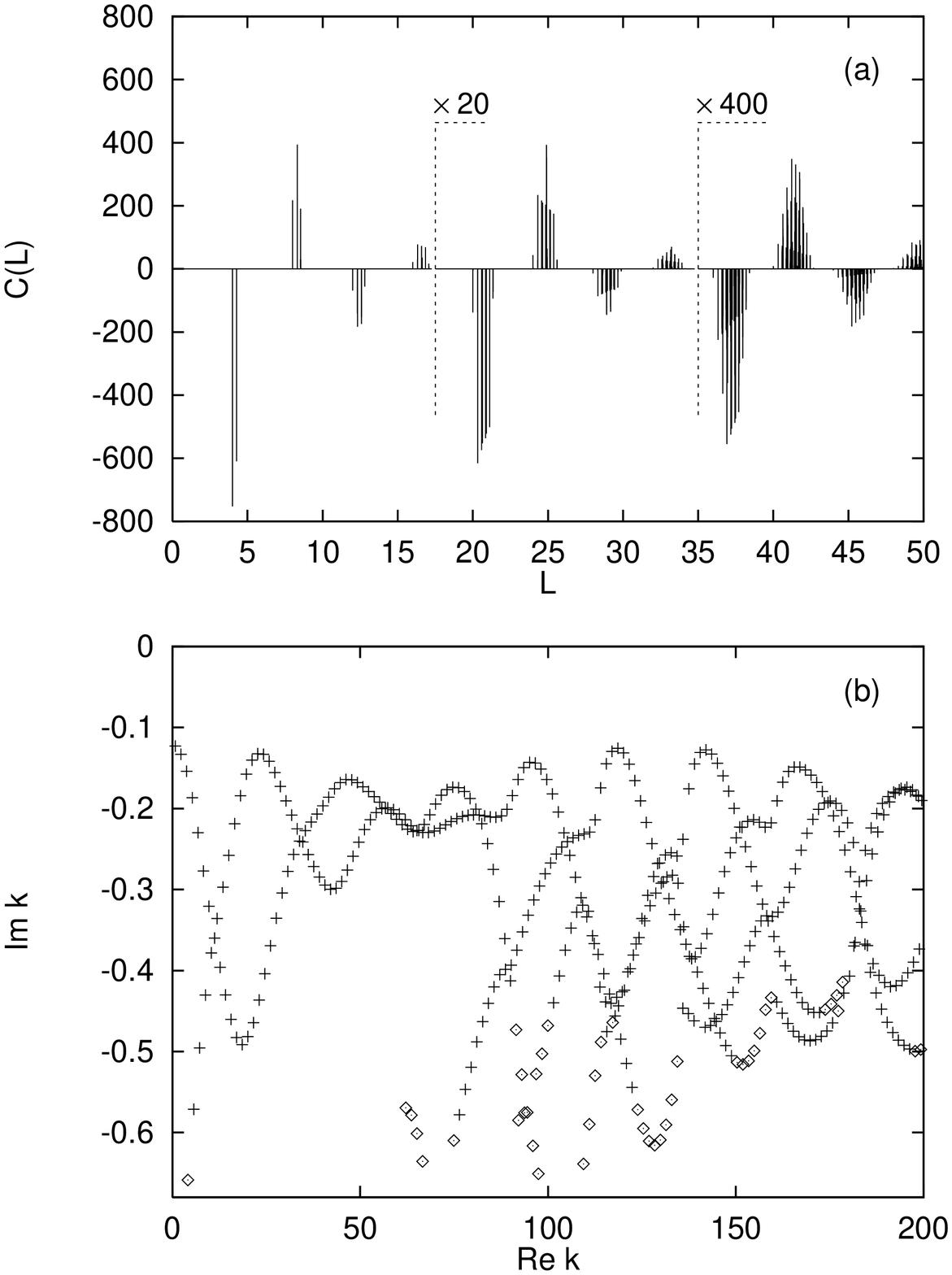}
\caption{\label{fig28} 
Three disk scattering system ($A_1$ subspace) with $R=1$, $d=6$.
(a) Periodic orbit recurrence function, $C(L)$.
The signal has been convoluted with a Gaussian function of width 
$\sigma=0.0015$.
(b) Semiclassical resonances. 
(From Ref.\ [82].)
}
\end{figure}

We now demonstrate the usefulness of the harmonic inversion technique 
for the semiclassical quantization of the three disk scattering system.
In contrast to the cycle expansion we will not make use of the symbolic
code of the periodic orbits and the approximate relations (\ref{po_relations})
between the periodic orbit quantities.
As in the previously discussed examples for billiard systems the scaled action
$s$ is given by the length $L$ of orbits $(s=L)$ and the quantized parameter 
is the absolute value of the wave vector $k=|{\bf k}|=\sqrt{2mE}/\hbar$.
For the three disk system the periodic orbit signal $C(L)$ reads
\begin{equation}
 C(L) = -i \sum_p\sum_{r=1}^\infty {\cal N}_p
  {L_p e^{-ir{\pi\over 2}\mu_p} \over \sqrt{|\det(M_p^r-I)|}} \,
  \delta(L-rL_p) \; ,
\label{C_3disk}
\end{equation}
with $L_p$, $M_p$, and $\mu_p$ the geometrical length, monodromy matrix, 
and Maslov index of the symmetry reduced primitive periodic orbit $p$, 
respectively.
The weight factors ${\cal N}_p$ depend on the irreducible subspace
($A_1$, $A_2$, and $E$), where resonances are calculated, and the
symmetries of the orbits.
Periodic orbits without symmetries, with rotational symmetry, and with
symmetry under reflection are characterized by $e$, $C_3,C_3^2$, and 
$\sigma_v$, respectively.
The weight factors ${\cal N}_p$ are the same as for the cycle expansion, 
and are given in Table \ref{table7}.
For details of the symmetry decomposition see Ref.\ \cite{Cvi93}.
In the following we calculate resonances in the irreducible subspace $A_1$,
i.e., ${\cal N}_p=1$ in (\ref{C_3disk}) for all orbits.
We first apply harmonic inversion to the case $R:d=1:6$ studied before.
Fig.\ \ref{fig28}a shows the periodic orbit recurrence function, i.e., 
the trace of the semiclassical propagator $C^{\rm sc}(L)$.
The groups with oscillating sign belong to periodic orbits with adjacent
cycle lengths.
To obtain a smooth function on an equidistant grid, which is required for
the harmonic inversion method, the $\delta$-functions in (\ref{C_3disk})
have been convoluted with a Gaussian function of width $\sigma=0.0015$.
As explained in Section \ref{Riemann_ansatz} this does not change the 
underlying spectrum.
The results of the harmonic inversion analysis of this signal are
presented in Fig.\ \ref{fig28}b.
The crosses in Fig.\ \ref{fig28}b represent semiclassical poles, for which 
the amplitudes $d_k$ are very close to $1$, mostly within one percent.
Because the amplitudes converge much slower than the frequencies these
resonance positions can be assumed to be very accurate within the 
semiclassical approximation.
For some broad resonances marked by diamonds in Fig.\ \ref{fig28}b
the $d_k$ deviate strongly from $1$, within 5 to maximal 50 percent.
It is not clear whether these strong deviations are due to numerical 
effects, such as convergence problems caused by too short a signal, or 
if they are a consequence of the semiclassical approximation.
A direct comparison between the semiclassical resonances $k^{\rm hi}$
obtained by harmonic inversion of the periodic orbit sum, and results
of Wirzba \cite{Cvi97,Wir97,Wir98}, i.e., the cycle expansion resonances 
$k^{\rm ce}$ calculated up to $12^{\rm th}$ order in the curvature expansion
of the Gutzwiller-Voros zeta function, and the exact quantum resonances 
$k^{\rm qm}$ is given in 
Table \ref{table8} for resonances with ${\rm Re}~k<30$ and in 
Table \ref{table9} for $120<{\rm Re}~k<132$.
Apart from a few resonances with large imaginary parts the differences
$|k^{\rm hi}-k^{\rm ce}|$ between the semiclassical resonances obtained by 
harmonic inversion and cycle expansion are by several orders of magnitude 
smaller than the semiclassical error, i.e., the differences 
$|k^{\rm hi}-k^{\rm qm}|$.

\begin{figure}[b]
\vspace{9.5cm}
\includegraphics{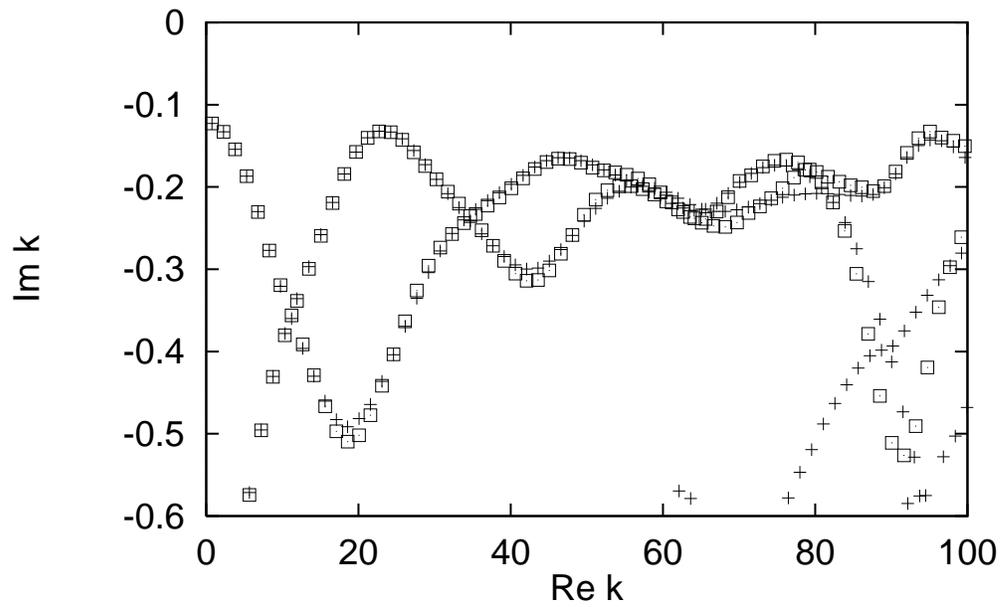}
\caption{\label{fig29} 
Three disk scattering system ($A_1$ subspace) with $R=1$, $d=6$.
Semiclassical resonances obtained by harmonic inversion of the 
periodic orbit recurrence function $C(L)$ with $L\le 24$ (squares) and 
$L\le 50$ (crosses).
}
\end{figure}

To compare the efficiency of the harmonic inversion and the cycle expansion
method we calculated the $A_1$ resonances of the three disk system with
distance $d=6$ from a short signal $C(L)$ with $L\le 24$, which includes
the recurrences of all periodic orbits with cycle length $n\le 5$.
The resonances obtained by harmonic inversion of the short signal are
presented as squares in Fig.\ \ref{fig29}.
The resonances of the two bands closest to the real axis qualitatively
agree with the correct semiclassical resonances 
(crosses in Fig.\ \ref{fig29}).
The accuracy is similar to the accuracy obtained by the cycle expansion up
to $3^{\rm rd}$ order in the curvature expansion \cite{Wir97}, which 
includes all periodic orbits with cycle length $n\le 4$.
The reason for the somewhat higher efficiency of the cycle expansion 
compared to harmonic inversion is probably that the basic requirement
(\ref{po_relations}) for the cycle expansion is a very good approximation
at the large distance $d=6$ between the disks.

\begin{figure}[b]
\vspace{18.0cm}
\includegraphics{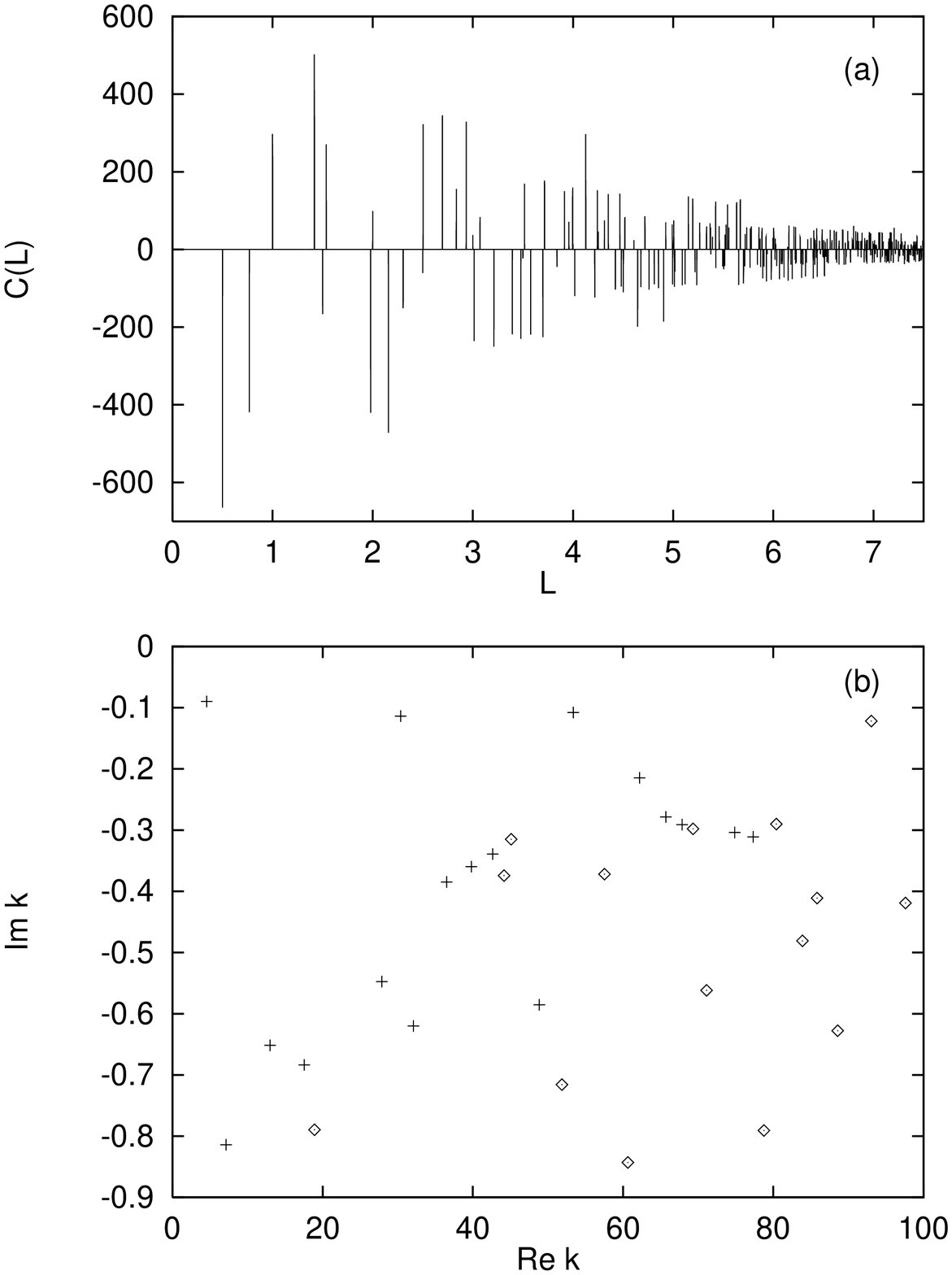}
\caption{\label{fig30} 
Three disk scattering system ($A_1$ subspace) with $R=1$, $d=2.5$.
(a) Periodic orbit recurrence function, $C(L)$.
The signal has been convoluted with a Gaussian function of width 
$\sigma=0.0003$.
(b) Semiclassical resonances. 
(From Ref.\ [82].)
}
\end{figure}

We now study the three disk scattering system with a short distance
ratio $d/R=2.5$.
The signal $C(L)$ is constructed from 356 periodic orbits with geometrical 
length $L\le 7.5$ (see Fig.\ \ref{fig30}a).
For large $L$ groups of orbits with the same cycle length of the symbolic 
code strongly overlap and cannot be recognized in Fig.\ \ref{fig30}a.
Note that the signal contains complete sets of orbits up to topological 
length (cycle length) $n=9$ only.
The resonances obtained by harmonic inversion of the signal $C(L)$ are
presented in Fig.\ \ref{fig30}b.
A comparison between the semiclassical resonances $k^{\rm hi}$
obtained by harmonic inversion, the cycle expansion resonances $k^{\rm ce}$ 
calculated up to $9^{\rm th}$ order in the curvature expansion of the 
Gutzwiller-Voros zeta function, and the exact quantum resonances $k^{\rm qm}$ 
is given in Table \ref{table10}.
The cycle expansion and exact quantum calculations have been performed
by Wirzba \cite{Wir98}.
The results of both semiclassical methods are in good agreement, although
a detailed comparison reveals that for some resonances towards the end of 
Table \ref{table10} the values of the cycle expansion are somewhat closer 
to the exact quantum mechanical results than those values obtained by 
harmonic inversion.

\subsection{Systems with mixed regular-chaotic dynamics}
\label{mixed_sec}
In Section \ref{zeta_sec} we have applied harmonic inversion for the
calculation of zeros of Riemann's zeta function as a mathematical model of 
a strongly chaotic bound system, and in Section \ref{three_disk_sec} we have 
used harmonic inversion for the periodic orbit quantization of the three disk
scattering system.
Both systems have been solved with other especially designed methods, i.e.,
the Riemann zeta function with the help of the Riemann-Siegel formula and 
the three disk system by application of cycle expansion techniques.
However, none of the special methods, which were designed to overcome the
convergence problems of the semiclassical trace formula (see, e.g., Refs.\
\cite{Cvi89,Ber92,Bog92b,Aur92a}), has succeeded so far in correctly 
describing generic dynamical systems with {\em mixed regular-chaotic} 
phase spaces.
In this section we want to demonstrate the universality of periodic orbit
quantization by harmonic inversion by investigating generic dynamical systems.
It will be the objective to contribute to solving the longstanding problem 
of semiclassical quantization of nonintegrable systems in the mixed 
regular-chaotic regime.

\begin{figure}[b]
\vspace{17.5cm}
\includegraphics{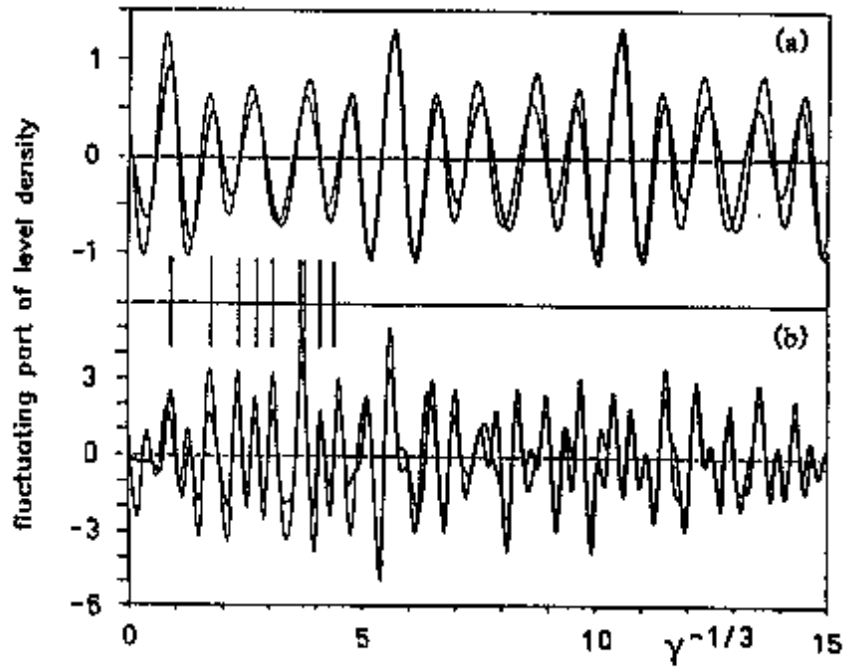}
\caption{\label{fig31} 
Smoothed fluctuating part of the density of states for the hydrogen atom in
a magnetic field at scaled energy $\tilde E=-0.2$.
The quantum results (thick lines) are smoothed over the first 100 eigenstates.
Semiclassical results (thin lines) are obtained by including (a) two orbits
(three contributions including repetitions) with scaled actions $s/2\pi<1.33$,
and (b) 13 orbits (19 including repetitions) with $s/2\pi<3$. 
The lowest quantum eigenvalues are marked as vertical bars.
(From Ref.\ [78].)
}
\end{figure}

A first step towards the periodic orbit quantization of mixed systems has
been done by Wintgen \cite{Win88} on the hydrogen atom in a magnetic field.
From Gutzwiller's truncated periodic orbit sum he obtained the smoothed part
of the density of states presented in Fig.\ \ref{fig31} at scaled energy
$\tilde E=-0.2$ which is in the mixed regular-chaotic regime.
The resolution of the smoothed spectra depends on the cutoff value of 
the periodic orbit sum which is $s_{\rm max}/2\pi=1.33$ and 
$s_{\rm max}/2\pi=3.00$ in Fig.\ \ref{fig31}a and \ref{fig31}b, respectively.
Nine approximate eigenvalues have been estimated from the low-resolution 
truncated periodic orbit sum in Fig.\ \ref{fig31}b (for comparison the exact 
quantum eigenvalues are marked by vertical bars in Fig.\ \ref{fig31}), but 
neither the accuracy nor the number of semiclassical eigenvalues could be 
improved by increasing the number of periodic orbits because of the 
non-convergence property of Gutzwiller's trace formula.
The results of Ref.\ \cite{Win88} therefore clearly demonstrate that
methods to overcome the convergence problems of the semiclassical periodic
orbit sum are of crucial importance to obtain highly resolved semiclassical
spectra in the mixed regime.
Note also that Gutzwiller's trace formula is not valid for non-isolated 
periodic orbits on invariant tori in the regular part of the phase space.
This fact has not been considered in Ref.\ \cite{Win88}.

The limiting cases of mixed systems are strongly chaotic and integrable
systems.
As has been proven by Gutzwiller \cite{Gut90}, for systems with complete 
chaotic (hyperbolic) classical dynamics the density of states can be 
expressed as an infinite sum over all (isolated) periodic orbits.
On the other extreme of complete integrability, it is well known that the 
semiclassical energy values can be obtained by EBK torus quantization 
\cite{Ein17}.
This requires the knowledge of all the constants of motion, which are 
not normally given in explicit form, and therefore practical EBK quantization
based on the direct or indirect numerical construction of the constants of 
motion turns out to be a formidable task \cite{Per77}.
As an alternative, EBK quantization was recast as a sum over all periodic 
orbits of a given topology on respective tori by Berry and Tabor 
\cite{Ber76,Ber77}.
The Berry-Tabor formula circumvents the numerical construction of the 
constants of motion but usually suffers from the convergence problems of the
infinite periodic orbit sum.
 
The extension of the Berry-Tabor formula into the near-integrable (KAM) 
regime was outlined by Ozorio de Almeida \cite{Alm88} and elaborated, at 
different levels of refinement, by Tomsovic et al.\ \cite{Tom95} and Ullmo 
et al.\ \cite{Ull96}.
These authors noted that in the near-integrable regime, according to the 
Poincar\'e-Birkhoff theorem, two periodic orbits survive the destruction
of a rational torus with similar actions, one stable and one hyperbolic
unstable, and worked out the ensuing modifications of the Berry-Tabor 
formula.
In this section we go one step further by noting that, with increasing 
perturbation, the stable orbit turns into an inverse hyperbolic one 
representing, together with its unstable companion with similar action, 
a remnant torus.
We include the contributions of these pairs of inverse hyperbolic and
hyperbolic orbits in the Berry-Tabor formula and demonstrate for a
system with mixed regular-chaotic dynamics that this procedure yields 
excellent results even in the deep mixed regular-chaotic regime.
As in Ref.\ \cite{Win88} we choose the hydrogen atom in a magnetic field, 
which is a real physical system and has served extensively as a prototype 
for the investigation of ``quantum chaos'' \cite{Fri89,Has89,Wat93}.

The fundamental obstacle bedeviling the semiclassical quantization of systems
with mixed regular-chaotic dynamics is that the periodic orbits are neither 
sufficiently isolated, as is required for Gutzwiller's trace formula 
\cite{Gut90}, nor are they part of invariant tori, as is necessary for the 
Berry-Tabor formula \cite{Ber76,Ber77}.
However, as will become clear below, it is the Berry-Tabor formula which 
lends itself in a natural way for an extension of periodic orbit quantization 
to mixed systems.
As previously, we consider scaling systems where the classical action $S$ 
scales as $S=sw$ with $w=\hbar_{\rm eff}^{-1}$ the scaling parameter and 
$s$ the scaled action.
For scaling systems with two degrees of freedom, which we will focus on,
the Berry-Tabor formula for the fluctuating part of the level density reads
\begin{equation}
 \varrho(w) = {1\over\pi}\, {\rm Re} \,
   \sum_{\bf M} {w^{1/2}s_{\bf M}\over M_2^{3/2}|g''_E|^{1/2}} \,
   e^{i(s_{\bf M}w-{\pi\over2}\eta_{\bf M}-{\pi\over 4})} \; ,
\label{BT_eq}
\end{equation}
with ${\bf M}=(M_1,M_2)$ pairs of integers specifying the individual periodic 
orbits on the tori (numbers of rotations per period, $M_2/M_1$ rational), and
$s_{\bf M}$ and $\eta_{\bf M}$ the scaled action and Maslov index of the 
periodic orbit ${\bf M}$.
The function $g_E$ in (\ref{BT_eq}) is obtained by inverting the Hamiltonian, 
expressed in terms of the actions $(I_1,I_2)$ of the corresponding torus,
with respect to $I_2$, viz.\ $H(I_1,I_2=g_E(I_1))=E$ \cite{Boh93}.
The calculation of $g''_E$ from the actions $(I_1,I_2)$ can be rather
laborious even for integrable and near-integrable systems, and, by definition,
becomes impossible for mixed systems in the chaotic part of the phase space.
Here we will adopt the method of Refs.\ \cite{Tom95,Ull96} and calculate 
$g''_E$, for given ${\bf M}=(\mu_1,\mu_2)$, with $(\mu_1,\mu_2)$ coprime 
integers specifying the primitive periodic orbit, directly from the parameters
of the two periodic orbits (stable (s) and hyperbolic unstable (h)) that 
survive the destruction of the rational torus $\bf M$, viz.
\begin{equation}
 g''_E = {2\over\pi \mu_2^3\Delta s}
   \left({1\over\sqrt{\det(M_{\rm s}-I)}}
       + {1\over\sqrt{-\det(M_{\rm h}-I)}}\right)^{-2} \, ,
\label{g2_eq}
\end{equation}
with
\begin{equation}
 \Delta s = {1\over 2} (s_{\rm h} - s_{\rm s})
\label{Ds_eq}
\end{equation}
the difference of the scaled actions, and $M_{\rm s}$ and $M_{\rm h}$ the 
monodromy matrices of the two orbits.
The action $s_{\bf M}$ in (\ref{BT_eq}) is to replaced with the mean action
\begin{equation}
 \bar s = {1\over 2} (s_{\rm h} + s_{\rm s}) \; .
\label{s_mean_eq}
\end{equation}
Eq.\ \ref{g2_eq} is an approximation which becomes exact in the limit
of an integrable system.

It is a characteristic feature of systems with mixed regular-chaotic dynamics
that with increasing nonintegrability the stable orbits turn into inverse 
hyperbolic unstable orbits in the chaotic part of the phase space.
These orbits, although embedded in the fully chaotic part of phase space,
are remnants of broken tori.
It is therefore natural to assume that Eqs.\ \ref{BT_eq} and \ref{g2_eq} can 
even be applied when these pairs of inverse hyperbolic and hyperbolic orbits 
are taken into account, i.e., more deeply in the mixed regular-chaotic regime.

It should be noted that the difference $\Delta s$ between the actions of the 
two orbits is normally still small, and it is therefore more appropriate 
to start from the Berry-Tabor formula for semiclassical quantization in that 
regime than from Gutzwiller's trace formula, which assumes well-isolated 
periodic orbits.
It is also important to note that the Berry-Tabor formula does not require
an extensive numerical periodic orbit search.
The periodic orbit parameters $s/M_2$ and $g''_E$ are smooth functions
of the rotation number $M_2/M_1$, and can be obtained for arbitrary periodic 
orbits with coprime integers $(M_1,M_2)$ by interpolation between ``simple''
rational numbers $M_2/M_1$.

\subsubsection{Hydrogen atom in a magnetic field}
\begin{figure}[b]
\vspace{14.0cm}
\includegraphics{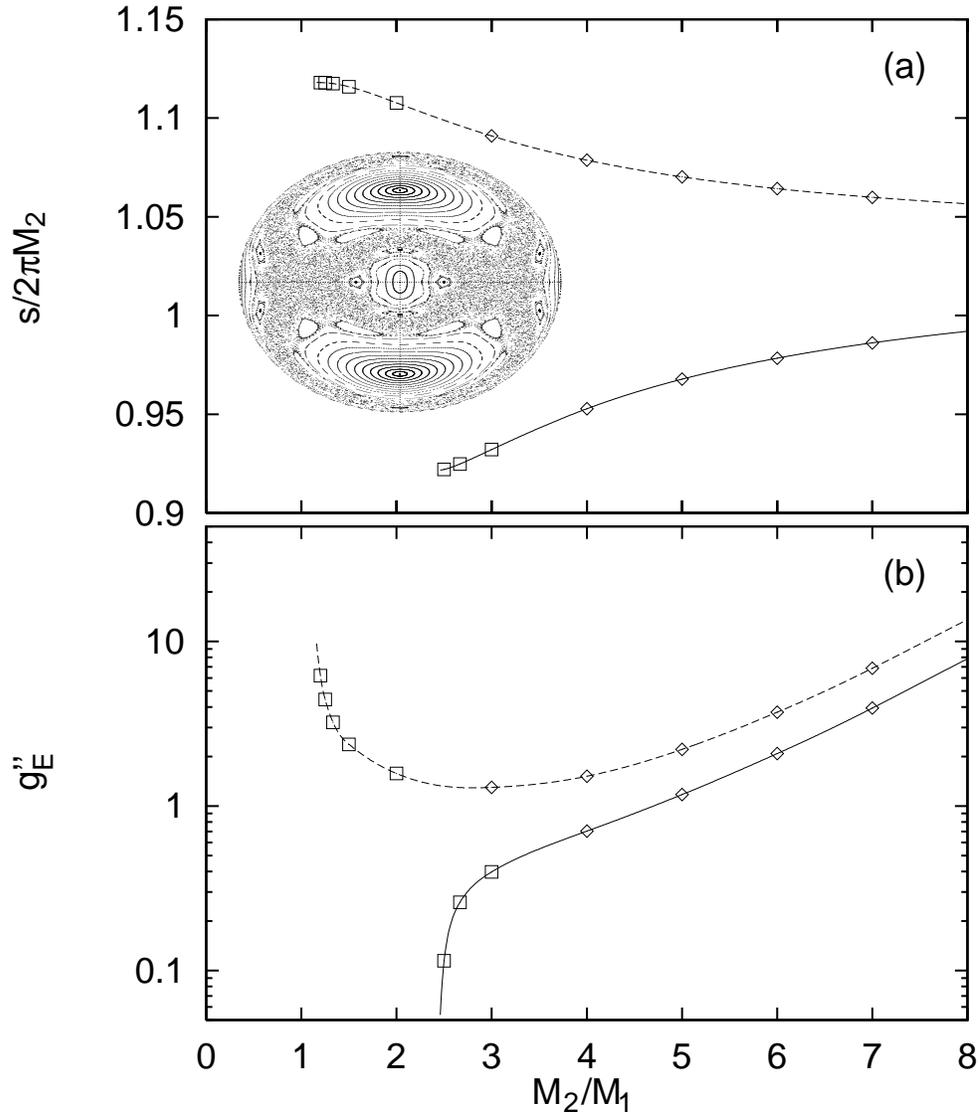}
\caption{\label{fig32} 
(a) Action $s/2\pi M_2$ and (b) second derivative $g''_E$ as a function of 
the frequency ratio $M_2/M_1$ for the rotator (solid lines) and vibrator 
(dashed lines) type motion of the hydrogen atom in a magnetic field at 
scaled energy $\tilde E=-0.4$.
Inset: Poincar\'e surface of section in semiparabolical coordinates
$(\mu,p_\mu;\nu=0)$.
(From Ref.\ [84].)
}
\end{figure}

We now demonstrate the high quality of the extension of Eqs.\ \ref{BT_eq} 
and \ref{g2_eq} to pairs of inverse hyperbolic and hyperbolic periodic orbits 
for a physical system that undergoes a transition from regularity to chaos, 
namely the hydrogen atom in a magnetic field.
This is a scaling system, with $w=\gamma^{-1/3}=\hbar_{\rm eff}^{-1}$ the 
scaling parameter and $\gamma=B/(2.35\times 10^5\, {\rm T})$ the magnetic 
field strength in atomic units.
The classical Hamiltonian and the scaling procedure have been discussed in
Section \ref{FT_spectra}.
At low scaled energies $\tilde E=E\gamma^{-2/3}<-0.6$ a Poincar\'e surface 
of section analysis \cite{Has89} of the classical Hamiltonian 
(Eq.\ \ref{Hamfkt_scal}) exhibits two different torus structures related to 
a ``rotator'' and ``vibrator'' type motion.
The separatrix between these tori is destroyed at a scaled energy of
$\tilde E \approx -0.6$, and the chaotic region around the separatrix grows
with increasing energy.
At $\tilde E=-0.127$ the classical phase space becomes completely chaotic.
We investigate the system at scaled energy $\tilde E=-0.4$, where about
40\% of the classical phase space volume is chaotic 
(see inset in Fig.\ \ref{fig32}), i.e.\ well in the region of mixed dynamics.
We use 8 pairs of periodic orbits to describe the rotator type motion in 
both the regular and chaotic region.
The results for the periodic orbit parameters $s/2\pi M_2$ and $g''_E$ are
presented as solid lines in Fig.\ \ref{fig32}.
The squares on the solid lines mark parameters obtained by pairs of stable
and unstable periodic orbits in the regular region of the phase space.
The diamonds mark parameters obtained by pairs of two unstable (inverse
hyperbolic and hyperbolic) periodic orbits in the chaotic region of phase 
space.
The cutoff is related to the winding angle $\phi=1.278$ of the fixed point of 
the rotator type motion, i.e., the orbit perpendicular to the magnetic field 
axis, $(M_2/M_1)_{\rm cutoff}=\pi/\phi=2.458$.
The solid lines have been obtained by spline interpolation of the data points.
In the same way the periodic orbit parameters for the vibrator type motion
have been obtained from 11 pairs of periodic orbits (see the dashed lines
in Fig.\ \ref{fig32}).
The cutoff at $M_2/M_1=\pi/\phi=1.158$ is related to the winding 
angle $\phi=2.714$ of the fixed point of the vibrator type motion, i.e.,
the orbit parallel to the field axis.

With the data of Fig.\ \ref{fig32} we have all the ingredients at hand to 
calculate the semiclassical density of states $\varrho(w)$ in Eq.\ \ref{BT_eq}.
The periodic orbit sum includes for both the rotator and vibrator type motion
the orbits with $M_2/M_1>(M_2/M_1)_{\rm cutoff}$.
For each orbit the action and the function $g''_E$ is obtained from the 
spline interpolations.
The Maslov indices are $\eta_{\bf M}=4M_2-M_1$ for the rotator and
$\eta_{\bf M}=4M_2+2M_1-1$ for the vibrator type orbits.
However, the problem is to extract the semiclassical eigenenergies from
Eq.\ \ref{BT_eq} because the periodic orbit sum does not converge.
We now adopt the method discussed in Section \ref{po_quant_sec} and
adjust the semiclassical recurrence signal, i.e., the Fourier transform
of the weighted density of states $w^{-1/2}\varrho(w)$
(Eq.\ \ref{BT_eq})
\begin{equation}
 C^{\rm sc}(s) = \sum_{\bf M} {\cal A}_{\bf M} \delta(s-s_{\bf M}) \; ,
\label{C_sc_BT}
\end{equation}
with the amplitudes being determined exclusively by periodic orbit quantities,
\begin{equation}
 {\cal A}_{\bf M} = {s_{\bf M}\over M_2^{3/2}|g''_E|^{1/2}} \, 
   e^{-i{\pi\over2}\eta_{\bf M}} \; ,
\end{equation}
to the functional form of its quantum mechanical analogue
\begin{equation}
 C^{\rm qm}(s) = -i \sum_k d_k e^{-iw_k s} \; ,
\label{C_qm_BT}
\end{equation}
where the $w_k$ are the quantum eigenvalues of the scaling parameter, 
and the $d_k$ are the multiplicities of the eigenvalues 
($d_k=1$ for nondegenerate states).
The frequencies obtained from this procedure are interpreted as the 
semiclassical eigenvalues $w_k$.
The technique used to adjust Eq.\ (\ref{C_sc_BT}) to the functional form of
Eq.\ (\ref{C_qm_BT}) is harmonic inversion \cite{Mai97d,Mai98b}.

\begin{figure}[b]
\vspace{14.0cm}
\includegraphics{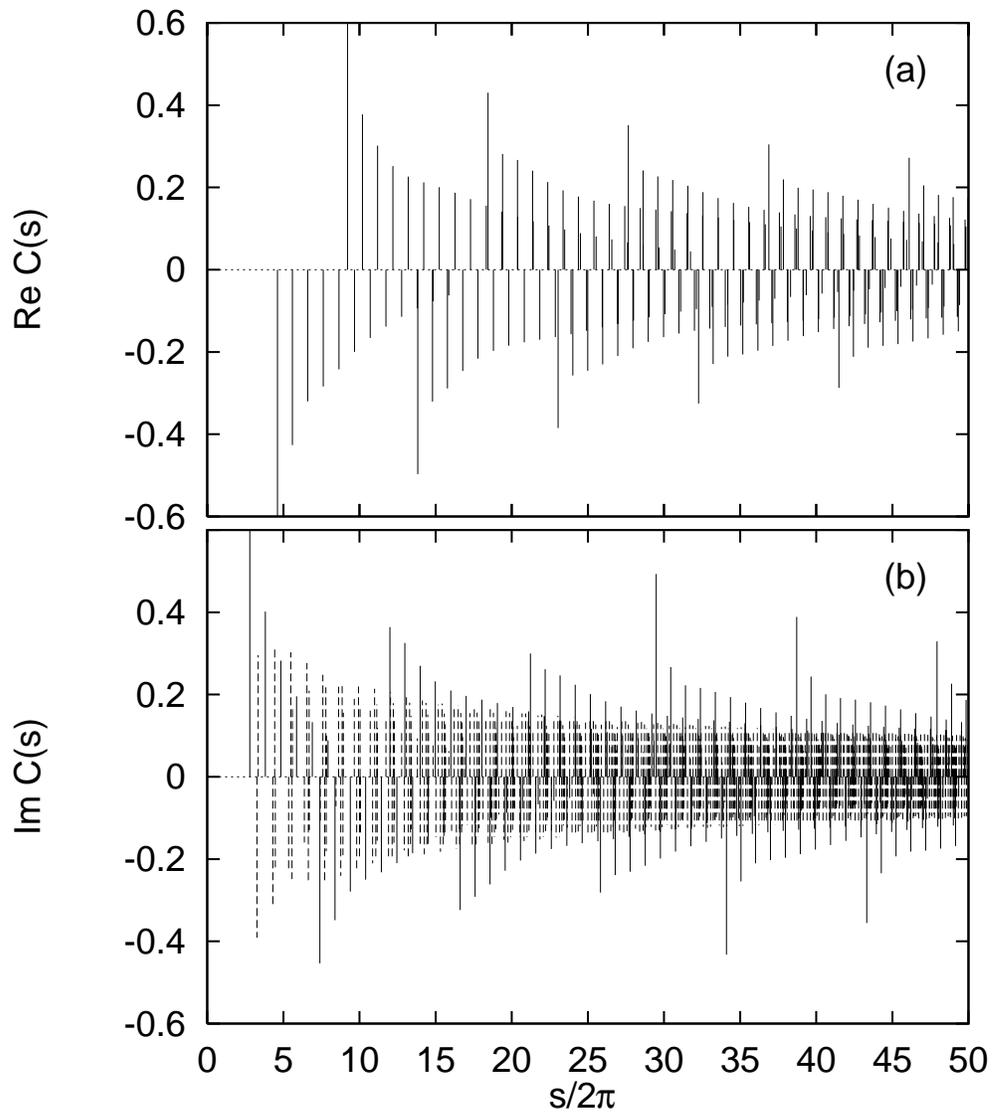}
\caption{\label{fig33} 
Semiclassical recurrence signal $C^{\rm sc}(s)$ for the hydrogen atom in a 
magnetic field at scaled energy $\tilde E=-0.4$.
Solid and dashed sticks: Signal from the rotator and vibrator type motion,
respectively.
(From Ref.\ [84].)
}
\end{figure}

For the hydrogen atom in a magnetic field part of the semiclassical 
recurrence signal $C^{\rm sc}(s)$ at scaled energy $\tilde E=-0.4$ is 
presented in Fig.\ \ref{fig33}.
The solid and dashed peaks mark the recurrencies of the rotator and vibrator
type orbits, respectively.
Note that $C^{\rm sc}(s)$ can be easily calculated even for long periods $s$
with the help of the spline interpolation functions in Fig.\ \ref{fig32}.
By contrast, the construction of the recurrence signal for Gutzwiller's 
trace formula usually requires an exponentially increasing effort for the
numerical periodic orbit search with growing period.

\begin{figure}[b]
\vspace{14.0cm}
\includegraphics{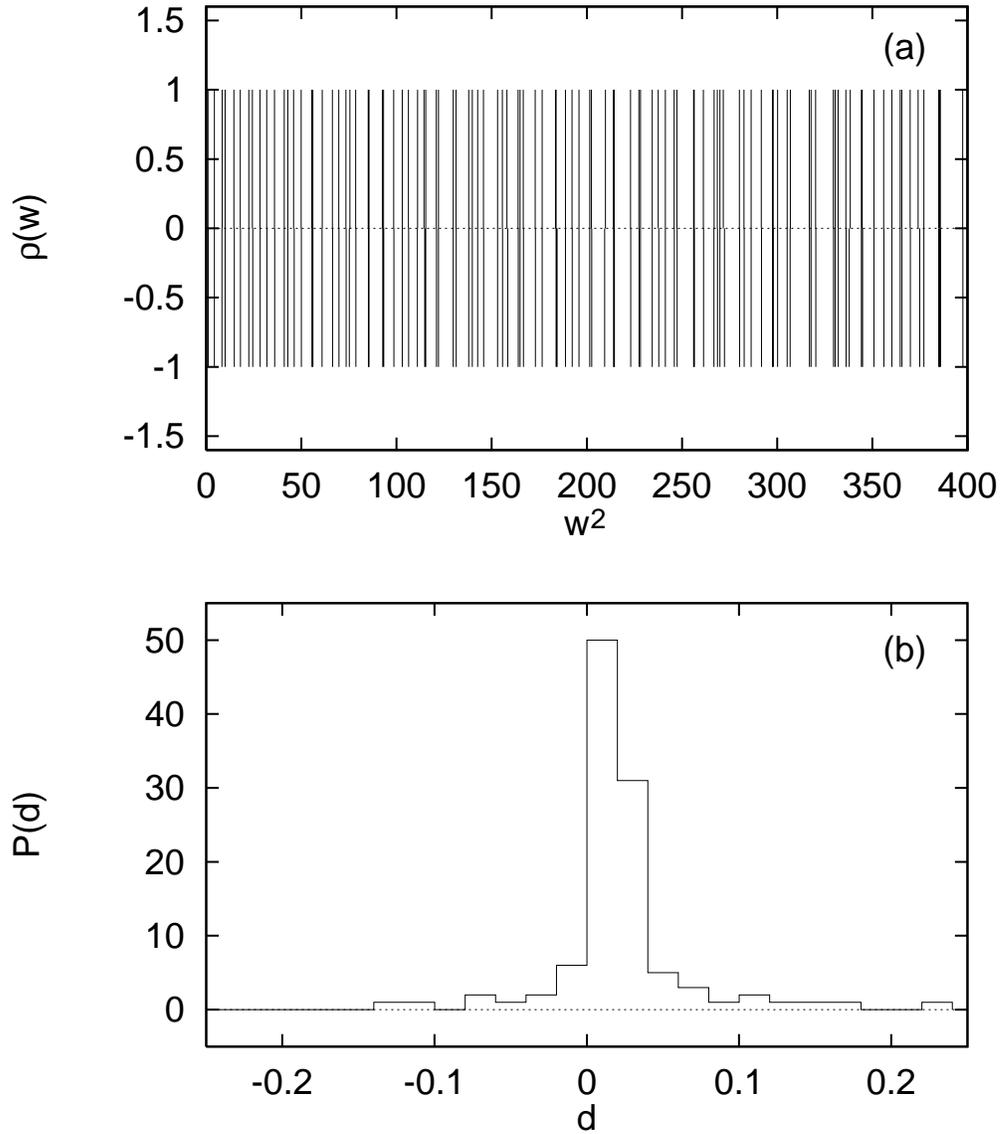}
\caption{\label{fig34} 
(a) Semiclassical and quantum mechanical spectrum of the hydrogen atom 
in a magnetic field at scaled energy $\tilde E=-0.4$.
(b) Distribution $P(d)$ of the semiclassical error in units of the mean 
level spacing,
$d=(w_{\rm qm}-w_{\rm sc})/\Delta w_{\rm av}$ for the lowest 106 eigenvalues.
(From Ref.\ [84].)
}
\end{figure}

We have analyzed $C^{\rm sc}(s)$ by the harmonic inversion technique 
in the region $0<s/2\pi<200$.
The resulting semiclassical spectrum of the lowest 106 states with
eigenvalues $w<20$ is shown in the upper part of Fig.\ \ref{fig34}a.
For graphical purposes the spectrum is presented as a function of the squared
scaling parameter $w^2$, which is equivalent to unfolding the spectrum
to constant mean level spacing.
For comparison the lower part of Fig.\ \ref{fig34}a shows the exact quantum
spectrum.
The semiclassical and quantum spectrum are seen to be in excellent agreement,
and deviations are less than the stick widths for nearly all states.
The distribution $P(d)$ of the semiclassical error with 
$d=(w_{\rm qm}-w_{\rm sc})/\Delta w_{\rm av}$ the error in units of the
mean level spacing, $\Delta w_{\rm av}=1.937/w$, is presented in 
Fig.\ \ref{fig34}b.
For most levels the semiclassical error is less than 4\% of the mean level
spacing, which is typical for a system with two degrees of freedom 
\cite{Boa94}.

The accuracy of the results presented in Fig.\ \ref{fig34} seems to be
surprising for two reasons.
First, we have not exploited the mean staircase function $\bar{\cal N}(w)$,
i.e., the number of eigenvalues $w_k$ with $w_k<w$, which is a basic
requirement of some other semiclassical quantization techniques for bound 
chaotic systems \cite{Aur92a,Ber92}.
Second, as mentioned before, Eq.\ \ref{g2_eq} has been derived for 
near-integrable systems, and is only an approximation, in particular, for 
mixed systems.
We have not taken into account any more refined extensions of the 
Berry-Tabor formula (\ref{BT_eq}) as discussed, e.g., in Refs.\ 
\cite{Tom95,Ull96}.
The answer to the second point is that the splitting of scaled actions of 
the periodic orbit pairs used in Fig.\ \ref{fig32} does not exceed 
$\Delta s=0.022$, and therefore for states with $w<20$ the phase shift 
between the two periodic orbit contributions is $w\Delta s=0.44$, at most.
For small phase shifts the extension of the Berry-Tabor formula to 
near-integrable systems results in a damping of the {\em amplitudes} of the 
periodic orbit recurrence signal in Fig.\ \ref{fig33} but seems not to effect 
the {\em frequencies}, i.e., the semiclassical eigenvalues $w_k$ obtained by 
the harmonic inversion of the function $C^{\rm sc}(s)$.

To summarize, in this section we have presented a solution to the fundamental 
problem of semiclassical quantization of nonintegrable systems in the mixed 
regular-chaotic regime.
We have demonstrated the excellent quality of our procedure for the hydrogen 
atom in a magnetic field at a scaled energy $\tilde E=-0.4$, where about 
40\% of the phase space volume is chaotic.
The lowest 106 semiclassical and quantum eigenenergies have been shown
to agree within a few percent of the mean level spacings.
The same method can be applied straightforwardly to other systems with 
mixed dynamics.

\subsection{Harmonic inversion of cross-correlated periodic orbit sums}
\label{cross_corr_po_sums}
In the previous sections we have introduced harmonic
inversion of semiclassical signals as a powerful and universal technique
for the problem of periodic orbit quantization in that it does not depend on 
special properties of the system such as being bound and ergodic, or the 
existence of a symbolic dynamics.
The method only requires the knowledge of periodic orbits and their physical
quantities up to a certain maximum period, which depends on the average local 
density of states.
Unfortunately, this method is not free of the general drawback of
most semiclassical approaches, which suffer from a rapid proliferation of 
periodic orbits with their period, which in turn requires an enormous number 
of orbits to be taken into account.
As discussed in Section \ref{Req_signal_len} the required signal length for
harmonic inversion is at least two times the Heisenberg period.
This is by a factor of four longer than the signal length required for the
application of the Riemann-Siegel look-alike formula \cite{Ber90,Kea92,Ber92}.
In chaotic systems, where the number of periodic orbits grows exponentially
with the period, the reduction of the required signal length by a factor 
of four implies that, e.g., instead of one million periodic orbits a reduced 
set of about 32 orbits is sufficient for the semiclassical quantization.
This example clearly illustrates that a shortening of the required signal
length is highly desirable.

In this section we want to introduce harmonic inversion of cross-correlated
periodic orbit sums as a method to reduce the required amount of periodic
orbit data \cite{Mai98c}.
The idea is that the informational content of a $D\times D$ cross-correlated
time signal is increased roughly by a factor of $D$ as compared to a 
$1\times 1$ signal.
The cross-correlated signal is constructed by introducing a set of
$D$ smooth and linearly independent operators.
Numerically, the harmonic inversion of the cross-correlated periodic orbit 
sum is based on an extension of the filter-diagonalization method to the
case of time cross-correlation functions \cite{Wal95,Nar97,Man98}.
This extended method provides highly resolved spectra even in situations 
of nearly degenerate states, as well as the diagonal matrix elements for 
the set of operators chosen.
The power of the method will be demonstrated in Section \ref{circle_billiard}
for the circle billiard, as an example of a completely integrable system.

Consider a quantum Hamiltonian $\hat H$ whose eigenvalues are 
$w_n$ and eigenstates $|n\rangle$. 
[As previously, we consider scaling systems with $w$ the scaling parameter,
which is not necessarily the energy, and with linear scaling of the classical
action, $S_{\rm po}=ws_{\rm po}$.]
We introduce a cross-correlated response function $(\alpha,\beta=1,2,\dots,D)$
\begin{equation}
   g_{\alpha\beta}(w)
 = \sum_n {b_{\alpha n}b_{\beta n} \over w-w_n+i0}\ ,
\label{g_ab_qm}
\end{equation}
where $b_{\alpha n}$ and $b_{\beta n}$ are the diagonal matrix elements 
of two operators $\hat A_\alpha$ and $\hat A_\beta$, respectively, i.e.\
\begin{equation}
 b_{\alpha n} = \langle n|\hat A_\alpha|n\rangle \; .
\end{equation}
Later we have to find a semiclassical approximation to Eq.\ \ref{g_ab_qm}.
In this context it is important to note that $g_{\alpha\beta}(w)$ can only 
be written as a trace formula,
\[
   g_{\alpha\beta}(w)
 = \mbox{tr}\, \left\{\hat A_\alpha\hat G^+(w)\hat A_\beta\right\}
\]
with the Green Function 
\[
 \hat G^+(w) = {1\over w-\hat H +i0} \; ,
\]
if either $\hat A_\alpha$ or $\hat A_\beta$ commutes with $\hat H$.
The weighted density of states is given by
\begin{equation}
 \varrho_{\alpha\beta}(w) = -{1\over\pi} \, {\rm Im} \, g_{\alpha\beta}(w)\ .
\end{equation}
Let us assume that the semiclassical approximation 
$g_{\alpha\beta}^{\rm sc}(w)$ to the quantum expression (\ref{g_ab_qm}) 
is given.
The general procedure of harmonic inversion as described in Section
\ref{po_quant_sec} would then be to adjust the Fourier transform of
$g_{\alpha\beta}^{\rm sc}(w)$ to the functional form of the quantum expression
\begin{equation}
   C_{\alpha\beta}^{\rm qm}(s)
 = {1\over 2\pi}\int_{-\infty}^{+\infty} g_{\alpha\beta}^{\rm qm}(w)e^{-isw}dw
 = -i \sum_n b_{\alpha n}b_{\beta n} e^{-iw_ns} \; .
\label{C_ab_qm}
\end{equation}
The conventional harmonic inversion problem is formulated as a nonlinear 
fit of the signal $C(s)$ by the sum of sinusoidal terms 
(see Section \ref{Circumvent} and Appendix \ref{harm_inv}),
\[
 C(s)=\sum_n d_n e^{-iw_ns} \; , 
\]
with the set of, in general, complex variational parameters
$\{w_n,d_n=-ib_{\alpha n}b_{\beta n}\}$.
As already discussed in Section \ref{Req_signal_len}, simple information 
theoretical considerations then yield an estimate for the required signal 
length, $s_{\rm max}\sim 4\pi\bar\varrho(w)$, for poles $w_n\le w$ which 
can be extracted by this method.
When a periodic orbit approximation of the quantum signal $C(s)$ is used, 
this estimate results sometimes in a very unfavorable scaling because of a 
rapid (exponential for chaotic systems) proliferation of periodic orbits 
with increasing period.
Let us consider now a generalized harmonic inversion problem, which assumes 
that the whole $s$-dependent $D\times D$ signal $C_{\alpha\beta}(s)$ is 
adjusted simultaneously to the form of Eq.\ \ref{C_ab_qm}, with 
$b_{\alpha n}$ and $w_n$ being the variational parameters.
The advantage of using the cross-correlation approach \cite{Wal95,Nar97,Man98}
is based on the simple argument that the total amount of independent 
information contained in the $D\times D$ signal is $D(D+1)$ multiplied by 
the length of the signal, while the total number of unknowns 
(here $b_{\alpha n}$ and $w_n$) is $(D+1)$ times the total number of poles 
$w_n$. 
Therefore the informational content of the $D\times D$ signal per unknown 
parameter is increased (compared to the case of Eq.\ \ref{C_sc}) by a factor 
of $D$. 
[Of course, this scaling holds only approximately and for sufficiently small 
numbers $D$ of operators $\hat A_{\alpha}$ chosen.]
Thus we have the result that, provided we are able to obtain a periodic orbit 
approximation for $C_{\alpha\beta}(s)$, with this procedure we can extract 
more information from the same set of periodic orbits.

We now have to find a semiclassical approximation for $C_{\alpha\beta}(s)$
in Eq.\ \ref{C_ab_qm}.
The problem has been solved in the literature only for special cases, i.e.,
if one or both operators $\hat A_{\alpha}$ and $\hat A_{\beta}$ are the
identity or, somewhat more general, if at least one of the operators commutes
with the Hamiltonian $\hat H$.
For the identity operator $\hat A_1=I$ the element $C_{11}^{\rm sc}(s)$ is 
the Fourier transform of Gutzwiller's trace formula \cite{Gut67,Gut90} for 
chaotic systems, and of the Berry-Tabor formula \cite{Ber76,Ber77} for 
regular systems, i.e.
\begin{equation}
   C_{11}^{\rm sc}(s) 
 = \sum_{\rm po} {\cal A}_{\rm po} \delta\left(s-s_{\rm po}\right) \; ,
\label{C_11_sc}
\end{equation}
where $s_{\rm po}$ are the periods of the orbits and ${\cal A}_{\rm po}$
the amplitudes (recurrence strengths) of the periodic orbit contributions 
including phase information.
For $\hat A_1=I$ and an arbitrary smooth operator $\hat A_\alpha$ the elements
$C_{\alpha 1}^{\rm sc}(s)$ are obtained from a semiclassical approximation to 
the generalized trace formula 
\begin{equation}
   g_{\alpha 1}(w)
 = {\rm tr} \, \left\{\hat G^+(w)\hat A_\alpha\right\} \; ,
\end{equation}
which has been investigated in detail in Refs.\ \cite{Wil88,Eck92}.
The result is that the amplitudes ${\cal A}_{\rm po}$ in (\ref{C_11_sc})
have to be multiplied by the classical average of the observable $A_\alpha$
along the periodic orbit,
\begin{equation}
 a_{\alpha,{\rm po}} = {1\over s_{\rm po}} \int_0^{s_{\rm po}}
 A_\alpha({\bf q}(s),{\bf p}(s)) ds \; ,
\label{a_po}
\end{equation}
with $A_\alpha({\bf q},{\bf p})$ the Wigner transform of the operator
$\hat A_\alpha$, i.e., the signal $C_{\alpha 1}^{\rm sc}(s)$ reads
\begin{equation}
   C_{\alpha 1}^{\rm sc}(s) 
 = \sum_{\rm po} a_{\alpha,{\rm po}}
     {\cal A}_{\rm po} \delta\left(s-s_{\rm po}\right) \; .
\label{C_a1_sc}
\end{equation}
If at least one of the operators $\hat A_\alpha$ and $\hat A_\beta$ commutes
with $\hat H$, Eq.\ \ref{g_ab_qm} can still be written as a trace formula,
\[
   g_{\alpha\beta}(w)
 = {\rm tr}\, \left\{\hat A_\alpha\hat G^+(w)\hat A_\beta\right\}
 = {\rm tr}\, \left\{\hat G^+(w)\left(\hat A_\alpha\hat A_\beta\right)\right\}
 \; ,
\]
and Eq.\ \ref{C_a1_sc} can be applied to the product 
$\hat A_\alpha\hat A_\beta$.
However, we do not want to restrict the operators to those commuting with 
$\hat H$, which obviously would be a severe restriction especially for 
chaotic systems, and
the problem is now to find a semiclassical approximation to Eq.\ \ref{C_ab_qm}
for the general case of two arbitrary smooth operators $\hat A_\alpha$ and 
$\hat A_\beta$.
A reasonable assumption is that the amplitudes ${\cal A}_{\rm po}$ in 
(\ref{C_11_sc}) have to be multiplied by the product of the classical 
averages, $a_{\alpha,{\rm po}}a_{\beta,{\rm po}}$, of these two observables,
i.e.\
\begin{equation}
   C_{\alpha\beta}^{\rm sc}(s)
 = \sum_{\rm po} a_{\alpha,{\rm po}} a_{\beta,{\rm po}} {\cal A}_{\rm po}
   \delta\left(s-s_{\rm po}\right) \; .
\label{C_ab_sc}
\end{equation}
Although no rigorous mathematical proof of Eq.\ \ref{C_ab_sc} will be given
here, we have strong numerical evidence, from the high resolution analysis 
of quantum spectra, part of which will be given below, that the conjecture
of Eq.\ \ref{C_ab_sc} is correct.
Details on semiclassical non-trace type formulae like Eq.\ \ref{C_ab_sc} 
are given in Ref.\ \cite{Mai98e}.
Eq.\ \ref{C_ab_sc} is the starting point for the following application of 
harmonic inversion of cross-correlation functions.
Note that all quantities in (\ref{C_ab_sc}) are obtained from the classical
periodic orbits.

The idea of periodic orbit quantization by harmonic inversion of
cross-correlated periodic orbit sums is to fit the semiclassical functions 
$C_{\alpha\beta}^{\rm sc}(s)$ given in a finite range $0<s<s_{\rm max}$ 
to the functional form of the quantum expression (\ref{C_ab_qm}).
As for the harmonic inversion of a one-dimensional signal (see Section
\ref{po_quant_sec}) the frequencies of the harmonic inversion analysis are 
then identified with the semiclassical eigenvalues $w_n$.
The amplitudes $b_{\alpha n}$ are identified with the semiclassical 
approximations to the diagonal matrix elements 
$\langle n|\hat A_\alpha|n\rangle$.
Here we only give a brief description how the harmonic inversion method is
extended to cross-correlation functions.
The details of the numerical procedure of solving the generalized harmonic 
inversion problem (\ref{C_ab_qm}) are presented in Refs.\ 
\cite{Wal95,Nar97,Man98} and in Appendix \ref{hi_cross_corr}.
As for the harmonic inversion of a single function the idea is to recast 
the nonlinear fit problem as a linear algebraic problem \cite{Wal95}. 
This is done by associating the signal $C_{\alpha\beta}(s)$
(to be inverted) with a time cross-correlation function between an initial 
state $\Phi_{\alpha}$ and a final state $\Phi_{\beta}$,
\begin{equation}
   C_{\alpha\beta}(s)
 = \left(\Phi_{\beta},e^{-i\hat\Omega s}\Phi_{\alpha}\right) \; ,
\label{C_ab_ansatz}
\end{equation}
where the fictitious quantum dynamical system is described by an effective 
Hamiltonian $\hat\Omega$.
The latter is defined implicitly by relating its spectrum to the set of 
unknown spectral parameters $w_n$ and $b_{\alpha n}$.
Diagonalization of $\hat\Omega$ would yield the desired $w_n$ and 
$b_{\alpha n}$. 
This is done by introducing an appropriate basis set in which the matrix 
elements of $\hat\Omega$ are available only in terms of the known
signals $C_{\alpha\beta}(s)$. 
The Hamiltonian $\hat\Omega$ is assumed to be complex symmetric even
in the case of a bound system. This makes the harmonic inversion stable 
with respect to ``noise'' due to the imperfections of the semiclassical 
approximation.
The most efficient numerical and practical implementation of the
harmonic inversion method with all relevant formulae can be found in 
Refs.\ \cite{Nar97,Man98} and Appendix \ref{hi_cross_corr}.

The method of harmonic inversion of cross-correlated periodic orbit sums
will be applied in Section \ref{circle_billiard} to the circle billiard.
As will be shown, for a given number of periodic orbits the accuracy of
semiclassical spectra can be significantly improved with the help of the
cross-correlation approach, or, alternatively, spectra with similar accuracy
can be obtained from a periodic orbit cross-correlation signal 
with significantly reduced signal length.

\subsection{$\hbar$ expansion for the periodic orbit quantization by
harmonic inversion}
\label{hbar_sec2}
Semiclassical spectra can be obtained for both regular and chaotic systems
in terms of the periodic orbits of the system.
For chaotic dynamics the semiclassical trace formula was derived by
Gutzwiller \cite{Gut67,Gut90}, and for integrable systems the Berry-Tabor 
formula \cite{Ber76,Ber77} is well known to be precisely equivalent to 
the EBK torus quantization \cite{Ein17,Bri26,Kel58}.
However, as already has been discussed in Section \ref{hbar_sec},
the semiclassical trace formulae are exact only in exceptional
cases, e.g., the geodesic motion on the constant negative curvature surface.
In general, they are just the leading order terms of an infinite series in
powers of the Planck constant and the accuracy of semiclassical quantization
is still an object of intense investigation \cite{Pro93,Boa94,Pri98}.
Methods for the calculation of the higher order periodic orbit 
contributions were developed in Refs.\ \cite{Gas93,Alo93,Vat96}.
In Section \ref{hbar_sec} we have demonstrated how the periodic orbit
quantities of the $\hbar$ expanded trace formula can be extracted from
the quantum and semiclassical spectra.
It is an even more fundamental problem to obtain semiclassical eigenenergies 
beyond the Gutzwiller and Berry-Tabor approximation directly from the 
$\hbar$ expanded periodic orbit sum.
Note that the $\hbar$ expansion of the periodic orbit sum does not solve
the general problem of the construction of the analytic continuation of
the trace formula, which is already a fundamental problem when only the
leading order terms of the $\hbar$ expansion is considered.
Up to now the $\hbar$ expansion for periodic orbit quantization is restricted
to systems with known symbolic dynamics, like the three disk scattering
problem, where cycle expansion techniques can be applied \cite{Alo93,Vat96},
and semiclassical eigenenergies beyond the Gutzwiller and Berry-Tabor
approximation cannot be calculated, e.g., for bound systems with the help
of Riemann-Siegel type formulae \cite{Kea92,Ber92} or surface of section
techniques \cite{Bog92a,Bog92b}.
In this section we extend the method of periodic orbit quantization by
harmonic inversion to the analysis of the $\hbar$ expansion of the
periodic orbit sum.
The accuracy of semiclassical eigenvalues can be improved by one to several 
orders of magnitude, as will be shown in Section \ref{circle_billiard}
by way of example of the circle billiard.

As in Section \ref{hbar_sec} we consider systems with a scaling property, 
i.e., where the classical action scales as $S_{\rm po}=ws_{\rm po}$, and
the scaling parameter $w\equiv\hbar_{\rm eff}^{-1}$ plays the role of an 
inverse effective Planck constant.
The $\hbar$ expansion of the periodic orbit sum is given (see Eq.\ 
\ref{g_hbar_series}) as a power series in $w^{-1}$,
\begin{equation}
   g(w)
 = \sum_{n=0}^\infty g_n(w)
 = \sum_{n=0}^\infty {1\over w^n} \sum_{\rm po} 
   {\cal A}_{\rm po}^{(n)} e^{is_{\rm po}w} \; .
\label{g_hbar_series2}
\end{equation}
The complex amplitudes ${\cal A}_{\rm po}^{(n)}$ of the $n^{\rm th}$
order periodic orbit contributions include the phase information from the 
Maslov indices.
For periodic orbit quantization the zeroth order contributions 
${\cal A}_{\rm po}^{(0)}$ are usually considered only.
The Fourier transform of the principal periodic orbit sum
\begin{equation}
  C_0(s) = {1\over 2\pi}\int_{-\infty}^{+\infty} g_0(w) e^{-isw}dw
        = \sum_{\rm po}{\cal A}_{\rm po}^{(0)} \delta(s-s_{\rm po})
\label{C0_sc}
\end{equation}
is adjusted by application of the harmonic inversion technique (see Section
\ref{po_quant_sec}) to the functional form of the exact quantum expression
\begin{equation}
   C(s)
 = {1\over 2\pi}\int_{-\infty}^{+\infty} \sum_k{d_k\over w-w_k+i0} e^{-iws}dw
 = -i\sum_k d_k e^{-iw_ks} \; ,
\label{C_qm0}
\end{equation}
with $\{w_k,d_k\}$ the eigenvalues and multiplicities.
The frequencies $w_{k,0}$ obtained by harmonic inversion of Eq.\ \ref{C0_sc} 
are the zeroth order $\hbar$ approximation to the semiclassical eigenvalues.
We will now demonstrate how the higher order correction terms to the 
semiclassical eigenvalues can be extracted from the periodic orbit sum 
(\ref{g_hbar_series2}).
We first remark that the asymptotic expansion (\ref{g_hbar_series2}) of the 
semiclassical response function suffers, for $n\ge 1$, from the singularities 
at $w=0$, and it is therefore not appropriate to harmonically invert the 
Fourier transform of (\ref{g_hbar_series2}), although the Fourier transform 
formally exists.
This means that the method of periodic orbit quantization by harmonic 
inversion cannot straightforwardly be extended to the $\hbar$ expansion of
the periodic orbit sum.
Instead we will calculate the correction terms to the semiclassical 
eigenvalues separately, order by order, as described in the following.

Let us assume that the $(n-1)^{\rm st}$ order approximations $w_{k,n-1}$ 
to the semiclassical eigenvalues are already obtained and the $w_{k,n}$
are to be calculated.
The difference between the two subsequent approximations to the quantum
mechanical response function reads \cite{Mai98d}
\begin{eqnarray}
     g_{n}(w)
 &=& \sum_k \left({d_k\over w-w_{k,n}+i0}
          - {d_k\over w-w_{k,n-1}+i0}\right)
     \nonumber \\
 &\approx& \sum_k{d_k\Delta w_{k,n}\over (w-\bar w_{k,n}+i0)^2} \; ,
\label{g_n}
\end{eqnarray}
with $\bar w_{k,n}=(w_{k,n}+w_{k,n-1})/2$ and 
$\Delta w_{k,n}=w_{k,n}-w_{k,n-1}$.
Integration of Eq.\ \ref{g_n} and multiplication by $w^n$ yields
\begin{equation}
 {\cal G}_{n}(w) = w^n \int g_{n}(w)dw
 = \sum_k {-d_kw^n\Delta w_{k,n}\over w-\bar w_{k,n}+i0} \; ,
\label{g_int_qm}
\end{equation}
which has the functional form of a quantum mechanical response function but
with residues proportional to the $n^{\rm th}$ order corrections 
$\Delta w_{k,n}$ to the semiclassical eigenvalues.
The semiclassical approximation to Eq.\ \ref{g_int_qm} is obtained from 
the term $g_{n}(w)$ in the periodic orbit sum (\ref{g_hbar_series2}) by 
integration and multiplication by $w^n$, i.e.
\begin{equation}
   {\cal G}_{n}(w)
 = w^n \int g_{n}(w)dw
 = -i\sum_{\rm po} {1\over s_{\rm po}}
   {\cal A}_{\rm po}^{(n)} e^{iws_{\rm po}}
   + {\cal O}\left(1\over w\right)  \; .
\label{g_int_sc}
\end{equation}
We can now Fourier transform both Eqs.\ \ref{g_int_qm} and \ref{g_int_sc},
and obtain ($n\ge 1$)
\begin{eqnarray}
\label{Cn_qm}
   C_{n}(s) &\equiv&
   {1\over 2\pi}\int_{-\infty}^{+\infty}{\cal G}_{n}(w)e^{-iws}dw
 = i\sum_k d_k (w_k)^n\Delta w_{k,n} e^{-iw_ks} \\
\label{Cn_sc}
 &\stackrel{\rm h.i.}{=}&
   -i\sum_{\rm po}{1\over s_{\rm po}}{\cal A}_{\rm po}^{(n)}
   \delta(s-s_{\rm po})  \; .
\end{eqnarray}
Eqs.\ \ref{Cn_qm} and \ref{Cn_sc} are the main result of this section.
They imply that the $\hbar$ expansion of the semiclassical eigenvalues can 
be obtained, order by order, by harmonic inversion (h.i.) of the periodic
orbit signal in Eq.\ \ref{Cn_sc} to the functional form of Eq.\ \ref{Cn_qm}.
The frequencies of the periodic orbit signal (\ref{Cn_sc}) are the 
semiclassical eigenvalues $w_k$.
Note that the accuracy of the semiclassical eigenvalues does not necessarily
increase with increasing order $n$.
We indicate this in Eq.\ \ref{Cn_qm} by omitting the index $n$ at the 
eigenvalues $w_k$.
The corrections $\Delta w_{k,n}$ to the eigenvalues are obtained from
the {amplitudes}, $d_k(w_k)^n\Delta w_{k,n}$, of the periodic orbit
signal.

The method requires as input the periodic orbits of the classical system 
up to a maximum period (scaled action), $s_{\rm max}$, determined by the 
average density of states \cite{Mai97d,Mai98b}.
The amplitudes ${\cal A}_{\rm po}^{(0)}$ are obtained from Gutzwiller's
trace formula \cite{Gut67,Gut90} and the Berry-Tabor formula 
\cite{Ber76,Ber77} for chaotic and regular systems, respectively.
For the next order correction ${\cal A}_{\rm po}^{(1)}$ explicit formulae
were derived by Gaspard and Alonso for chaotic systems with smooth potentials
\cite{Gas93} and in Refs.\ \cite{Alo93,Vat96} for billiards.
With appropriate modifications \cite{Wei99a,Wei99b} the formulae can be used 
for regular systems as well.
As an example we investigate the $\hbar$ expansion of the periodic orbit sum
for the circle billiard in the next section.

\subsection{The circle billiard}
\label{circle_billiard}
In Section \ref{cross_corr_po_sums} (see also \cite{Mai98c}) we have 
introduced harmonic inversion of cross-correlated periodic orbit sums as 
a method to significantly reduce the required number of periodic orbits for 
semiclassical quantization, and in Section \ref{hbar_sec2} (see also 
\cite{Mai98d}) we have discussed the $\hbar$ expansion of the periodic
orbit sum and the calculation of semiclassical eigenenergies beyond the
Gutzwiller \cite{Gut90} and Berry-Tabor \cite{Ber76,Ber77} approximation.
We now demonstrate both methods by way of example of the circle billiard.
The circle billiard is a regular system and has been chosen here for the
following reasons.
\begin{enumerate}
\item
The nearest neighbor level statistics of integrable systems is a Poisson 
distribution, with a high probability for nearly degenerate states.
The conventional method for periodic orbit quantization by harmonic inversion
requires very long signals to resolve the nearly degenerate states.
We will demonstrate the power of harmonic inversion of cross-correlated
periodic orbit sums by fully resolving those nearly degenerate states
with a significantly reduced set of orbits.
\item
All relevant physical quantities, i.e., the quantum and semiclassical
eigen\-energies, the matrix elements of operators, the periodic orbits and
their zeroth and first order amplitudes of the $\hbar$ expanded periodic
orbit sum, and the periodic orbit averages of classical observables can 
easily be obtained.
\item
The semiclassical quantization of the circle billiard as an example of an
{\em integrable} system demonstrates the universality and wide applicability 
of periodic orbit quantization by harmonic inversion, i.e., the method is not
restricted to systems with hyperbolic dynamics like, e.g., pinball scattering.
\end{enumerate}
The circle billiard has already been introduced in Section 
\ref{circle_analysis}.
The exact quantum mechanical eigenvalues $E=\hbar^2k^2/2M$ are given as 
zeros of Bessel functions $J_{|m|}(kR)=0$, where $m$ is the angular 
momentum quantum number and $R$, the radius of the circle.
In the following we choose $R=1$.
The semiclassical eigenvalues are obtained from the EBK quantization
condition, Eq.\ \ref{circle_EBK},
\[
 kR\sqrt{1-(m/kR)^2} - |m|\arccos{|m|\over kR} = \pi\left(n+{3\over 4}\right)
 \; ,
\]
and the $\hbar$ expansion of the periodic orbit sum reads 
(see Eq.\ \ref{g_hbar_series_circ})
\[
   {1\over\sqrt{k}} g(k)
 = {1\over\sqrt{k}} \sum_{n=0}^\infty g_n(k)
 = \sum_{n=0}^\infty {1\over k^{n}} \sum_{\rm po} 
   {\cal A}_{\rm po}^{(n)} e^{i\ell_{\rm po}k} \; ,
\]
with (see Eqs.\ \ref{A0_circ} and \ref{A1_circ})
\begin{eqnarray*}
   {\cal A}_{\rm po}^{(0)}
&=&\sqrt{\pi\over 2}{\ell_{\rm po}^{3/2}\over m_r^2} \,
   e^{-i({\pi\over 2}\mu_{\rm po}+{\pi\over 4})}  \; , \nonumber \\
   {\cal A}_{\rm po}^{(1)}
&=&{1\over 2}\sqrt{\pi m_r} \, {5-2\sin^2\gamma\over 3\sin^{3/2}\gamma} \,
   e^{-i({\pi\over 2}\mu_{\rm po}-{\pi\over 4})}  \; .
\end{eqnarray*}
The angle $\gamma$ is defined as $\gamma\equiv\pi m_\phi/m_r$, with
$m_\phi=1,2,\dots$ the number of turns of the periodic orbit around the 
origin, and $m_r=2m_\phi,2m_\phi+1,\dots$ the number of reflections at the 
boundary of the circle.
$\ell_{\rm po} = 2m_r \sin \gamma$ and $\mu_{\rm po}=3m_r$ are the
geometrical length and Maslov index of the orbits, respectively.\\

\subsubsection{Harmonic inversion of the cross-correlated periodic orbit sum}
\begin{figure}[b]
\vspace{18.0cm}
\includegraphics{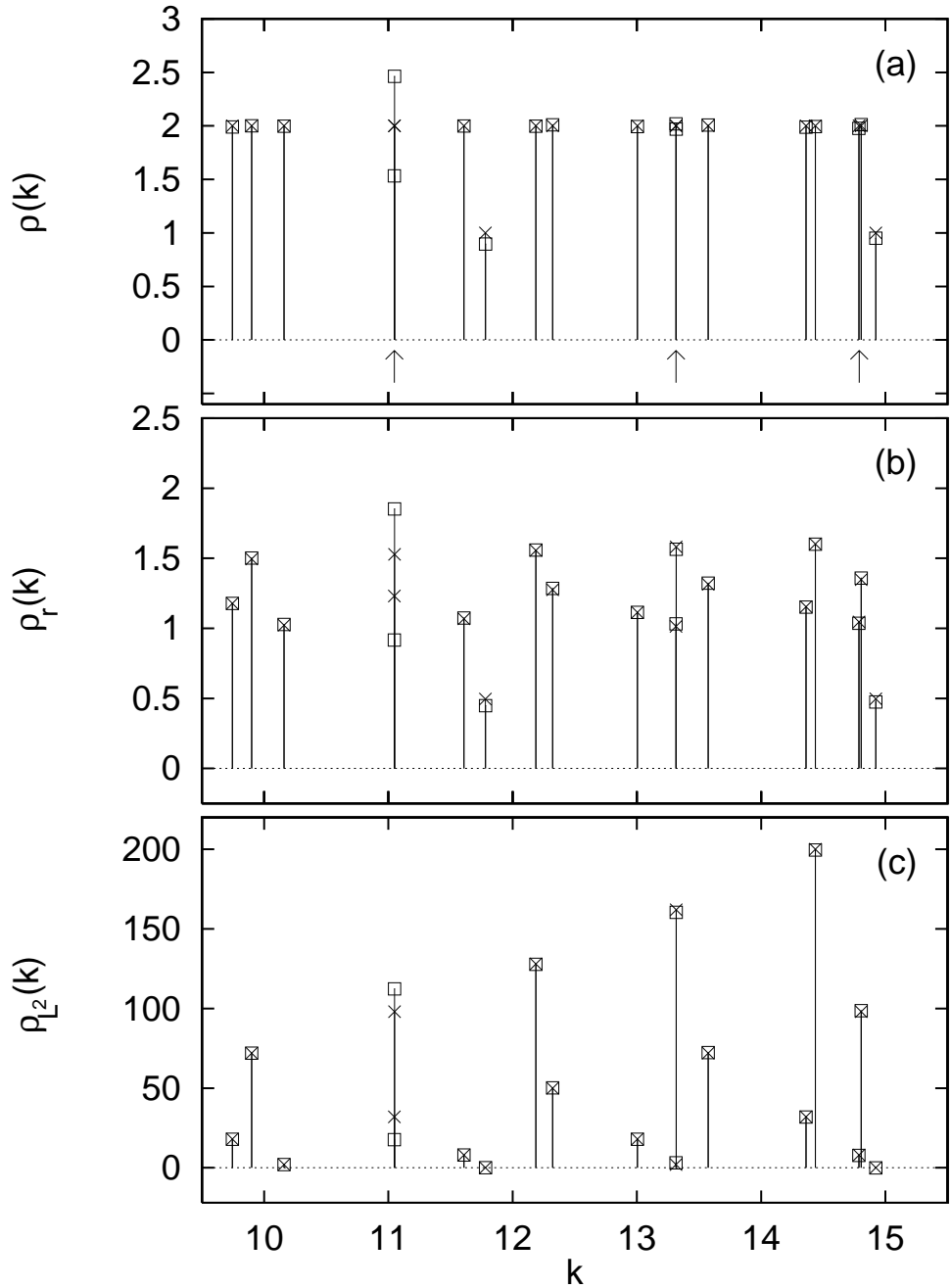}
\caption{\label{fig35} 
Density of states weighted with the diagonal matrix elements of the
operators (a) $\hat A=I$, (b) $\hat A=r$, (c) $\hat A=L^2$ for the
circle billiard with radius $R=1$. 
Crosses: EBK eigenvalues and quantum matrix elements.
Squares: Eigenvalues and matrix elements obtained by harmonic inversion
of cross-correlated periodic orbit sums.
Three nearly degenerate states are marked by arrows.
(From Ref.\ [85].)
}
\end{figure}

We now calculate the semiclassical eigenenergies of the circle billiard
by harmonic inversion of the cross-correlated periodic orbit sum 
(\ref{C_ab_sc}) with ${\cal A}_{\rm po}={\cal A}_{\rm po}^{(0)}$ 
(Eq.\ \ref{A0_circ}) the amplitudes of the Berry-Tabor formula 
\cite{Ber76,Ber77}, i.e., the lowest order $\hbar$ approximation.
To construct the periodic orbit cross-correlation signal
$C_{\alpha\beta}^{\rm sc}(\ell)$ we choose 
three different operators, 
\[
\hat A_1=I
\]
the identity,
\[
\hat A_2=r
\]
the distance from the origin, and
\[
\hat A_3=(L/k)^2
\]the square of the scaled angular momentum.
For these operators the classical weights $a_{\alpha,{\rm po}}$ 
(Eq.\ \ref{a_po}) are obtained as
\begin{eqnarray}
 a_{1,{\rm po}} &=& 1 \nonumber \\
\label{a_circ}
 a_{2,{\rm po}} &=& {1\over 2}\left(1+{\cos\gamma\over 
                    \tan\gamma} {\rm arsinh} \, \tan\gamma\right) \\
 a_{3,{\rm po}} &=& \cos^2\gamma \; . \nonumber
\end{eqnarray}
Once all the ingredients of Eq.\ \ref{C_ab_sc} for the circle billiard are 
available, the $3\times 3$ periodic orbit cross-correlation signal 
$C_{\alpha\beta}^{\rm sc}(\ell)$ can easily be constructed and
inverted by the generalized filter-diagonalization method.
Results obtained from the periodic orbits with maximum length $s_{\rm max}=100$
are presented in Fig.\ \ref{fig35}.
Fig.\ \ref{fig35}a is part of the density of states, $\varrho(k)$,
Figs.\ \ref{fig35}b and \ref{fig35}c are the density of states weighted with 
the diagonal matrix elements of the operators $\hat A=r$ and $\hat A=L^2$, 
respectively.
The squares are the results from the harmonic inversion of the periodic
orbit cross-correlation signals.
For comparison the crosses mark the matrix elements obtained by exact
quantum calculations at positions $k^{\rm EBK}$ obtained from the EBK
quantization condition (\ref{circle_EBK}).
In this section we do not compare with the exact zeros of the Bessel 
functions because Eq.\ \ref{C_ab_sc} is correct only to first order in 
$\hbar$ and thus the harmonic inversion of $C_{\alpha\beta}^{\rm sc}(s)$ 
cannot provide the exact quantum mechanical eigenvalues.
The calculation of eigenenergies beyond the Berry-Tabor approximation will 
be discussed in Section \ref{po_quant_beyond_BT_app}.
However, the perfect agreement between the eigenvalues $k^{\rm HI}$
obtained by harmonic inversion and the EBK eigenvalues $k^{\rm EBK}$ is
remarkable, and this is even true for nearly degenerate states marked by 
arrows in Fig.\ \ref{fig35}a.
The eigenvalues of some nearly degenerate states are presented in Table
\ref{table11}.
It is important to emphasize that these states with level splittings of, e.g.,
$\Delta k=6\times 10^{-4}$ cannot be resolved by the originally proposed 
method of periodic orbit quantization by harmonic inversion 
(see Section \ref{po_quant_sec})
with a periodic orbit signal length $s_{\rm max}=100$.
To resolve the two levels at $k\approx 11.049$ (see Table \ref{table11})
a signal length of at least $s_{\rm max}\approx 500$ is required if a single 
periodic orbit function $C^{\rm sc}(s)$ is used instead of a cross-correlation
function.
The method presented in Section \ref{cross_corr_po_sums} can therefore 
be used to significantly reduce the required signal length and thus the 
required number of periodic orbits for periodic orbit quantization by 
harmonic inversion.
As such the part of the spectrum shown in Fig.\ \ref{fig35} can even be
resolved, apart from the splittings of the nearly degenerate states marked 
by the arrows, from a short cross-correlation signal with $s_{\rm max}=30$, 
which is about the Heisenberg period $s_H=2\pi\bar\varrho(k)$, i.e.\
half of the signal length required for the harmonic inversion
of a $1\times 1$ signal.
With five operators and a $5\times 5$ cross-correlation signal highly excited
states around $k=130$ have even been obtained with a signal length
$s_{\rm max}\approx 0.7 s_H$ \cite{Wei99a,Wei99b}, which is close to the
signal length $s_{\rm max}\approx 0.5 s_H$ required for the Riemann-Siegel
type quantization \cite{Ber90,Kea92,Ber92}.
The reduction of the signal length is especially important if the periodic 
orbit parameters are not given analytically, as in our example of the circle 
billiard, but must be obtained from a numerical periodic orbit search.
How small can $s_{\rm max}$ get as one uses more and more operators in the
method?
It might be that half of the Heisenberg period is a fundamental barrier
for bound systems with chaotic dynamics in analogy to the Riemann-Siegel
formula \cite{Ber92} while for regular systems an even further reduction
of the signal length should in principle be possible.
However, further investigations are necessary to clarify this point.

\subsubsection{Periodic orbit quantization beyond the Berry-Tabor approximation}
\label{po_quant_beyond_BT_app}
The semiclassical eigenvalues obtained by harmonic inversion of a
cross-corre\-lated or a sufficiently long single signal are in excellent
agreement with the results of the EBK torus quantization, Eq.\ 
\ref{circle_EBK}.
However, they deviate from the exact quantum mechanical eigenenergies,
i.e., the zeros of the Bessel functions because the Berry-Tabor formula 
\cite{Ber76,Ber77} is only the lowest order $\hbar$ approximation of the 
periodic orbit sum.
We now demonstrate the $\hbar$ expansion of the periodic orbit sum and apply 
the technique discussed in Section \ref{hbar_sec2} to the circle billiard.

\begin{figure}[t]
\vspace{9.0cm}
\includegraphics{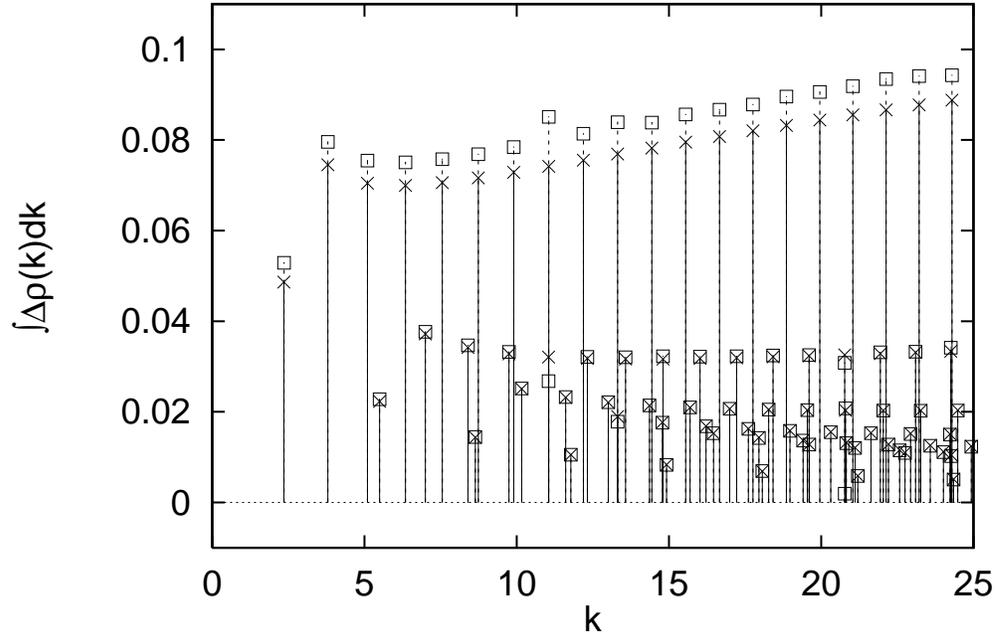}
\caption{\label{fig36} 
Integrated difference of the density of states, $\int\Delta\varrho(k)dk$,
for the circle billiard with radius $R=1$. 
Crosses: $\Delta\varrho(k)=\varrho^{\rm ex}(k)-\varrho^{\rm EBK}(k)$.
Squares: $\Delta\varrho(k)=\varrho^{(1)}(k)-\varrho^{(0)}(k)$
obtained from the $\hbar$ expansion of the periodic orbit signal.
}
\end{figure}

The first order corrections to the semiclassical eigenvalues,
$\Delta k=k^{(1)}-k^{(0)}$
are obtained by harmonic inversion of the periodic orbit signal
$C_1(\ell)$ (see Eq.\ \ref{Cn_sc}),
\begin{equation}
   C_1(\ell)
 = -i\sum_{\rm po}{1\over \ell_{\rm po}}{\cal A}_{\rm po}^{(1)}
     \delta(\ell-\ell_{\rm po})
 \stackrel{\rm h.i.}{=}
 i\sum_j d_j k_j^{(0)} \left[k_j^{(1)}-k_j^{(0)}\right] \, e^{-ik_j\ell}  \; ,
\label{C1_sc}
\end{equation}
with $d_j$ the multiplicities of states.
The signal $C_1(\ell)$ in Eq.\ \ref{C1_sc} can be inverted as a single
function as has been done in Ref.\ \cite{Mai98d}, where the accuracy of the
eigenenergies was improved by one to several orders of magnitude,
apart from the nearly degenerate states marked by arrows in Fig.\ 
\ref{fig35}.
Here we go one step further and use Eq.\ \ref{C1_sc} as part of a $3\times 3$
cross-correlation signal.
For the two other diagonal components of the cross-correlation matrix
we use the identity operator, $\hat A_1=I$, and the distance from the origin,
$\hat A_2=r$.
By applying the cross-correlation technique of Section \ref{cross_corr_po_sums}
we obtain both the zeroth and first order $\hbar$ expansion of the 
eigenenergies and the diagonal matrix elements of the chosen operators
simultaneously from one and the same harmonic inversion procedure.
Furthermore, we can now even resolve the nearly degenerate states.
The spectrum of the integrated differences of the density of states 
$\int\Delta\varrho(k)dk$ obtained by harmonic inversion of the $3\times 3$
cross-correlation matrix with signal length $\ell_{\rm max}=150$ is shown 
in Fig.\ \ref{fig36}.
The squares mark the spectrum for 
$\Delta\varrho(k)=\varrho^{\rm (1)}(k)-\varrho^{\rm (0)}(k)$
obtained from the harmonic inversion of the signal $C_1(s)$.
For comparison the crosses present the same spectrum but for the difference 
$\Delta\varrho(k)=\varrho^{\rm ex}(k)-\varrho^{\rm EBK}(k)$ between the 
exact quantum mechanical and the EBK-spectrum.
The deviations between the peak heights exhibit the contributions of terms
of the $\hbar$ expansion series beyond the first order approximation.
The peak heights of the levels in Fig.\ \ref{fig36} (solid lines and crosses)
are, up to a multiplicity factor for the degenerate states, the shifts 
$\Delta k$ between the zeroth and first order semiclassical approximations 
to the eigenvalues $k$.
The zeroth and first order eigenvalues, $k^{(0)}$ and 
$k^{(1)}=k^{(0)}+\Delta k$ are presented in Table \ref{table12} for the
40 lowest eigenstates.
The zeroth order eigenvalues, $k^{(0)}$, agree within the numerical accuracy 
with the results of the torus quantization, Eq.\ \ref{circle_EBK} 
(see eigenvalues $k^{\rm EBK}$ in Table \ref{table12}).
However, the semiclassical eigenvalues deviate significantly, especially
for states with low radial quantum numbers $n$, from the exact quantum
mechanical eigenvalues $k^{\rm ex}$ in Table \ref{table12}.
By contrast, the semiclassical error of the first order eigenvalues, $k^{(1)}$,
is by orders of magnitude reduced compared to the lowest order approximation.

\begin{figure}[b]
\vspace{9.0cm}
\includegraphics{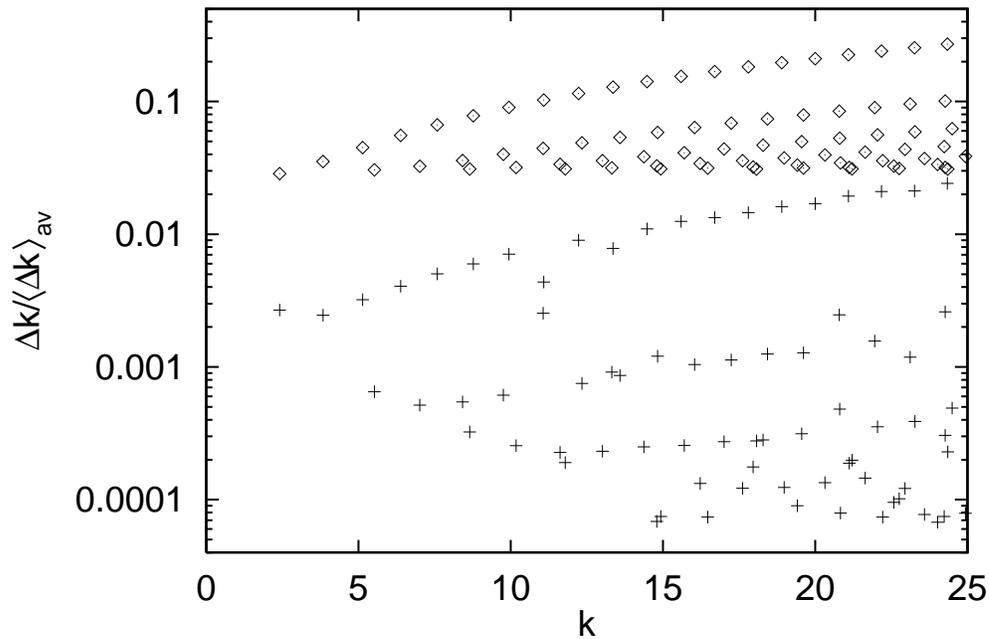}
\caption{\label{fig37} 
Semiclassical error $|k^{\rm (0)}-k^{\rm ex}|$ (diamonds) and 
$|k^{\rm (1)}-k^{\rm ex}|$ (crosses) in units of the average level spacing
$\langle\Delta k\rangle_{\rm av}\approx 4/k$.
}
\end{figure}

An appropriate measure for the accuracy of semiclassical eigenvalues is 
the deviation from the exact quantum eigenvalues in units of the average
level spacings, $\langle\Delta k\rangle_{\rm av}=1/\bar\varrho(k)$.
Fig.\ \ref{fig37} presents the semiclassical error in units of the average 
level spacings $\langle\Delta k\rangle_{\rm av}\approx 4/k$ for the zeroth
order (diamonds) and first order (crosses) approximations to the eigenvalues.
In the zeroth order approximation the semiclassical error for the low lying 
states is about 3 to 10 percent of the mean level spacing.
This error is reduced in the first order approximation by at least one order
of magnitude for the least semiclassical states with radial quantum number
$n=0$.
The accuracy of states with $n\ge 1$ is improved by two or more orders of 
magnitude.

Finally, we want to note that the small splittings between the nearly
degenerate states are extremely sensitive to the higher order $\hbar$
corrections.
E.g., in the zeroth order approximation the splitting between the two 
states around $k\approx 11.05$ is $\Delta k^{(0)}=6\times 10^{-4}$.
In the first order approximation the splitting between the same states 
is $\Delta k^{(1)}=0.0242$, which is very close to the exact splitting
$\Delta k^{\rm ex}=0.0217$.
The accuracy obtained here for the first order approximations to the nearly
degenerate states goes beyond the results presented in Ref.\ \cite{Mai98d},
and is achieved by the combined application of the methods introduced 
in Sections \ref{cross_corr_po_sums} and \ref{hbar_sec2}, i.e., the 
harmonic inversion of cross-correlation functions and the analysis of 
the $\hbar$ expanded periodic orbit sum, respectively.

\subsection{Semiclassical calculation of transition matrix elements 
for atoms in external fields}
\label{photo_sec}
The interpretation of photoabsorption spectra of atoms in external fields 
is a fundamental problem of atomic physics.
Although the ``exact'' quantum mechanics accurately describes the
energies and transition strengths of individual levels it has completely
failed to present a simple physical picture of the long ranged modulations
which have been observed in early low resolution spectra of barium atoms 
in a magnetic field \cite{Gar69} and later in the Fourier transform 
recurrence spectra of the magnetized hydrogen atom \cite{Hol86,Mai86}.
However, the long ranged modulations of the quantum photoabsorption spectra 
can be naturally interpreted in terms of the periods of classical closed 
orbits starting at and returning back to the nucleus where the initial 
state is localized.
The link between the quantum spectra and classical trajectories is given
by {\em closed orbit theory} \cite{Du87,Du88,Bog88a,Bog89} which describes
the photoabsorption cross section as the sum of a smooth part and the 
superposition of sinusoidal modulations.
The frequencies, amplitudes, and phases of the modulations are directly
obtained from the quantities of the closed orbits.
When the photoabsorption spectra are Fourier transformed or analyzed with
a high resolution method (see Sections \ref{FT_spectra} to \ref{prec_check})
the sinusoidal modulations result in sharp peaks in the Fourier transform 
recurrence spectra, and closed orbit theory has been most successful to 
explain quantum mechanical recurrence spectra qualitatively and 
even quantitatively in terms of the closed orbits of the underlying 
classical system \cite{Mai91,Mai94a,Vel93,Cou94}.

However, up to now practical applications of closed orbit theory have always
been restricted to the semiclassical calculation of {\em low resolution} 
photoabsorption spectra for the following two reasons.
First, the closed orbit sum requires, in principle, the knowledge of all
orbits up to infinite length, which are usually not available from a
numerical closed orbit search, and second, the infinite closed orbit sum 
suffers from fundamental convergence problems \cite{Du87,Du88,Bog88a,Bog89}.
It is therefore commonly accepted that the calculation of {\em individual} 
transition matrix elements $\langle\phi_i|D|\psi_f\rangle$ of the dipole 
operator $D$, which describe the transition strengths from the initial state 
$|\phi_i\rangle$ to final states $|\psi_f\rangle$, is a problem beyond
the applicability of the semiclassical closed orbit theory, i.e., is the 
domain of quantum mechanical methods.

In this section we disprove this common believe and demonstrate that 
individual eigenenergies and transition matrix elements can be directly 
extracted from the quantities of the classical closed orbits.
To that end, we slightly generalize closed orbit theory to the semiclassical
calculation of cross-correlated recurrence functions.
We then adopt the cross-correlation approach introduced in Section 
\ref{cross_corr_po_sums} to harmonically invert the cross-correlated 
recurrence signal and to extract the semiclassical eigenenergies and 
transition matrix elements.
Results will be presented for the photo excitation of the hydrogen atom in
a magnetic field.

The oscillator strength $f$ for the photo excitation of atoms in external 
fields can be written as
\begin{equation}
 f(E) = -{2\over\pi} (E-E_i) \:
   {\rm Im}\, \langle\phi_i|D G_E^+ D|\phi_i\rangle \; ,
\label{fE:eq}
\end{equation}
where $|\phi_i\rangle$ is the initial state at energy $E_i$, $D$ is the
dipole operator, and $G_E^+$ the retarded Green's function of the atomic 
system.
The basic steps for the derivation of closed orbit theory are to replace
the quantum mechanical Green's function in (\ref{fE:eq}) with its
semiclassical Van Vleck-Gutzwiller approximation and to carry out the
overlap integrals with the initial state $|\phi_i\rangle$.
Here we go one step further by introducing a cross-correlation matrix
\begin{equation}
 g_{\alpha\alpha'} = \langle\phi_\alpha|DG_E^+D|\phi_{\alpha'}\rangle
\label{g_def:eq}
\end{equation}
with $|\phi_\alpha\rangle$, $\alpha=1,2,\dots,L$ a set of independent initial 
states.
As will be shown below the use of cross-correlation matrices can considerably
improve the convergence properties of the semiclassical procedure.
In the following we will concentrate on the hydrogen atom in a magnetic field
with $\gamma=B/(2.35 \times 10^5\,{\rm T})$ the magnetic field strength
in atomic units.
As discussed in Section \ref{FT_spectra} the system has a scaling property, 
i.e., the shape of periodic orbits does not depend on the scaling parameter, 
$w=\gamma^{-1/3}=\hbar_{\rm eff}^{-1}$,
and the classical action scales as $S=sw$ with $s$ the scaled action.
As, e.g., in Ref.\ \cite{Mai94a} we consider scaled photoabsorption spectra 
at constant scaled energy $\tilde E=E\gamma^{-2/3}$ as a function of the 
scaling parameter $w$.
We choose dipole transitions between states with magnetic quantum number
$m=0$.
Note that the following ideas can be applied in an analogous way to atoms
in electric fields. 
Following the derivation of Refs.\ \cite{Du88,Bog89} the semiclassical
approximation to the fluctuating part of $g_{\alpha\alpha'}$ in Eq.\
\ref{g_def:eq} reads
\begin{eqnarray}
     g_{\alpha\alpha'}^{\rm sc}(w)
 &=& w^{-1/2} \sum_{{\rm co}} {-(2\pi)^{5/2}\over\sqrt{|m_{12}^{\rm co}|}}
    \sqrt{\sin\vartheta_i^{\rm co}\sin\vartheta_f^{\rm co}} \nonumber \\
 &\times&
    {\cal Y}_\alpha(\vartheta_i^{\rm co})
    {\cal Y}_{\alpha'}(\vartheta_f^{\rm co}) \,
    e^{i\left(s_{\rm co}w-{\pi\over 2}\mu_{\rm co}+{\pi\over 4}\right)} \; ,
\label{g_sc:eq}
\end{eqnarray}
with $s_{\rm co}$ and $\mu_{\rm co}$ the scaled action and Maslov index
of the closed orbit (co), $m_{12}^{\rm co}$ an element of the monodromy
matrix, and $\vartheta_i^{\rm co}$ and $\vartheta_f^{\rm co}$ the initial
and final angle of the trajectory with respect to the magnetic field axis.
The angular functions ${\cal Y}_\alpha(\vartheta)$ depend on the states
$|\phi_\alpha\rangle$ and the dipole operator $D$ and are given as a
linear superposition of Legendre polynomials,
${\cal Y}_\alpha(\vartheta)=\sum_l{\cal B}_{l\alpha} P_l(\cos\vartheta)$ 
with usually only few nonzero coefficients ${\cal B}_{l\alpha}$ with low $l$.
Explicit formulae for the calculation of the coefficients can be found in
Refs.\ \cite{Du88,Bog89} and in Appendix \ref{Ym_app}.
The problem is now to extract the semiclassical eigenenergies and transition
matrix elements from Eq.\ \ref{g_sc:eq} because the closed orbit sum does
not converge.
The Fourier transformation of $w^{1/2}g_{\alpha\alpha'}^{\rm sc}(w)$
yields the cross-correlated recurrence signals
\begin{equation}
   C_{\alpha\alpha'}^{\rm sc}(s)
 = \sum_{{\rm co}} {\cal A}_{\alpha\alpha'}^{\rm co} \delta(s - s_{\rm co})\; ,
\label{C_sc:eq}
\end{equation}
with the amplitudes
\begin{equation}
   {\cal A}_{\alpha\alpha'}^{\rm co}
 = {-(2\pi)^{5/2}\over\sqrt{|m_{12}^{\rm co}|}}
   \sqrt{\sin\vartheta_i^{\rm co}\sin\vartheta_f^{\rm co}} \,
   {\cal Y}_{\alpha }(\vartheta_i^{\rm co})
   {\cal Y}_{\alpha'}(\vartheta_f^{\rm co}) \,
   e^{i\left(-{\pi\over 2}\mu_{\rm co}+{\pi\over 4}\right)}
\label{ampl:eq}
\end{equation}
being determined exclusively by closed orbit quantities.
The corresponding quantum mechanical cross-correlated recurrence functions, 
i.e., the Fourier transforms of $w^{1/2}g_{\alpha\alpha'}^{\rm qm}(w)$ read
\begin{equation}
   C^{\rm qm}_{\alpha\alpha'}(s)
 = -i \sum_k b_{\alpha k} b_{\alpha' k} \, e^{-iw_ks} \; ,
\label{C_qm:eq}
\end{equation}
with $w_k$ the eigenvalues of the scaling parameter, and
\begin{equation}
 b_{\alpha k} = w_k^{1/4} \langle\phi_\alpha|D|\psi_k\rangle
\end{equation}
proportional to the transition matrix element for the transition from the 
initial state $|\phi_\alpha\rangle$ to the final state $|\psi_k\rangle$.

\begin{figure}[b]
\vspace{18.0cm}
\includegraphics{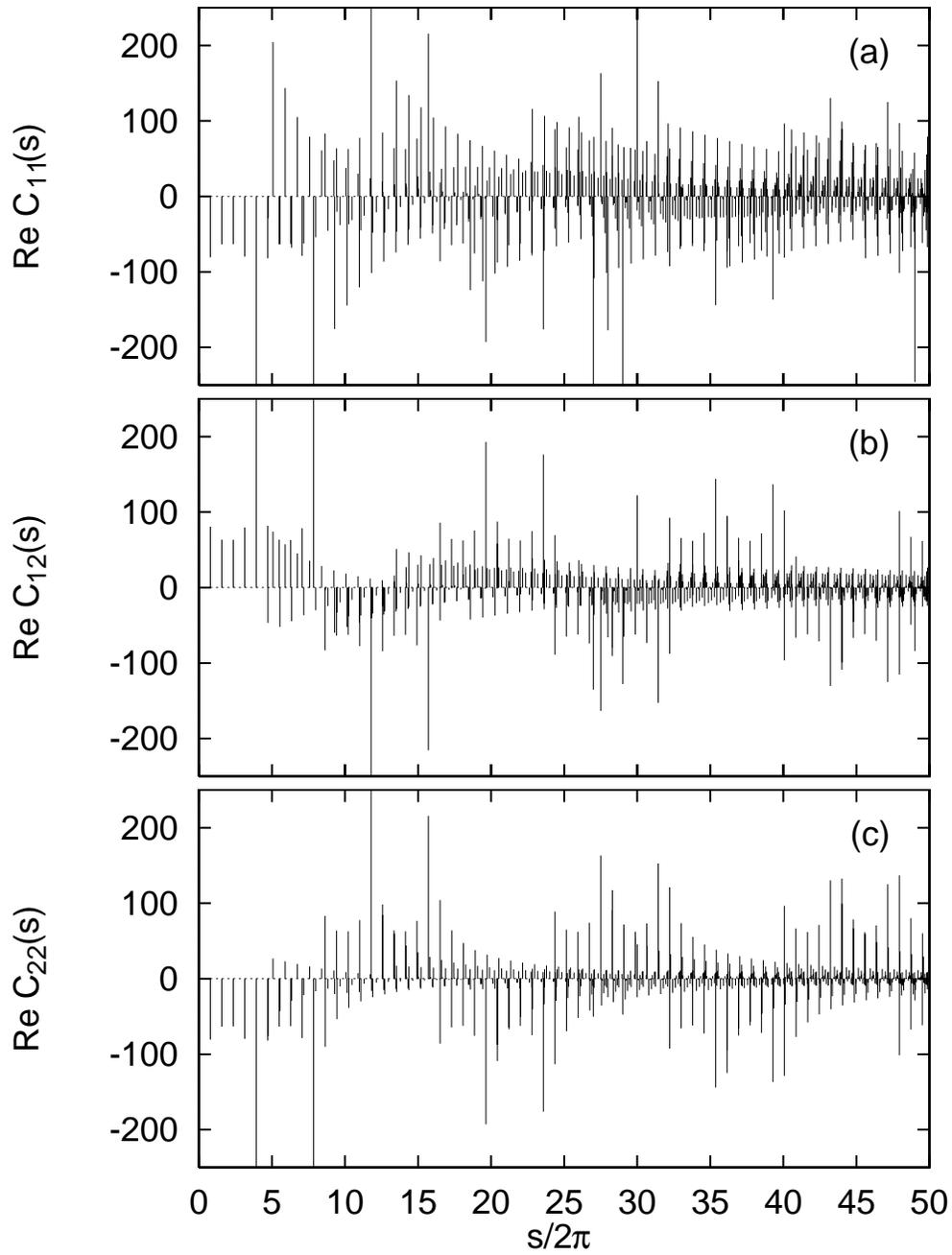}
\caption{\label{fig38} 
Real parts of the cross-correlated recurrence functions for the hydrogen 
atom in a magnetic field at constant scaled energy $\tilde E=-0.7$.
(From Ref.\ [89].)
}
\end{figure}

\begin{figure}[b]
\vspace{18.0cm}
\includegraphics{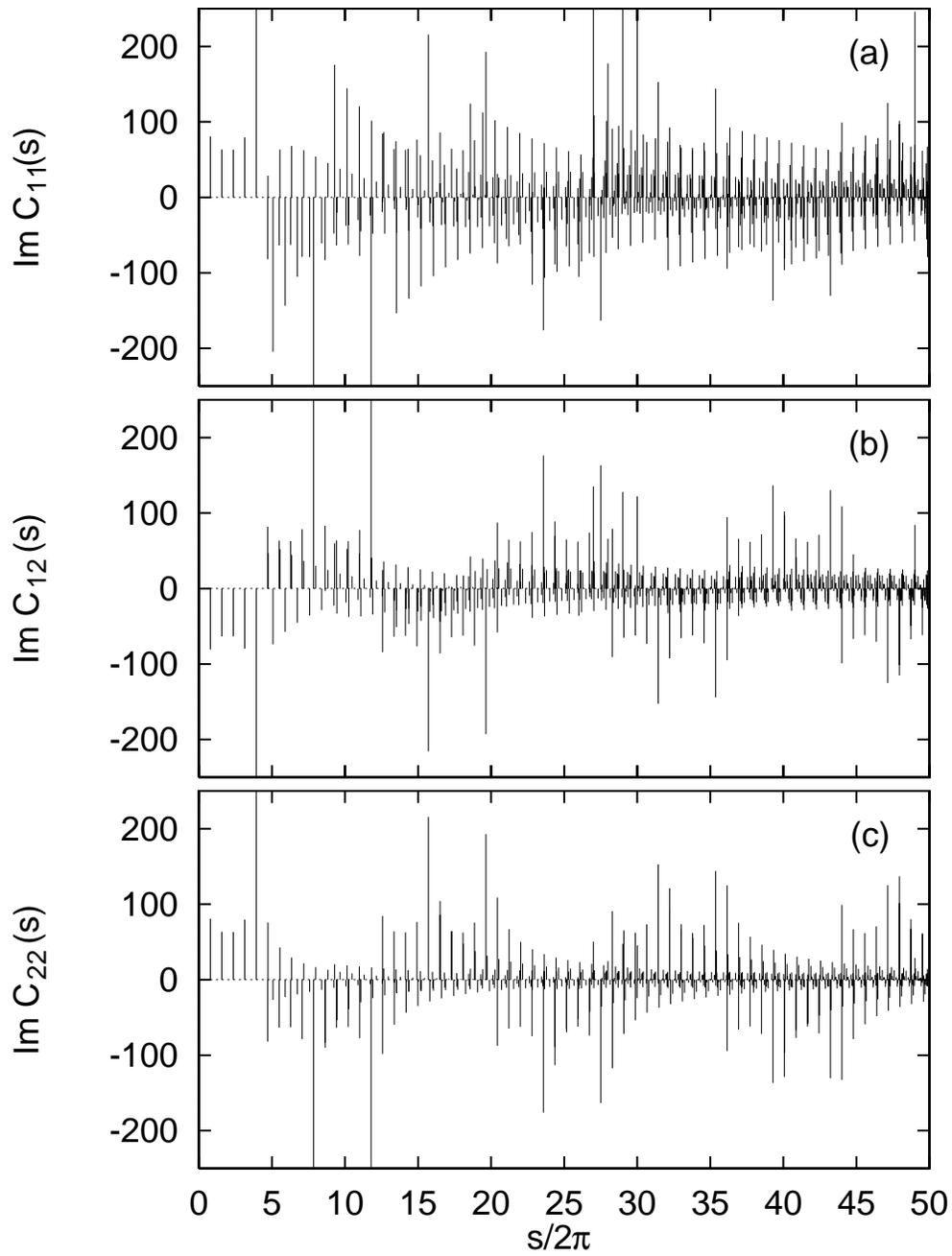}
\caption{\label{fig39} 
Same as Fig.\ \ref{fig38} but for the imaginary parts of the cross-correlated 
recurrence functions.
}
\end{figure}

The method to adjust Eq.\ (\ref{C_sc:eq}) to the functional form of
Eq.\ (\ref{C_qm:eq}) for a set of initial
states $|\phi_\alpha\rangle$, $\alpha=1,2,\dots,L$ is the harmonic inversion of
cross-correlation functions as discussed in Section \ref{cross_corr_po_sums}
and Appendix \ref{hi_cross_corr}.
We now demonstrate the method of harmonic inversion of the cross-correlated
closed orbit recurrence functions (\ref{C_sc:eq}) for the example of the
hydrogen atom in a magnetic field at constant scaled energy $\tilde E=-0.7$.
This energy was also chosen for detailed experimental investigations on
the helium atom \cite{Vel93}.
We investigate dipole transitions from the initial state 
$|\phi_1\rangle = |2p0\rangle$ with light polarized parallel to the 
magnetic field axis to final states with magnetic quantum number $m=0$.
For this transition the angular function in Eq.\ \ref{ampl:eq} reads
(see Appendix \ref{Ym_app})
${\cal Y}_1(\vartheta)=(2\pi)^{-1/2}2^7e^{-4}(4\cos^2\vartheta-1)$.
For the construction of a $2 \times 2$ cross-correlated recurrence signal
we use for simplicity as a second transition formally an outgoing $s$-wave, 
i.e., $D|\phi_2\rangle \propto Y_{0,0}$, and, thus, 
${\cal Y}_2(\vartheta)=\mbox{const}$.
A numerical closed orbit search yields 1395 primitive closed orbits 
(2397 orbits including repetitions) with scaled action $s/2\pi<100$.
With the closed orbit quantities at hand it is straightforward to calculate
the cross-correlated recurrence functions in (\ref{C_sc:eq}).
The real and imaginary parts of the complex functions $C_{11}^{\rm sc}(s)$, 
$C_{12}^{\rm sc}(s)$, and $C_{22}^{\rm sc}(s)$ with $s/2\pi<50$ are 
presented in Figs.\ \ref{fig38} and \ref{fig39}, respectively.
Note that for symmetry reasons $C_{21}^{\rm sc}(s)=C_{12}^{\rm sc}(s)$.

\begin{figure}[b]
\vspace{18.0cm}
\includegraphics{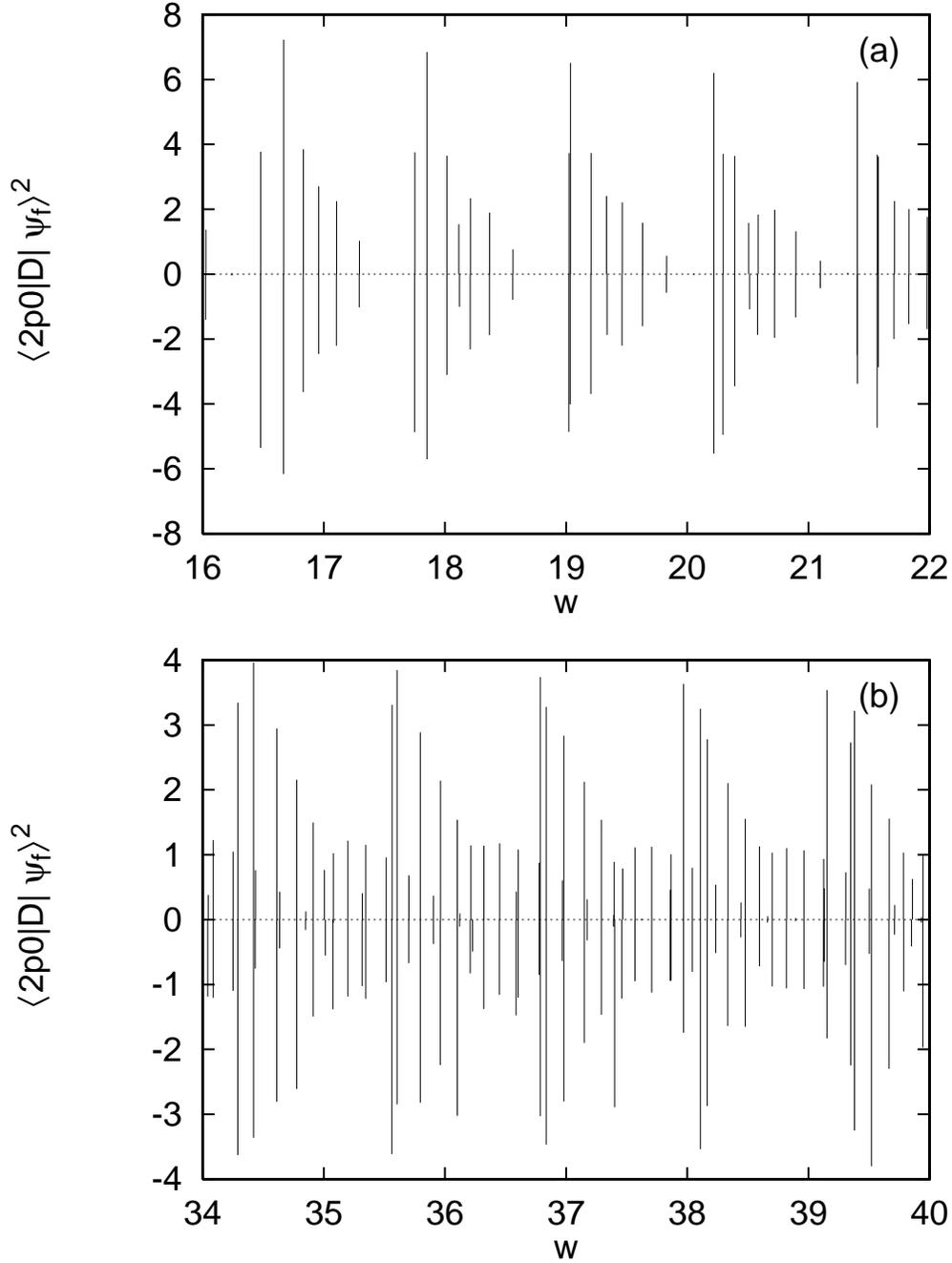}
\caption{\label{fig40} 
Quantum (upper part) and semiclassical (lower part) photoabsorption spectra
of the hydrogen atom in a magnetic field at scaled energy $\tilde E=-0.7$.
Transition matrix elements $\langle 2p0|D|\psi_f\rangle^2$ for dipole
transitions with light polarized parallel to the magnetic field axis.
(From Ref.\ [89].)
}
\end{figure}

We have inverted the $2 \times 2$ cross-correlated recurrence functions
in the region $0<s/2\pi<100$.
The resulting semiclassical photoabsorption spectrum is compared 
with the exact quantum spectrum in Fig.\ \ref{fig40}a for the region $16<w<21$
and in Fig.\ \ref{fig40}b for the region $34<w<40$.
The upper and lower parts in Fig.\ \ref{fig40} show the exact quantum spectrum 
and the semiclassical spectrum, respectively.
Note that the region of the spectrum presented in Fig.\ \ref{fig40}b belongs
well to the experimentally accessible regime with laboratory field strengths
$B=6.0 \, {\rm T}$ to $B=3.7 \,{\rm T}$.
The overall agreement between the quantum and semiclassical spectrum is 
impressive, even though a line by line comparison still reveals small 
differences for a few matrix elements. 
It is important to note that the high quality of the semiclassical spectrum
could only be achieved by our application of the cross-correlation approach.
For example, the two nearly degenerate states at $w=36.969$ and $w=36.982$
cannot be resolved and the very weak transition at $w=38.894$ with
$\langle2p0|D|\psi_f\rangle^2=0.028$ is not detected with a single 
$(1 \times 1)$ recurrence signal of the same length.
However, these hardly visible details are indeed present in the semiclassical 
spectrum in Fig.\ \ref{fig40}b obtained from the harmonic inversion of the 
$2 \times 2$ cross-correlated recurrence functions.

To summarize, we have demonstrated that closed orbit theory is not 
restricted to describe long ranged modulations in quantum mechanical
photoabsorption spectra of atoms in external fields but can well be 
applied to extract individual eigenenergies and transition matrix 
elements from the closed orbit quantities.
This is achieved by the high resolution spectral analysis (harmonic inversion)
of cross-correlated closed orbit recurrence signals.
For the hydrogen atom in a magnetic field we have obtained individual
transition matrix elements between low lying and highly excited
Rydberg states solely from the classical closed orbit data.

\section{Conclusion}
\label{conclusion}
\setcounter{equation}{0}
In summary, we have shown that harmonic inversion is a powerful tool
for the analysis of quantum spectra, and is the foundation for a novel
and universal method for periodic orbit quantization of both regular
and chaotic systems.

The high resolution analysis of finite range quantum spectra allows to 
circumvent the restrictions imposed by the uncertainty principle of the 
conventional Fourier transformation.
Therefore physical phenomena can directly be revealed in the
quantum spectra which previously were unattainable.
Topical examples are the study of quantum manifestations of periodic
orbit bifurcations and catastrophe theory, and the uncovering of symmetry
breaking effects.
The investigation of these phenomena provides a deeper understanding of
the relation between quantum mechanics and the dynamics of the underlying
classical system.
The high resolution technique is demonstrated in this work for numerically
calculated quantum spectra of, e.g., the hydrogen atom in external fields,
three disk pinball scattering, and the circle billiard.
Theoretical spectra are especially suited for the high resolution analysis
because of the very high accuracy of most quantum computational methods.
In principle, harmonic inversion may be applied to experimental spectra 
as well, e.g., to study atoms in external fields measured with the technique
of {\em scaled-energy spectroscopy} 
\cite{Hol88,Mai91,Mai94a,Vel93,Del94,Rai94,Cou94,Cou95a,Cou95b,Neu97,Spe97}.
However, the exact requirements on the precision of experimental data
to achieve high resolution recurrence spectra beyond the limitations of 
the uncertainty principle are not yet known.
It will certainly be a challenge for future experimental work to verify
the quantum manifestations of bifurcations, which have been extracted here
from theoretically computed spectra, using real systems in the laboratory.

We have also introduced harmonic inversion as a new and general tool for 
semiclassical periodic orbit quantization.
Here we briefly recall the highlights of our technique.
The method requires the complete set of periodic orbits up to a given maximum
period as input but does not depend on special properties of the orbits,
as, e.g., the existence of a symbolic code or a functional equation.
The universality and wide applicability has been demonstrated 
by applying it to systems with completely different properties, namely the 
zeros of the Riemann zeta function, the three disk scattering problem, and
the circle billiard.
These systems have been treated before by separate efficient methods, 
which, however, are restricted to bound ergodic systems, systems with a 
complete symbolic dynamics, or integrable systems.
The harmonic inversion technique allows to solve all these problems with 
one and the same method.
The method has furthermore been successfully applied to the hydrogen atom in 
a magnetic field as a prototype example of a system with mixed regular-chaotic
dynamics.
The efficiency of the method can be improved if additional semiclassical
information obtained from a set of linearly independent observables is used
to construct a cross-correlated periodic orbit sum, which can then be 
inverted with a generalized harmonic inversion technique.
The cross-correlated periodic orbit sum allows the calculation of 
semiclassical eigenenergies from a significantly reduced set of orbits.
Eigenenergies beyond the Gutzwiller and Berry-Tabor approximation are
obtained by the harmonic inversion of the $\hbar$ expansion of the periodic
orbit sum.
When applied, e.g., to the circle billiard the semiclassical accuracy is
improved by at least one to several orders of magnitude.
The combination of closed orbit theory with the cross-correlation approach 
and the harmonic inversion technique also allows the semiclassical calculation
of individual quantum transition strengths for atoms in external fields.

Periodic orbit quantization by harmonic inversion has been applied in this
work to systems with scaling properties, i.e., systems where the classical 
actions of periodic orbits depend linearly on a scaling parameter, $w$.
However, the method can even be used for the semiclassical quantization
of systems with non-homogeneous potentials such as the potential surfaces 
of molecules.
The basic idea is to introduce a generalized scaling technique with the 
inverse Planck constant $w\equiv 1/\hbar$ as the new formal scaling
parameter.
The generalized scaling technique can be applied, e.g., to the analysis of 
the rovibrational dynamics of the HO$_2$ molecule \cite{Mai97e}.
By varying the energy of the system, the harmonic inversion method yields
the semiclassical eigenenergies in the $(E,w)$ plane.
For non-scaling systems the semiclassical spectra can then be
compared along the line with the true physical Planck constant,
$w=1/\hbar=1$, with experimental measurements in the laboratory.

\begin{ack}
The author thanks H.\ S.\ Taylor and V.\ A.\ Mandelshtam for their 
collaboration and kind hospitality during his stay at the University of 
Southern California, where part of this work was initiated.
Financial support from the Feodor Lynen program of the Alexander von Humboldt 
foundation is gratefully acknowledged.
Stimulating discussions with
M.\ V.\ Berry,
E.\ B.\ Bogomolny,
D.\ Boos\'e,
J.\ Briggs,
D.\ Delande,
J.\ B.\ Delos,
B.\ Eckhardt,
S.\ Freund,
M.\ Gutzwiller,
F.\ Haake,
M.\ Haggerty,
E.\ Heller,
C.\ Jung,
J.\ Keating,
D.\ Kleppner,
M.\ Ku\'s,
C.\ Neumann,
B.\ Mehlig,
K.\ M\"uller,
H.\ Schomerus,
N.\ Spellmeyer,
F.\ Steiner,
G.\ Tanner,
T.\ Uzer,
K.\ H.\ Welge,
A.\ Wirzba, and
J.\ Zakrzewski
are also gratefully acknowledged.
The author thanks V.\ A.\ Mandelshtam for his program on 
filter-diagonalization and B.\ Eckhardt and A.\ Wirzba for supplying 
him with numerical data of the three disk system.

Part of the present work has been done in collaboration with K.\ Weibert 
and K.\ Wilmesmeyer, whose PhD thesis \cite{Wei99b} and diploma thesis 
\cite{Wil97} are prepared under the author's guidance.

The author is indebted to G.\ Wunner for his permanent support and his 
interest in the progress of this work.

The work was supported by the Deutsche Forschungsgemeinschaft via a 
Habilitandenstipendium (Grant No.\ Ma 1639/3) and the Sonderforschungsbereich
237 ``Unordnung und gro{\ss}e Fluktuationen''.

\end{ack}

\begin{appendix}
\section{Harmonic inversion by filter-diagonalization}
\label{harm_inv_app}
In the following we give details about the numerical method of harmonic 
inversion by filter-diagonalization.
We begin with the harmonic inversion of a single function and then extend
the method to the harmonic inversion of cross-correlation functions.

\subsection{Harmonic inversion of a single function}
\label{harm_inv}
The harmonic inversion problem can be formulated as a nonlinear fit 
(see, e.g., Ref.\ \cite{Mar87}) of the signal $C(s)$ defined on
an equidistant grid,
\begin{equation}
c_n\equiv C(n\tau)=\sum_k d_ke^{-in\tau w_k}\ ,\ \ 
n=0,1,2, \dots N,\label{eq:complexCt}
\end{equation}
with the set of generally complex variational parameters $\{w_k,d_k\}$.
(In this context the Discrete Fourier Transform scheme would 
correspond to a linear fit with $N$ amplitudes $d_k$ and 
fixed real frequencies $w_k=2\pi k/N\tau,\ k=1,2, \dots N$. The latter
implies the so called ``uncertainty principle'', i.e., the resolution, 
defined by the Fourier grid spacing, $\Delta w$, is inversely 
proportional to the length, $s_{\rm max}=N\tau$, of the signal C(s).)
The ``high resolution'' property associated with Eq.\ \ref{eq:complexCt} 
is due to the fact that there is no restriction for the closeness of the 
frequencies $w_k$ as they are variational parameters.
In Ref.\ \cite{Wal95} it was shown how this nonlinear fitting problem can be
recast as a linear algebraic one using the filter-diagonalization procedure.
The essential idea is to associate the signal $c_n$ with an autocorrelation 
function of a suitable dynamical system, 
\begin{equation}
c_n=\left(\Phi_0,\hat U^n \Phi_0\right), 
\label{eq:Neuhc_n}
\end{equation}
where $(\ \cdot\ ,\ \cdot\ )$ defines a complex symmetric inner product 
(i.e., no complex conjugation).
The evolution operator can be defined implicitly by
\begin{equation}
   \hat U\equiv e^{-i\tau\hat\Omega}
 = \sum_{k=1}^K e^{-i\tau w_k} |\Upsilon_k)(\Upsilon_k| \; ,
\label{eq:U}
\end{equation}
where the set of eigenvectors $\{\Upsilon_k\}$ is associated with 
an arbitrary orthonormal basis set and the eigenvalues of $\hat U$ are 
$u_k\equiv e^{-i\tau w_k}$ (or equivalently the eigenvalues of 
$\hat \Omega$ are $ w_k$).
Inserting Eq.\ \ref{eq:U} into Eq.\ \ref{eq:Neuhc_n} we obtain Eq.\ 
\ref{eq:complexCt} with 
\begin{equation}
d_k=(\Upsilon_k,\Phi_0)^2,\label{eq:dk_def}
\end{equation}
which also implicitly
defines the ``initial state'' $\Phi_0$.
This construction establishes an equivalence between
the problem of extracting spectral information from the signal with the one of
diagonalizing the evolution operator $\hat U=e^{-i\tau\hat \Omega}$ 
(or the Hamiltonian $\hat \Omega$) of the
fictitious underlying dynamical system. 
The filter-diagonalization method is then used for extracting the eigenvalues 
of the Hamiltonian $\hat \Omega$ in any chosen small energy window. 
Operationally this is done by solving a small generalized eigenvalue problem
whose eigenvalues yield the frequencies in a chosen window.
The knowledge of the operator $\hat \Omega$ itself is not required,
as for a properly chosen basis the matrix elements of $\hat \Omega$ can be 
expressed only in terms of $c_n$.
The advantage of the  filter-diagonalization procedure 
is its numerical stability with respect to both the length
and complexity 
(the number and density of  the contributing frequencies) of the signal.
Here we apply the method of Refs.\ \cite{Man97a,Man97b} which is an 
improvement of the filter-diagonalization method of Ref.\ \cite{Wal95} in 
that it allows to significantly reduce the required length of the signal by 
implementing a different Fourier-type basis
with an efficient rectangular filter. Such a basis
is defined by choosing a small set of values $\varphi_j$ in the frequency 
interval of interest,
$\tau w_{min}<\varphi_j<\tau w_{max},\ j=1,2,...,J$, and
the maximum order, $M$, of the Krylov vectors, 
$\Phi_n = e^{-in\tau\hat\Omega}\Phi_0$, used in the Fourier series,
\begin{equation} 
\Psi_j\equiv\Psi(\varphi_j)=\sum_{n=0}^M e^{in\varphi_j}\Phi_n
\equiv\sum_{n=0}^M e^{in(\varphi_j-\tau\hat\Omega)}\Phi_0.
\label{eq:Psi_j}
\end{equation} 
It is convenient to introduce the notations, 
\begin{equation}
U^{(p)}_{jj'}\equiv  U^{(p)}(\varphi_j,\varphi_{j'})= 
\left(\Psi(\varphi_j),e^{-ip\tau\hat\Omega}\Psi(\varphi_{j'})\right), 
\label{eq:Up}
\end{equation}
for the matrix elements of the operator $e^{-ip\tau\hat\Omega}$,
and $\mbox{\bf U}^{(p)}$, for the corresponding small $J\times J$ complex 
symmetric matrix.
As such $\mbox{\bf U}^{(1)}$ denotes the
matrix representation of the operator $\hat U$ itself and $\mbox{\bf U}^{(0)}$,
the overlap matrix with elements $(\Psi(\varphi_j),\Psi(\varphi_{j'}))$,
which is required as the vectors $\Psi(\varphi_j)$ are not generally 
orthonormal.
Now using these definitions  we can set up a generalized eigenvalue problem,
\begin{equation}
\mbox{\bf U}^{(p)}\mbox{\bf B}_k=e^{-ip\tau w_k} 
\mbox{\bf U}^{(0)}\mbox{\bf B}_k, 
\label{eq:generalized}
\end{equation}
for the eigenvalues $e^{-ip\tau w_k}$ 
of the operator $e^{-ip\tau\hat\Omega}$. 
The column vectors $\mbox{\bf B}_k$ with elements $B_{jk}$ define 
the eigenvectors $\Upsilon_k$ in terms of the basis functions $\Psi_j$ as
\begin{equation}
\Upsilon_k = \sum_{j=1}^J B_{jk}\Psi_j, \label{eq:Bk}
\end{equation}
assuming that the $\Psi_j$'s form a locally complete basis.

The matrix elements (\ref{eq:Up}) can be expressed in terms of the signal 
$c_n$, the explicit knowledge of the auxiliary objects 
$\hat \Omega$, $\Upsilon_k$ or $\Phi_0$ is not needed.
Indeed, insertion of Eq.\ \ref{eq:Psi_j} into Eq.\ \ref{eq:Up},
use of the symmetry property,
$(\Psi,\hat U\Phi)=(\hat U\Psi,\Phi)$, and
the definition of $c_n$, Eq.\ \ref{eq:Neuhc_n}, gives after some arithmetics
\begin{eqnarray}
\label{eq:Upjj}
   U^{(p)} (\varphi,\varphi')
 &=& \big(e^{-i\varphi}-e^{-i\varphi'}\big)^{-1} 
     \Big[e^{-i\varphi}\sum_{n=0}^M e^{in\varphi'} c_{n+p}\\
 &-& e^{-i\varphi'}\sum_{n=0}^M e^{in\varphi}c_{n+p}
   -e^{iM\varphi}\sum_{n=M+1}^{2M}e^{i(n-M-1)\varphi'}c_{n+p}\nonumber\\
 &+& e^{iM\varphi'}\sum_{n=M+1}^{2M}e^{i(n-M-1)\varphi}c_{n+p}\Big],\ 
     \varphi\neq\varphi'\ ,\nonumber \\
   U^{(p)} (\varphi,\varphi)
 &=& \sum_{n=0}^{2M}(M-|M-n|+1)e^{in\varphi}c_{n+p}.\nonumber
\end{eqnarray}
(Note that the evaluation of {\bf U}$^{(p)}$ requires the knowledge of 
$c_n$ for $n=p,p+1, \dots ,N=2M+p$.)

The generalized eigenvalue problem (\ref{eq:generalized}) can be solved 
by a singular value decomposition of the matrix {\bf U}$^{(0)}$, or more 
accurately by application of the QZ algorithm \cite{Wil79}, which is 
implemented, e.g., in the NAG library.
Each value of $p$ yields a set of frequencies $w_k$ and,
due to Eqs.\ \ref{eq:dk_def}, \ref{eq:Psi_j} and \ref{eq:Bk}, amplitudes, 
\begin{equation}
d_k=\left(\sum_{j=1}^J B_{jk}\sum_{n=0}^M c_ne^{in\varphi_j}\right)^2. 
\label{eq:dk}
\end{equation}
Note that Eq.\ \ref{eq:dk} is a functional of the half signal $c_n, 
n=1,2,\dots,M$.
An even better expression for the coefficients $d_k$ 
(see Ref.\ \cite{Man97b}) reads
\begin{eqnarray}
d_k & = & \left[{1\over M+1}\sum_{j=1}^{J}B_{jk}
{\Big(}\Psi(\varphi_j),\Psi(w_k){\Big)}\right]^2\nonumber\\
& \equiv & \left[{1\over M+1}\sum_{j=1}^{J}B_{jk}U^{(0)}
  (\varphi_j,w_k)\right]^2 \; ,
\label{eq:bestdk}
\end{eqnarray}
with $U^{(0)}(\varphi_j,w_k)$ defined by Eq.\ \ref{eq:Upjj}.
Eq.\ \ref{eq:bestdk} is a functional of the whole available signal $c_n$,
$n=0,1,\dots,2M$ and therefore sometimes provides more precise results than 
Eq.\ \ref{eq:dk}.

The converged $w_k$ and $d_k$ should not depend on $p$. 
This condition allows us to identify spurious or non-converged frequencies 
by comparing the results with different values of $p$ 
(e.g., with $p=1$ and $p=2$).
We can define the simplest error estimate $\varepsilon$ as the difference 
between the frequencies $w_k$ obtained from diagonalizations with $p=1$ and 
$p=2$, i.e.\
\begin{equation}
 \varepsilon = | w_k^{(p=1)} - w_k^{(p=2)} | \; .
\label{eps_def}
\end{equation}

\subsection{Harmonic inversion of cross-correlation functions}
\label{hi_cross_corr}
We now consider a cross-correlation signal, i.e., a $D\times D$ matrix
of signals defined on an equidistant grid $(\alpha,\alpha'=1,2,\dots,D)$:
\begin{equation}
   c_{\alpha\alpha'}(n) \equiv C_{\alpha\alpha'}(n\tau)
 = \sum_k b_{\alpha,k}b_{\alpha',k} e^{-in\tau w_k} \; , \quad
 n = 0, 1, 2, \dots , N \; .
\label{c_cross_corr}
\end{equation}
[We choose $\tau=1$ for simplicity in what follows.]
Each component of the signal $c_{\alpha\alpha'}(n)$ contains the same
set of frequencies $w_k$, and the amplitudes belonging to each frequency
are correlated, i.e., $d_{\alpha\alpha',k}=b_{\alpha,k}b_{\alpha',k}$ with
only $D$ (instead of $D^2$) independent parameters $b_{\alpha,k}$.
As for the harmonic inversion of a single function the nonlinear problem 
of adjusting the parameters $\{w_k,b_{\alpha,k}\}$ can be recast as a
linear algebra one using the filter-diagonalization procedure
\cite{Wal95,Nar97,Man98}.
The cross-correlation signal (\ref{c_cross_corr}) is associated with the
cross-correlation function of a suitable dynamical system,
\begin{equation}
   c_{\alpha\alpha',n}
 = \left(\Phi_\alpha, \hat U^n \Phi_{\alpha'}\right) \; ,
\label{c_cross_corr_U}
\end{equation}
with the same complex symmetric inner product as in Eq.\ \ref{eq:Neuhc_n},
and the evolution operator $\hat U$ defined implicitly by Eq.\ \ref{eq:U}.
The extension from Eq.\ \ref{eq:Neuhc_n} to Eq.\ \ref{c_cross_corr_U}
is that the autocorrelation function $(\Phi_0,\hat U^n \Phi_0)$ built of
a single state $\Phi_0$ is replaced with the cross-correlation function 
$(\Phi_\alpha,\hat U^n \Phi_{\alpha'})$ built of a set of $D$ different 
states $\Phi_\alpha$.
Inserting (\ref{eq:U}) into Eq.\ (\ref{c_cross_corr_U}) we obtain Eq.\
(\ref{c_cross_corr}) with
\begin{equation}
 b_{\alpha,k} = (\Upsilon_k,\Phi_\alpha) \; ,
\end{equation}
which now implicitly defines the states $\Phi_\alpha$.
After choosing a basis set in analogy to Eq.\ \ref{eq:Psi_j},
\begin{equation} 
   \Psi_{\alpha j}
 = \Psi_\alpha(\varphi_j)
 = \sum_{n=0}^M e^{in(\varphi_j-\hat\Omega)}\Phi_\alpha \; ,
\label{eq:Psi_aj}
\end{equation} 
and introducing the notations
\begin{equation}
   \mbox{\bf U}^{(p)}
 \equiv U^{(p)}_{\alpha j,\alpha'j'}
 = U_{\alpha\alpha'}^{(p)}(\varphi_j,\varphi_{j'})
 = \left(\Psi_\alpha(\varphi_j),
     e^{-ip\hat\Omega}\Psi_{\alpha'}(\varphi_{j'})\right) \; ,
\label{eq:Up_generalized}
\end{equation}
for the small matrix of the operator $e^{-ip\hat\Omega}$ in the basis
set (\ref{eq:Psi_aj}), we can set up a generalized eigenvalue problem,
\[
\mbox{\bf U}^{(p)}\mbox{\bf B}_k=e^{-ip w_k} 
\mbox{\bf U}^{(0)}\mbox{\bf B}_k \; , 
\]
for the eigenvalues $e^{-ip w_k}$ of the operator $e^{-ip\hat\Omega}$, 
which is formally identical with Eq.\ \ref{eq:generalized}.
The matrix elements (\ref{eq:Up_generalized}) can be expressed in terms
of the signal $c_{\alpha\alpha',n}$.
The following expression for the matrix elements of $\mbox{\bf U}^{(p)}$ is 
derived in complete analogy with Refs.\ \cite{Man97a,Man97b} with 
the additional indices $\alpha,\alpha'$ being the only difference,
\begin{eqnarray}
U^{(p)}_{\alpha\alpha'} (\varphi,\varphi')  &=&  \big(e^{-i\varphi}
- e^{-i\varphi'}\big)^{-1} 
\Big[e^{-i\varphi}\sum_{n=0}^M e^{in\varphi'} c_{\alpha\alpha'}(n+p)\nonumber\\
&&- e^{-i\varphi'} \sum_{n=0}^M e^{in\varphi}c_{\alpha\alpha'}(n+p)\nonumber\\
&&- e^{iM\varphi} \sum_{n=M+1}^{2M}e^{i(n-M-1)\varphi'}
c_{\alpha\alpha'}(n+p)\nonumber \\
&&+ e^{iM\varphi'} 
  \sum_{n=M+1}^{2M}e^{i(n-M-1)\varphi}c_{\alpha\alpha'}(n+p)\Big] \; , \;
\varphi\neq\varphi' \; , \\
U^{(p)}_{\alpha\alpha'}, (\varphi,\varphi) &=&
 \sum_{n=0}^{2M}(M-|M-n|+1)e^{in\varphi}c_{\alpha\alpha'}(n+p) \; .
\end{eqnarray}
Given the cross-correlation signal $c_{\alpha\alpha'}(n)$, the solution of 
the generalized eigenvalue problem (\ref{eq:generalized}) yields the 
eigenfrequencies $w_k$ and the eigenvectors $\mbox{\bf B}_k$.
The latter can be used to compute the amplitudes,
\begin{equation}
b_{\alpha k}={1-e^{-\gamma}\over 1-e^{-(M+1)\gamma} }
\sum_{j=1}^{J}\sum_{\alpha'=1}^D B_{\alpha' j,k}
U^{(0)}_{\alpha\alpha'}(\varphi_j, w_k+i\gamma) \; ,
\label{eq:bestbk}
\end{equation}
where the adjusting parameter $\gamma$ is chosen so that
$U^{(0)}(\varphi_j, w_k+i\gamma)$ is numerically stable \cite{Man98}.
One correct choice is 
$\gamma=-\mbox{Im}\ w_k$ for $\mbox{Im}\ w_k<0$ and 
$\gamma=0$ for $\mbox{Im}\ w_k>0$.

\section{Angular function ${\cal Y}_{\lowercase{m}}(\vartheta)$}
\label{Ym_app}
In Section \ref{bfly_sec} we have presented Eq.\ \ref{f_osc} as the final
result of closed orbit theory for the semiclassical photoabsorption spectrum
of the hydrogen atom in a magnetic field.
Here we define explicitly the angular function ${\cal Y}_m(\vartheta)$
in Eq.\ \ref{f_osc}.

The angular function ${\cal Y}_m(\vartheta)$ solely depends on the initial
state $\psi_i$ and the dipole operator $D$ and is a linear superposition
of spherical harmonics:
\begin{equation}
 {\cal Y}_m(\vartheta) = \sum_{\ell=|m|}^\infty (-1)^\ell {\cal B}_{\ell m}
                         Y_{\ell m}(\vartheta,0)  \; .
\end{equation}
The coefficients ${\cal B}_{\ell m}$ are defined by the overlap integrals
\begin{equation}
 {\cal B}_{\ell m} = \int d^3x (D\psi_i)({\bf x}) \sqrt{2/r} \,
   {\rm J}_{2\ell+1}\left(\sqrt{8r}\right) Y_{\ell m}^\ast(\vartheta,\varphi)
\end{equation}
(with ${\rm J}_\nu(x)$ the Bessel functions)
and can be calculated analytically \cite{Bog89}.
For excitations of the ground state $\psi_i=|1s0\rangle$ with $\pi$-polarized
light (i.e.\ dipole operator $D=z$) the explicit result is
\begin{equation}
 {\cal Y}_0(\vartheta) = -\pi^{-1/2} 2^3 e^{-2} \cos\vartheta \; ,
\end{equation}
and for $\psi_i=|2p0\rangle$, i.e., the initial state in many spectroscopic
measurements on hydrogen \cite{Hol88,Mai91,Mai94a} we obtain
\begin{equation}
 {\cal Y}_0(\vartheta) = (2\pi)^{-1/2} 2^7 e^{-4} 
    \left(4\cos^2\vartheta-1\right)  \; .
\end{equation}

\section{Catastrophe diffraction integrals}
\label{diffr_int_app}
Here we give some technical details about the numerical calculation
and the asymptotic expansion of catastrophe diffraction integrals for the
hyperbolic umbilic and the butterfly catastrophe.

\subsection{The hyperbolic umbilic catastrophe}
\label{hyp_umb_app}
The catastrophe diffraction integral of the hyperbolic umbilic
(Eq.\ \ref{Psi_def}) reads
\begin{equation}
 \Psi(x,y) = \int_{-\infty}^{+\infty}dp \int_{-\infty}^{+\infty}dq
   e^{i[p^3 + q^3 + y(p+q)^2 + x(p+q)]} \; .
\label{Psi_def_app}
\end{equation}
By substituting
\begin{eqnarray*}
 p &=& s_1 - {y\over 3} \\
 q &=& s_2 - {y\over 3}
\end{eqnarray*}
we obtain
\begin{equation}
 \Psi(x,y) = e^{i\left[{2\over 3}y\left({5\over 9}y^2-x\right)\right]}
  \Phi\left(x-y^2,2y\right) \; ,
\end{equation}
with
\begin{equation}
 \Phi(\xi,\eta) = \int_{-\infty}^{+\infty}ds_1 \int_{-\infty}^{+\infty}ds_2
   e^{i[s_1^3 + s_2^3 + \xi(s_1+s_2) + \eta s_1 s_2]} \; .
\end{equation}
The integral $\Phi(\xi,\eta)$ can be expanded into a Taylor series around
$\eta=0$.
Using
\begin{eqnarray}
     {\partial^n \Phi(\xi,\eta)\over\partial\eta^n}\bigg|_{\eta=0}
 &=& \int_{-\infty}^{+\infty}ds_1 \int_{-\infty}^{+\infty}ds_2
     i^n (s_1s_2)^n e^{i[s_1^3 + s_2^3 + \xi(s_1+s_2)]} \nonumber \\
 &=& i^n \left(\int_{-\infty}^{+\infty}ds\, s^n e^{i(s^3+\xi s)}\right)^2 \; ,
\end{eqnarray}
and solving the one-dimensional integrals \cite{Gra65}
\begin{eqnarray}
 &&  \int_{-\infty}^{+\infty}ds s^n e^{i(s^3+\xi s)} \nonumber \\
 &=& {1\over 3} \sum_{k=0}^\infty {(-\xi)^k\over k!}
     \Gamma\left({k+n+1\over 3}\right)
     \left[e^{i(n+1-2k)\pi/6}+(-1)^n e^{-i(n+1-2k)\pi/6}\right] \; ,
\end{eqnarray}
we finally obtain
\begin{eqnarray}
     \Phi(\xi,\eta)
 &=& {1\over 9}\sum_{n=0}^\infty{(i\eta)^n\over n!}
 \Bigg\{\sum_{k=0}^\infty {(-\xi)^k\over k!}
     \Gamma\left({k+n+1\over 3}\right)
 \nonumber \\  &\times& 
     \left[e^{i(n+1-2k)\pi/6}+(-1)^n e^{-i(n+1-2k)\pi/6}\right]\Bigg\}^2 \, ,
\end{eqnarray}
which is a convergent series for all $\xi$ and $\eta$.

\subsection{The butterfly catastrophe}
\label{bfly_app}
The uniform phase integral $\Psi(x,y)$ of the butterfly catastrophe
is expanded in a two-parametric Taylor series around $x=y=0$:
\begin{eqnarray}
 \Psi(x,y) & \equiv & \int_{-\infty}^{+\infty}
           \exp{\left[-i(x t^2 + y t^4 + t^6)\right]} dt  \nonumber \\
\label{Psidef}
 & = & \sum_{n=0}^\infty \sum_{m=0}^\infty {1 \over {i^{n+m}}}
     {x^n y^m \over n!\,m!} \int_{-\infty}^{+\infty} t^{2n+4m}
                             \exp{\left[-it^6\right]} dt  \; .
\end{eqnarray}
With the substitution $z=t^{2n+4m+1}$ we obtain \cite{Gra65}
\begin{eqnarray*}
 &   &  \int_{-\infty}^{+\infty} t^{2n+4m} \exp{\left[-it^6\right]} dt \\
 & = &  {2 \over {2n+4m+1}} \int_0^{\infty}
                               \exp{\left[-iz^{6/(2n+4m+1)}\right]} dz \\
 & = &  {1 \over 3} \exp{\left[-i{2n+4m+1 \over 12}\pi\right]} ~
        \Gamma \left( {2n+4m+1 \over 6} \right) \; ,
\end{eqnarray*}
and finally
\begin{eqnarray}
 \Psi(x,y) & = & {1 \over 3} \exp{\left[-i{\pi\over12}\right]}
   \sum_{n=0}^\infty \sum_{m=0}^\infty
   {1 \over n!\,m!} \Gamma  \left( {2n+4m+1 \over 6} \right)   \nonumber \\
 & \times &  
   \left( x \exp{\left[-i{2\over3}\pi\right]} \right) ^n
   \left( y \exp{\left[-i{5\over6}\pi\right]} \right) ^m \; ,
\end{eqnarray}
which is a convergent series for all $x$ and $y$.

The asymptotic behavior of $\Psi(x,y)$ for $x\to\pm\infty$ is obtained
in a stationary phase approximation to Eq.\ (\ref{Psidef}) with the
stationary points $t_0$ being defined by
\begin{equation}
 t_0 (6t_0^4 + 4yt_0^2 + 2x) = 0 \; .   \label{t_0_poly}
\end{equation}
a) $x\to-\infty$:

\noindent
There are three real stationary points given by
\[
  t_0=0 ~~ {\rm and} ~~
  t_0^2=-{1\over3}y + \sqrt{{1\over9}y^2-{1\over3}x}
\]
and we obtain
\begin{eqnarray}
  \Psi(x,y) & \stackrel{x\to-\infty}{\longrightarrow} &
  \int_{-\infty}^{+\infty} \exp{\left[-ixt^2\right]} dt 
   + 2\exp{\left[-i(xt_0^2+yt_0^4+t_0^6)\right]} \nonumber \\
 & \times & \int_{-\infty}^{+\infty}
             \exp{\left[-i(x+6yt_0^2+15t_0^4)\,t^2\right]} dt \nonumber \\
 & = & \sqrt{\pi\over -x} \exp{\left[i{\pi\over4}\right]}
 + \sqrt{\pi\over{-x+{1\over3}y^2-y\sqrt{{1\over9}y^2-{1\over3}x}}} \nonumber \\
 & \times & \exp{\left\{i\left[2\left({1\over9}y^2-{1\over3}x\right)^{3/2}
       -{1\over3}y\left({2\over9}y^2-x\right)-{\pi\over4}\right]\right\}} \; .
\label{psi_asy_minus}
\end{eqnarray}
b) $x\to+\infty$:

\noindent
The only real solution of (\ref{t_0_poly}) is $t_0=0$ and $\Psi(x,y)$
is approximated as
\begin{equation}
   \Psi(x,y) \stackrel{x\to+\infty}{\longrightarrow}
   \int_{-\infty}^{+\infty} \exp{\left[-ixt^2\right]} dt 
 = \sqrt{\pi\over x} \, \exp{\left[-i{\pi\over4}\right]}  \; .
\label{Psi_right_asy}
\end{equation}
In the case of $y<0$ and $x \stackrel{>}{\sim} {1\over3}y^2$ the complex zeros 
\[
  t_0= \pm \sqrt{-{1\over3}y \pm i \sqrt{{1\over3}x-{1\over9}y^2}}
\]
of Eq.\ (\ref{t_0_poly}) are situated close to the real axis.
When one considers these complex zeros in a stationary phase approximation
Eq.\ (\ref{Psi_right_asy}) is modified by an additional term exponentially
damped for large $x$:
\begin{eqnarray}
  \Psi(x,y) & \stackrel{x\to+\infty}{\longrightarrow} &
   \sqrt{\pi\over x} \exp{\left[-i{\pi\over4}\right]}
 + \sqrt{\pi\over{x-{1\over3}y^2-iy\sqrt{{1\over3}x-{1\over9}y^2}}} \nonumber \\
 & \times & \exp{\left\{ -2\left({1\over3}x-{1\over9}y^2\right)^{3/2} \right\}}
     \exp{\left\{-i\left[{1\over3}y\left({2\over9}y^2-x\right)-{\pi\over4}
      \right] \right\}} \; . \nonumber \\
\label{psi_asy_plus}
\end{eqnarray}

\end{appendix}

%

\newpage
\setcounter{section}{4}
\section{Figures}
\newpage
\phantom{}
\newpage
\phantom{}
\newpage
\phantom{}
\newpage
\phantom{}
\newpage
\phantom{}
\newpage
\phantom{}
\newpage
\phantom{}
\newpage
\phantom{}
\newpage
\phantom{}
\newpage
\phantom{}
\newpage
\phantom{}
\newpage
\phantom{}
\newpage
\phantom{}
\newpage
\phantom{}
\newpage
\phantom{}
\newpage
\phantom{}
\newpage
\phantom{}
\newpage
\phantom{}
\newpage
\phantom{}
\newpage
\phantom{}
\newpage
\phantom{}
\newpage
\phantom{}
\newpage
\phantom{}
\newpage
\phantom{}
\newpage
\phantom{}
\newpage
\phantom{}
\newpage
\phantom{}
\newpage
\phantom{}
\newpage
\phantom{}
\newpage
\phantom{}
\newpage
\phantom{}
\newpage
\phantom{}
\newpage
\phantom{}
\newpage
\phantom{}
\newpage
\phantom{}
\newpage
\phantom{}
\newpage
\phantom{}

\newpage
\section{Tables}
\begin{table}[h]
\caption{\label{table1}
Hydrogen atom in a magnetic field at scaled energy $\tilde E=-0.1$.
$s_{\rm po}^{\rm qm}$ and $|A^{\rm qm}|$: Actions and absolute values of
amplitudes obtained by harmonic inversion of the quantum spectrum.
$s_{\rm po}^{\rm cl}$ and $|A^{\rm cl}|$: Actions and absolute values of
amplitudes obtained by periodic orbit theory.
}

\bigskip
\begin{center}
\begin{tabular}[t]{rrr|rr}
  \multicolumn{1}{c}{${\rm Re}~s_{\rm po}^{\rm qm}/2\pi$} &
  \multicolumn{1}{c}{${\rm Im}~s_{\rm po}^{\rm qm}/2\pi$} &
  \multicolumn{1}{c|}{$|A^{\rm qm}|$} &
  \multicolumn{1}{c}{$s_{\rm po}^{\rm cl}/2\pi$} &
  \multicolumn{1}{c}{$|A^{\rm cl}|$} \\
\hline
   0.67746349 &   0.00000000 &   0.29925998 &    0.67746283 &    0.29929657 \\
   1.09456040 &   0.00003414 &   1.07896385 &    1.09457049 &    1.14785331 \\
   1.11461451 &  -0.00000723 &   0.51757197 &    1.11457036 &    0.51991351 \\
   1.35492782 &   0.00000089 &   0.28203634 &    1.35492566 &    0.28236677 \\
   1.56500143 &   0.00000054 &   0.42062528 &    1.56499821 &    0.42053218 \\
   1.69802779 &  -0.00000017 &   0.39860976 &    1.69802585 &    0.39857924 \\
   1.79106306 &   0.00000075 &   0.34636062 &    1.79106067 &    0.34672899 \\
   1.87643070 &   0.00000051 &   0.34352848 &    1.87642962 &    0.34375997 \\
   1.93352314 &   0.00000177 &   0.31816675 &    1.93352213 &    0.31888913 \\
   1.99328138 &   0.00000352 &   0.32173573 &    1.99328294 &    0.32238331 \\
   2.03199050 &  -0.00017651 &   0.31129071 &    2.03194819 &    0.30888967 \\
   2.07515409 &   0.00002170 &   0.31251061 &    2.07517790 &    0.31598033 \\
   2.10239191 &   0.00003926 &   0.30768786 &    2.10234679 &    0.30947859 \\
   2.13389568 &   0.00007623 &   0.33987795 &    2.13380984 &    0.31999751 \\
   2.15304246 &   0.00035849 &   0.34510366 &    2.15300805 &    0.31929721 \\
   2.17505591 &  -0.00021399 &   0.24582893 &    2.17556880 &    0.33461616 \\
   2.18921873 &   0.00034303 &   0.79129103 &    2.18914098 &    0.84694565 \\
   2.20287712 &  -0.00045692 &   0.35778573 &    2.20439817 &    0.36428325 \\
   2.30616383 &  -0.00000277 &   0.47937694 &    2.30615714 &    0.47918792 \\
   2.40955313 &   0.00000007 &   0.35374810 &    2.40954609 &    0.35398381 \\
   2.58910596 &  -0.00002565 &   0.57871391 &    2.58913149 &    0.57621849 \\
   2.60055850 &  -0.00002093 &   0.94842381 &    2.60051951 &    0.94785988 \\
   2.62069577 &  -0.00001446 &   0.39565368 &    2.62066666 &    0.39496450 \\
   2.70985681 &  -0.00000436 &   0.09049970 &    2.70985132 &    0.09040051 \\
   2.76685370 &  -0.00008167 &   0.70094383 & \multicolumn{2}{c}{ghost orbit}\\
   2.84036264 &  -0.01311947 &   0.00065741 & \multicolumn{2}{c}{ghost orbit}\\
   2.87476026 &  -0.00973465 &   0.00390219 & \multicolumn{2}{c}{ghost orbit}\\
   2.95857109 &  -0.00000045 &   0.31520915 &    2.95857293 &    0.31514488 \\

\end{tabular}
\end{center}
\end{table}
\begin{table}
\caption{\label{table2}
Hydrogen atom in a magnetic field at scaled energy $\tilde E=-0.1$.
Relative deviations between the quantum mechanical and classical actions 
and amplitudes of the three recurrence peaks around $s/2\pi\approx 2.6$.
}

\bigskip
\begin{center}
\begin{tabular}[t]{ccc}
  \multicolumn{1}{c}{$s^{\rm cl}/2\pi$} &
  \multicolumn{1}{c}{$|s^{\rm qm}-s^{\rm cl}|/s^{\rm cl}$} &
  \multicolumn{1}{c}{$|A^{\rm qm}-A^{\rm cl}|/|A^{\rm cl}|$} \\
\hline
   2.589131 &   $9.86 \times 10^{-6}$  &  $0.00433$ \\
   2.600520 &   $1.50 \times 10^{-5}$  &  $0.00060$ \\
   2.620667 &   $1.11 \times 10^{-5}$  &  $0.00174$ \\

\end{tabular}
\end{center}
\end{table}
\begin{table}
\caption{\label{table3}
Periodic orbit quantities for the three disk scattering system with
$R=1$, $d=6$.
$P$: symbolic code; 
$L_P$ and ${\cal A}^{(1)}/{\cal A}^{(0)}$: Periods and ratio of zeroth 
and first order amplitudes obtained by harmonic inversion of spectra (hi)
and by the classical calculations (cl) of Alonso and Gaspard in
Ref.\ [23].
}

\bigskip
\begin{center}
\begin{tabular}[t]{l|rr|cc}
  \multicolumn{1}{c|}{$P$} &
  \multicolumn{1}{c}{$L_P^{\rm hi}$} &
  \multicolumn{1}{c|}{$L_P^{\rm cl}$} &
  \multicolumn{1}{c}{$|{\cal A}^{(1)}/{\cal A}^{(0)}|^{\rm hi}$} &
  \multicolumn{1}{c}{$|{\cal A}^{(1)}/{\cal A}^{(0)}|^{\rm cl}$} \\
\hline
 0     &    4.000498 &   4.000000 &   0.3135 &    0.31250 \\
 1     &    4.268069 &   4.267949 &   0.5554 &    0.56216 \\
 01    &    8.316597 &   8.316529 &   1.0172 &    1.01990 \\
 001   &   12.321782 &  12.321747 &   1.3465 &    1.35493 \\
 011   &   12.580837 &  12.580808 &   1.5558 &    1.55617 \\
 0001  &   16.322308 &  16.322276 &   1.6584 &    1.67009 \\
 0011  &   16.585261 &  16.585243 &   1.8514 &            \\
 0111  &   16.849133 &  16.849072 &   2.1172 &    2.12219 \\
 00001 &   20.322343 &  20.322330 &   1.9961 &            \\
 00011 &   20.585725 &  20.585690 &   2.1682 &            \\
 00101 &   20.638284 &  20.638238 &   2.4061 &            \\
 00111 &   20.853593 &  20.853572 &   2.4387 &    2.45127 \\
 01011 &   20.897413 &  20.897369 &   2.6127 &            \\
 01111 &   21.117009 &  21.116994 &   2.6764 &            \\

\end{tabular}
\end{center}
\end{table}
\begin{table}
\caption{\label{table4}
Non-trivial zeros $w_k$, multiplicities $d_k$, and error estimate 
$\varepsilon$ for the Riemann zeta function.}

\bigskip
\begin{center}
\begin{tabular}[t]{r|rr|rr|r}
  \multicolumn{1}{c|}{$k$} &
  \multicolumn{1}{c}{${\rm Re}~w_k$} &
  \multicolumn{1}{c|}{${\rm Im}~w_k$} &
  \multicolumn{1}{c}{${\rm Re}~d_k$} &
  \multicolumn{1}{c|}{${\rm Im}~d_k$} &
  \multicolumn{1}{c}{$\varepsilon$} \\ \hline
  1 &   14.13472514 &  4.05E-12 &     1.00000011 & -5.07E-08 &   3.90E-13 \\
  2 &   21.02203964 & -2.23E-12 &     1.00000014 &  1.62E-07 &   9.80E-13 \\
  3 &   25.01085758 &  1.66E-11 &     0.99999975 & -2.64E-07 &   5.20E-12 \\
  4 &   30.42487613 & -6.88E-11 &     0.99999981 & -1.65E-07 &   1.90E-12 \\
  5 &   32.93506159 &  7.62E-11 &     1.00000020 &  5.94E-08 &   7.10E-13 \\
  6 &   37.58617816 &  1.46E-10 &     1.00000034 &  5.13E-07 &   1.00E-12 \\
  7 &   40.91871901 & -3.14E-10 &     0.99999856 &  1.60E-06 &   4.90E-11 \\
  8 &   43.32707328 &  1.67E-11 &     1.00000008 &  3.29E-07 &   1.90E-12 \\
  9 &   48.00515088 &  4.35E-11 &     0.99999975 & -1.35E-07 &   1.40E-12 \\
 10 &   49.77383248 &  7.02E-11 &     1.00000254 & -4.59E-07 &   1.10E-10 \\
 11 &   52.97032148 &  1.92E-10 &     1.00000122 &  7.31E-07 &   6.00E-11 \\
 12 &   56.44624770 & -1.30E-10 &     0.99999993 &  4.51E-07 &   5.50E-12 \\
 13 &   59.34704400 &  5.40E-11 &     0.99999954 &  2.34E-06 &   2.30E-10 \\
 14 &   60.83177852 & -3.94E-10 &     1.00000014 &  1.11E-06 &   3.00E-11 \\
 15 &   65.11254406 & -4.98E-09 &     0.99998010 & -8.30E-06 &   2.70E-08 \\
 16 &   67.07981053 & -2.05E-10 &     0.99999892 & -8.04E-07 &   5.30E-11 \\
 17 &   69.54640171 &  2.51E-11 &     0.99999951 &  9.45E-07 &   6.80E-12 \\
 18 &   72.06715767 & -5.74E-10 &     0.99999974 &  8.63E-06 &   4.30E-10 \\
 19 &   75.70469070 &  3.93E-10 &     1.00000082 &  1.07E-06 &   3.20E-11 \\
 20 &   77.14484007 & -2.70E-12 &     0.99999979 &  1.25E-06 &   1.30E-11 \\
 21 &   79.33737502 & -3.58E-11 &     1.00000086 &  3.03E-07 &   9.20E-12 \\
 22 &   82.91038085 &  1.56E-10 &     0.99999912 & -8.58E-07 &   1.60E-11 \\
 23 &   84.73549298 &  3.34E-10 &     0.99999940 & -7.09E-07 &   2.50E-11 \\
 24 &   87.42527461 &  1.20E-09 &     0.99999866 &  1.39E-06 &   6.70E-11 \\
 25 &   88.80911121 & -9.42E-10 &     1.00000101 &  1.49E-06 &   4.80E-11 \\
 26 &   92.49189927 & -4.11E-09 &     0.99999761 & -1.93E-06 &   1.50E-10 \\
 27 &   94.65134404 & -7.11E-09 &     1.00000520 & -1.27E-06 &   6.80E-10 \\
 28 &   95.87063426 &  8.06E-09 &     0.99999001 & -1.15E-05 &   5.20E-09 \\
 29 &   98.83119422 & -1.78E-11 &     0.99999936 &  5.70E-07 &   3.10E-12 \\
 30 &  101.31785101 &  4.22E-11 &     0.99999969 & -4.73E-07 &   4.50E-12 \\
\end{tabular}
\end{center}
\end{table}
\begin{table}
\caption{\label{table5}
Non-trivial zeros $w_k$, multiplicities $d_k$, and error estimate 
$\varepsilon$ for the Riemann zeta function.}

\bigskip
\begin{center}
\begin{tabular}[t]{r|rr|rr|r}
  \multicolumn{1}{c|}{$k$} &
  \multicolumn{1}{c}{${\rm Re}~w_k$} &
  \multicolumn{1}{c|}{${\rm Im}~w_k$} &
  \multicolumn{1}{c}{${\rm Re}~d_k$} &
  \multicolumn{1}{c|}{${\rm Im}~d_k$} &
  \multicolumn{1}{c}{$\varepsilon$} \\ \hline
 2532 & 3063.43508648 & -1.64E-09 &     0.99999901 &  1.34E-06 &   5.50E-11 \\
 2533 & 3065.28655558 &  1.15E-09 &     1.00000107 & -3.63E-07 &   2.40E-11 \\
 2534 & 3066.32025039 & -1.66E-10 &     1.00000231 &  1.00E-06 &   1.20E-10 \\
 2535 & 3067.07132023 &  3.68E-09 &     1.00000334 &  2.73E-07 &   2.20E-10 \\
 2536 & 3068.01350133 & -1.51E-09 &     1.00000291 &  1.46E-06 &   2.10E-10 \\
 2537 & 3068.98426618 & -5.92E-09 &     1.00000205 &  2.94E-06 &   2.60E-10 \\
 2538 & 3069.78290477 & -4.40E-09 &     1.00000237 &  2.51E-06 &   2.40E-10 \\
 2539 & 3070.54262154 & -7.71E-10 &     1.00000169 &  9.57E-07 &   7.90E-11 \\
 2540 & 3072.00099337 & -6.44E-11 &     0.99999908 &  2.17E-07 &   2.00E-11 \\
 2541 & 3073.18523777 &  9.17E-11 &     0.99999942 & -1.07E-06 &   3.00E-11 \\
 2542 & 3074.52349428 &  6.73E-09 &     1.00000391 & -6.51E-07 &   3.50E-10 \\
 2543 & 3075.03387288 & -1.22E-08 &     1.00000117 & -5.69E-06 &   7.30E-10 \\
 2544 & 3075.83347924 & -3.13E-09 &     1.00000013 & -3.86E-06 &   3.10E-10 \\
 2545 & 3077.42747330 &  5.76E-10 &     1.00000561 &  4.69E-06 &   1.10E-09 \\
 2546 & 3078.28622690 &  1.34E-08 &     1.00001283 &  1.10E-06 &   3.80E-09 \\
 2547 & 3078.89737915 &  1.61E-09 &     1.00000487 & -8.04E-06 &   2.10E-09 \\
 2548 & 3079.87139464 &  1.70E-09 &     1.00000275 & -2.32E-06 &   3.00E-10 \\
 2549 & 3080.85638233 &  8.67E-10 &     1.00000159 & -4.12E-07 &   5.90E-11 \\
 2550 & 3082.16316375 & -5.88E-10 &     1.00000013 &  8.44E-07 &   1.70E-11 \\
 2551 & 3083.36135798 &  8.43E-10 &     0.99999923 &  3.45E-07 &   1.50E-11 \\
 2552 & 3084.83845150 &  2.72E-09 &     1.00000057 & -2.86E-06 &   1.80E-10 \\
 2553 & 3085.37726898 & -1.37E-08 &     0.99999576 & -2.88E-06 &   5.50E-10 \\
 2554 & 3085.96552225 &  6.39E-09 &     0.99999667 &  1.50E-06 &   2.80E-10 \\
 2555 & 3087.01881535 &  3.46E-11 &     0.99999845 & -3.63E-07 &   5.20E-11 \\
 2556 & 3088.08343703 & -3.89E-10 &     0.99999931 & -8.44E-07 &   2.40E-11 \\
 2557 & 3089.22230894 & -3.31E-10 &     1.00000017 & -9.21E-07 &   1.80E-11 \\
 2558 & 3090.28219490 &  2.97E-10 &     1.00000069 & -7.17E-07 &   2.10E-11 \\
 2559 & 3091.15446969 &  1.10E-09 &     1.00000052 & -6.59E-07 &   1.50E-11 \\
 2560 & 3092.68766704 &  2.25E-09 &     1.00000033 &  1.45E-06 &   5.20E-11 \\
 2561 & 3093.18544571 & -2.33E-09 &     1.00000168 & -1.50E-07 &   6.40E-11 \\
\end{tabular}
\end{center}
\end{table}
\begin{table}
\caption{\label{table6}
Trivial zeros and pole of the Riemann zeta function.}

\bigskip
\begin{center}
\begin{tabular}[t]{rr|rr|r}
  \multicolumn{1}{c}{${\rm Re}~w_k$} &
  \multicolumn{1}{c|}{${\rm Im}~w_k$} &
  \multicolumn{1}{c}{${\rm Re}~d_k$} &
  \multicolumn{1}{c|}{${\rm Im}~d_k$} &
  \multicolumn{1}{c}{$\varepsilon$} \\ \hline
     0.00000000 &  0.50000000 &    -1.00000002 & -4.26E-08 &   1.80E-14 \\
    -0.00000060 & -2.49999941 &     0.99992487 & -3.66E-05 &   1.80E-07 \\
    -0.00129915 & -4.49987911 &     1.00069939 & -3.25E-03 &   4.40E-05 \\
    -0.09761173 & -6.53286064 &     1.07141445 & -1.49E-01 &   1.70E-03 \\
\end{tabular}
\end{center}
\end{table}
\begin{table}
\caption{\label{table7}
Weight factors ${\cal N}_p$ for the symmetry decomposition of the three
disk scattering system.
}

\smallskip
\begin{center}
\begin{tabular}[t]{l|rrr}
  \multicolumn{1}{c|}{$C_{3v}$} &
  \multicolumn{1}{c}{$A_1$} &
  \multicolumn{1}{c}{$A_2$} &
  \multicolumn{1}{c}{$E$} \\
\hline
 $e$         &  1 &  1 &  2 \\
 $C_3,C_3^2$ &  1 &  1 & -1 \\
 $\sigma_v$  &  1 & -1 &  0 \\
\end{tabular}
\end{center}
\end{table}
\begin{table}
\caption{\label{table8}
Semiclassical and exact quantum mechanical resonances for the three disk 
scattering problem ($A_1$ subspace) with $R=1$, $d=6$. 
Resonances $k^{\rm hi}$ have been obtained by harmonic inversion of the 
periodic orbit sum, resonances $k^{\rm ce}$ (cycle expansion) and the exact 
quantum resonances $k^{\rm qm}$ have been calculated by Wirzba. 
}

\bigskip
\begin{center}
\begin{tabular}[t]{rr|rr|rr}
  \multicolumn{1}{c}{${\rm Re}~k^{\rm hi}$} &
  \multicolumn{1}{c|}{${\rm Im}~k^{\rm hi}$} &
  \multicolumn{1}{c}{${\rm Re}~k^{\rm ce}$} &
  \multicolumn{1}{c|}{${\rm Im}~k^{\rm ce}$} &
  \multicolumn{1}{c}{${\rm Re}~k^{\rm qm}$} &
  \multicolumn{1}{c}{${\rm Im}~k^{\rm qm}$} \\
\hline
    0.7583139 &   -0.1228222 &    0.7583139 &  -0.1228222 &   0.6979958 &  -0.0750137 \\
    2.2742786 &   -0.1330587 &    2.2742786 &  -0.1330587 &   2.2396014 &  -0.1187664 \\
    3.7878768 &   -0.1541274 &    3.7878768 &  -0.1541274 &   3.7626868 &  -0.1475455 \\
    4.1456898 &   -0.6585397 &    4.1474774 &  -0.6604761 &   4.1316606 &  -0.6170418 \\
    5.2960678 &   -0.1867873 &    5.2960678 &  -0.1867873 &   5.2756666 &  -0.1832203 \\
    5.6814976 &   -0.5713721 &    5.6820274 &  -0.5715543 &   5.6694976 &  -0.5534079 \\
    6.7936365 &   -0.2299221 &    6.7936365 &  -0.2299221 &   6.7760661 &  -0.2275078 \\
    7.2240580 &   -0.4954243 &    7.2242175 &  -0.4954066 &   7.2152706 &  -0.4856243 \\
    8.2763906 &   -0.2770805 &    8.2763906 &  -0.2770805 &   8.2611376 &  -0.2749083 \\
    8.7792134 &   -0.4302561 &    8.7791917 &  -0.4302718 &   8.7724709 &  -0.4241019 \\
    9.7476329 &   -0.3208170 &    9.7476329 &  -0.3208170 &   9.7345075 &  -0.3188052 \\
   10.3442257 &   -0.3781988 &   10.3442254 &  -0.3781988 &  10.3381881 &  -0.3737056 \\
   11.2134778 &   -0.3599639 &   11.2134779 &  -0.3599639 &  11.2021099 &  -0.3582265 \\
   11.9134496 &   -0.3357346 &   11.9134496 &  -0.3357346 &  11.9075971 &  -0.3322326 \\
   12.6775319 &   -0.3961154 &   12.6775320 &  -0.3961159 &  12.6675941 &  -0.3946675 \\
   13.4826489 &   -0.2969478 &   13.4826489 &  -0.2969477 &  13.4769269 &  -0.2941108 \\
   14.1424136 &   -0.4300604 &   14.1424117 &  -0.4300584 &  14.1337039 &  -0.4288264 \\
   15.0473050 &   -0.2578357 &   15.0473050 &  -0.2578357 &  15.0416935 &  -0.2555072 \\
   15.6114431 &   -0.4603838 &   15.6113293 &  -0.4604377 &  15.6037211 &  -0.4592793 \\
   16.6025598 &   -0.2188731 &   16.6025599 &  -0.2188731 &  16.5970551 &  -0.2170025 \\
   17.0875557 &   -0.4826796 &   17.0876372 &  -0.4827914 &  17.0809957 &  -0.4815364 \\
   18.1465009 &   -0.1842319 &   18.1465009 &  -0.1842319 &  18.1411380 &  -0.1827959 \\
   18.5733904 &   -0.4913642 &   18.5731865 &  -0.4914087 &  18.5673580 &  -0.4899438 \\
   19.6808375 &   -0.1575927 &   19.6808375 &  -0.1575927 &  19.6756560 &  -0.1565422 \\
   20.0679755 &   -0.4814959 &   20.0685648 &  -0.4842307 &  20.0633947 &  -0.4825586 \\
   21.2080634 &   -0.1403086 &   21.2080634 &  -0.1403086 &  21.2030727 &  -0.1395793 \\
   21.5736413 &   -0.4643351 &   21.5736471 &  -0.4643077 &  21.5689872 &  -0.4625282 \\
   22.7296581 &   -0.1323433 &   22.7296581 &  -0.1323433 &  22.7248396 &  -0.1318732 \\
   23.0872357 &   -0.4363434 &   23.0872405 &  -0.4363396 &  23.0829816 &  -0.4345679 \\
   24.2458779 &   -0.1331136 &   24.2458779 &  -0.1331136 &  24.2411967 &  -0.1328421 \\
   24.6079881 &   -0.4038977 &   24.6079949 &  -0.4038962 &  24.6040788 &  -0.4022114 \\
   25.7560497 &   -0.1415265 &   25.7560497 &  -0.1415265 &  25.7514701 &  -0.1413856 \\
   26.1353684 &   -0.3694453 &   26.1353709 &  -0.3694420 &  26.1317670 &  -0.3679001 \\
   27.2592433 &   -0.1556272 &   27.2592433 &  -0.1556272 &  27.2547515 &  -0.1555416 \\
   27.6694386 &   -0.3353429 &   27.6694384 &  -0.3353414 &  27.6661128 &  -0.3339924 \\
   28.7553526 &   -0.1727506 &   28.7553526 &  -0.1727506 &  28.7509680 &  -0.1726604 \\
   29.2098235 &   -0.3041527 &   29.2098230 &  -0.3041533 &  29.2067213 &  -0.3030272 \\

\end{tabular}
\end{center}
\end{table}
\begin{table}
\caption{\label{table9}
Same as Table \ref{table8} but for resonances with 
$120<{\rm Re}~k<132$.
}

\bigskip
\begin{center}
\begin{tabular}[t]{rr|rr|rr}
  \multicolumn{1}{c}{${\rm Re}~k^{\rm hi}$} &
  \multicolumn{1}{c|}{${\rm Im}~k^{\rm hi}$} &
  \multicolumn{1}{c}{${\rm Re}~k^{\rm ce}$} &
  \multicolumn{1}{c|}{${\rm Im}~k^{\rm ce}$} &
  \multicolumn{1}{c}{${\rm Re}~k^{\rm qm}$} &
  \multicolumn{1}{c}{${\rm Im}~k^{\rm qm}$} \\
\hline
  120.0966075 &   -0.1313240 &  120.0966093 &  -0.1313207 & 120.0956703 &  -0.1313419 \\
  120.3615795 &   -0.4242281 &  120.3624892 &  -0.4246464 & 120.3617516 &  -0.4244457 \\
  120.8941439 &   -0.5145734 &  120.8959933 &  -0.5143028 & 120.8949118 &  -0.5143053 \\
  121.2644970 &   -0.4017485 &  121.2643338 &  -0.4018042 & 121.2633401 &  -0.4017384 \\
  121.6158735 &   -0.1451572 &  121.6158734 &  -0.1451572 & 121.6149357 &  -0.1452113 \\
  121.9157839 &   -0.3979469 &  121.9158422 &  -0.3978279 & 121.9151103 &  -0.3976081 \\
  122.3933221 &   -0.5441615 &  122.3962246 &  -0.5427722 & 122.3951716 &  -0.5427575 \\
  122.7533968 &   -0.3809547 &  122.7533176 &  -0.3809838 & 122.7523307 &  -0.3809219 \\
  123.1345930 &   -0.1656546 &  123.1345929 &  -0.1656547 & 123.1336504 &  -0.1657315 \\
  123.4680089 &   -0.3672776 &  123.4679489 &  -0.3672571 & 123.4672237 &  -0.3670296 \\
  123.8916697 &   -0.5717434 &  123.8969484 &  -0.5675655 & 123.8959339 &  -0.5675273 \\
  124.2423970 &   -0.3593830 &  124.2423480 &  -0.3593825 & 124.2413639 &  -0.3593262 \\
  124.6520739 &   -0.1905948 &  124.6520738 &  -0.1905950 & 124.6511239 &  -0.1906812 \\
  125.0191421 &   -0.3358345 &  125.0191284 &  -0.3358584 & 125.0184111 &  -0.3356413 \\
  125.3924617 &   -0.5950322 &  125.4000681 &  -0.5858492 & 125.3990978 &  -0.5857818 \\
  125.7306095 &   -0.3386874 &  125.7305801 &  -0.3386762 & 125.7295973 &  -0.3386268 \\
  126.1681278 &   -0.2172657 &  126.1681277 &  -0.2172664 & 126.1671733 &  -0.2173496 \\
  126.5700003 &   -0.3071799 &  126.5700090 &  -0.3071895 & 126.5692997 &  -0.3069994 \\
  126.8986333 &   -0.6105834 &  126.9067485 &  -0.5956839 & 126.9058213 &  -0.5955889 \\
  127.2175968 &   -0.3201029 &  127.2175825 &  -0.3200895 & 127.2166018 &  -0.3200477 \\
  127.6830865 &   -0.2434140 &  127.6830866 &  -0.2434159 & 127.6821330 &  -0.2434894 \\
  128.1211609 &   -0.2838964 &  128.1211675 &  -0.2838934 & 128.1204656 &  -0.2837384 \\
  128.4113722 &   -0.6157741 &  128.4171380 &  -0.5966404 & 128.4162452 &  -0.5965218 \\
  128.7033407 &   -0.3044266 &  128.7033374 &  -0.3044145 & 128.7023607 &  -0.3043815 \\
  129.1973295 &   -0.2678886 &  129.1973310 &  -0.2678924 & 129.1963819 &  -0.2679547 \\
  129.6731970 &   -0.2671784 &  129.6731961 &  -0.2671733 & 129.6725028 &  -0.2670549 \\
  129.9292721 &   -0.6091832 &  129.9304559 &  -0.5891942 & 129.9295889 &  -0.5890492 \\
  130.1879622 &   -0.2922354 &  130.1879664 &  -0.2922273 & 130.1869953 &  -0.2922052 \\
  130.7109808 &   -0.2904524 &  130.7109850 &  -0.2904576 & 130.7100422 &  -0.2905099 \\
  131.2271782 &   -0.2573647 &  131.2271733 &  -0.2573650 & 131.2264935 &  -0.2572819 \\
  131.4488921 &   -0.5905439 &  131.4451877 &  -0.5734808 & 131.4443391 &  -0.5733029 \\
  131.6713958 &   -0.2842971 &  131.6714040 &  -0.2842951 & 131.6704389 &  -0.2842886 \\

\end{tabular}
\end{center}
\end{table}
\begin{table}
\caption{\label{table10}
Semiclassical and exact quantum mechanical resonances for the three disk 
scattering problem ($A_1$ subspace) with $R=1$, $d=2.5$. 
Resonances $k^{\rm hi}$ have been obtained by harmonic inversion of the 
periodic orbit sum, resonances $k^{\rm ce}$ (cycle expansion) and the exact 
quantum resonances $k^{\rm qm}$ have been calculated by Wirzba. 
}

\bigskip
\begin{center}
\begin{tabular}[t]{rr|rr|rr}
  \multicolumn{1}{c}{${\rm Re}~k^{\rm hi}$} &
  \multicolumn{1}{c|}{${\rm Im}~k^{\rm hi}$} &
  \multicolumn{1}{c}{${\rm Re}~k^{\rm ce}$} &
  \multicolumn{1}{c|}{${\rm Im}~k^{\rm ce}$} &
  \multicolumn{1}{c}{${\rm Re}~k^{\rm qm}$} &
  \multicolumn{1}{c}{${\rm Im}~k^{\rm qm}$} \\
\hline
     4.5811788 &  -0.0899874 &   4.5811768 &    -0.0899911 &   4.4692836 &   -0.0015711 \\
     7.1436576 &  -0.8112421 &   7.1441170 &    -0.8107256 &   7.0917132 &   -0.7207876 \\
    13.0000861 &  -0.6516382 &  13.0000510 &    -0.6516077 &  12.9503201 &   -0.6282388 \\
    17.5688087 &  -0.6848798 &  17.5699350 &    -0.6845598 &  17.5042296 &   -0.6352583 \\
    18.9210923 &  -0.7860616 &  18.9266454 &    -0.7836853 &  18.9254527 &   -0.7662881 \\
    26.6126422 &  -1.8066664 &  26.5430240 &    -1.8607912 &  26.5316680 &   -1.8273425 \\
    27.8887512 &  -0.5431988 &  27.8881874 &    -0.5431833 &  27.8577850 &   -0.5499256 \\
    30.3885091 &  -0.1134324 &  30.3884554 &    -0.1134504 &  30.3528872 &   -0.1056653 \\
    32.0961916 &  -0.6218604 &  32.0966982 &    -0.6223710 &  32.0693735 &   -0.6077432 \\
    33.7905462 &  -1.9974557 &  33.9721550 &    -2.0780146 &  33.9323425 &   -2.0379172 \\
    36.5066867 &  -0.3845559 &  36.5066384 &    -0.3846400 &  36.4822805 &   -0.3839172 \\
    39.8138808 &  -0.3582253 &  39.8139242 &    -0.3580251 &  39.7859698 &   -0.3508678 \\
    42.6585345 &  -0.3514612 &  42.6556455 &    -0.3492807 &  42.6312420 &   -0.3403609 \\
    44.3271565 &  -0.3596874 &  44.2457223 &    -0.4072838 &  44.2295828 &   -0.4015643 \\
    45.1114372 &  -0.4173349 &  45.0602989 &    -0.3453843 &  45.0400958 &   -0.3421851 \\
    48.8432026 &  -0.5925064 &  48.8417782 &    -0.5913485 &  48.8203073 &   -0.5853389 \\
    51.9171049 &  -0.6951700 &  51.9146021 &    -0.6791551 &  51.8983106 &   -0.6721309 \\
    53.3788306 &  -0.1062924 &  53.3766455 &    -0.1005604 &  53.3584390 &   -0.1005951 \\
    60.5907039 &  -0.8405420 &  60.6204084 &    -0.8126455 &  60.6075221 &   -0.8098089 \\
    62.2000118 &  -0.2161506 &  62.2004010 &    -0.2137105 &  62.1832903 &   -0.2111191 \\
    64.8371533 &  -2.1940372 &  64.1528609 &    -1.6713984 &  64.1457569 &   -1.6638733 \\
    65.6804780 &  -0.2791339 &  65.6804671 &    -0.2737728 &  65.6638670 &   -0.2725830 \\
    67.8612092 &  -0.2865945 &  67.8688350 &    -0.2881904 &  67.8515132 &   -0.2865564 \\
    69.3238832 &  -0.2988268 &  69.3443564 &    -0.3123706 &  69.3334558 &   -0.3092481 \\
    71.1163719 &  -0.5635882 &  71.0822786 &    -0.5381921 &  71.0672666 &   -0.5353411 \\
    74.8566924 &  -0.3058255 &  74.8552672 &    -0.3022484 &  74.8405304 &   -0.2994128 \\
    77.3118119 &  -0.3223758 &  77.3193218 &    -0.3129301 &  77.3088118 &   -0.3107104 \\
    78.5807841 &  -0.9179862 &  78.9236428 &    -0.9433778 &  78.9042646 &   -0.9416087 \\
    80.3395505 &  -0.3052621 &  80.4173794 &    -0.3670163 &  80.4002169 &   -0.3628903 \\
    83.9450649 &  -0.4910890 &  83.8740934 &    -0.5035157 &  83.8631132 &   -0.5005356 \\
    85.8041790 &  -0.3988778 &  85.8000989 &    -0.4147646 &  85.7918872 &   -0.4152898 \\
    88.5979364 &  -0.6089526 &  88.4703014 &    -0.6739426 &  88.4561398 &   -0.6778186 \\
    93.2377427 &  -0.1707857 &  93.0234303 &    -0.1231810 &  93.0113364 &   -0.1220480 \\
    94.4188748 &  -0.5644403 &  94.4284207 &    -0.5013008 &  94.4351896 &   -0.5267138 \\
    97.5331975 &  -0.4015302 &  97.5576754 &    -0.4167467 &  97.5478240 &   -0.4181066 \\
\end{tabular}
\end{center}
\end{table}
\begin{table}
\caption{\label{table11}
Nearly degenerate semiclassical states of the circle billiard.
$k^{\rm EBK}$: Results from EBK-quantization.
$k^{\rm HI}$: Eigenvalues obtained by harmonic inversion of cross-correlated
periodic orbit sums. States are labeled by the radial and angular momentum
quantum numbers $(n,m)$.}

\bigskip
\begin{center}
\begin{tabular}[t]{r|r|r|r}
  \multicolumn{1}{c|}{$n$} &
  \multicolumn{1}{c|}{$m$} &
  \multicolumn{1}{c|}{$k^{\rm EBK}$} &
  \multicolumn{1}{c}{$k^{\rm HI}$} \\ 
\hline
   1 &  4 &  11.048664 &   11.048569 \\
   0 &  7 &  11.049268 &   11.049239 \\
\hline
   3 &  1 &  13.314197 &   13.314216 \\
   0 &  9 &  13.315852 &   13.315869 \\
\hline
   3 &  2 &  14.787105 &   14.787036 \\
   1 &  7 &  14.805435 &   14.805345 \\
\hline
   1 & 11 &  19.599795 &   19.599863 \\
   5 &  1 &  19.609451 &   19.608981 \\
\hline
   1 & 15 &  24.252501 &   24.252721 \\
   6 &  2 &  24.264873 &   24.264887 \\
\end{tabular}
\end{center}
\end{table}
\begin{table}
\caption{\label{table12}
The 40 lowest eigenstates of the circle billiard with radius $R=1$.
$n,m$: Radial and angular momentum quantum numbers;
$k^{\rm EBK}$: Results from EBK-quantization;
$k^{\rm (0)}$ and $k^{\rm (1)}$: Zeroth and first order semiclassical
eigenvalues obtained by harmonic inversion of the periodic orbit signal;
$k^{\rm ex}$: Exact eigenvalues, i.e., zeros of the Bessel functions 
$J_m(kR)=0$.}

\smallskip
\begin{center}
\begin{tabular}[t]{r|r|r|r|r|r}
  \multicolumn{1}{c|}{$n$} &
  \multicolumn{1}{c|}{$m$} &
  \multicolumn{1}{c|}{$k^{\rm EBK}$} &
  \multicolumn{1}{c|}{$k^{\rm (0)}$} &
  \multicolumn{1}{c|}{$k^{\rm (1)}$} &
  \multicolumn{1}{c}{$k^{\rm ex}$} \\ 
\hline
   0 &   0 &    2.356194 &    2.356230 &    2.409288 &    2.404826 \\
   0 &   1 &    3.794440 &    3.794440 &    3.834267 &    3.831706 \\
   0 &   2 &    5.100386 &    5.100382 &    5.138118 &    5.135622 \\
   1 &   0 &    5.497787 &    5.497816 &    5.520550 &    5.520078 \\
   0 &   3 &    6.345186 &    6.345182 &    6.382709 &    6.380162 \\
   1 &   1 &    6.997002 &    6.997006 &    7.015881 &    7.015587 \\
   0 &   4 &    7.553060 &    7.553055 &    7.590990 &    7.588342 \\
   1 &   2 &    8.400144 &    8.400145 &    8.417503 &    8.417244 \\
   2 &   0 &    8.639380 &    8.639394 &    8.653878 &    8.653728 \\
   0 &   5 &    8.735670 &    8.735672 &    8.774213 &    8.771484 \\
   1 &   3 &    9.744628 &    9.744627 &    9.761274 &    9.761023 \\
   0 &   6 &    9.899671 &    9.899660 &    9.938954 &    9.936110 \\
   2 &   1 &   10.160928 &   10.160949 &   10.173568 &   10.173468 \\
   1 &   4 &   11.048664 &   11.048635 &   11.063791 &   11.064709 \\
   0 &   7 &   11.049268 &   11.049228 &   11.087943 &   11.086370 \\
   2 &   2 &   11.608251 &   11.608254 &   11.619919 &   11.619841 \\
   3 &   0 &   11.780972 &   11.780993 &   11.791599 &   11.791534 \\
   0 &   8 &   12.187316 &   12.187302 &   12.228037 &   12.225092 \\
   1 &   5 &   12.322723 &   12.322721 &   12.338847 &   12.338604 \\
   2 &   3 &   13.004166 &   13.004167 &   13.015272 &   13.015201 \\
   3 &   1 &   13.314197 &   13.314192 &   13.323418 &   13.323692 \\
   0 &   9 &   13.315852 &   13.315782 &   13.356645 &   13.354300 \\
   1 &   6 &   13.573465 &   13.573464 &   13.589544 &   13.589290 \\
   2 &   4 &   14.361846 &   14.361846 &   14.372606 &   14.372537 \\
   0 &  10 &   14.436391 &   14.436375 &   14.478531 &   14.475501 \\
   3 &   2 &   14.787105 &   14.787091 &   14.795970 &   14.795952 \\
   1 &   7 &   14.805435 &   14.805457 &   14.821595 &   14.821269 \\
   4 &   0 &   14.922565 &   14.922572 &   14.930938 &   14.930918 \\
   0 &  11 &   15.550089 &   15.550084 &   15.593060 &   15.589848 \\
   2 &   5 &   15.689703 &   15.689701 &   15.700239 &   15.700174 \\
   1 &   8 &   16.021889 &   16.021888 &   16.038034 &   16.037774 \\
   3 &   3 &   16.215041 &   16.215047 &   16.223499 &   16.223466 \\
   4 &   1 &   16.462981 &   16.462982 &   16.470648 &   16.470630 \\
   0 &  12 &   16.657857 &   16.657846 &   16.701442 &   16.698250 \\
   2 &   6 &   16.993489 &   16.993486 &   17.003884 &   17.003820 \\
   1 &   9 &   17.225257 &   17.225252 &   17.241482 &   17.241220 \\
   3 &   4 &   17.607830 &   17.607831 &   17.615994 &   17.615966 \\
   0 &  13 &   17.760424 &   17.760386 &   17.804708 &   17.801435 \\
   4 &   2 &   17.952638 &   17.952662 &   17.959859 &   17.959819 \\
   5 &   0 &   18.064158 &   18.064201 &   18.071125 &   18.071064 \\
\end{tabular}
\end{center}
\end{table}

\end{document}